\documentclass[a4paper,onecolumn,superscriptaddress,11pt,unpublished,accepted=2023-01-27]{quantumarticle}
\pdfoutput=1
\usepackage{cite}
\usepackage{lineno}
\usepackage{graphicx}
\usepackage[toc,page]{appendix}
\usepackage{makeidx} 
\usepackage{color}
\usepackage{xcolor}
\usepackage{amsmath}
\usepackage{amssymb}
\usepackage{hyperref}
\usepackage{cleveref}
\usepackage{mathrsfs}
\usepackage{tikz}
\usepackage{enumitem}
\usepackage{slashbox}
\usepackage{diagbox}
\usepackage{boldline, makecell, booktabs}
\usepackage{multirow}
\hypersetup{colorlinks, citecolor=magenta, filecolor=blue, linkcolor=blue, urlcolor=blue}
\newcolumntype{?}{!{\vrule width 1pt}}
\usepackage[normalem]{ulem}

\RequirePackage{etex}

\usepackage{pgfplots}
\pgfplotsset{compat=1.6}

\usepackage{lipsum, babel}

\def\duzomniejsze{<\kern-.7mm<}
\def\duzowieksze{>\kern-.7mm>}

\def\textbf#1{{\bf #1}}
\def\beq{\begin{equation}}
\def\eeq{\end{equation}}
\def\be{\begin{equation}}
\def\ee{\end{equation}}
\def\ben{\begin{eqnarray}}
\def\een{\end{eqnarray}}
\def\beqa{\begin{eqnarray}}
\def\eeqa{\end{eqnarray}}
\def\eea{\end{array}}
\def\bea{\begin{array}}
\newcommand{\bei}{\begin{itemize}}
\newcommand{\eei}{\end{itemize}}
\newcommand{\bee}{\begin{enumerate}}
\newcommand{\eee}{\end{enumerate}}
\newlength{\Oldarrayrulewidth}
\newcommand{\Cline}[2]{%
  \noalign{\global\setlength{\Oldarrayrulewidth}{\arrayrulewidth}}%
  \noalign{\global\setlength{\arrayrulewidth}{#1}}\cline{#2}%
  \noalign{\global\setlength{\arrayrulewidth}{\Oldarrayrulewidth}}}

\definecolor{indiagreen}{rgb}{0.07, 0.53, 0.03}

\def\ce{\mathscr{E}}
\def\cep{\mathscr{E}_{pure}}
\def\cm{\mathscr{M}}

\def\tr{{\rm Tr}}

\def\>{\rangle}
\def\<{\langle}

\def\bed{\begin{definition}}
\def\eed{\end{definition}}
\def\bel{\begin{lemma}}
\def\eel{\end{lemma}}
\def\bet{\begin{theorem}}
\def\eet{\end{theorem}}

\DeclareMathAlphabet{\pazocal}{OMS}{zplm}{m}{n}

\usepackage{rotating}
\usepackage{floatpag}
\rotfloatpagestyle{empty}
\renewcommand{\theenumi}{\arabic{enumi}}
\usepackage{hyperref}
\usepackage{dsfont}
\usepackage{xspace,epsfig}%color,
\usepackage{graphicx}
\graphicspath{{.}{./figures/}}
\usepackage{marvosym}

\usepackage{tikzfig}
\usepackage{stmaryrd}
\usepackage{docmute}
\usepackage{keycommand}

\newkeycommand{\pointmap}[style=point][1]{\,\tikz{\node[style=\commandkey{style}] (x) at (0,-0.3) {$#1$};\node [style=none] (3) at (0, 1.0) {};\node [style=none] (3) at (0, -1.0) {};\draw (x)--(0,0.7);}\,}

\newkeycommand{\typedkpoint}[style=kpoint,edgestyle=][2]{%
\,\begin{tikzpicture}
  \begin{pgfonlayer}{nodelayer}
    \node [style=none] (0) at (0, 1) {};
    \node [style=\commandkey{style}] (1) at (0, -0.75) {$#2$};
    \node [style=right label] (2) at (0.25, 0.5) {$#1$};
  \end{pgfonlayer}
  \begin{pgfonlayer}{edgelayer}
    \draw [\commandkey{edgestyle}] (1) to (0.center);
  \end{pgfonlayer}
\end{tikzpicture}\,}

\newkeycommand{\typedkpointdag}[style=kpoint adjoint,edgestyle=][2]{%
\,\begin{tikzpicture}
  \begin{pgfonlayer}{nodelayer}
    \node [style=none] (0) at (0, -1) {};
    \node [style=\commandkey{style}] (1) at (0, 0.75) {$#2$};
    \node [style=right label] (2) at (0.25, -0.5) {$#1$};
  \end{pgfonlayer}
  \begin{pgfonlayer}{edgelayer}
    \draw [\commandkey{edgestyle}] (0.center) to (1);
  \end{pgfonlayer}
\end{tikzpicture}\,}

\newcommand{\bistateadj}[1]{\,\begin{tikzpicture}[yshift=-3mm]
  \begin{pgfonlayer}{nodelayer}
    \node [style=kpointadj, minimum width=1 cm, inner sep=2pt] (0) at (0, 0.5) {$\psi$};
    \node [style=none] (1) at (-0.75, 0.25) {};
    \node [style=none] (2) at (0.75, 0.25) {};
    \node [style=none] (3) at (-0.75, -0.5) {};
    \node [style=none] (4) at (0.75, -0.5) {};
  \end{pgfonlayer}
  \begin{pgfonlayer}{edgelayer}
    \draw (1.center) to (3.center);
    \draw (2.center) to (4.center);
  \end{pgfonlayer}
\end{tikzpicture}\,}

\newkeycommand{\pointketbra}[style1=copoint,style2=point][2]{\,%
\begin{tikzpicture}
  \begin{pgfonlayer}{nodelayer}
    \node [style=none] (0) at (0, -1.5) {};
    \node [style=\commandkey{style1}] (1) at (0, -0.75) {$#1$};
    \node [style=\commandkey{style2}] (2) at (0, 0.75) {$#2$};
    \node [style=none] (3) at (0, 1.5) {};
  \end{pgfonlayer}
  \begin{pgfonlayer}{edgelayer}
    \draw (0.center) to (1);
    \draw (2) to (3.center);
  \end{pgfonlayer}
\end{tikzpicture}\,}

\newkeycommand{\twocopointketbra}[style1=copoint,style2=copoint,style3=point][3]{\,%
\begin{tikzpicture}
  \begin{pgfonlayer}{nodelayer}
    \node [style=none] (0) at (-0.75, -1.5) {};
    \node [style=none] (1) at (0.75, -1.5) {};
    \node [style=\commandkey{style1}] (2) at (-0.75, -0.75) {$#1$};
    \node [style=\commandkey{style2}] (3) at (0.75, -0.75) {$#2$};
    \node [style=\commandkey{style3}] (4) at (0, 0.75) {$#3$};
    \node [style=none] (5) at (0, 1.5) {};
  \end{pgfonlayer}
  \begin{pgfonlayer}{edgelayer}
    \draw (0.center) to (2);
    \draw (1.center) to (3);
    \draw (4) to (5.center);
  \end{pgfonlayer}
\end{tikzpicture}\,}

\newkeycommand{\twopointketbra}[style1=copoint,style2=point,style3=point][3]{\,%
\begin{tikzpicture}
  \begin{pgfonlayer}{nodelayer}
    \node [style=none] (0) at (0, -1.5) {};
    \node [style=none] (1) at (-0.75, 1.5) {};
    \node [style=\commandkey{style1}] (2) at (0, -0.75) {$#1$};
    \node [style=\commandkey{style2}] (3) at (-0.75, 0.75) {$#2$};
    \node [style=\commandkey{style3}] (4) at (0.75, 0.75) {$#3$};
    \node [style=none] (5) at (0.75, 1.5) {};
  \end{pgfonlayer}
  \begin{pgfonlayer}{edgelayer}
    \draw (0.center) to (2);
    \draw (1.center) to (3);
    \draw (4) to (5.center);
  \end{pgfonlayer}
\end{tikzpicture}\,}

\newkeycommand{\pointbraket}[style1=point,style2=copoint][2]{\,%
\begin{tikzpicture}
  \begin{pgfonlayer}{nodelayer}
    \node [style=\commandkey{style1}] (0) at (0, -0.5) {$#1$};
    \node [style=\commandkey{style2}] (1) at (0, 0.5) {$#2$};
  \end{pgfonlayer}
  \begin{pgfonlayer}{edgelayer}
    \draw (0) to (1);
  \end{pgfonlayer}
\end{tikzpicture}\,}

\newkeycommand{\kpointbraket}[style1=kpoint,style2=kpoint adjoint][2]{\,%
\begin{tikzpicture}
  \begin{pgfonlayer}{nodelayer}
    \node [style=\commandkey{style1}] (0) at (0, -0.75) {$#1$};
    \node [style=\commandkey{style2}] (1) at (0, 0.75) {$#2$};
  \end{pgfonlayer}
  \begin{pgfonlayer}{edgelayer}
    \draw (0) to (1);
  \end{pgfonlayer}
\end{tikzpicture}\,}

\newkeycommand{\kpointmap}[style1=kpoint,style2=map][2]{\,%
\begin{tikzpicture}
  \begin{pgfonlayer}{nodelayer}
    \node [style=\commandkey{style1}] (0) at (0, -1.1) {$#1$};
    \node [style=\commandkey{style2}] (1) at (0, 0.2) {$#2$};
    \node [style=none] (2) at (0, 1.5) {};
  \end{pgfonlayer}
  \begin{pgfonlayer}{edgelayer}
    \draw (0) to (1);
    \draw (1) to (2.center);
  \end{pgfonlayer}
\end{tikzpicture}%
\,}

\newkeycommand{\sandwichtwo}[style1=point,style2=map,style3=map,style4=copoint][4]{\,%
\begin{tikzpicture}
  \begin{pgfonlayer}{nodelayer}
    \node [style=\commandkey{style2}] (0) at (0, -0.75) {$#2$};
    \node [style=\commandkey{style3}] (1) at (0, 0.75) {$#3$};
    \node [style=\commandkey{style1}] (2) at (0, -2) {$#1$};
    \node [style=\commandkey{style4}] (3) at (0, 2) {$#4$};
  \end{pgfonlayer}
  \begin{pgfonlayer}{edgelayer}
    \draw (2) to (0);
    \draw (0) to (1);
    \draw (1) to (3);
  \end{pgfonlayer}
\end{tikzpicture}\,}

\newkeycommand{\longbraket}[style1=point,style2=copoint][2]{\,%
\begin{tikzpicture}
  \begin{pgfonlayer}{nodelayer}
    \node [style=\commandkey{style1}] (0) at (0, -2) {$#1$};
    \node [style=\commandkey{style2}] (1) at (0, 2) {$#2$};
  \end{pgfonlayer}
  \begin{pgfonlayer}{edgelayer}
    \draw (0) to (1);
  \end{pgfonlayer}
\end{tikzpicture}\,}

\newkeycommand{\onb}[style=point]{\ensuremath{\left\{\tikz{\node[style=\commandkey{style}] (x) at (0,-0.3) {$j$};\draw (x)--(0,0.7);}\right\}}\xspace}

\newkeycommand{\copointmap}[style=copoint][1]{\,\tikz{\node[style=\commandkey{style}] (x) at (0,0.3) {$#1$};\draw (x)--(0,-0.7);}\,}

\newcommand{\bra}[1]{\ensuremath{\left\langle #1 \right|}}
\newcommand{\ket}[1]{\ensuremath{\left|  #1 \right\rangle}}

\newcommand{\ketbra}[2]{\ensuremath{\ket{#1}\!\bra{#2}}}

\tikzstyle{cdiag}=[matrix of math nodes, row sep=3em, column sep=3em, text height=1.5ex, text depth=0.25ex,inner sep=0.5em]
\tikzstyle{arrow above}=[transform canvas={yshift=0.5ex}]
\tikzstyle{arrow below}=[transform canvas={yshift=-0.5ex}]

\usetikzlibrary{shapes.multipart}

\tikzstyle{bwSpider}=[
       rectangle split,
       rectangle split parts=2,
       rectangle split part fill={black,white},
 minimum size=3.6 mm, inner sep=-2mm, draw=black,scale=0.5,rounded corners=0.8 mm
       ]
 \tikzstyle{wbSpider}=[
       rectangle split,
       rectangle split parts=2,
       rectangle split part fill={white,black},
 minimum size=3.6 mm, inner sep=-2mm, draw=black,scale=0.5,rounded corners=0.8 mm
       ]
\tikzstyle{qWire}=[line width = 1pt, color=quantumviolet]
\tikzstyle{cWire}=[color=quantumgray,thin]%[densely dotted, thick]
\tikzstyle{env}=[copoint,regular polygon rotate=0,minimum width=0.2cm, fill=black]

\tikzstyle{probs}=[shape=semicircle,fill=white,draw=black,shape border rotate=180,minimum width=1.2cm]

\tikzstyle{every picture}=[baseline=-0.25em,scale=0.5]
\tikzstyle{dotpic}=[] % for backwards-compatibility
\tikzstyle{diredges}=[every to/.style={diredge}]
\tikzstyle{math matrix}=[matrix of math nodes,left delimiter=(,right delimiter=),inner sep=2pt,column sep=1em,row sep=0.5em,nodes={inner sep=0pt},text height=1.5ex, text depth=0.25ex]

\tikzstyle{inline text}=[text height=1.5ex, text depth=0.25ex,yshift=0.5mm]
\tikzstyle{label}=[font=\footnotesize,text height=1.5ex, text depth=0.25ex,yshift=0.5mm]
\tikzstyle{left label}=[label,anchor=east,xshift=1.5mm]
\tikzstyle{right label}=[label,anchor=west,xshift=-1mm]

\tikzstyle{braceedge}=[decorate,decoration={brace,amplitude=2mm,raise=-1mm}]
\tikzstyle{small braceedge}=[decorate,decoration={brace,amplitude=1mm,raise=-1mm}]

\tikzstyle{doubled}=[line width=1.6pt] % set the line width for all doubled (quantum) maps/wires
\tikzstyle{boldedge}=[doubled,shorten <=-0.17mm,shorten >=-0.17mm]
\tikzstyle{boldedgegray}=[doubled,gray,shorten <=-0.17mm,shorten >=-0.17mm]
\tikzstyle{singleedgegray}=[gray]%,shorten <=-0.1mm,shorten >=-0.1mm]

\tikzstyle{semidoubled}=[line width=1.4pt] % set the line width for all doubled (quantum) maps/wires
\tikzstyle{semiboldedgegray}=[semidoubled,gray,shorten <=-0.17mm,shorten >=-0.17mm]

\tikzstyle{boxedge}=[semiboldedgegray]

\tikzstyle{boldedgedashed}=[very thick,dashed,shorten <=-0.17mm,shorten >=-0.17mm]
\tikzstyle{vboldedgedashed}=[doubled,dashed,shorten <=-0.17mm,shorten >=-0.17mm]
\tikzstyle{left hook arrow}=[left hook-latex]
\tikzstyle{right hook arrow}=[right hook-latex]
\tikzstyle{sembracket}=[line width=0.5pt,shorten <=-0.07mm,shorten >=-0.07mm]

\tikzstyle{causal edge}=[->,thick,gray]
\tikzstyle{causal nondir}=[thick,gray]
\tikzstyle{timeline}=[thick,gray, dashed]

\tikzstyle{cedge}=[<->,thick,gray!70!white]

\tikzstyle{empty diagram}=[draw=gray!40!white,dashed,shape=rectangle,minimum width=1cm,minimum height=1cm]
\tikzstyle{empty diagram small}=[draw=gray!50!white,dashed,shape=rectangle,minimum width=0.6cm,minimum height=0.5cm]

\tikzstyle{dot}=[inner sep=0mm,minimum width=2mm,minimum height=2mm,draw,shape=circle]
\tikzstyle{leak}=[white dot, shape=regular polygon, minimum size=3.3 mm, regular polygon sides=3, outer sep=-0.2mm, regular polygon rotate=270]
\tikzstyle{proj}=[draw,fill=white,chamfered rectangle,chamfered rectangle angle=30, minimum width=2mm, minimum height= 1mm,scale=0.5, outer sep=-0.2mm]
\tikzstyle{wide proj}=[draw,fill=white,chamfered rectangle,chamfered rectangle angle=30, minimum width=15mm,minimum height=1mm,scale=0.5, outer sep=-0.2mm]
\tikzstyle{very wide proj}=[draw,fill=white,chamfered rectangle,chamfered rectangle angle=30, minimum width=25mm,minimum height=1mm,scale=0.5, outer sep=-0.2mm]
\tikzstyle{very very wide proj}=[draw,fill=white,chamfered rectangle,chamfered rectangle angle=30, minimum width=35mm,minimum height=1mm,scale=0.5, outer sep=-0.2mm]
\tikzstyle{preleak}=[proj]

\tikzstyle{split proj out}=[regular polygon,regular polygon sides=3,draw,scale=0.75,inner sep=-0.5pt,minimum width=3.3mm,fill=white,regular polygon rotate=180]
\tikzstyle{split proj in}=[regular polygon,regular polygon sides=3,draw,scale=0.75,inner sep=-0.5pt,minimum width=3.3mm,fill=white]
\tikzstyle{Vleak}=[white dot, shape=regular polygon, minimum size=3.3 mm, regular polygon sides=3, outer sep=-0.2mm, regular polygon rotate=90]
\tikzstyle{dleak}=[white dot, line width=1.6pt, shape=regular polygon, minimum size=3.3 mm, regular polygon sides=3, outer sep=-0.2mm, regular polygon rotate=270]

\tikzstyle{Wsquare}=[white dot, shape=regular polygon, rounded corners=0.8 mm, minimum size=3.3 mm, regular polygon sides=3, outer sep=-0.2mm]
\tikzstyle{Wsquareadj}=[white dot, shape=regular polygon, rounded corners=0.8 mm, minimum size=3.3 mm, regular polygon sides=3, outer sep=-0.2mm, regular polygon rotate=180]
\tikzstyle{ddot}=[inner sep=0mm, doubled, minimum width=2.5mm,minimum height=2.5mm,draw,shape=circle]

\tikzstyle{black dot}=[dot,fill=black]
\tikzstyle{white dot}=[dot,fill=white,,text depth=-0.2mm]
\tikzstyle{white Wsquare}=[Wsquare,fill=gray,,text depth=-0.2mm]
\tikzstyle{white Wsquareadj}=[Wsquareadj,fill=white,,text depth=-0.2mm]
\tikzstyle{green dot}=[white dot] % for backwards-compatibility
\tikzstyle{gray dot}=[dot,fill=gray!40!white,,text depth=-0.2mm]
\tikzstyle{red dot}=[gray dot] % for backwards-compatibility

\tikzstyle{black ddot}=[ddot,fill=black]
\tikzstyle{white ddot}=[ddot,fill=white]
\tikzstyle{gray ddot}=[ddot,fill=gray!40!white]

\tikzstyle{gray edge}=[gray!60!white]

\tikzstyle{small dot}=[inner sep=0.5mm,minimum width=0pt,minimum height=0pt,draw,shape=circle]

\tikzstyle{small black dot}=[small dot,fill=black]
\tikzstyle{small white dot}=[small dot,fill=white]
\tikzstyle{small gray dot}=[small dot,fill=gray!40!white]

\tikzstyle{causal dot}=[inner sep=0.4mm,minimum width=0pt,minimum height=0pt,draw=white,shape=circle,fill=gray!40!white]

\tikzstyle{phase dimensions}=[minimum size=5mm,font=\footnotesize,rectangle,rounded corners=2.5mm,inner sep=0.2mm,outer sep=-2mm]
\tikzstyle{dphase dimensions}=[minimum size=5mm,font=\footnotesize,rectangle,rounded corners=2.5mm,inner sep=0.2mm,outer sep=-2mm]
\tikzstyle{white phase dot}=[dot,fill=white,phase dimensions]
\tikzstyle{white phase ddot}=[ddot,fill=white,dphase dimensions]

\tikzstyle{white rect ddot}=[draw=black,fill=white,doubled,minimum size=5mm,font=\footnotesize,rectangle,rounded corners=2.5mm,inner sep=0.2mm]
\tikzstyle{gray rect ddot}=[draw=black,fill=gray!40!white,doubled,minimum size=6mm,font=\footnotesize,rectangle,rounded corners=3mm]

\tikzstyle{gray phase dot}=[dot,fill=gray!40!white,phase dimensions]
\tikzstyle{gray phase ddot}=[ddot,fill=gray!40!white,dphase dimensions]
\tikzstyle{grey phase dot}=[gray phase dot]
\tikzstyle{grey phase ddot}=[gray phase ddot]

\tikzstyle{small phase dimensions}=[minimum size=4mm,font=\tiny,rectangle,rounded corners=2mm,inner sep=0.2mm,outer sep=-2mm]
\tikzstyle{small dphase dimensions}=[minimum size=4mm,font=\tiny,rectangle,rounded corners=2mm,inner sep=0.2mm,outer sep=-2mm]

\tikzstyle{small gray phase dot}=[dot,fill=gray!40!white,small phase dimensions]
\tikzstyle{small gray phase ddot}=[ddot,fill=gray!40!white,small dphase dimensions]

\tikzstyle{small map}=[draw,shape=rectangle,minimum height=4mm,minimum width=4mm,fill=white]

\tikzstyle{cnot}=[fill=white,shape=circle,inner sep=-1.4pt]

\tikzstyle{asym hadamard}=[fill=white,draw,shape=NEbox,inner sep=0.6mm,font=\footnotesize,minimum height=4mm]
\tikzstyle{asym hadamard conj}=[fill=white,draw,shape=NWbox,inner sep=0.6mm,font=\footnotesize,minimum height=4mm]
\tikzstyle{asym hadamard dag}=[fill=white,draw,shape=SEbox,inner sep=0.6mm,font=\footnotesize,minimum height=4mm]

\tikzstyle{hadamard}=[fill=white,draw,inner sep=0.6mm,font=\footnotesize,minimum height=4mm,minimum width=4mm]
\tikzstyle{small hadamard}=[fill=white,draw,inner sep=0.6mm,minimum height=1.5mm,minimum width=1.5mm]
\tikzstyle{small hadamard rotate}=[small hadamard,rotate=45]
\tikzstyle{dhadamard}=[hadamard,doubled]
\tikzstyle{small dhadamard}=[small hadamard,doubled]
\tikzstyle{small dhadamard rotate}=[small hadamard rotate,doubled]
\tikzstyle{antipode}=[white dot,inner sep=0.3mm,font=\footnotesize]

\tikzstyle{scalar}=[diamond,draw,inner sep=0.5pt,font=\small]
\tikzstyle{dscalar}=[diamond,doubled, draw,inner sep=0.5pt,font=\small]

\tikzstyle{small box}=[rectangle,inline text,fill=white,draw,minimum height=5mm,yshift=-0.5mm,minimum width=5mm,font=\small]
\tikzstyle{small gray box}=[small box,fill=gray!30]
\tikzstyle{medium box}=[rectangle,inline text,fill=white,draw,minimum height=5mm,yshift=-0.5mm,minimum width=10mm,font=\small]
\tikzstyle{square box}=[small box] % for backwards-compatibility
\tikzstyle{medium gray box}=[small box,fill=gray!30]
\tikzstyle{semilarge box}=[rectangle,inline text,fill=white,draw,minimum height=5mm,yshift=-0.5mm,minimum width=12.5mm,font=\small]
\tikzstyle{large box}=[rectangle,inline text,fill=white,draw,minimum height=5mm,yshift=-0.5mm,minimum width=15mm,font=\small]
\tikzstyle{large gray box}=[small box,fill=gray!30]

\tikzstyle{Bayes box}=[rectangle,fill=black,draw, minimum height=3mm, minimum width=3mm]

\tikzstyle{gray square point}=[small box,fill=gray!50]

\tikzstyle{dphase box white}=[dhadamard]
\tikzstyle{dphase box gray}=[dhadamard,fill=gray!50!white]
\tikzstyle{phase box white}=[hadamard]
\tikzstyle{phase box gray}=[hadamard,fill=gray!50!white]

\tikzstyle{point}=[regular polygon,regular polygon sides=3,draw,scale=0.75,inner sep=-0.5pt,minimum width=9mm,fill=white,regular polygon rotate=180]
\tikzstyle{point nosep}=[regular polygon,regular polygon sides=3,draw,scale=0.75,inner sep=-2pt,minimum width=9mm,fill=white,regular polygon rotate=180]
\tikzstyle{copoint}=[regular polygon,regular polygon sides=3,draw,scale=0.75,inner sep=-0.5pt,minimum width=9mm,fill=white]
\tikzstyle{dpoint}=[point,doubled]
\tikzstyle{dcopoint}=[copoint,doubled]

\tikzstyle{pointgrow}=[shape=cornerpoint,kpoint common,scale=0.75,inner sep=3pt]
\tikzstyle{pointgrow dag}=[shape=cornercopoint,kpoint common,scale=0.75,inner sep=3pt]

\tikzstyle{wide copoint}=[fill=white,draw,shape=isosceles triangle,shape border rotate=90,isosceles triangle stretches=true,inner sep=0pt,minimum width=1.5cm,minimum height=6.12mm]
\tikzstyle{wide point}=[fill=white,draw,shape=isosceles triangle,shape border rotate=-90,isosceles triangle stretches=true,inner sep=0pt,minimum width=1.5cm,minimum height=6.12mm,yshift=-0.0mm]
\tikzstyle{wide point plus}=[fill=white,draw,shape=isosceles triangle,shape border rotate=-90,isosceles triangle stretches=true,inner sep=0pt,minimum width=1.74cm,minimum height=7mm,yshift=-0.0mm]

\tikzstyle{wide dpoint}=[fill=white,doubled,draw,shape=isosceles triangle,shape border rotate=-90,isosceles triangle stretches=true,inner sep=0pt,minimum width=1.5cm,minimum height=6.12mm,yshift=-0.0mm]

\tikzstyle{tinypoint}=[regular polygon,regular polygon sides=3,draw,scale=0.55,inner sep=-0.15pt,minimum width=6mm,fill=white,regular polygon rotate=180]

\tikzstyle{white point}=[point]
\tikzstyle{white dpoint}=[dpoint]
\tikzstyle{green point}=[white point] % for backwards-compatibility
\tikzstyle{white copoint}=[copoint]
\tikzstyle{gray point}=[point,fill=gray!40!white]
\tikzstyle{gray dpoint}=[gray point,doubled]
\tikzstyle{red point}=[gray point] % for backwards-compatibility
\tikzstyle{gray copoint}=[copoint,fill=gray!40!white]
\tikzstyle{gray dcopoint}=[gray copoint,doubled]

\tikzstyle{white point guide}=[regular polygon,regular polygon sides=3,font=\scriptsize,draw,scale=0.65,inner sep=-0.5pt,minimum width=9mm,fill=white,regular polygon rotate=180]

\tikzstyle{black point}=[point,fill=black,font=\color{white}]
\tikzstyle{black copoint}=[copoint,fill=black,font=\color{white}]

\tikzstyle{tiny gray point}=[tinypoint,fill=gray!40!white]

\tikzstyle{diredge}=[->]
\tikzstyle{ddiredge}=[<->]
\tikzstyle{rdiredge}=[<-]
\tikzstyle{thickdiredge}=[->, very thick]
\tikzstyle{pointer edge}=[->,very thick,gray]
\tikzstyle{pointer edge part}=[very thick,gray]
\tikzstyle{dashed edge}=[dashed]
\tikzstyle{thick dashed edge}=[very thick,dashed]
\tikzstyle{thick gray dashed edge}=[thick dashed edge,gray!40]
\tikzstyle{thick map edge}=[very thick,|->]

\makeatletter
\newcommand{\boxshape}[3]{%
\pgfdeclareshape{#1}{
\inheritsavedanchors[from=rectangle] % this is nearly a rectangle
\inheritanchorborder[from=rectangle]
\inheritanchor[from=rectangle]{center}
\inheritanchor[from=rectangle]{north}
\inheritanchor[from=rectangle]{south}
\inheritanchor[from=rectangle]{west}
\inheritanchor[from=rectangle]{east}
\backgroundpath{% this is new
\southwest \pgf@xa=\pgf@x \pgf@ya=\pgf@y
\northeast \pgf@xb=\pgf@x \pgf@yb=\pgf@y

\@tempdima=#2
\@tempdimb=#3

\pgfpathmoveto{\pgfpoint{\pgf@xa - 5pt + \@tempdima}{\pgf@ya}}
\pgfpathlineto{\pgfpoint{\pgf@xa - 5pt - \@tempdima}{\pgf@yb}}
\pgfpathlineto{\pgfpoint{\pgf@xb + 5pt + \@tempdimb}{\pgf@yb}}
\pgfpathlineto{\pgfpoint{\pgf@xb + 5pt - \@tempdimb}{\pgf@ya}}
\pgfpathlineto{\pgfpoint{\pgf@xa - 5pt + \@tempdima}{\pgf@ya}}
\pgfpathclose
}
}}

\boxshape{NEbox}{0pt}{3pt}
\boxshape{SEbox}{0pt}{-3pt}
\boxshape{NWbox}{5pt}{0pt}
\boxshape{SWbox}{-5pt}{0pt}
\boxshape{EBox}{-3pt}{3pt}
\boxshape{WBox}{3pt}{-3pt}
\makeatother

\tikzstyle{cloud}=[shape=cloud,draw,minimum width=1.5cm,minimum height=1.5cm]

\tikzstyle{map}=[draw,shape=NEbox,inner sep=1pt,minimum height=4mm,fill=white]
\tikzstyle{dashedmap}=[draw,dashed,shape=NEbox,inner sep=2pt,minimum height=6mm,fill=white]
\tikzstyle{mapdag}=[draw,shape=SEbox,inner sep=1pt,minimum height=4mm,fill=white]
\tikzstyle{mapadj}=[draw,shape=SEbox,inner sep=2pt,minimum height=6mm,fill=white]
\tikzstyle{maptrans}=[draw,shape=SWbox,inner sep=2pt,minimum height=6mm,fill=white]
\tikzstyle{mapconj}=[draw,shape=NWbox,inner sep=2pt,minimum height=6mm,fill=white]

\tikzstyle{medium map}=[draw,shape=NEbox,inner sep=2pt,minimum height=6mm,fill=white,minimum width=7mm]
\tikzstyle{medium map dag}=[draw,shape=SEbox,inner sep=2pt,minimum height=6mm,fill=white,minimum width=7mm]
\tikzstyle{medium map adj}=[draw,shape=SEbox,inner sep=2pt,minimum height=6mm,fill=white,minimum width=7mm]
\tikzstyle{medium map trans}=[draw,shape=SWbox,inner sep=2pt,minimum height=6mm,fill=white,minimum width=7mm]
\tikzstyle{medium map conj}=[draw,shape=NWbox,inner sep=2pt,minimum height=6mm,fill=white,minimum width=7mm]
\tikzstyle{semilarge map}=[draw,shape=NEbox,inner sep=2pt,minimum height=6mm,fill=white,minimum width=9.5mm]
\tikzstyle{semilarge map trans}=[draw,shape=SWbox,inner sep=2pt,minimum height=6mm,fill=white,minimum width=9.5mm]
\tikzstyle{semilarge map adj}=[draw,shape=SEbox,inner sep=2pt,minimum height=6mm,fill=white,minimum width=9.5mm]
\tikzstyle{semilarge map dag}=[draw,shape=SEbox,inner sep=2pt,minimum height=6mm,fill=white,minimum width=9.5mm]
\tikzstyle{semilarge map conj}=[draw,shape=NWbox,inner sep=2pt,minimum height=6mm,fill=white,minimum width=9.5mm]
\tikzstyle{large map}=[draw,shape=NEbox,inner sep=2pt,minimum height=6mm,fill=white,minimum width=12mm]
\tikzstyle{large map conj}=[draw,shape=NWbox,inner sep=2pt,minimum height=6mm,fill=white,minimum width=12mm]
\tikzstyle{very large map}=[draw,shape=NEbox,inner sep=2pt,minimum height=6mm,fill=white,minimum width=17mm]

\tikzstyle{medium dmap}=[draw,doubled,shape=NEbox,inner sep=2pt,minimum height=6mm,fill=white,minimum width=7mm]
\tikzstyle{medium dmap dag}=[draw,doubled,shape=SEbox,inner sep=2pt,minimum height=6mm,fill=white,minimum width=7mm]
\tikzstyle{medium dmap adj}=[draw,doubled,shape=SEbox,inner sep=2pt,minimum height=6mm,fill=white,minimum width=7mm]
\tikzstyle{medium dmap trans}=[draw,doubled,shape=SWbox,inner sep=2pt,minimum height=6mm,fill=white,minimum width=7mm]
\tikzstyle{medium dmap conj}=[draw,doubled,shape=NWbox,inner sep=2pt,minimum height=6mm,fill=white,minimum width=7mm]
\tikzstyle{semilarge dmap}=[draw,doubled,shape=NEbox,inner sep=2pt,minimum height=6mm,fill=white,minimum width=9.5mm]
\tikzstyle{semilarge dmap trans}=[draw,doubled,shape=SWbox,inner sep=2pt,minimum height=6mm,fill=white,minimum width=9.5mm]
\tikzstyle{semilarge dmap adj}=[draw,doubled,shape=SEbox,inner sep=2pt,minimum height=6mm,fill=white,minimum width=9.5mm]
\tikzstyle{semilarge dmap dag}=[draw,doubled,shape=SEbox,inner sep=2pt,minimum height=6mm,fill=white,minimum width=9.5mm]
\tikzstyle{semilarge dmap conj}=[draw,doubled,shape=NWbox,inner sep=2pt,minimum height=6mm,fill=white,minimum width=9.5mm]
\tikzstyle{large dmap}=[draw,doubled,shape=NEbox,inner sep=2pt,minimum height=6mm,fill=white,minimum width=12mm]
\tikzstyle{large dmap conj}=[draw,doubled,shape=NWbox,inner sep=2pt,minimum height=6mm,fill=white,minimum width=12mm]
\tikzstyle{large dmap trans}=[draw,doubled,shape=SWbox,inner sep=2pt,minimum height=6mm,fill=white,minimum width=12mm]
\tikzstyle{large dmap adj}=[draw,doubled,shape=SEbox,inner sep=2pt,minimum height=6mm,fill=white,minimum width=12mm]
\tikzstyle{large dmap dag}=[draw,doubled,shape=SEbox,inner sep=2pt,minimum height=6mm,fill=white,minimum width=12mm]
\tikzstyle{very large dmap}=[draw,doubled,shape=NEbox,inner sep=2pt,minimum height=6mm,fill=white,minimum width=19.5mm]

\tikzstyle{muxbox}=[draw,shape=rectangle,minimum height=3mm,minimum width=3mm,fill=white]
\tikzstyle{dmuxbox}=[muxbox,doubled]

\tikzstyle{box}=[draw,shape=rectangle,inner sep=2pt,minimum height=6mm,minimum width=6mm,fill=white]
\tikzstyle{dbox}=[draw,doubled,shape=rectangle,inner sep=2pt,minimum height=6mm,minimum width=6mm,fill=white]
\tikzstyle{dmap}=[draw,doubled,shape=NEbox,inner sep=2pt,minimum height=6mm,fill=white]
\tikzstyle{dmapdag}=[draw,doubled,shape=SEbox,inner sep=2pt,minimum height=6mm,fill=white]
\tikzstyle{dmapadj}=[draw,doubled,shape=SEbox,inner sep=2pt,minimum height=6mm,fill=white]
\tikzstyle{dmaptrans}=[draw,doubled,shape=SWbox,inner sep=2pt,minimum height=6mm,fill=white]
\tikzstyle{dmapconj}=[draw,doubled,shape=NWbox,inner sep=2pt,minimum height=6mm,fill=white]

\tikzstyle{ddmap}=[draw,doubled,dashed,shape=NEbox,inner sep=2pt,minimum height=6mm,fill=white]
\tikzstyle{ddmapdag}=[draw,doubled,dashed,shape=SEbox,inner sep=2pt,minimum height=6mm,fill=white]
\tikzstyle{ddmapadj}=[draw,doubled,dashed,shape=SEbox,inner sep=2pt,minimum height=6mm,fill=white]
\tikzstyle{ddmaptrans}=[draw,doubled,dashed,shape=SWbox,inner sep=2pt,minimum height=6mm,fill=white]
\tikzstyle{ddmapconj}=[draw,doubled,dashed,shape=NWbox,inner sep=2pt,minimum height=6mm,fill=white]

\boxshape{sNEbox}{0pt}{3pt}
\boxshape{sSEbox}{0pt}{-3pt}
\boxshape{sNWbox}{3pt}{0pt}
\boxshape{sSWbox}{-3pt}{0pt}
\tikzstyle{smap}=[draw,shape=sNEbox,fill=white]
\tikzstyle{smapdag}=[draw,shape=sSEbox,fill=white]
\tikzstyle{smapadj}=[draw,shape=sSEbox,fill=white]
\tikzstyle{smaptrans}=[draw,shape=sSWbox,fill=white]
\tikzstyle{smapconj}=[draw,shape=sNWbox,fill=white]

\tikzstyle{dsmap}=[draw,dashed,shape=sNEbox,fill=white]
\tikzstyle{dsmapdag}=[draw,dashed,shape=sSEbox,fill=white]
\tikzstyle{dsmaptrans}=[draw,dashed,shape=sSWbox,fill=white]
\tikzstyle{dsmapconj}=[draw,dashed,shape=sNWbox,fill=white]

\boxshape{mNEbox}{0pt}{10pt}
\boxshape{mSEbox}{0pt}{-10pt}
\boxshape{mNWbox}{10pt}{0pt}
\boxshape{mSWbox}{-10pt}{0pt}
\tikzstyle{mmap}=[draw,shape=mNEbox]
\tikzstyle{mmapdag}=[draw,shape=mSEbox]
\tikzstyle{mmaptrans}=[draw,shape=mSWbox]
\tikzstyle{mmapconj}=[draw,shape=mNWbox]

\tikzstyle{mmapgray}=[draw,fill=gray!40!white,shape=mNEbox]
\tikzstyle{smapgray}=[draw,fill=gray!40!white,shape=sNEbox]

\makeatletter

\pgfdeclareshape{cornerpoint}{
\inheritsavedanchors[from=rectangle] % this is nearly a rectangle
\inheritanchorborder[from=rectangle]
\inheritanchor[from=rectangle]{center}
\inheritanchor[from=rectangle]{north}
\inheritanchor[from=rectangle]{south}
\inheritanchor[from=rectangle]{west}
\inheritanchor[from=rectangle]{east}
\backgroundpath{% this is new
\southwest \pgf@xa=\pgf@x \pgf@ya=\pgf@y
\northeast \pgf@xb=\pgf@x \pgf@yb=\pgf@y

\pgfmathsetmacro{\pgf@shorten@left}{\pgfkeysvalueof{/tikz/shorten left}}
\pgfmathsetmacro{\pgf@shorten@right}{\pgfkeysvalueof{/tikz/shorten right}}

\pgfpathmoveto{\pgfpoint{0.5 * (\pgf@xa + \pgf@xb)}{\pgf@ya - 5pt}}
\pgfpathlineto{\pgfpoint{\pgf@xa - 8pt + \pgf@shorten@left}{\pgf@yb - 1.5 * \pgf@shorten@left}}
\pgfpathlineto{\pgfpoint{\pgf@xa - 8pt + \pgf@shorten@left}{\pgf@yb}}
\pgfpathlineto{\pgfpoint{\pgf@xb + 8pt - \pgf@shorten@right}{\pgf@yb}}
\pgfpathlineto{\pgfpoint{\pgf@xb + 8pt - \pgf@shorten@right}{\pgf@yb - 1.5 * \pgf@shorten@right}}
\pgfpathclose
}
}

\pgfdeclareshape{cornercopoint}{
\inheritsavedanchors[from=rectangle] % this is nearly a rectangle
\inheritanchorborder[from=rectangle]
\inheritanchor[from=rectangle]{center}
\inheritanchor[from=rectangle]{north}
\inheritanchor[from=rectangle]{south}
\inheritanchor[from=rectangle]{west}
\inheritanchor[from=rectangle]{east}
\backgroundpath{% this is new
\southwest \pgf@xa=\pgf@x \pgf@ya=\pgf@y
\northeast \pgf@xb=\pgf@x \pgf@yb=\pgf@y

\pgfmathsetmacro{\pgf@shorten@left}{\pgfkeysvalueof{/tikz/shorten left}}
\pgfmathsetmacro{\pgf@shorten@right}{\pgfkeysvalueof{/tikz/shorten right}}

\pgfpathmoveto{\pgfpoint{0.5 * (\pgf@xa + \pgf@xb)}{\pgf@yb + 5pt}}
\pgfpathlineto{\pgfpoint{\pgf@xa - 8pt + \pgf@shorten@left}{\pgf@ya + 1.5 * \pgf@shorten@left}}
\pgfpathlineto{\pgfpoint{\pgf@xa - 8pt + \pgf@shorten@left}{\pgf@ya}}
\pgfpathlineto{\pgfpoint{\pgf@xb + 8pt - \pgf@shorten@right}{\pgf@ya}}
\pgfpathlineto{\pgfpoint{\pgf@xb + 8pt - \pgf@shorten@right}{\pgf@ya + 1.5 * \pgf@shorten@right}}
\pgfpathclose
}
}

\makeatother

\pgfkeyssetvalue{/tikz/shorten left}{0pt}
\pgfkeyssetvalue{/tikz/shorten right}{0pt}

\tikzstyle{kpoint common}=[draw,fill=white,inner sep=1pt,minimum height=4mm]
\tikzstyle{kpoint sc}=[shape=cornerpoint,kpoint common]
\tikzstyle{kpoint adjoint sc}=[shape=cornercopoint,kpoint common]
\tikzstyle{kpoint}=[shape=cornerpoint,shorten left=5pt,kpoint common]
\tikzstyle{kpoint adjoint}=[shape=cornercopoint,shorten left=5pt,kpoint common]
\tikzstyle{kpoint conjugate}=[shape=cornerpoint,shorten right=5pt,kpoint common]
\tikzstyle{kpoint transpose}=[shape=cornercopoint,shorten right=5pt,kpoint common]
\tikzstyle{kpoint symm}=[shape=cornerpoint,shorten left=5pt,shorten right=5pt,kpoint common]

\tikzstyle{wide kpoint sc}=[shape=cornerpoint,kpoint common, minimum width=1 cm]
\tikzstyle{wide kpointdag sc}=[shape=cornercopoint,kpoint common, minimum width=1 cm]

\tikzstyle{black kpoint}=[shape=cornerpoint,shorten left=5pt,kpoint common,fill=black,font=\color{white}]

\tikzstyle{black kpoint sm}=[shape=cornerpoint,shorten left=5pt,kpoint common,fill=black,font=\color{white},scale=0.75]

\tikzstyle{black kpoint adjoint}=[shape=cornercopoint,shorten left=5pt,kpoint common,fill=black,font=\color{white}]
\tikzstyle{black kpointadj}=[shape=cornercopoint,shorten left=5pt,kpoint common,fill=black,font=\color{white}]

\tikzstyle{black kpointadj sm}=[shape=cornercopoint,shorten left=5pt,kpoint common,fill=black,font=\color{white},scale=0.75]

\tikzstyle{black dkpoint}=[shape=cornerpoint,shorten left=5pt,kpoint common,fill=black, doubled,font=\color{white}]
\tikzstyle{black dkpoint adjoint}=[shape=cornercopoint,shorten left=5pt,kpoint common,fill=black, doubled,font=\color{white}]
\tikzstyle{black dkpointadj}=[shape=cornercopoint,shorten left=5pt,kpoint common,fill=black, doubled,font=\color{white}]

\tikzstyle{black dkpoint sm}=[shape=cornerpoint,shorten left=5pt,kpoint common,fill=black, doubled,font=\color{white},scale=0.75]
\tikzstyle{black dkpointadj sm}=[shape=cornercopoint,shorten left=5pt,kpoint common,fill=black, doubled,font=\color{white},scale=0.75]

\tikzstyle{kpointdag}=[kpoint adjoint]
\tikzstyle{kpointadj}=[kpoint adjoint]
\tikzstyle{kpointconj}=[kpoint conjugate]
\tikzstyle{kpointtrans}=[kpoint transpose]

\tikzstyle{big kpoint}=[kpoint, minimum width=1.2 cm, minimum height=8mm, inner sep=4pt, text depth=3mm]

\tikzstyle{wide kpoint}=[kpoint, minimum width=1 cm, inner sep=2pt]%, text depth=-0.7 mm]
\tikzstyle{wide kpointdag}=[kpointdag, minimum width=1 cm, inner sep=2pt]%, text depth=0.7 mm]
\tikzstyle{wide kpointconj}=[kpointconj, minimum width=1 cm, inner sep=2pt]%, text depth=-0.7 mm]
\tikzstyle{wide kpointtrans}=[kpointtrans, minimum width=1 cm, inner sep=2pt]%, text depth=0.7 mm]

\tikzstyle{wider kpoint}=[kpoint, minimum width=1.25 cm, inner sep=2pt]%, text depth=-0.7 mm]
\tikzstyle{wider kpointdag}=[kpointdag, minimum width=1.25 cm, inner sep=2pt]%, text depth=0.7 mm]
\tikzstyle{wider kpointconj}=[kpointconj, minimum width=1.25 cm, inner sep=2pt]%, text depth=-0.7 mm]
\tikzstyle{wider kpointtrans}=[kpointtrans, minimum width=1.25 cm, inner sep=2pt]%, text depth=0.7 mm]

\tikzstyle{gray kpoint}=[kpoint,fill=gray!50!white]
\tikzstyle{gray kpointdag}=[kpointdag,fill=gray!50!white]
\tikzstyle{gray kpointadj}=[kpointadj,fill=gray!50!white]
\tikzstyle{gray kpointconj}=[kpointconj,fill=gray!50!white]
\tikzstyle{gray kpointtrans}=[kpointtrans,fill=gray!50!white]

\tikzstyle{gray dkpoint}=[kpoint,fill=gray!50!white,doubled]
\tikzstyle{gray dkpointdag}=[kpointdag,fill=gray!50!white,doubled]
\tikzstyle{gray dkpointadj}=[kpointadj,fill=gray!50!white,doubled]
\tikzstyle{gray dkpointconj}=[kpointconj,fill=gray!50!white,doubled]
\tikzstyle{gray dkpointtrans}=[kpointtrans,fill=gray!50!white,doubled]

\tikzstyle{white label}=[draw,fill=white,rectangle,inner sep=0.7 mm]
\tikzstyle{gray label}=[draw,fill=gray!50!white,rectangle,inner sep=0.7 mm]
\tikzstyle{black label}=[draw,fill=black,rectangle,inner sep=0.7 mm]

\tikzstyle{dkpoint}=[kpoint,doubled]
\tikzstyle{wide dkpoint}=[wide kpoint,doubled]
\tikzstyle{dkpointdag}=[kpoint adjoint,doubled]
\tikzstyle{wide dkpointdag}=[wide kpointdag,doubled]
\tikzstyle{dkcopoint}=[kpoint adjoint,doubled]
\tikzstyle{dkpointadj}=[kpoint adjoint,doubled]
\tikzstyle{dkpointconj}=[kpoint conjugate,doubled]
\tikzstyle{dkpointtrans}=[kpoint transpose,doubled]

\tikzstyle{kscalar}=[kpoint common, shape=EBox, inner xsep=-1pt, inner ysep=3pt,font=\small]
\tikzstyle{kscalarconj}=[kpoint common, shape=WBox, inner xsep=-1pt, inner ysep=3pt,font=\small]

\tikzstyle{spekpoint}=[kpoint sc,minimum height=5mm,inner sep=3pt]
\tikzstyle{spekcopoint}=[kpoint adjoint sc,minimum height=5mm,inner sep=3pt]

\tikzstyle{dspekpoint}=[spekpoint,doubled]
\tikzstyle{dspekcopoint}=[spekcopoint,doubled]

 \tikzstyle{upground}=[circuit ee IEC,thick,ground,rotate=90,scale=2.5]
 \tikzstyle{downground}=[circuit ee IEC,thick,ground,rotate=-90,scale=2.5]
 \tikzstyle{bigground}=[regular polygon,regular polygon sides=3,draw=gray,scale=0.50,inner sep=-0.5pt,minimum width=10mm,fill=gray]
\tikzstyle{arrs}=[-latex,font=\small,auto]
\tikzstyle{arrow plain}=[arrs]
\tikzstyle{arrow dashed}=[dashed,arrs]
\tikzstyle{arrow bold}=[very thick,arrs]
\tikzstyle{arrow hide}=[draw=white!0,-]
\tikzstyle{arrow reverse}=[latex-]
\tikzstyle{cdnode}=[]

\let\olddagger\dagger
\renewcommand{\dagger}{\ensuremath{\olddagger}\xspace}

\usepackage{color}
\def\bR{\begin{color}{red}}
\def\bB{\begin{color}{blue}}
\def\bM{\begin{color}{magenta}}
\def\bC{\begin{color}{cyan}}
\def\bW{\begin{color}{white}}
\def\bBl{\begin{color}{black}}
\def\bG{\begin{color}{green}}
\def\bY{\begin{color}{yellow}}
\def\e{\end{color}\xspace}
\def\jR{\begin{color}{DarkRed}}
\def\jB{\begin{color}{MidnightBlue}}
\def\jM{\begin{color}{magenta}}
\def\jC{\begin{color}{cyan}}
\def\jW{\begin{color}{white}}
\def\jBl{\begin{color}{black}}
\def\jG{\begin{color}{green}}
\def\jY{\begin{color}{yellow}}

\def\cR{\begin{color}{Crimson}}
\def\cB{\begin{color}{cyan}}
\def\cM{\begin{color}{magenta}}
\def\cC{\begin{color}{cyan}}
\def\cW{\begin{color}{white}}
\def\cBl{\begin{color}{black}}
\def\cG{\begin{color}{green}}
\def\cY{\begin{color}{yellow}}
\definecolor{quantumviolet}{RGB}{79, 4, 134}
\definecolor{quantumgray}{RGB}{134,79,4}

\usepackage{amsthm}
\newtheorem{theorem}{Theorem}[]
\newtheorem{lemma}[theorem]{Lemma}
\newtheorem{proposition}[theorem]{Proposition}

\newtheorem{definition}[theorem]{Definition}
\newtheorem{definition_SM}[theorem]{Definition}
\newtheorem{observation}[theorem]{Observation}

\newtheorem{rem}[theorem]{Remark}
\newtheorem{corollary}[theorem]{Corollary}

\newtheorem{conjecture}[theorem]{Conjecture}

\def\john{\begin{color}{blue}}
\def\blk{\end{color}\xspace}

\def\karol{\begin{color}{brown}}
\def\endkarol{\end{color}\xspace}

\definecolor{GrT}{rgb}{0.13, 0.55, 0.13}

\begin{document}
%\runningpagewiselinenumbers
%\linenumbers

\title{Complete extension: the non-signalling analog of quantum purification}
%\date{\today} 

\author{Marek Winczewski}
\affiliation{International Centre for Theory of Quantum Technologies,
University of Gda{\'ns}k, Wita Stwosza 63, 80-308 Gda{\'n}sk, Poland }
\affiliation{Institute of Theoretical Physics and Astrophysics and National Quantum Information Centre in Gda\'nsk, University of Gda\'nsk, 80--952 Gda\'nsk, Poland}

\author{Tamoghna Das}
\affiliation{International Centre for Theory of Quantum Technologies,
University of Gda{\'ns}k, Wita Stwosza 63, 80-308 Gda{\'n}sk, Poland }
\affiliation{Department of Physics, Indian Institute of Technology Kharagpur, Kharagpur-721302, India}

\author{John H. Selby}
\affiliation{International Centre for Theory of Quantum Technologies,
University of Gda{\'ns}k, Wita Stwosza 63, 80-308 Gda{\'n}sk, Poland }

\author{Karol Horodecki}
\affiliation{Institute of Informatics and National Quantum Information Centre in Gda\'nsk, Faculty of Mathematics, Physics and Informatics, University of Gda\'nsk, 80--952 Gda\'nsk, Poland}
\affiliation{International Centre for Theory of Quantum Technologies,
University of Gda{\'ns}k, Wita Stwosza 63, 80-308 Gda{\'n}sk, Poland }

\author{Pawe\l{} Horodecki} 
\affiliation{International Centre for Theory of Quantum Technologies,
University of Gda{\'ns}k, Wita Stwosza 63, 80-308 Gda{\'n}sk, Poland }
\affiliation{Faculty of Applied Physics and Mathematics, Gda\'nsk University of Technology, 80--233 Gda\'nsk, Poland}
\affiliation{National Quantum Information Centre, University of Gda{\'ns}k, ul. Jana Ba{\.z}y{\'n}skiego 8, 80-309 Gda{\'n}sk, Poland}

\author{\L{}ukasz Pankowski} 
\affiliation{VOICELAB.AI, Al. Grunwaldzka 135A; 80-264 Gdańsk, Poland,}

\author{Marco Piani}
\affiliation{evolutionQ Inc., Waterloo, Ontario, N2L 3L3, Canada}
\affiliation{{\it SUPA} and Department of Physics, University of Strathclyde, Glasgow, G4 0NG, UK}

\author{Ravishankar Ramanathan}
\affiliation{Department of Computer Science, The University of Hong Kong, Pokfulam Road, Hong Kong}

\begin{abstract}
Deriving quantum mechanics from information-theoretic postulates is a recent research direction 
taken, in part, with the view of finding a beyond-quantum theory; once the postulates are
clear, we can consider modifications to them. A key postulate is the purification postulate, which we propose to replace by a more generally applicable postulate that we call the complete extension postulate (CEP), i.e.,  the existence of an extension of a physical system from which one can generate any other extension.  This new concept leads to a plethora of open questions and research directions in the study of  general  theories satisfying  the  CEP (which may include a theory that hyper-decoheres to quantum theory). For example, we show that the CEP implies the impossibility of bit-commitment. This is exemplified by a case study of the theory of non-signalling behaviors which we show satisfies the CEP. We moreover show that in certain cases the complete extension will not be pure, highlighting the key divergence from the purification postulate. 
\end{abstract}

\maketitle

\tableofcontents

\section{Introduction}\label{sec:intro}

Quantum mechanics appears to be one of the best validated physical  theories and, at the same time, lacking a  single agreed-on interpretation  \cite{Peres2002}. It can be expressed  via several mathematical axioms \cite{Dirac,vn,vNeumann} and leads to numerous phenomena, which have been confirmed experimentally,  and now form the basis for many of our everyday technologies.
However, it cannot explain some of the visible (or detectable) parts of the universe -- 
the key example being the realm of the theory of gravity. For this reason,  an important scientific effort is 
underway to find a beyond-quantum theory which would explain phenomena beyond quantum theory \cite{PR,WBertramI,Bertram2018II,hardy2001quantum,zyczkowski2008quartic,smolin2006could}. However, it is not clear which  axiom  
of quantum mechanics should be replaced or modified in order to keep  the theory as expressive as quantum mechanics whilst explaining phenomena, such as gravity, which seem to be beyond quantum theory. 

  It is known that the standard axioms of quantum mechanics are not independent from one another \cite{stone1932one,gleason1975measures,masanes2019measurement}, i.e., they can be used to derive one another. It is therefore difficult to modify the standard postulates of quantum theory as they cannot be independently modified.   In order to examine modifications of quantum theory,  it is therefore useful to first  re-express quantum mechanics in the form of a new set of postulates  which can be independently modified.  Recently, there has been a great deal of interest  in such \emph{reformulations},  particularly,  via   information theoretic  postulates within the framework of so-called Generalised Probabilistic Theories (GPTs).  This idea led to quite a  number of reformulations of quantum theory \cite{DiakicBruckner,hardy2001quantum,Chiribella2010,Hardy2,Clifton2003,Goyal2008,MasanesMuller2011,BarnumMullerUdudec2014,Wilce16,Hhn2017,BudiyonoRohrlich2017,selbyReconstruction,STull,Wetering2019,Nakahira2020}  in which sets of axioms are proposed which single out quantum mechanics from the space of all GPTs.

  One of the key insights in the reconstruction programme came from \cite{Chiribella2011} which introduced the purification postulate. 
This was an essential postulate, in that it was the sole postulate that distinguished quantum from classical theory. Since it's introduction the purification postulate has been used extensively in the literature to prove many results, for example, pertaining to computation \cite{lee2016generalised,lee2016deriving,barnum2018oracles}, cryptography \cite{chiribella2010probabilistic,sikora2018simple}, thermodynamics \cite{chiribella2015entanglement,chiribella2017microcanonical}, and interference \cite{barnum2017ruling}.
Essentially the purification postulate is a generalisation of the notion of purification within quantum theory to arbitrary GPTs: 
 \begin{itemize}
    \item[] {\bf Quantum Purifications}
    {\it In quantum theory, for any state $\rho_A$ there exists a system $B$ and a bipartite pure state $\ket{\psi}_{AB}$,  which satisfies $\mathsf{tr}_B(\ketbra{\psi}{\psi})=\rho_A$. The bipartite state $\ket{\psi}_{AB}$ is known as a purification of $\rho_A$, and these are essentially unique, in that there is always an isometry mapping between any pair of purifications.} 
\end{itemize}
We formally introduce purifications in Sec.~\ref{sup:GPTs} and the formalism of GPTs necessary to talk about it in Sec.~\ref{subsec:GPT}. At this point, however, let us observe that there are 
  various issues with the purification postulate that may lead one to believe that it will have to be modified in order to find a more fundamental theory of nature.
 \begin{enumerate}
     \item There is a no-go theorem proven in Ref.~\cite{lee2018no} which shows that if one believes that a more fundamental theory should be causal, and should reduce to quantum theory by a decoherence-like mechanism, then one must give up on the purification postulate.
     \item We prove a theorem in Sec.~\ref{sec:no-go-discrete} which shows that the purification postulate must fail in any discrete theory, whilst in Refs.~\cite{Buniy_2005,Muller_2009, palmer2020discretisation} ideas coming from quantum gravity are used to argue that on a fundamental level the quantum state space will become discrete.
     \item The purification postulate also fails to hold in super-selected quantum systems \cite{selbyReconstruction,Nakahira2020,westerbaan2022computer}, and so if we believe in any fundamental superselection rules then the postulate must be modified in some way.
     \item There are also questions about whether the purification postulate is suitable in a theory which permits indefinite causal structure \cite{hardy2005probability,Hardy_gravity,oreshkov2012quantum,chiribella2013quantum}, as it is shown in Ref.~\cite{araujo2017purification} that a purification-like postulate fails to hold for all quantum process matrices.  That a theory should permit indefinite causal structure can again be motivated by ideas coming from quantum gravity. 
 \end{enumerate}
 For all of these reasons  one can then ask if the purification postulate  can be weakened or replaced by another  postulate, resulting in a 
physical theory  with  more explanatory power. In this manuscript, we propose  such 
a replacement for the purification postulate  called the Complete Extension Postulate (CEP)  and study its consequences in various contexts. 

 In order to  understand the complete extension  postulate  we  begin by observing that,  within quantum mechanics, purifications are examples of complete extensions  \cite[Exercise 2.8.2]{Nielsen-Chuang}. 
\begin{itemize}
    \item[] {\bf Quantum Complete Extensions}
        {\it A complete extension for a state $\rho_A$ in quantum theory, is an extension $\sigma^*_{AB}$ of $\rho_A$, that is is satisfies $\mathsf{tr}_B(\sigma^*_{AB})= \rho_A$, where, moreover, any other extension $\sigma$ can be reached via local operations on $B$, and any ensemble for $\rho_A$ can be reached via a measurement on $B$.} 
\end{itemize}
It can be straightforwardly seen that, within quantum theory, all quantum purifications are quantum complete extensions but not vice versa. It is precisely these complete extensions which we generalise to GPTs in order to define the complete extension postulate in Sec.~\ref{sec:ConnectionToPP}. As this is a weaker notion within quantum theory it is therefore plausible that the associated postulate will apply more broadly and avoid some of the aforementioned issues with the purification postulate.
Indeed, we show that this is the case for the first three issues raised above:
 \begin{enumerate}
     \item In Sec.~\ref{sec:Hyper} we show the proof of the no-go theorem for hyperdecoherence provided in \cite{lee2018no} no longer holds if we use the complete extension postulate.
     \item We show that the complete extension postulate can hold in discrete theories, in particular, in Sec.~\ref{sec:BoxWorld} we demonstrate this for the widely studied theory of non-signalling behaviours \cite{PR,barrett2007information,Bell-nonlocality} colloquially known as \emph{boxworld}.
     \item In Sec.~\ref{sec:ConnectionToPP} we show that the complete extension postulate holds in super-selected quantum systems and within classical systems.
 \end{enumerate}

This may therefore lead one to wonder whether perhaps the complete extension postulate is too weak and does not have any meaningful consequences. We show that this is not the case. In particular, we show that there are theories which have been studied in the literature, such as those found in \cite{barnum2008nonclassicality}, which fail to satisfy the CEP. This means that it is not a trivial postulate and does offer explanatory power. Indeed, there are two key reasons that a given GPT may fail to satisfy the CEP. On the one hand, it may not have sufficient bipartite states, in which case it would not admit of an extension that enables access to all ensembles. On the other hand, it may not have sufficient dynamics, as would be the case for highly asymmetric state spaces. Namely, in the latter case, even if an extension enables access to all ensembles, it may not give a possibility for the generation of any other extension from it by this dynamics. Moreover, in Sec.~\ref{sec:ProofNBC} We show that any theory which satisfies the CEP cannot allow for the crytographic primitive of bit-commitment, and, moreover, following the techniques of \cite{sikora2018simple}, we find a non-trivial lower bound on the success probability for any protocol. As an added benefit, this unifies the proof of impossibility of bit-commitment for the quantum and classical cases which previously were treated separately.

We also study the complete extensions within the theory of non-signalling behaviours in some depth in Sec.~\ref{sec:BoxWorld} and in App.~\ref{app:Examples}. In particular, we show how a complete extension can be constructed within this theory, and that the dimensions of these complete extensions are always finite. In this way, we show that CEP is not an empty postulate in general theories, as the complete extensions can actually be constructed. Furthermore, we characterise the minimal ensembles of the non-local isotropic behaviours, which was already proved useful in the context of the non-signalling
device independent secure key agreement \cite{WDH}.

The introduction of a new postulate opens up many doors for future exploration. We touch on just a few of them in this paper and point out many others as future research directions as we go. In particular, we collect together many of these in Sec.~\ref{sec:conclusions}.

\subsection{Generalised probabilistic theories}\label{subsec:GPT}

   The purification postulate and complete extension postulate that we introduce are expressed in the language of generalised probabilistic theories (GPTs), hence, in order to understand these we first provide a brief introduction to this formalism. For a more in-depth introduction see, for example, Refs.~\cite{plavala2021general,muller2021probabilistic,lami2018non} 
 
The formalism of GPTs \cite{hardy2001quantum,barrett2007information} is a formalism for describing the operational predictions of essentially arbitrary conceivable theories of physics. In the case of quantum theory, this operational description is isomorphic to the standard quantum informatic approach to quantum theory (that is, of completely positive trace preserving maps, density matrices, and positive operator value measures). 

The primitive building blocks of any GPT, $\mathcal{G}$ are the \emph{systems}, denoted $A, B,...\in \textsf{Syst}[\mathcal{G}]$, these come equipped with an associative binary composition rule $\otimes:\textsf{Syst}[\mathcal{G}]\times \textsf{Syst}[\mathcal{G}] \to \textsf{Syst}[\mathcal{G}]$ which allow us to build composite systems out of simpler system. In quantum theory these systems can be labeled by natural numbers, i.e., by their dimension, and then these labels compose simply by multiplication, as the dimension of a tensor product of two Hilbert spaces is the product of the individual dimensions.
%, where the natural number represents the dimension of the quantum system.

Each system $A$ corresponds to a finite dimensional real vector space $V_A$. The fact that it is finite can either be viewed as simply a choice for technical convenience, or can be motivated operationally as the idea that in practice we can only ever perform a finite number of measurements to do tomography on a system (even if in principle there are an infinite number of degrees of freedom). In quantum theory these real vector spaces are the spaces of Hermitian operators on some finite dimensional Hilbert space, whilst in classical theory these real vector spaces are the function spaces $\mathds{R}^\Lambda$ for some finite sample space $\Lambda$. It is important to note at this point that $V_{A\otimes B}$ is not necessarily equal to $V_A\otimes V_B$. These are equal only under the assumption of tomographic locality \cite{hardy2001quantum} -- that every multipartite state can be characterised by means of local measurements on its parts. Note that both quantum and classical theory are  tomographically local theories.

Within these vector spaces $V_A$ there is a specified convex set $\Omega_A \subset V_A$ which describes the \emph{state space} for the system. This convex set must be compact and closed, and, moreover, must have an affine dimension one less than the linear dimension of the vector space such that the affine span of $\Omega_A$ does not intersect the origin. We can also define the convex state cone $K_A$ which includes also subnormalised and supernormalised states as $K_A:=\{r s | r\in \mathds{R}^+, s\in \Omega_A\}$. In quantum theory this state space corresponds to the space of density matrices and the state cone to positive semidefinite operators.

Within the dual vector space $V_A^*$ -- that is, the vector space of linear functionals on $V_A$ -- there is also a specified convex set $\mathcal{E}_A \subset V_A^*$ which describes the \emph{effect space} for the system. This convex set must also be compact and closed, however, it will be full dimensional and contain the origin. The effects $e\in\mathcal{E}_A$ assign probabilities to measurement outcomes when the measurement is performed on an arbitrary state in $\Omega_A$, they therefore must satisfy $e(\Omega_A) \subseteq [0,1]$. The effect space $\mathcal{E}_A$ must also contain the unique unit effect $u_A \in \mathcal{E}_A$ which is defined as $u_A(\Omega_A)=\{1\}$. The uniqueness of this effect is built into the setup of our framework, however, there are modifications to the framework in which this is instead an additional principle which captures the notion of causality \cite{Chiribella2010}. In quantum theory this effect space corresponds to $\mathsf{tr}(\sigma \cdot)$ where $\sigma$ is a POVM element and the unit corresponds to $\sigma=\mathds{1}$. 

Like the systems themselves, these state and effect spaces have an associative composition rule, $\Omega_A\otimes \Omega_B := \Omega_{A\otimes B}$, $\mathcal{E}_A\otimes \mathcal{E}_B:=\mathcal{E_{A\otimes B}}$ and $u_A\otimes u_B := u_{A\otimes B}$. Note that in general $\otimes$ is not related to the tensor product of vector spaces, the symbol is used in analogy to the role of the tensor product in composing quantum systems. This composition must be bilinear, and satisfy $e\otimes f(s\otimes t)= e(s) f(t)$ for all $e\in \mathcal{E}_A$, $f\in \mathcal{E}_B$, $s\in\Omega_A$ and $t\in\Omega_B$. These conditions, in particular, ensure that the unit effect $u_B$ gives a way to uniquely define a kind of ``partial trace'' and hence marginal states -- this in turn ensures that the theory does not permit signalling without sending a physical system \cite{Chiribella2010,coecke2014terminality,kissinger2017equivalence}.

Next we consider transformations between systems. For a pair of systems $A$ and $B$ there is a space of transformations $\mathcal{T}_A^B$ from system $A$ to system $B$. These again form a closed compact and convex set. In this case, there are two relevant associative composition rules. The first, parallel composition, $\mathcal{T}_A^B \otimes \mathcal{T}_C^D := \mathcal{T}_{A\otimes C}^{B\otimes D}$ and the second, sequential composition, $\mathcal{T}_B^C \circ \mathcal{T}_A^B:= \mathcal{T}_A^C$. These are both bilinear and must satisfy the condition that $(T_1 \otimes T_2) \circ (T_3\otimes T_4) = (T_1\circ T_3) \otimes (T_2\circ T_4)$. Note that states (and effects) can be viewed as particular kinds of transformations, namely those that have a trivial input (resp. output) system. Denoting this trivial system by $\star$ and noting that $\star$ is a unit of system composition ($A\otimes \star = A = \star \otimes A$) we can therefore write that $\Omega_A = \mathcal{T}_\star^A$ and $\mathcal{E}_A=\mathcal{T}_A^\star$. Note that for every system, $A$, there is an identity transformation $\mathds{1}_A$ which  for every $T:A\to B$ must satisfy $\mathds{1_B}\circ T = T = T\circ \mathds{1}_A$. One can then observe that what we are defining here is nothing but a monoidal category where the objects are the GPT systems and the morphisms are the GPT transformations. 

It is convenient to work with the convention whereby every GPT contains classical systems, denoted $\Delta_I$, in which measurement outcomes can be encoded, and which can act as control variables in descriptions of experiments. See, for example, \cite{gogioso2017categorical,selbyReconstruction}. In particular, a demolition measurement of system $A$ with outcome set $I$ can then be viewed as a transformation in the set $\mathcal{T}_A^{\Delta_I}$\footnote{Nondemolition measurements would live in the space $\mathcal{T}_A^{A\otimes \Delta_I}$}. In order to see the connection between this and the description of measurements in terms of effects, we first set up some notation for describing classical systems.

Consider a classical system $\Delta_I$ which is a system corresponding to, for example, some outcome degree of freedom on some measurement device where $I$ labels the set of possible outcomes. The vector space $V_{\Delta_I}$ will correspond to the real vector space of real functions from $I\to \mathds{R}$, denoted $\mathds{R}^I$. The state space $\Omega_{\Delta_I}$ is the space of probability distributions over $I$, that is, real functions $p:I\to \mathds{R}$ such that $p(i) \in [0,1]$ and $\sum_i p(i)=1$ for all $i\in I$. Note that geometrically this is a simplex with vertices labelled by the elements of $I$. These vertices correspond to delta function distributions which we denote as $\delta_i$. The effect space $\mathcal{E}_{\Delta_I}$ lives in the dual vector space, however, here we make use of the Riesz representation theorem via the inner product $\sum_{i\in I} f(i) g(i)$ in order to view the effects as living in $\mathds{R}^I$. Using this representation then the effects correspond to functions $e:I\to \mathds{R}$ such that $e(i) \in [0,1]$. Geometrically these form a hypercube which contains the simplex of states. The vertices of the simplex, when interpreted as effects via the Riesz representation will be denoted as $\epsilon_i$ such that $\epsilon_i(\delta_j) = \sum_{k\in I} \delta_i(k)\delta_j(k) = \delta_{ij}$. These systems compose via $\Delta_I\otimes \Delta_J := \Delta_{I\times J}$. Transformations between classical systems correspond to stochastic linear maps.    

Now, note an important property of classical theory, that identity transformations $\mathds{1}_{\Delta_I}$ can be decomposed as $\sum_{i\in I} \delta_i \circ \epsilon_i$. Hence, any measurement $M:A\to\Delta_I$ satisfies $M = \mathds{1}_{\Delta_I} \circ M = \sum_{i\in I}  \delta_i \circ \epsilon_i \circ M$ then, noting that $e_i:=\epsilon_i \circ M \in \mathcal{T}_A^\star = \mathcal{E}_A$ we can rewrite this as $\sum_{i\in I} \delta_i \circ e_i$. One can then, recalling that $\epsilon_i(\delta_j)=\delta_{ij}$, see that it is possible to construct an isomorphism between measurements as transformations to a classical system and measurements as a collection of effects.

\section{The purification postulate}\label{sup:GPTs}

The purification postulate \cite{Chiribella2010} has been widely studied within the literature on generalised probabilistic theories, from the study of thermodynamics \cite{chiribella2015entanglement,chiribella2017microcanonical}, cryptography \cite{chiribella2010probabilistic,sikora2018simple}, and computation \cite{lee2016generalised,lee2016deriving,barnum2018oracles} through to higher-order interference \cite{barnum2017ruling} and reconstructions of quantum theory \cite{Chiribella2011}. One of the reasons for the wide use of this principle is the fact that it is mathematically very powerful. For instance, one immediate corollary is that the state space is transitive \cite{Chiribella2010} and there is a continuum of pure states (see Theorem~\ref{thm:nogo} below).

 Despite the utility of the purification postulate, as we discussed in Sec.~\ref{sec:intro}, there are several criticisms which can be made of it. This motivates the search for alternative postulates such as the complete extension postulate that we propose here. Nonetheless, the purification postulate served as the inspiration behind the complete extension so it is useful to introduce it in some detail here. 

 The purification postulate takes the key ideas behind quantum purification and recasts them in the language of GPTs such that it can be taken as a postulate. We will slowly build up the relevant GPT concepts here before presenting the postulate itself.

\begin{definition}[Pure states]
A state $s$ of a system $A$ is said to be \emph{pure} if and only if it is extremal in $\Omega_A$.
\end{definition}
In the case of quantum theory this coincides with the usual notion of purity, namely that of rank-$1$ density matrices. More generally, pure states are thought of as states of maximal knowledge whilst mixed states can be thought of as describing classical uncertainty of knowledge about which pure state has been prepared. 

We can then define the notion of a purification of a given state: 
\begin{definition}[Purifications of a state]
A purification of a state $\omega_A$ of system $A$
%, where the system $A$ may be arbitrarily complex, 
is a non-signalling extension to system $B$. That is, a state $\varepsilon_{AB}$ of the composite system $A\otimes B$, satisfying
	\begin{enumerate}[label=(\alph*)]
		\item $[\mathds{1}_A\otimes u_{B}](\varepsilon_{AB}) =\omega_A$.
		\label{newproof:tr}
		\item $\varepsilon_{AB}$ is a pure state.
		\label{newproof:vertex}
	\end{enumerate}
	If $s$ is pure by itself, we assume that it is one of its own purifications, by taking $B$ system to be trivial. We denote the set of purifications of a given state $\omega_A$ as $\mathbf{Purif}[\omega_A]$.
	\label{def:purification}
\end{definition}

%For mathematical simplicity of the proof we did not restrict ourselves to the definition of purification which is minimal in any sense.

	According to Definition \ref{def:purification}, a state may possess more than one state that purifies it. That is, $\mathbf{Purif}[s]$ typically is not a singleton set. In quantum mechanics,  for example  $|\psi_{AB}\rangle$ and $|\psi_{AB}\rangle \otimes |\phi_{B^\prime}\rangle$ purify the same state, i.e., $\mathsf{tr}_B(\ketbra{\psi_{AB}}{\psi_{AB}})$. In quantum mechanics, however, purifications are unique up to isometries on the purifying system. This motivates the following:
	
\begin{definition}[Essential uniqueness of purifications]\label{def:EU}
The purifications of a given state $s$ are said to be essentially unique, if and only if the elements in $\mathbf{Purif}[\omega_A]$ with the same purifying system can be related by reversible transformations on the purifying systems. 
\end{definition}

With these ideas in place we are then in the position that we can succinctly define the purification postulate:
\begin{definition}[Purification Postulate]\label{def:purificationpostulate}\label{def:theorywithpurification}
A GPT $\mathcal{G}$ satisfies the \emph{Purification Postulate} if and only if for all systems $A$ and all states $\omega_A\in \Omega_A$, there exists purifications~(Def.~\ref{def:purification}) which are \emph{essentially unique}~(Def.~\ref{def:EU}).
% Essential uniqueness means that, if we are given any two purifications $\sigma^p_1,\sigma^p_2 \in \mathbf{Ext}_P[s]$ with the same extending system $E$, then there exists a reversible transformation $T_{1\to 2}:E\to E$ such that:
% \beq
% \mathds{1}_A\otimes T_{1\to 2}(\sigma^p_1) = \sigma^p_2
% \eeq
\end{definition}

\subsection{No-go theorem for purification in discrete theories}
\label{sec:no-go-discrete} 

In this subsection, we will prove that there cannot exist purifications of arbitrary states in any discrete theory. Later on, we will see that this implies that in such theories the complete extensions that we have introduced will generically not be pure. 
This fact has been proven in case of classical theory and for the theory of non-signalling behaviours in \cite{Chiribella2010,Chiribella2011}. We show below a different, direct argument, which holds in general for all non-signalling, convex discrete theories.
%, including the theory of contextuality \cite{Kochen-Specker}. 
In fact, the result developed here is valid in any convex, non-signalling theory, provided that in each dimension number of different systems' types (state spaces with distinct shapes) is countable. We show that there is no single finite dimensional discrete theory \cite{Pfister-Wehner-phys-theories,CzekajHorodzie2017,PolygonTheories2011}, which vertices can purify all the states from a theory of a smaller dimension.  We recall below the formal definition of a discrete theory: 

\begin{definition}[Discrete theory]\label{def:discretetheory}
 A GPT $\mathcal{G}$ is said to be discrete, if and only if, for each system $T\in \mathsf{Syst}[\mathcal{G}]$ there is a discrete number of pure states, that is, where each state space, $\Omega_T$, is a polytope. 
\end{definition}

The demand for the theory to be convex presumes that the state space  lives inside a vector space, the dimension of which is also the dimension of a theory. These assumptions together with the non-signalling condition allow for the well-definiteness of the partial trace as a linear map $\mathrm{Tr}_B\left(\cdot\right): \Omega_A \otimes \Omega_B \equiv \Omega_{A\otimes B} \mapsto \Omega_A$\footnote{The theory of  composition in GPTs was developed in various places, for example, Refs.~\cite{barnum2009ordered,Chiribella2010,janotta2013generalized}.}, which for normalized states (in a sense of probabilistic outcomes) $\rho_{A}, \sigma_B$ satisfies $\mathrm{Tr}_B\left(\rho_A \otimes \sigma_B\right)=(\mathds{1}_A\otimes u_B)(\rho_A \otimes \sigma_B)=\rho_A$, where $u_B$ is the unit effect on the system B. %Before we provide the proof of no-go theorem we should focus on what do we mean by {\em purification} and discuss some of it's basic properties.  
%We assume all states to be normalized, and also that the set of possible measurements is not trivial.

\begin{lemma}
	In any convex (nontrivial) theory, the cardinality of the set of states is at least of power of the continuum $\mathfrak{c}$. 
	\label{lemma:continuum}
\end{lemma}
\begin{proof}
Let us take $\omega_A \neq \omega_A^\prime$ in $\Omega_A$. Since the theory is convex, any state of the form $p\omega_A + (1-p) \omega_A^\prime$, for $p\in[0,1]$ is still a state in $\Omega_A$. 
The set of states $\left\{p\omega_A + (1-p) \omega_A^\prime\right\}_{p=0}^{p=1}$, forms an interval in $\Omega_A$. Since there is a bijection between any interval and the set of real numbers that has cardinality of continuum, the state space has at least cardinality of continuum.
\end{proof}

We are ready to state the main theorem of this section now.
\begin{theorem}\label{thm:NoGo}\label{thm:nogo}
In any discrete theory (Def.~\ref{def:discretetheory}) there are no (non-trivial) systems that have purifications for all states. Hence, theories with purifications (Def.~\ref{def:theorywithpurification}) cannot be discrete.
\end{theorem}
\begin{proof}
As the theory is a discrete theory, any system, $T$ within the theory 
contains only finite number of $v$, vertices,
and so the vertices can only purify a finite number of states $f$ (as each pure state purifies a single mixed state). On the other hand  according to Lemma   \ref{lemma:continuum}, for all (nontrivial) systems, $T'$, in the theory, the cardinality of the set of states is at least $\mathfrak{c}$. So according to set-theoretic fact \cite{Kuratowski} $f <\aleph_0<\mathfrak{c}$, in the system $T$ there are not enough vertices to purify all states from the system $T'$. As it is true for all systems $T$ in the theory, there exist no 
discrete theory, that can purify all states.
\end{proof}

 Note that the proof of theorem \ref{thm:nogo} does not strictly rely on convexity, but just that the cardinality of the state space is greater than $\aleph_0$. One could therefore replace the assumption of convexity simply with the assumption that the number of mixed states in a considered discrete theory is greater than $\aleph_0$. 
 
As a consequence of Theorem~\ref{thm:nogo}, we have the following Proposition. Note that this result does not rely on the essential uniqueness of purifications, only their existence.
\begin{proposition}
For any theory (with a countable number of system types) where purifications exist for all mixed states, there must be at least one system with a continuum of pure states.
\end{proposition}
\proof
In any non-trivial theory there are state spaces which correspond to convex sets other than the singleton set. Such convex sets necessarily contain a continuum of mixed states, i.e., a continuum of non-extremal states. If purifications exist then each of these mixed states must purify to a distinct pure state in the theory (as marginalisation is described by a mapping from the bipartite states to the local states).

Now suppose, for the sake of contradiction, that there is a countable number of system types, that the state spaces of these systems live in finite dimensional vector spaces (as per the definition of a GPT) , and, that these state spaces all have a discrete number of vertices (i.e., pure states). Then, the theory as a whole has a countable number of pure states, and hence there are not enough pure states to purify the continuum of mixed states -- in contradiction to our initial assumption. 

If we want to hold on to the countable number of system types, the finite dimensional vector spaces, and the existence of purifications, it must therefore be the case that there exists a system with a continuum of pure states.
\endproof

\section{The complete extension postulate} 
\label{sec:ConnectionToPP}

The idea of the complete extension  is that it should be taken as a competitor to the purification postulate and is, prima facie, a mathematically weaker postulate.  Indeed, one cannot derive either transitivity of the state space or the necessity for a continuum of pure states from this postulate  whereas these are  consequences of the purification postulate. Consequently, the complete extension postulate  has the potential to be much more broadly applicable in scope and, hence, any results which can be derived using this postulate will hold in a broader class of theories than those derived using the purification postulate.

In this section we will formalise the definition of the complete extension postulate in the language of GPTs introduced in Sec.~\ref{sup:GPTs}. We will work towards this by first defining ensembles, extensions and the principles of $\textsc{access}$ and $\textsc{generation}$ within this language.

To begin, suppose we have some state, $s$, a vector in the convex set of states $\Omega_A$ of a GPT system $A$. An \emph{ensemble} for $s$, is some set of pairs $\{(p_i,s_i)\}_{i\in I}$ such that $s_i \in \Omega_A$ and $\{p_i\}$ define a probability distribution ($p_i \in \mathds{R}$, $p_i \geq 0$, and $\sum_i p_i =1$), satisfying:
\beq
s=\sum_{i\in I} p_i s_i
\eeq
in other words, it is some convex decomposition of $s$ within the state space. We will denote the set of all possible ensembles for a state $s$ as $\mathbf{Ens}[s]$.

The set of states which appear in some ensemble in $\mathbf{Ens}[s]$ is, will be, a face of the state space $\Omega_A$, we denote this as $\mathbf{Face}[s]$. A state is \emph{pure} if and only if $\mathbf{Face}[s]=\{s\}$ and is said to be an \emph{interior} state if and only if $\mathbf{Face}[s]=\Omega_A$.

We can then define a \emph{pure ensemble} for $s$ as an ensemble $\{(p_i,s_i)\}_{i\in I} \in \mathbf{Ens}[s]$ in which all of the $s_i$ are pure, that is, $\mathbf{Face}[s_i]=\{s_i\}\ \forall i\in I$. Let us denote the set of such pure ensembles as $\mathbf{Ens}_{P}[s] \subseteq \mathbf{Ens}[s]$, this is necessarily non-empty as we can always decompose a point inside a   compact and closed   convex set in terms of the vertices of the convex set.

An \emph{extension} of a state $s\in\Omega_A$ of system $A$ is a state $\sigma\in\Omega_{A\otimes E}$ of a bipartite system $A\otimes E$ for which we have:
\beq
s = [\mathds{1}_A\otimes u_E](\sigma)
\eeq
we will denote the set of extensions, $\sigma$, of a state $s$ as $\mathbf{Ext}[s]$, note that, in general, there will be extensions with many different extending systems $E$. A \emph{purification}, should it exist, is simply some $\sigma\in \mathbf{Ext}[s]$ which is pure, that is, $\mathbf{Purif}[s] \subseteq \mathbf{Ext}[s]$.
%(note, this may be an empty set).

Now we can define what we mean for an extension to satisfy \textsc{generation}. An extension $\sigma^*\in\mathbf{Ext}[s]$ with extending system $E^*$ is said to be \emph{generating} if and only if for every $\sigma\in\mathbf{Ext}[s]$ with arbitrary extending system $E$ there is some transformation $T_\sigma:E^*\to E$ such that:
\beq
\mathds{1}_A\otimes T_\sigma (\sigma^*) = \sigma
\eeq

Next we show an equivalence between the notion of an ensemble and a particular class of extensions.
This particular class of extensions are extensions in which the extending system, $E$, is taken to be a \emph{classical} system, $\Delta_d$ for some $d\in \mathds{N}$. We will denote this class of extensions for a state $s$ as $\mathbf{Ext}_{class}[s]$ and show that there is an isomorphism between $\mathbf{Ext}_{class}[s]$ and $\mathbf{Ens}[s]$.

\begin{proposition}
The space of extensions with classical extending system is isomorphic to the space of ensembles, i.e., $\mathbf{Ext}_{class}[s] \cong \mathbf{Ens}[s]$
\end{proposition}
\proof
Starting with an ensemble:
\beq
\left\{\left(p_i, s_i
\right)\right\}_{i\in I}
\eeq
we can construct an extension:
\beq
\sum_{i\in I}p_i s_i\otimes \delta_i \in \Omega_A\otimes\Delta_I
\eeq
where $\delta_i\in \mathsf{Vert}[\Delta_I]$.

Whilst, on the other hand, if we start with an extension 
\beq
\sigma \in \Omega_A\otimes \Delta_J
\eeq
we can construct an ensemble:
\beq
\left\{\left(
u_A\otimes \epsilon_j (\sigma)
\ \ ,\ \
\frac{1}{u_A\otimes \epsilon_j (\sigma)}
\mathds{1}_A\otimes \epsilon_j (\sigma)
\right)\right\}_{j\in J}
\eeq
It is then simple to verify that these two constructions are the inverse of one another and hence establishes an isomorphism between the two sets.
\endproof

Using this isomorphism we can easily express the property of \textsc{access} in a way which makes it manifestly a special case of \textsc{generation}. That is, an extension $\sigma^*\in \mathbf{Ext}[s]$ with extending system $E^*$ is an \emph{extension with access} if and only if, for any ensemble $\sigma \in \mathbf{Ext}_{class}[s]$ with extending system $\Delta_I$ we can find a measurement $M_\sigma:E^*\to \Delta_I$ such that:
\beq
\mathds{1}_A\otimes M_\sigma (\sigma^*) = \sigma
\eeq

Then it is clear that:
\begin{proposition}
If an extension $\sigma$ has the property of \textsc{generation} then it necessarily has the property of \textsc{access}.
\label{thm:gen_implies_access}
\end{proposition}
\begin{proof}
This immediately follows from inspection of the two definitions, and noting that $\mathbf{Ens}[s] \cong \mathbf{Ext}_{class}[s] \subseteq \mathbf{Ext}[s]$.
\end{proof}

The converse however will not always be true, that is, from \textsc{access} we cannot generally derive \textsc{generation}.
\begin{proposition}
If an extension $\sigma$ has the property of \textsc{access} then it does \emph{not} necessarily have the property of \textsc{generation}.
\end{proposition}
\proof
We prove this by providing an example of a GPT which has extensions which allow for \textsc{access} but do not allow for generation. 

Specifically, consider the GPT which has the same states and measurements as quantum theory. Then, the static part of the purification postulate, namely the existence of purifications, is satisfied. Hence, we have that any purification of a state allows us to access the ensembles of that state -- that is they have the property \textsc{access}.

On the other hand, let us assume that the set of possible dynamics is restricted relative to quantum theory, such that  transformations between non-trivial systems are necessarily noisy,  that is,
mixed with some non-zero parameter $\epsilon$ of the totally depolarising channel. Then we fail the dynamical, ``essential uniqueness'' part of the purification postulate and so these extensions will not have the property of \textsc{generation}.
\endproof

We are now in a position such that we can formally define the complete extension postulate within the language of GPTs.
\begin{definition}[Complete Extension Postulate]
A GPT $\mathcal{G}$ satisfies the \emph{Complete Extension Postulate} (CEP) iff: for all systems $A$ and all states $s\in\Omega_A$, there exists an extension $\sigma^* \in \mathbf{Ext}[s]$ which is generating, that is, which has the \textsc{generation} property.
\end{definition}

With this definition in place, we can now compare this to the Purification Postulate \cite{Chiribella2010}, Def.~\ref{def:purificationpostulate}. 
The essential uniqueness property in the purification postulate is intuitively very closely related to the \textsc{generation} property of the complete extension postulate. However, note an important subtlety -- essential uniqueness is a property for extensions on the same extending system, whilst \textsc{generation} is a property for extensions on arbitrary extending systems. In order to derive \textsc{generation} from essential uniqueness we must therefore invoke one additional, relatively innocuous looking, assumption about the GPT. Specifically, that the parallel composition of pure states is necessarily pure.

\begin{proposition}\label{prop:PPtoCEP}
For the set of GPTs in which the product of pure states is pure,  the purification postulate implies the complete extension postulate.
\end{proposition}
\proof
See App.~\ref{proof:PPtoCEP}.
\endproof

 An immediate corollary of this is that quantum theory satisfies the CEP, as it is a GPT satisfying the purification postulate and in which the product of pure states is pure. More specifically,  in the case of quantum theory, a purification $\ket{\psi}$ of a state $\rho$ is a complete extension. In particular, it allows for \textsc{generation} of arbitrary extensions and \textsc{access} of arbitrary ensembles.

It can also easily be shown that classical probability theory satisfies the CEP. That is, the complete extension  of a probability distribution with finite support over some set $X$ has a complete extension given by simply copying the variable $X$. See, for example, Ex. 3.4 in \cite{cho2017disintegration} for details.  
 Specifically, a  general  extension of a probability distribution $p(x)$ is a  simply a  distribution $q(x,y)$ which marginalises to $p(x)$, that is, such that $\sum_{y}q(x,y) = p(x)$. It is then straightforward to verify that the bipartite probability distribution $p_{ce}(x,x'):= p(x) \delta(x,x')$ (obtained by copying $x$) is a complete extension, as any other extension can be generated by applying a stochastic map to $x'$.  This is in stark contrast to the purification postulate, which does not hold in the case of classical theory. In this sense, CEP is a far more natural postulate to consider; indeed, it can be thought of as salvaging the most intuitive part of the purification postulate, which is common to both quantum and classical theory (even though the full purification postulate does not hold in classical theory).   We also show in Sec.~\ref{sec:BoxWorld},  that the complete extension postulate holds in the GPT of non-signalling boxes (colloquially known as Boxworld) which is also known to fail the purification postulate. 

Another proposal to modify the purification postulate was put forward in Ref.~\cite{selbyReconstruction} which provides a time-symmetric version of the purification postulate. It can be shown that this time-symmetric purification postulate, together with another postulate from Ref.~\cite{selbyReconstruction}, namely, the existence of cups \& caps, suffices to derive the CEP. More specifically, if we take a time-symmetric purification of some state and then turn the input via a cup, this defines a complete extension of the state.
Since both quantum and classical theory satisfy all of the postulates of Ref.~\cite{selbyReconstruction}, it therefore immediately follows that they both    have complete extensions. Moreover, superselected quantum systems also satsify the time-symmetric purification postulate and have cups \& caps, hence they also satisfy CEP.

 We now consider two results which have been proven using the purification postulate, and consider whether or not they can be reproven using the complete extension postulate instead. 

\subsection{Impossibility of bit-commitment}\label{sec:ProofNBC}

The task of \emph{bit-commitment} is a two-party cryptographic task that can be used as a primitive in building up other important cryptographic protocols such as coin-flipping~\cite{Blu81} and zero-knowledge proofs~\cite{GMR89}. It is well known that bit-commitment and its generalisation to \emph{integer-commitment} is impossible in both quantum and classical theory \cite{May97, LC97a}. However, this is not true in all GPTs, indeed in \cite{barnum2008nonclassicality} they show that any GPT which is non-classical but does not have entanglement allows for bit-commitment. It is therefore interesting to study what are the features of quantum theory which make this task impossible.

There have been various works on this subject including \cite[Corollary 45]{Chiribella2010} and \cite{sikora2018simple}. The former demonstrates the impossibility of bit-commitment under a set of physical postulates including the purification postulate. The latter goes beyond a strict impossibility proof and provides an analytic lower bound on the product of the two relevant cheating probabilities, again, however, relying on the purification postulate in the process. In this section we will show that we can derive exactly the same analytic lower bound but using the complete extension rather than the purification postulate.
%For the full formalisation of all of this within the framework of GPTs we direct readers to \cite{sikora2018simple}, here we will simply state the result and show how  the proof in \cite{sikora2018simple} must be modified to obtain the following result:

 Before getting to this result, however, let us first introduce the protocol.
 For honest parties this task  consists of two phases: i) the commit phase, in which Alice  chooses an integer $j\in \{1,...,n\}$  via a  uniform random distribution and ``commits'' it to Bob by passing him a ``token''. After doing so it should be impossible for Alice to change the value $j$ but it should also be impossible for Bob to learn the value $j$. ii) the reveal phase, in which Alice communicates the integer $j$ to Bob and sends him a second token to verify the integer. Bob must be able to check that the integer that Alice revealed was indeed the integer that she committed to in the first phase.  Succinctly, for a protocol to be secure, it must be the case that Bob cannot learn anything about the integer $j$ prior to the reveal phase and that Alice cannot change $j$ after the commit phase.  
The extent to which they can deviate from this ideal is characterised by the cheating probabilities $p_B^*$ and $p_A^*$, respectively. 

 More formally,  the protocol can be described as follows: (i) In the commit phase,  Alice chooses an integer $j\in\{1,...,n\}$ via a uniform random distribution and creates a corresponding bipartite state $s^j \in \Omega_{A\otimes B}$ of the bipartite system $A\otimes B$. This is her ``commitment'' to the integer $j$. She then sends the system $B$ to Bob to complete this phase. The idea behind this being that is now having access to only one part of the state will prevent Alice from being able to change her commitment, whilst at the same time, the fact that Bob only has access to one part of the state will prevent him from being able to learn the integer value. (ii) In the reveal phase, Alice communicates the integer $j$ to Bob and sends him the system $A$, Bob applies a two-outcome measurement $(e^j_{accept},e^j_{reject})$ to the bipartite system $AB$ to check that it is in fact in state $s^j$ to verify that the integer that Alice reveals is indeed the integer that she originally committed to.

The cheating probabilities can then be described as follows. $p_B^*$ is the maximum probability with which Bob can learn the value $j$ prior to the reveal phase, that is, by performing some measurement on the reduced states on system $B$ alone. In the case of quantum theory, this can be computed as a semi-definite program, whilst, as shown in \cite{sikora2018simple}, in the case of generic GPTs, they are computed via cone programs (a natural generalisation of SDPs). $p_A^*$ is the maximum probability with which Alice can reveal an integer other than the value $j$ that she committed to in the commit phase. There are many ways in which Alice can attempt to cheat but in order to find a lower bound on $p_A^*$ we focus on the following particular cheating strategy:  she applies some transformation to her part of the shared state between the two phases of the task thereby transforming $s_j$ into some other state $s'_{j'}$. Alice's success probability with this strategy is then the probability which Bob's measurement $(e^{j'}_{accept},e^{j'}_{reject})$ will give the accept outcome on state~$s'_{j'}$.   
A perfect protocol would be one such that $p_B^*=p_A^*=0$, that is, in which there is no probability that Bob can learn anything about $j$ prior to the reveal phase, and there is no probability that Alice can influence the value after the commit phase. In the case that the GPT satisfies the complete extension postulate and an additional assumption called the no-restriction hypothesis \cite{Chiribella2010}, then it can be shown that such a perfect protocol is impossible. The following theorem encapsulates the result:

\begin{theorem}
In any GPT satisfying the complete extension postulate and the no-restriction hypothesis \cite{Chiribella2010}, and in any integer-commitment protocol within that GPT, Alice and Bob's cheating probabilities satisfy

\beq
p_A^*\cdot p_B^* \geq \frac{\alpha}{n}>\frac{1}{2n}
\eeq
\end{theorem}
\proof[Proof. (Adapted from \cite{sikora2018simple})]
The proof is identical up to the paragraph containing eq.~14. At which point we must rewrite the proof as follows:

Since $\rho^j$ is the marginal of $s^j$, we call $s^j$ a \emph{extension}
let $t^j$ be an extension of $r^j$. Then
\begin{equation} \label{chi}
\chi^j := \frac{1}{u_\mathcal{B}[x] \cdot n} (s^j + t^j) \in K_{\mathcal{A} \otimes \mathcal{B}}
\end{equation}
is an extension of $x'$ for each $j$.

Now, consider a complete extension of $x'$, denoted $\tilde{\chi} \in K_{B \otimes E^*}$. Then, the \textsc{generation} property implies that, for each $j$ there exists $T_j:E^*\to A$ such that $\mathds{1}_A \otimes T_j (\tilde{\chi})=\chi^j$.

We now have the following cheating strategy for Alice: Alice prepares the state $\tilde{\chi}$ and passes system $B$ to Bob keeping hold of the system $E^*$.
Then to reveal $j$, she `steers' the state $\tilde{\chi}$ to $\chi^j$ using $T_j$ 
before sending $j$ and the system $A$ to Bob.

The remainder of the proof is identical relying on the use of the tool from convex-optimisation known as \emph{cone programming} \cite{BV}. This is the natural generalisation of the notion of semi-definite programs, which are ubiquitous in quantum information theory, to the setting of GPTs. To date there have only been a handful of papers which utilise this tool either in quantum theory \cite{GSU13, BCJRWY14, LP15, NST16, SW17} or in GPTs \cite{fiorini2014generalized, JP17,bae2016structure,TensorProduct,sikora2018simple,sikora2019impossibility,selby2018make}. 
\endproof

 In the above, the no-restriction hypothesis \cite{Chiribella2010} means that any logically possible measurement  compatible with the state space of theory is physically realisable within the GPT. This is satisfied by both quantum and classical theory and is well motivated in adversarial scenarios when one does not want to make assumptions on the capabilities of the adversaries.  
Formally this means that, for every system $A$ in the GPT, that the effect space $\mathcal{E}_A$ is the dual of the state space $\Omega$. That is:
\beq
\mathcal{E}_A = \Omega_A^* := \{ e \in V_A^* | e(s) \in [0,1] \forall s\in \Omega_A \subset V_A\}
\eeq
That is, any convex-linear functional which gives valid probabilities for all states corresponds to a physically realisable effect. 

\subsection{No no-go for hyperdecoherence}\label{sec:Hyper}

\begin{figure}[t]
\begin{center}
\includegraphics[scale=0.6,angle =0]{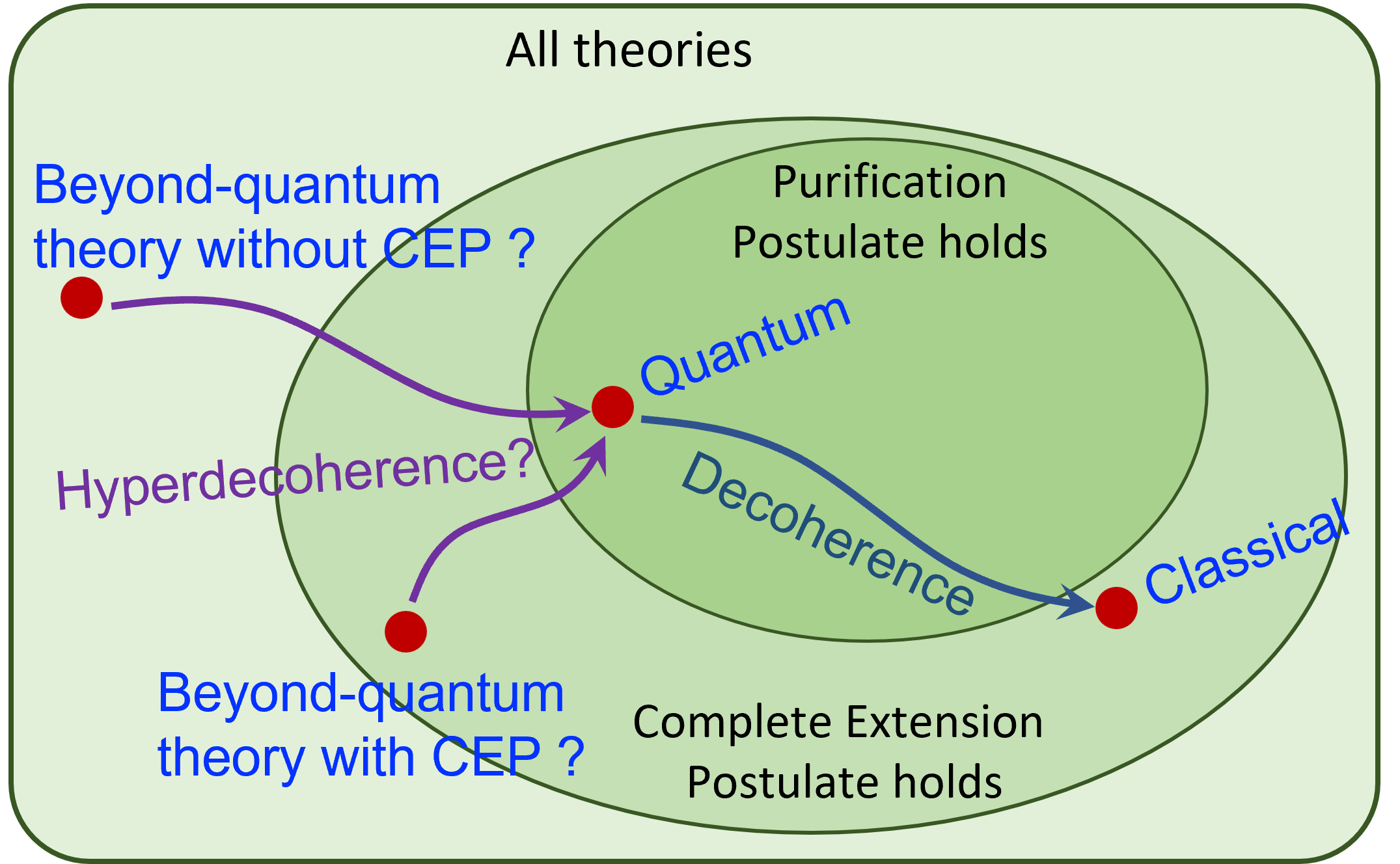}
\caption{Schematic depiction of the relationship between the set of all theories, those are satisfying the complete extension postulate and those are satisfying  the purification postulate. Also shown  is the hyperdecoherence mechanism (purple arrows) relating hypothetical beyond quantum theories and quantum theory in analogy to the relationship between quantum and classical theory by a decoherence mechanism (blue arrow). }
\label{fig:theories}
\end{center}
\end{figure}

The question that any beyond-quantum theory must answer is the following: why do we not observe beyond-quantum phenomena in all of our current experimental tests of quantum theory? That is, within the beyond-quantum theory there must be some mechanism which explains the emergence of the quantum world. We can gain intuition for this by considering the beyond-classical theory that is quantum theory. Within quantum theory, the mechanism that best explains the emergence of the classical world is decoherence. 

It has recently been understood \cite{coecke2017two,selby2017leaks} how decoherence is not simply a way in which quantum states can be made effectively classical (as diagonal density matrices are isomorphic to classical probability distributions) but that this can be lifted to a mechanism which shows how the entirety of quantum theory (i.e., including states, measurements, transformations, and so on) can be made effectively classical (i.e., the theory of classical stochastic dynamics). In \cite{lee2018no} this idea was generalised to define the notion of hyperdecoherence (a term coined in \cite{zyczkowski2008quartic}) which similarly describes how the entirety of a beyond-quantum theory can be made effectively quantum. 

The basic idea, is that for every beyond-quantum system, $A$, there exists some  hyperdecoherence process,
\beq \mathbf{H}_A:A\to A,
\eeq 
which will cause the system $A$ to behave essentially as quantum systems. These hyperdecoherence processes must satisfy  three  basic properties: 
\begin{enumerate} \item they must be unit-effect preserving, 
\beq u_A \circ \mathbf{H}_A = u_A,
\eeq
which is analogous to the trace-preservation condition for quantum decoherence processes; and \item that they must be idempotent, \beq\mathbf{H}_A\circ \mathbf{H}_A = \mathbf{H}_A, \eeq which means that once the beyond-quantum features have been lost, that hyperdecohering again does nothing. 
\item Moreover, they must be chosen compositionally, that is, such that \beq \mathbf{H}_{A\otimes B}=\mathbf{H}_A\otimes \mathbf{H}_B, \eeq
such that if two systems hyperdecohere independently then the global system will behave as a quantum system.
\end{enumerate}

We then can model the hyperdecoherence of a beyond-quantum theory by replacing the identity processes $\mathds{1}_A$ with hyperdecoherence processes $\mathbf{H}_A$, the intuition being that we are trying to describe a situation in which the hyperdecoherence happens on a time scale which is much shorter than any we can currently experimentally probe such that when we ``do nothing'' we are actually letting the system hyperdecohere. The conditions that we have imposed on the hyperdecoherence processes ensure that this can be described in  consistent manner and that the result of this is a valid physical theory. 

In order for such a procedure to describe the emergence of the quantum world from some beyond-quantum theory, it must be the case that the physical theory that we obtain by replacing the identities by hyperdecoherence processes in the beyond-quantum theory, is (isomorphic to) quantum theory. In \cite{lee2018no}, they demonstrate that any beyond-quantum theory which can hyperdecohere to quantum theory in this way, must violate either the causality principle -- that information can only flow forwards in time, or the purification postulate.  Letting go of purification seems like the more palatable option, and so this motivates the question of whether or not the complete extension postulate could be satisfied by a beyond-quantum theory which hyperdecoheres to quantum theory. 

By examination of the proof of \cite{lee2018no}, however, it becomes clear that one cannot obtain the same result using the complete extension postulate  -- at least, not with the same proof technique. One may have suspected that the proof would still go through as 
%the proof of the main theorem in \cite{lee2018no}, 
it is clear that at no point does it crucially rely on the fact that a purification is actually pure. Instead, what is needed is the fact, which can be derived from the purification postulate, that pure extensions are generating. Or, in other words, that the purification postulate tells us that pure extensions are necessarily complete extensions.  This fact, however, does not follow from the complete extension postulate alone, and so the proof does not go through. 
It therefore remains a possibility that there is a beyond-quantum theory satisfying the CEP, which moreover satisfies the causality principle and can hyperdecohere to quantum theory. An important direction for future research is therefore to try to formulate such a beyond-quantum theory. 

\section{Case study - theory of non-signalling behaviours}\label{sec:BoxWorld} 

As a case study, in this section, we focus on
the theory of non-signalling behaviours \cite{PR}. It exhibits stronger correlations between its subsystem than quantum theory,  for example,  it violates the Information Causality principle \cite{Info-Causal} and reaches the algebraic maximum for the CHSH inequality.   
It is also proved that, such behaviours can not be used for teleportation, and it does not allow for swapping of non-separable correlations \cite{ Barrett, no_ent_swap}  between two parties which were initially completely uncorrelated.
 This is not surprising as we already know that the theory of non-signalling behaviours violates the purification postulate (as it is a discrete theory), whilst in  case of quantum theory,  the ability to teleport an unknown state or entanglement swapping directly related to the existence of purifications \cite{teleportation, Zukowski1993}. 

In this section, we consider the lack of existence of purifications for the theory of non-signalling behaviours
 and try to bypass it with other entity fitted for the theory of non-signalling behaviours, namely, the complete extension that we introduced in Sec.~\ref{sec:ConnectionToPP}.
For a given behaviour $P_A$, we want to construct a non-signalling extension $P_{AE}$,  which, whilst not being a purification, still satisfies the two key properties of purifications  that we identified,  namely: 
\begin{enumerate}
\item[] \hspace{-1cm} \textsc{access}: The extension gives access to all ensembles of the extended system. 
\end{enumerate}
That is one can generate an arbitrary ensemble $\{(p_i, P_A^i)\}$ of $P_A$, (such that $\sum_i p_i P_A^i = P_A$ and $\sum_i p_i = 1$) from the extension $P_{AE}$.
\begin{enumerate}
\item[] \hspace{-1cm} \textsc{generation}: The 
extension can be transformed to any other extension of $P_A$. 
\label{qprop:exten}
\end{enumerate} 
That is, upon applying pre(post)-processing of inputs(outputs) (a classical channel) $\mathscr{P}_E$ on system $E$ of $P_{AE}$,  it should be transformed to an arbitrary extension $P'_{AE} = \mathscr{P}_E(P_{AE})$, e.g., $\text{tr}_E (P'_{AE}) = P_A$.

\begin{figure}[h]
\begin{center}
\includegraphics[scale=0.55,angle =0]{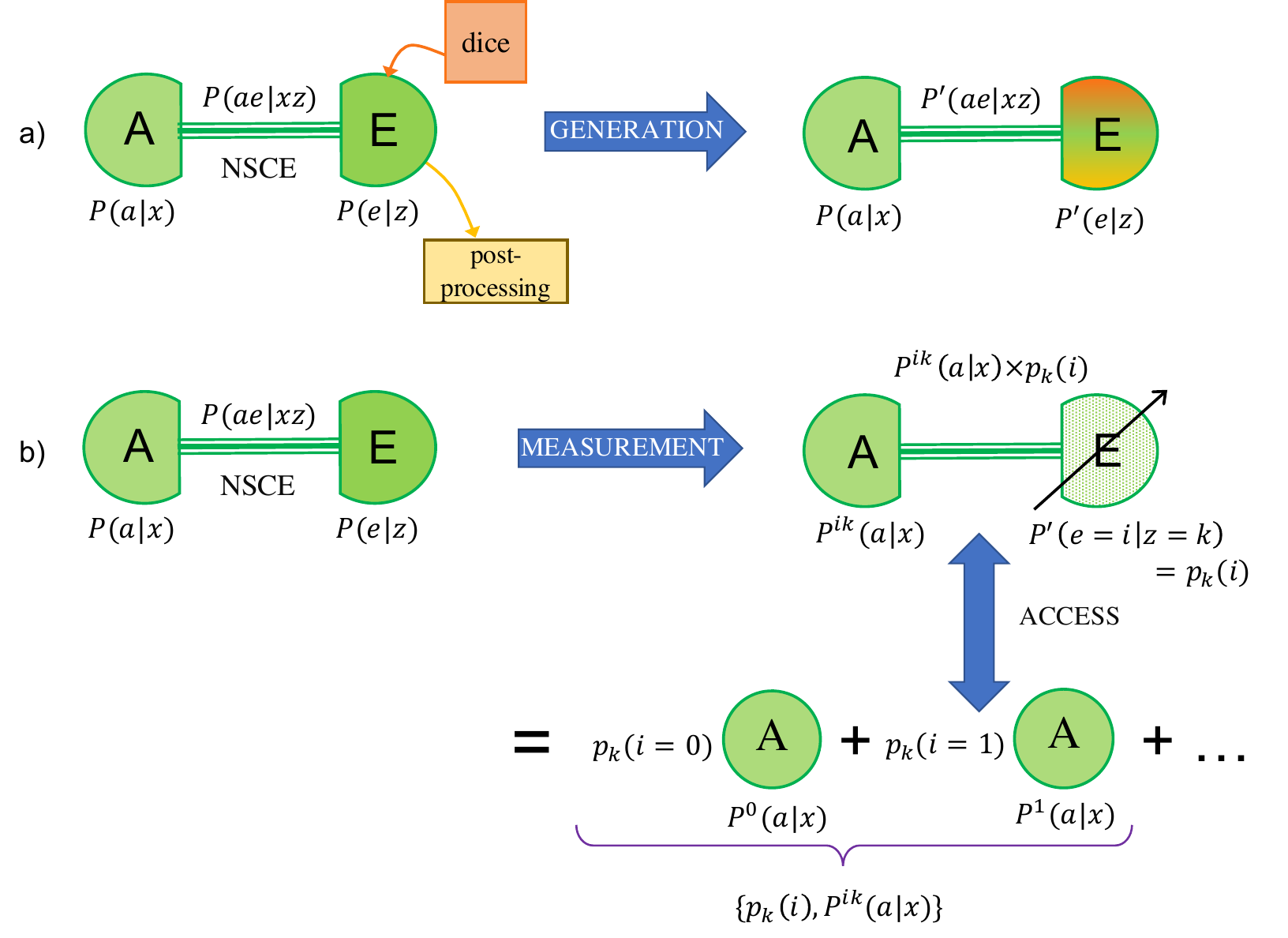}
\end{center}
\caption{Schematic diagram. 
 In any discrete convex theory, generic states  do not have an extension which is a vertex (aka purification). This is different to the quantum theory that is not discrete.
 However, in non-signalling theory, the Complete Extension postulate is satisfied. 
 a) From the Non-Signalling Complete Extension (NSCE)
  of a system, $A$ one can generate any other extension by means of the input randomizer and output post-processing.
 b) The extending system $E$ of the NSCE provides
 access to any ensemble $\{p_k(i),P^{ik}(a|x)\}$ of the system $A$,
 which can be generated upon measurement $z=k$ performed on the system $E$.
}\label{fig:NSCE_schematic}
\end{figure}

In the following, we show that this is indeed possible, and, hence,  the theory of non-signalling behaviours satisfies the Complete Extension Postulate. In this way we show that CEP is not an empty postulate and complete extensions can be actually constructed in certain theories. Moreover, we explore the properties of such extensions.  In particular, we show that there are  infinitely many ensembles of each behaviour, therefore each behaviour can be extended to infinite number of different extensions. Each ensemble corresponds then to certain choice of input in the extending system. 
As a consequence, these extensions might possess an arbitrarily large number of inputs. But in the theory of non-signalling behaviours, one can have behaviours lying only in a finite-dimensional polytope. This shows that some of the ensembles (inputs) are redundant, in the sense that they can be obtained from others via adequate operations of pre(post)-processing of inputs(outputs), and others can be equivalent under reversible operations, e.g., relabelling of inputs and outputs (see Fig. \ref{fig:NSCE_schematic}). In this section, we will show that corresponding to each behaviour, there exist many finite-dimensional non-signalling extensions with the property of \textsc{access} and \textsc{generation}. In fact, what we show, the above properties are equivalent, access to all possible ensembles is equivalent to generation of arbitrary extensions.

We also show that there is always a finite-dimensional complete extension, usually of a large dimension. We give an upper bound on this dimension as a function of the dimension of an extended system and the number of inputs and outputs.
Among these possible complete extensions, we give a prescription of how to construct (up to relabeling) the  minimal  one,  which we call the non-signalling extension with access (NSEA). This 
 lies in the lowest dimensional polytope among all possible extensions with the above properties.
The fact that access to
all possible ensembles of a non-signalling system, has been considered operationally {for the worst} case extension someone might have, in context of device
independent cryptography \cite{acin-2007-98, hanggi-2009} against non-signalling adversary by the Authors of Ref. \cite{Kent, Masanes, hanggi-2009, Hanggi-phd}, and more recently in private randomness \cite{Colbeck_randomness, Gallego_randomness, Mironowicz_randomness, Brandao_randomness, Ramanathan_randomness, Wojewodka_randomness} .
The NSEA which we define is the structure responsible for this fact: 
access to this special extension, gives the non-signalling eavesdropper an ultimate operational power. 
That is, the extending system of NSEA naturally represents the minimal memory of a non-signalling adversary  with maximal power in cryptography based on no-faster than light communication \cite{WDH}.

Interestingly,  whilst we know that not all NSEAs can be pure, we show that  some of them are. In particular, the aforementioned NSEA of
 a maximally mixed binary input binary output behaviour, i.e., the PR box \cite{PR} is a pure non-signalling behaviour. Observing that a PR box is an extension with access for a maximally mixed system can be viewed as a {\it derivation} of the PR box without referring to the notion of a Bell inequality -- this is in stark contrast to the original approach of Popescu and Rohrlich \cite{PR}. 
 The sense in which we derive the PR box is the following. We assume only a) the structure of single-partite systems, and b) that NSCE of any single-partite system is a valid state of the theory. In particular, we do \emph{not} presuppose a) that all non-signalling behaviours belong to the given theory, or b) a particular composition rule, or c) the no-restriction hypothesis. This means that if the PR box was not the NSCE of any single-partite behaviour, then it would be ``excluded'' from the state space built with the above prescription -- i.e., apriori, we do not know that the PR box is an allowed physical  behaviour. What we the show is that the PR box is the complete extension of the maximally mixed single-partite behaviour, and hence, by our assumptions, it is a valid  bipartite state.

\begin{figure}[t]
\begin{center}
\includegraphics[scale=0.65,angle =0]{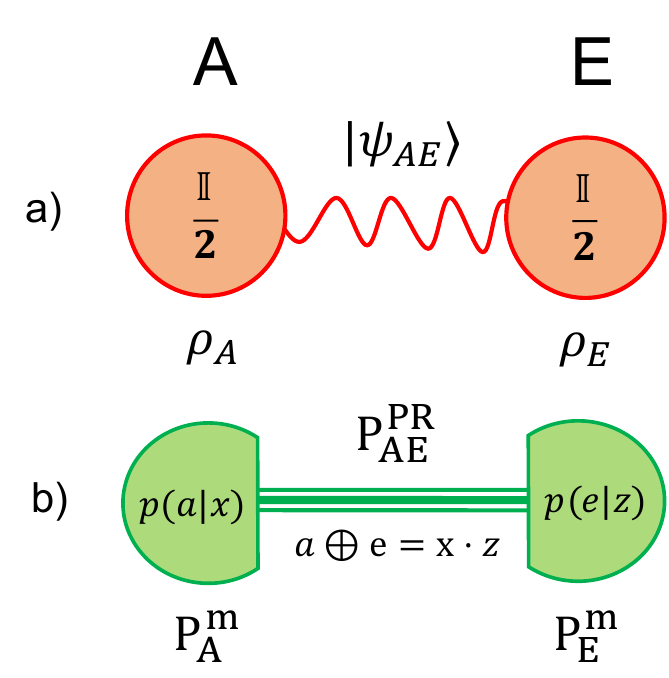}
\end{center}
\caption{Schematic diagram of purifications of system $A$ to an extended system $E$. a) Purification of any quantum state $\rho_A$, to $|\psi_{AE}\rangle$ in the extended state space. If the quantum state is a single qubit maximally mixed state $\frac{\mathbb{I}}{2}$, then its purification is maximally entangled Bell state.  b) Complete extension of a maximally mixed behaviour $P^m_{A}$, (given in Eq. (\ref{eqn:maximally_mixed})) is a pure behaviour - the well known PR box, which is however generically not the case.
}
\label{fig:schematic_quantum}
\end{figure}

Since an NSEA $P_{AE}$ is generically not extreme in the set of behaviours, it  can be non-trivially extended to a new system $E'$ giving a new NSEA $P_{AEE'}$. Note that an interesting corollary of  Theorem   \ref{thm:NoGo} is that for particular behaviours this process can be iterated indefinitely. Thus, identifying or constructing such behaviours and understanding their properties is an important direction for future research. 

From our proof technique (heavily based on convex
geometry) we expect that the complete extensions 
exemplified using the theory of non-signalling behaviours can also be defined in a
similar way in any discrete convex theory. Proving this, however,
needs further work.

In Sec.~\ref{sec:CE-def}, we provide the basic definition of the {\it Non-signalling extension with access} (NSEA)  and prove that it is a complete extension as it satisfies \textsc{access} and \textsc{generation}, and moreover, in Sec.~\ref{sec:limitedDim} that it is a minimal complete extension. 
In App.~\ref{app:Examples} we construct many explicit examples and study their properties. In particular in App.~\ref{sec:example-PR}, we prove that the NSEA of the maximally mixed single party binary input and binary output behaviour is the  Popescu-Rohrlich behaviour (See. Fig.~\ref{fig:schematic_quantum}). We also give an explicit example showing that the non-signalling complete extension has the property of \textsc{access} and \textsc{generation}.
In App. \ref{sec:Bell-Tsirelsen}, we find the NSCE for two party binary input and binary output behaviour in the isotropic line, which enclose the Bell-Tsirelsen {box} and the NSCE for Specker's triangle has been found in Sec. \ref{sec:spekkens}.

%\subsection{The non-signalling extension with access as  a   non-signalling complete extension}
\subsection{The complete extension in the theory of non-signalling behavious}
\label{sec:CE-def}

In this section, we provide a definition of the {\it non-signalling extension with access} (NSEA) in the theory of non-signalling behaviours (NS),  and show that NSEA satisfies the  property \textsc{access}. 
We then show that NSEA also satisfies the \textsc{generation} property, and,  moreover,  that in this theory these two properties are  actually equivalent. This will allow us to refer to NSEA
as  a  non-signalling complete extension (NSCE),  and demonstrates that NS satisfies the complete extension postulate (CEP).  What is interesting here is that it provides a constructive proof that NS satisfies the CEP, which proves useful for introducing explicitly the state of the system of the eavesdropper in the non-signalling device independent secure key agreement protocol \cite{WDH} .

Let us first fix the notation. 
We call a conditional probability distribution, $P_A = p_{\cal A|X}(a|x)$  a box  or behaviour , representing the state of system $A$, where ${\cal X}$ stands for all possible measurement choices (inputs) and the index $x \in \pazocal{X}$ is a particular  choice.  ${\cal A}_x$ 
is the set of measurement outcomes (outputs), corresponding to input $x$  and the index $a \equiv a_x \in {\cal A}_x$ represents one particular instance.  Note that if all of the outputs have the same outcome set then we will drop the subscript $x$ and write $\cal A$ instead.   As $P_A$ is a  conditional  probability distribution it must satisfy the positivity conditions, $0 \leq p_{\cal A|X}(a|x) \leq 1, ~\forall a\in {\cal A}_x,~ x\in {\cal X}$, and  the normalization condition $\sum_{a \in {\cal A}_x} p_{\cal A|X}(a|x) = 1, ~\forall x \in {\cal X}$\footnote{The summation over the output here is an analogue of partial trace.}.
 Let $P_{ AE}= p_{\cal AE|XZ}(ae|xz)$  be a bipartite conditional probability distribution,   where the system $E$ has associated input and output sets ${\cal Z}$ and ${\cal E}$ respectively. 
 We say that it is a non-signalling
extension of the behaviour $P_A$ 
if $P_A$  is its marginal distribution 
\ben
\sum_{e \in {\cal E}_z} p_{\cal AE|XZ}(ae|xz) = p_{\cal A|X}(a|x), ~\forall a \in {\cal A}_x,x \in {\cal X} ,z \in {\cal Z},
\een
and it satisfies non-signalling conditions
\begin{eqnarray}
\sum_{e \in {\cal E}_z} p_{\cal AE|XZ}(ae|xz) &=& \sum_{e \in {\cal E}_{z'}} p_{\cal AE|XZ}(ae|xz'), ~\forall a \in {\cal A}_x,x \in {\cal X}, ~ z,z'\in {\cal Z},\\
 \sum_{a\in {\cal A}_x} p_{\cal AE|XZ}(ae|xz) &=& \sum_{a\in {\cal A}_{x'}} p_{\cal AE|XZ}(ae|x'z), ~\forall  e \in {\cal E}_z,z \in {\cal Z}, ~x, x'\in {\cal X},
\end{eqnarray}
For the sake of simplicity we omit the subscript, and we assign $p_{\cal A|X}(a|x) \equiv P_A (a|x)$.
The set of all non-signalling behaviours of system $A$ is denoted by $\Omega_A$, mutatis mutandis for multipartite systems. 
 The $P_A$ can have inner non-signalling structure \cite{PR}, i.e., system $A$ can be compound (e.g., $A = A_1A_2$) itself, where the system $A_1$ can not signal to system $A_2$ and vice versa.  We will encounter this  later when we consider the extension of the binary two input two output behaviour.

All $n$-partite non-signalling behaviours, with the same cardinalities  of inputs and outputs satisfy a set of linear equations and inequalities (constraints). These constraints define a convex, bounded polyhedron (a polytope) that is a subset of $\mathbb{R}^N$  for some $N\in\mathds{N}$,  with a finite number $D$ of vertices. Each point within this polytope represents a different non-signalling behaviour. In  such  a polytope of behaviours,  we will distinguish those which are {\it extremal} (vertices), i.e., the behaviours which can not be expressed as a convex combination of other non-signalling behaviours, distinguishing them by subscript $\texttt{E}$: $P_\texttt{E}$.  

An ensemble $\{(p_i, P^i)\}_i$,  of a behaviour $P$, where $P = \sum_i p_i P^i$, we will denote by $\ce(P)$. In this paper, we will consider only those ensembles which consists of finite number of elements. 
We will also need the notion of the set of {\it members} of an ensemble $\ce(P) = \{(p_i, P^i)\}_i$, this is defined by $V(\ce(P))= V(\{(p_i, P^i)\}_i):= \{ P^i : p_i > 0\}$, and its {\it distribution}, which is $\{p_i\}$.  If \emph{all} the members, $P^i \in V(\ce)$, are extremal \textcolor{black}{(pure)}, $P^{i}_\texttt{E}$,  then we call the ensemble a {\it pure members ensemble} (PME) and denoted as $\cep(P)$. We say that an ensemble $\cep(P)$ is generated on system $A$ by measurement $\cal M$ on system $E$, if upon this measurement on the extending system, the outcome $e = i$ is obtained on $ E$ with probability $p_i$ and conditionally upon it, the state of the system $A$ is described by behaviour $P^i_\texttt{E}$. By $S_P$ we will denote the set of all PME of a behaviour $P$.

Our aim is to obtain the minimal  extension with properties \textsc{access} and \textsc{generation}. 
To begin with, however, as an intermediate step used to simplify further proofs, we will define an extension, satisfying properties \textsc{access} and \textsc{generation} yet of a larger dimension hence called {{\it overcomplete non-signalling extension with access} (ONSEA)}, defined as:
\begin{definition}[Overcomplete non-signalling extension with access - {ONSEA}] Given a behaviour $P_A: P_A(a|x)$, we say that a behaviour $P_{AE}: P_{AE}(ae|xz)$ is 
its  overcomplete non-signalling extension with access \textcolor{black}{(extension to system $E$ with access to system $A$)}, if for any input choice $z = k$ and outcome $e = i$ obtained in the extending part, there holds
\ben\label{eqn:overcomplete_expression}
P_{AE}(a,e=i|x,z=k) = 
P^{i,k}_\texttt{E}(a|x)p(e=i|z=k),
\een
such, that,  for each $k$,  the ensemble $\left\{\left(p(e=i|z=k),P^{i,k}_\texttt{E}(a|x)\right)\right\}_{i}$ is a pure members ensemble of the behaviour $P_{A}$, and corresponding to each pure members ensemble of $P_A$, there is exactly one input $z = k$, in the extending system which
generates it.
\label{def:overcomplete_ext}
\end{definition}

 It is simple to see that the overcomplete extension {ONSEA} of an arbitrary behaviour $P_A$ exists.  Indeed, $P_A$ belongs to a polytope. That is it belongs to a set with a finite number of pure behaviours. Any pure member ensemble  of $P_A$ is a subset of the set of pure behaviours. Hence, there is finite number of the latter ensembles and we can construct extension where for each of such ensemble we have $k$ such that $z=k$ generates it. This extension is precisely the ONSEA.  We will denote it  as $\tilde{\cal E}(P)_{AE}$. Notice that when a particular input $z = k$ is chosen in the extending part, an outcome $e = i$ occurs with the probability $p(e=i|z=k)$, resulting in a pure behaviour $P^{i,k}_\texttt{E}(a|x)$ in part of $A$. 
Eq. (\ref{eqn:overcomplete_expression}), expresses the partially measured behaviour with the probability times the conditional pure behaviour. Moreover, $\left\{\left(p(e=i|z=k),P^{i,k}_\texttt{E}(a|x)\right)\right\}_i$ is a pure members ensemble of $P_A$, for each $k$, as 
\ben
\sum_i p(e=i|z=k)P^{i,k}_\texttt{E}(a|x) = \sum_i  \tilde{\cal E}(P)_{AE}(a,e=i|x,z=k) = P_A(a|x),
\een
and given an overcomplete extension ({ONSEA}) there are  $|\mathcal{Z}|$ 
 pure members ensembles, where $\mathcal{Z} = \{z\}$, is the set of all input choices of the extending system.
 
The above definition of ONSEA satisfies the non-signalling condition for its both subsystems. For system $A$, it is by construction, and for system $E$ it holds due to the fact that for each input output pair $(z=k,e = i)$ of $E$, system $A$ holds a behaviour $P^{i,k}_\texttt{E}(a|x)$ according to Eq. (\ref{eqn:overcomplete_expression}), which gives $1$ when summed over $a$.

 We will see later that the ONSEA is actually a complete extension as it allows for \textsc{access} and \textsc{generation}. However, the dimension of the extending system is larger than necessary, we therefore seek to construct a minimal complete extension\footnote{Minimality will be proven in Prop.~\ref{prop:minimality} in the following subsection.}. To do so we  will now introduce a representative subset of all PMEs.
An ensemble $\cep(P) = \{p_i,P^i_\texttt{E}\}_i$,
of a behaviour $P = \sum_i p_i P^i_E$,  is called {\it minimal ensemble},  if any proper subset  $V' \subset V(\cep)$, with another choices of probabilities $\{p'_j\}$ is not an ensemble of the behaviour $P$, i.e., $P \neq \sum_{j: P^j_\texttt{E} \in V'} p_j' P^j_\texttt{E} $. Any minimal ensemble will be denoted as $\cm(P)$, to distinguish it from  an arbitrary  PME that is not necessarily minimal. 

Next, we will eliminate the redundant ensembles from the {ONSEA}, to obtain the {NSEA} that has the lowest number of inputs sufficient to satisfy \textsc{access} property. To achieve this we will show that having access to all minimal ensembles $\{\cm(P)\}$ of $P$, along with arbitrary randomness, one can generate any PME, $\cep(P)$. 

\begin{theorem} \textsc{access} to PMEs is equivalent to \textsc{access} to minimal ensembles. \label{thm:acesstoPMEequivaccesstoMinEns}
\end{theorem}
\proof See App.~\ref{proof:acesstoPMEequivaccesstoMinEns}. \endproof

\begin{rem}
    This theorem (Thm.~\ref{thm:acesstoPMEequivaccesstoMinEns}) can be generalized to any GPT that is a discrete theory (see Definition \ref{def:discretetheory}).
\end{rem} 

The above  theorem motivates  the following
definition of \emph{non-signalling extension with access} ({NSEA}),  which, like the extension defined in Definition \ref{def:overcomplete_ext}, also satisfies the properties of access and generation:

\begin{definition_SM}[Non-signalling  Extension with Access]\label{def:NSEA_SI}
Given a behaviour $P_A: P_A(a|x)$, we say that a behaviour $P_{AE}: P_{AE}(ae|xz)$ is 
its non-signalling extension with access extended to system $E$ if for any input choice $z = k$ an outcome  $e = i$ occurred in the extending system, there holds
\ben
P_{AE}(a,e=i|x,z=k) = 
P^{i,k}_\texttt{E}(a|x)p(e=i|z=k)
\een
such, that,  for each $k$,  the ensemble $\left\{\left(p(e=i|z=k),P^{i,k}_\texttt{E}(a|x)\right)\right\}_{i}$ is a minimal ensemble of the behaviour $P_{A}$. Moreover, corresponding to each minimal ensemble of $P_A$, there is exactly one input $z = k$, in part of the extending system which generates it.
\label{def:complete_ext}
\end{definition_SM}

\begin{figure}[t]
\begin{center}
\includegraphics[scale=0.6,angle =0]{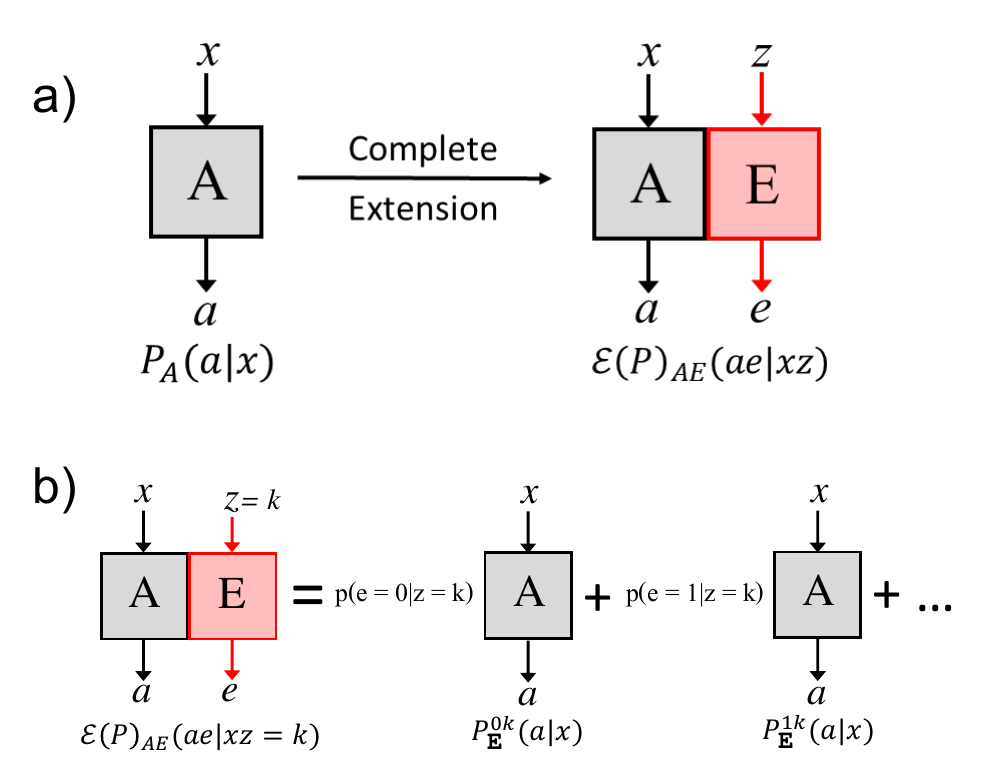}
\end{center}
\caption{Panel $a)$, schematic diagram of a {NSEA} of an arbitrary behaviour $P_{A}$ to ${\cal E}(P)_{AE}$. Panel $b)$, Corresponding to each input $z = k$ in part of the extending system, the composite behaviour partitioned to a minimal ensemble $\{p(e|z =k),P^{ek}_\texttt{E}(a|x)\}$ in part of $A$.
}
\label{fig:schematic_minimal}
\end{figure}

We will represent the {NSEA} of an arbitrary behaviour $P_A$ as ${\cal E}(P)_{AE}$. If there is more than one subsystem in the system $A$ satisfying non-signalling constraints, i.e., if $A\equiv A_1A_2\ldots A_N$, and none of the $k$ systems $\{A_{i_1},\ldots,A_{i_k}\}$ with $I=\{i_1,\ldots,i_k\}$ can signal to the remaining $N-k$ systems $\{A_{j_1,\ldots,j_{N-k}}\}$ with $j_l \in \{1,\ldots,N\}\setminus I$, for all $k \in \{1,2,\ldots, N-1\}$, then also our method of constructing {NSEA} holds, and it will be denoted by ${\cal E}(P)_{ A_1A_2\ldots A_NE}$.
 A schematic diagram of {NSEA} has been depicted in Fig. \ref{fig:schematic_minimal}. For {NSEA} defined above,  the inputs of the extending system correspond to the minimal ensembles of the given behaviour $P_A$. If there are $N$ number of minimal ensembles, then the total number of input choices in the extending part is $|\mathcal{Z}| = N$. 
 
 \begin{proposition}
    For each behavior $P_A$ of system $A$, its NSEA $\mathcal{E}(P)_{AE}$ is unique up to the local relabeling on the extending system $E$.
\end{proposition}
\begin{proof}
Let $P_{AE}$ and $P_{AE}'$ be two different
NSEA of a behaviour $P_A$. First note, that
$P_A$ determines the number of inputs $|E|$.
However the inputs of $P_{AE}$ can be 
labelled in different way than that of $P_{AE}'$. However, due to Definition \ref{def:NSEA_SI}, each input of $P_{AE}$
corresponds to a unique minimal ensemble
of $P_A$. The same holds for $P_{AE}'$. Thus for each input $e$ of $P_{AE}$
there exists input $e'$ of $P_{AE}'$ 
which corresponds to the same unique 
minimal ensemble. Hence for this pair
of inputs the distribution of outputs 
of $P_{AE}$ and $P_{AE}'$ are the 
same up to labeling of outputs. For this
reason, the two extensions differ only
by relabelling of inputs and outputs
of the extending system. 
Thus, relabellings establish an equivalence class over different behaviors being NCEA of behavior $P_A$.
\end{proof}

 Note that one can easily show that such an extension necessarily exists for any behaviour $P_A$. For example,
one  can check (see Section
\ref{sec:example-PR} 
of the Appendix) that the NSEA of the behaviour $P_A$,  the maximally mixed behaviour with a single binary
input and single binary output,   extended to system $E$, has the following structure:
\ben
\mathrm{P_{AE}}\left(ae|xz\right)=
\begin{array}{cc?cc?cc}
 &\multicolumn{1}{c}{x} & \multicolumn{2}{c}{0}	& \multicolumn{2}{c}{1} \\
z &$\diagbox[width=1.8em, height=1.8em, innerrightsep=0pt]{e}{a}$  & 0	& 1	&0  & 1\\
	\Cline{1pt}{2-6} \\ [-1em]
 \multirow{2}{*}{0 }  & 0	& 1/2	& 0				& 1/2		& 0\\ 
 & 1	& 0			& 1/2	&0				 	& 1/2 \\
 \Cline{1pt}{2-6}  \\ [-1em]
\multirow{2}{*}{1 }   &0 & 1/2	& 0				&0 				& 1/2\\
& 1 & 0			& 1/2	& 1/2 	& 0 
\end{array},
\label{eqn:PAX_purified_PR_maintext}
\een
which is nothing but the famous Popescu-Rohrlich ($PR$) behaviour  satisfying conditions $x.z=a\oplus e$ with $\oplus$ being addition modulo $2$. We have thus arrived at this structure without referring to the CHSH inequality \cite{CHSH} (in contrast to the way in which it was done in \cite{PR}).

We are in a position to show, that having access to NSEA we can generate any PME.

{\corollary The non-signalling extension with access (NSEA) of a behaviour P given in Definition \ref{def:complete_ext}, together with access to arbitrary local randomness, gives access to any pure members ensemble of a behaviour $P$. 
\label{th:min_puri+coin}
}\\
\proof See App.~\ref{proof:min_puri+coin}. \endproof

\begin{figure}[t]
\begin{center}
\includegraphics[scale=0.5,angle =0]{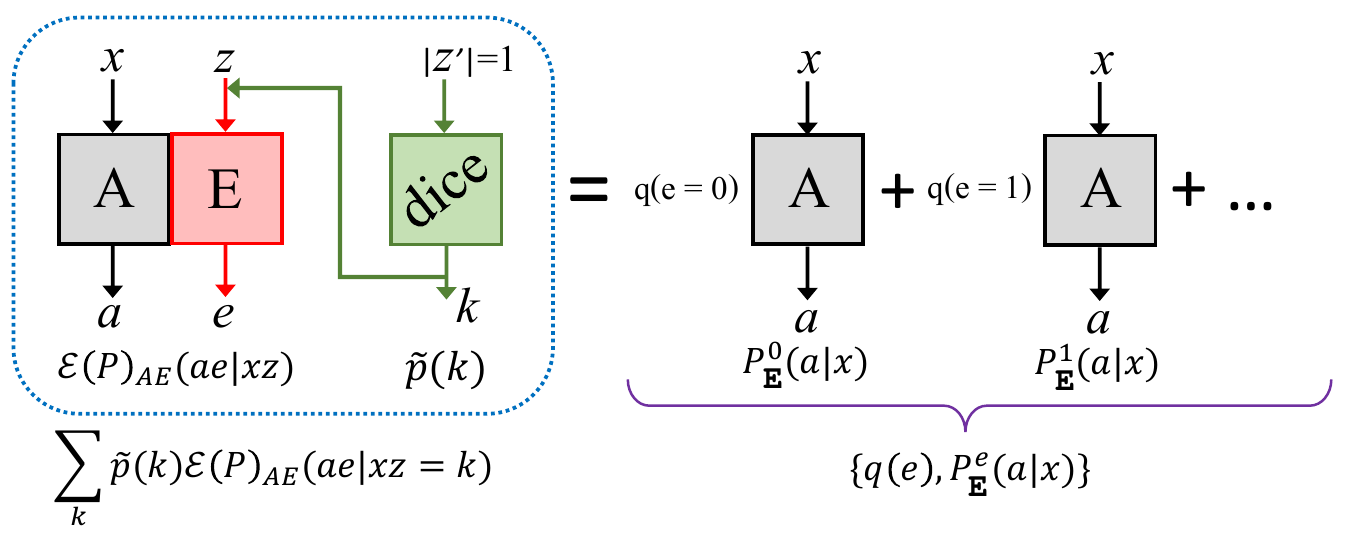}
\end{center}
\caption{Schematic diagram visualizing the  mixing the minimal ensembles $\{{\cal M}_k\}$ of $A$ with arbitrary randomness $p(k)$ in part of the extending system on the {NSEA} of the behaviour, which is obtained from the output of a dice (a local behaviour with unary input), results an arbitrary pure members ensemble $\left\{ p(k), \{(p(e|z=k),P^{ez=k}_\texttt{E}(a|x))\}\right\} = \{q(e), P^e_\texttt{E}(ab|xy) \}$, where $q(e) = \sum_{k} p(k)p(e|z=k)$.
}
\label{fig:schematic_pure_members_ensemble}
\end{figure}

From the above corollary it is clear that an arbitrary PME can be accessed from the NSEA by using a randomness generator $\{p(k)\}$, i.e., by using a dice(coin). 
{To access  all possible PME, one needs an access to  arbitrary} randomness. This can be done by setting the output $(k)$ of a dice with a distribution, $p(k|z')$, where $z'$ is the tuning parameter, as the input of the extending party of {NSEA}. Here $|\{k\}| = |\cal Z|$, and $|\{z'\}|$ will be equal to the possible number of PME one wants to generate.  The dice can be thought of as a  local behaviour with $z'$ being the input and $k$ as the output.  Different choices of the $z'$ can be considered as dices with different probabilities of outcome, actually led to different  PMEs. Accessing all possible PME has been pictorially depicted in Fig. \ref{fig:schematic_pure_members_ensemble}.

An explicit example of constructing an arbitrary pure members ensembles  has been given in Sec \ref{sec:any-PME}. Where we have chosen an arbitrary behaviour containing single binary input and single binary output.

\begin{theorem}  \textsc{access} to PMEs is equivalent to \textsc{access} to all ensembles. \label{thm:PME_mixed}  \end{theorem}
\proof
Let us consider the access of the mixed ensemble $\ce_{mix}(P) = \{(p_m, P^m)\}_m$,  which is the most general ensemble, from the set of all ensembles of a given behavior $P$. Now each $P^m$ lies in the same polytope as $P$, hence, all of them has a pure behaviour decomposition,
\ben
P^m = \sum_i q^m_i P^i_\texttt{E},
\een
where $\sum_i q^m_i = 1,~ \forall m$ and $ 0 \le q^m_i \le 1,~ \forall i,m$. Note that this decomposition is not unique, unless it is a minimal decomposition. Now $P = \sum_m p_m P^m = \sum_{m,i} p_m q^m_i P^i_\texttt{E} = \sum_i r_i P^i_\texttt{E}$, where $r_i = \sum_m p_m q^m_i $, implies that $ \{(r_i, P^i_\texttt{E})\}_i$ is also a PME of $P$. From Theorem \ref{th:min_puri+coin}, we know that by using an appropriate  randomness generator in the input of the extending system $E$ (input randomizer), such PME can be realized. To interpret the mixed ensembles, from $ \{(r_i, P^i_\texttt{E})\}_i$, the extending system send her output ($e = i$, of the NSEA along with an appropriate input randomizer) through a post-processing channel $p_c(m|e)$. 
The existence of this channel (\ref{channel}), $p_c(m|e)$ with $\sum_m p(m|e) = 1$, is guaranteed by the initial decomposition of the members of the mixed ensemble into pure behaviours.
\ben
\sum_i p_c(m|e = i) r_i P^i_E = \sum_i \frac{p_m q^m_i }{ r_i} r_i P^i_E = p_m P^m,
\label{channel}
\een 
where $p_c(m|e)$ take the form $\frac{p_m q^m_e}{r_e}$.
\endproof

\begin{corollary}\label{theorem:mixed} (NSEA  satisfies  \textsc{access}) The extending system of the {NSEA} gives access to any possible (even mixed) ensemble of the extended behaviour. 
\end{corollary}
\begin{proof}
Suppose the extending system wants to access an arbitrary mixed ensemble, once she obtained the NSEA. 
From Corollary \ref{th:min_puri+coin},  NSEA gives access to any PME, and from Theorem \ref{thm:PME_mixed}, it is clear that access to PMEs is equivalent to access to all ensembles. Hence, NSEA gives access to any possible (even mixed) ensemble of the extended behaviour. 
\end{proof}

\begin{figure}[t]
\begin{center}
\includegraphics[scale=0.5,angle =0]{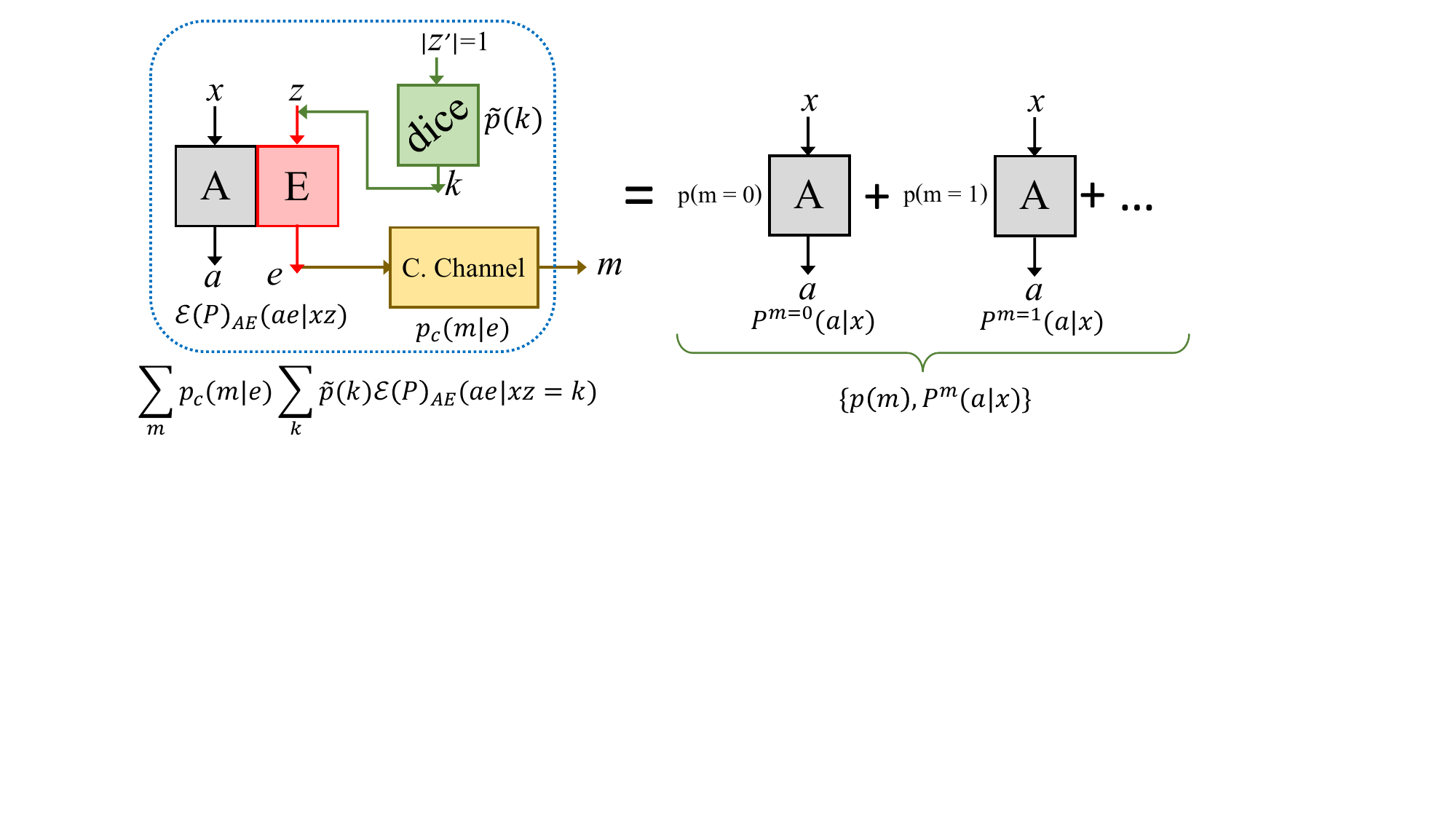}
\end{center}
\caption{Explanation of Theorem \ref{theorem:mixed} of accessing all possible ensembles even mixed in part of the extended system. By passing the output $e$ of the extending party's (Eve) behaviour, through a post-processing channel PC, Eve is able to interpret the behaviour shared by Alice and Bob as an ensemble of mixed bipartite behaviours $\{(p(m), P^m(ab|xy))\}$, where $P^m(ab|xy)$, are mixed behaviours and can be expanded as $P^m(ab|xy) = \sum_e q_e^m P^e_E(ab|xy)$. Now the post-processing channel $p_c(m|e)$, helps Eve to interpret the mixed behaviours as $\sum_e p_c(m|e)r(e)P^e_E(ab|xy) = p(m)P^m(ab|xy)$.
}
\label{fig:schematic_mixed_ensemble}
\end{figure}

From the above theorem, we observe that, corresponding to each mixed ensemble of $P$, there exists at least one PME of $P$, from which one can  obtain  the mixed ensemble by using 
an appropriate  post-processing channel. In this construction  the  NSEA plays an important role due to the fact that Theorem~\ref{th:min_puri+coin} certifies the existence of that particular PME. 
 In Sec. \ref{sec:mixen} of the Appendix, we exemplify Theorem~\ref{theorem:mixed} for a single party binary input binary output behaviour, by providing the  explicit  form of input randomizer and the post-processing channel.
A visualization of Theorem~\ref{theorem:mixed}, has been given in Fig.~\ref{fig:schematic_mixed_ensemble}. Note that the input randomizer $\tilde{p}(k)$, which generates the required PME of P,  is a dice with a unary input (trivial input), $\tilde{p}(k|z' = \text{fix})$. 
One can generate all possible ensembles by  appropriately   tuning the dice with the help of the dice input $z'$ and a post-processing channel (also dependent on $z'$) on part of the extending system.

 We have now seen that we can construct the NSEA which is an extension with \textsc{access}. However, in order for this to be a complete extension we need that it also satisfies \textsc{generation}. One can show that this is indeed the case by demonstrating that \textsc{access} and \textsc{generation} are equivalent conditions in the theory of non-signalling behaviours.  
\begin{theorem} In the theory of non-signalling behaviours \textsc{access} is equivalent to \textsc{generation}. \label{thm:accessequivgeneration}   \label{cor:GiffA}
\end{theorem}
\proof See App.~\ref{proof:accessequivgeneration}. \endproof

 From the above theorem, Theorem \ref{theorem:mixed}, and the fact that NSEAs exist for all behavious, we obtain the main result of this section: 
\begin{corollary}\label{thm:allensem_anyextension}
The Non-signalling Extension with Access (NSEA) has the properties of \textsc{access} and \textsc{generation}, that is, it is a Non-signalling Complete Extension (NSCE).  Hence, the theory of non-signalling behavious satisfies the Complete Extension Postulate (CEP). 
\end{corollary}

% From the results presented in this section we have the following statement. 
% \begin{corollary}
%     In the theory of non-signalling behaviours an extension has the property \textsc{generation} if and only if it has the property \textsc{access}.
%     \label{cor:GiffA}
% \end{corollary}
% \begin{proof}
%     If an extension has property \textsc{generation} one can generate NSEA as it is one of possible extensions. This implies \textsc{access} property by Theorem \ref{theorem:mixed}. Conversely if an extension has property \textsc{access}, then by lemma \ref{thm:cegacae} it has property \textsc{generation}.
% \end{proof}

% Another important corollary of Theorem~\ref{thm:allensem_anyextension} is the following:
% \begin{corollary}
% \john In particular, this means that the theory of non-signalling behaviours (NS) satisfies the complete extension postulate (CEP). \blk
% \end{corollary}
% \proof
%  Theorem~\ref{thm:allensem_anyextension} implies that NSEAs are complete extensions, and, as mentioned earlier, it is simple to see that these necessarily exist for arbitrary behaviours $P_A$.
% \endproof

\subsection{On the dimensionality of the complete extension}\label{sec:limitedDim}

In this subsection we are going to discuss the size of the non-signalling complete extension NSEA, and, what is most important, is that we are going to show that it is finite.
The restriction on the size of the non-signalling complete extension (NSEA) is particularly important from point of view of non-signalling Device Independent  cryptography \cite{Kent,hanggi-2009,DI_WDK}, and interesting on its own. We finish this section with another observation, namely we show that for any behaviour $P$, its NSEA $\mathcal{E}(P)$ has the lowest dimension (is minimal) among all non-signalling extension of $P$ having the property of \textsc{access}.

The following Theorem holds.

\begin{theorem}\label{thm:limitedDim}
     Let be $\mathcal{B}$ be a polytope of $n$-partite non-signalling behaviours, with $m_i$ inputs for parties and $v_{ij}$ outputs respectively. Then for each $P \in \mathcal{B}$, there exists a non-signalling polytope $\tilde{\mathcal{B}} \ni \mathcal{E}(P)$  which the NSEA lives in,  such that:
    \begin{align}
         &\dim \tilde{\mathcal{B}} <\left(\dim \mathcal{B}+1\right) \nonumber\\
         &\times \left(\binom{\binom{2t -\left \lfloor{t/2}\right \rfloor - \dim \mathcal{B} }{\left \lfloor{t/2}\right \rfloor}+ \binom{3t -\left \lfloor{t/2}\right \rfloor - \left(\dim \mathcal{B}+1\right) }{t-\left \lfloor{t/2}\right \rfloor-1}}{\dim \mathrm{B}+1}\dim \mathcal{B} +1\right), \label{eqn:DimUB}
    \end{align}
     where:
     \begin{align}
         &\dim \mathcal{B} = \prod_{i=1}^{n} \left(\sum_{j=1}^{m_i}(v_{ij}-1)+1\right)-1,~~t=\prod_{i=1}^{n} \sum_{j=1}^{m_i} v_{ij}.
     \end{align}
\end{theorem}
\begin{proof}
    The proof is the content of Section \ref{sec:app:dim} of the Appendix. 
\end{proof}

In Theorem \ref{thm:limitedDim} we were interested only in showing that the dimension of the complete extension is always finite and therefore the upper bound is very loose. For instance, the expression in equation \eqref{eqn:DimUB} in the case of $n=1$, $m_i=2$, $v_{ij}=2$  yields the dimension of $339$, for $n=2$, $m_i=2$, $v_{ij}=2$ we have c.a. $1.2 \times 10^{54}$, and for $n=2$, $m_i=3$, $v_{ij}=3$ we obtain c.a. $1.14\times 10^{1762}$. All exemplified results are far above any numerical predictions. The discussion about the possible improvements in the upper bound is left to the Appendix.
% \sout{In particular the number of candidates for a minimal ensemble is very inefficient upper bound on the number of minimal ensembles, although obtained with simple combinatorics.  Therefore, the first place for improvement is to find tighter upper bound on the number of minimal ensembles. A more rigorous treatment can show that many candidates lead to the same minimal ensemble and moreover that some of them are not valid. Furthermore, the upper bound on the number of vertices that we used is very general and does not incorporate, for example, symmetries of the non-signalling polytope. }
Importantly for us, however, is the following simple corollary of the existence of any upper bound, namely: 
\begin{corollary}
    The dimension of the NSCE  can always be  finite.  
\end{corollary}
% \begin{proof}
%     From the proof of Theorem \ref{thm:limitedDim} it is clear that in any sense the non-signalling complete extension can be taken to be a finite mathematical entity. \jnote{This should be clarified or removed}
% \end{proof}

%
A tempting question to ask about NSEA is whether it has the lowest dimension amongst all  NSCEs,    and therefore whether it is the minimal one. Similarly to earlier in this section, we refer to the dimension of the extension as the dimension of the behaviour polytope to which it belongs. The answer to this query is positive and therefore we finish this section with the following proposition.

\begin{proposition}\label{prop:minimality}
Among all non-signalling extensions of a behaviour $P$, having the property of \textsc{access}, NSEA $\mathcal{E}(P)$ is a minimal one.
\label{obs:nsea_is_minimal}
\end{proposition}
\begin{proof}
Let $P$ be a non-signalling behaviour in  $\mathcal{B}$,  and $\mathcal{E}(P) \in \tilde{\mathcal{B}}$ be its NSEA. The dimensions of non-signalling polytopes can be determined with equation (\ref{eq:dim_def}) as before. Suppose now, there exists another non-signalling extension $\hat{\mathcal{E}}(P) \in \hat{\mathcal{B}}$, having a property of \textsc{access}, such that $\dim \hat{\mathcal{B}} < \dim \tilde{\mathcal{B}}$.
The fact that ${\hat{\mathcal{E}}(P)}$ has the property of \textsc{access} implies that upon processing (possibly trivial) of inputs and outputs of the extending system with local randomness, ${\hat{\mathcal{E}}(P)}$ also has access also to all minimal ensembles of $P$. However, the minimal ensembles of $P$ are extremal points in the polytope of all possible ensembles (see Theorem \ref{thm:min-ext}) of the behaviour $P$, and so cannot be created via probabilistic processing of inputs and outputs in the extending system (class of operations considered in \textsc{access}). Therefore, ${\hat{\mathcal{E}}(P)}$ having a property of \textsc{access}, must have for each minimal ensemble of $P$, an input that generates it, like $\mathcal{E}(P)$ does. This implies that $\dim \hat{\mathcal{B}} \ge \dim \tilde{\mathcal{B}}$, and so proves by contradiction that NSEA $\mathcal{E}(P)$ is the minimal extension of $P$ having the property of \textsc{access}.
\end{proof}

\section{Conclusions}\label{sec:conclusions}

To summarize, our main contribution is a new
concept, which is the {\it complete extension}. We show that complete extensions are present
in classical theory,  quantum theory, super-selected quantum theory, the theory of non-signalling behaviours, and, moreover, any theory satisfying the purification postulate in which the product of pure states is pure.  We also postulate that it may  exist hypothetical beyond-quantum theories which could hyperdecohere to quantum theory.  In the case of quantum and classical theory, as well as the theory of non-signalling behaviours, we can explicitly construct these complete extensions. 
This notion implies a number of paths for research,  some of which 
we have exemplified in our case study on  the theory of non-signalling behaviors. 

The idea of the CEP sets a demarcation
line in the set of results obtained on the 
basis of the purification postulate. It divides them into those
that really require all of the purification postulate and those for which the CEP
suffices. We exemplify this by considering the possibility of bit commitment showing that the no-go for it is not specific to theories with the purification postulate.  This has the added benefit of giving a unified proof for the quantum and classical cases. 

The CEP may also be viewed as a razor for excluding theories that can not substitute quantum theory in the future. Indeed, a theory not satisfying CEP may not be physical, as (see Theorem \ref{thm:allensem_anyextension}) even the theory of non-signalling behaviours satisfies it.
The easiest way to obtain theories that do not satisfy CEP is to restrict state space or dynamics so that it is not possible to generate all extensions of some state. An example of a theory that does not satisfy CEP for this reason is given in \cite{barnum2008nonclassicality}. %Although mathematically well defined, it may be hard to find such a restricted theory as natural, and hence hard to consider it as a physical one.

We also show that an interesting ``mirror'' property of quantum purifications no longer holds for the case of non-signalling behaviours. 
 That is,  suppose that we have a purification $\ket{\psi_{AB}}$ of a quantum state $\rho_A$, and let us define $\sigma_B:=\tr_A(\ket{\psi_{AB}}\bra{\psi_{AB}})$. Then, any purification of $\sigma_B$ on system $A$, (which necessarily exists as $\ket{\psi_{AB}}$ is such a purification)) we denote as $\ket{\phi_{AB}}$ and call the ``{\it mirror}'' purification, is equal to $\ket{\psi_{AB}}$ up to a local unitary on $A$. That is $\ket{\psi_{AB}} = U_A\otimes \mathds{1}_B \ket{\phi_{AB}}$ for some unitary $U_A$.  In general, beyond-quantum theory, it need not be the case, as we have exemplified in the theory of non-signalling behaviours (see Sec. 
\ref{sec:CEconjugate} 
of the Appendix). A properly defined minimal distance between the complete extension and its ``mirror'' one, overall systems whose complete extension is not pure, can characterise to what extent a given theory departs from quantum mechanics.

Following the development of  quantum cryptography, one can ask if a post-quantum
theory should lead to secure communication.
As it is shown in \cite{PawelRavi},
this need not always be the case. In the quantum case, the system $E$ of the purification $\psi_{AE}$ describes the worst-case state of the knowledge of the quantum adversary that may have about a given system $A$ in many information processing protocols such as QKD.
Similarly, in the theory of non-signalling behaviours,  the system $E$ of the NSCE $\tau_{AE}$ represents the worst case knowledge of an adversary about system $A$.
The  minimal dimension of the system $E$, therefore, represents 
the maximal memory needed
by the adversary. We, therefore, enable the study of the worst-case adversary's capabilities in other post-quantum theories, see Ref.~\cite{WDH} for an application of this idea.

Significantly, one can use CEP to define composed systems
in a post-quantum theory. Namely given
a set of states of a single system $A$, one can
define as  valid  states of a joint system $AB$ only those that are  
 either (i) complete extensions of  the states of  $A$, or (ii) states obtained from  such  complete extensions by some operations valid in this theory.
We exemplify this approach by arriving 
at the structure of the PR box, assuming (i)
in the case  that $A$ is a local system from NS  with two binary inputs and two binary outputs. Together with (ii) the local transformations in NS (including e.g. relabelings of intputs and outputs), we can
reach all the non-local vertices of the non-signalling polytope of behaviours with two binary inputs and two binary outputs.
Given the fact, that the complete extension of a local deterministic behavior of system $A$ is a product of two deterministic behaviors on systems $A\otimes B$, via operation of mixing two or more behaviors we show that the state space of non-signalling behaviours includes the whole non-signalling polytope of behaviours with two binary inputs and two binary outputs. 

Interestingly, a post-quantum
theory that does not satisfy the purification postulate, but rather the CEP,
can have a property that there is an infinite
sequence of complete extensions of complete extensions as none of them
is pure. This  will always  hold unless the number
of pure states in the theory is of cardinality continuum at least.  This is indeed the case in both classical theory and the theory of non-signalling behaviours. 
 
Verifying if CEP holds in the case of other beyond-quantum theories and studying its consequences is an important direction to follow. 
 We formulate the hypothesis that
there exists a generalised probabilistic theory with complete extensions that may naturally hyper-decohere to  quantum mechanics. This is supported by demonstrating that  the proof of  an existing no-go result, which applies to generalised probabilistic theories with purifications, no longer holds. The confirmation of this hypothesis would open a new
arena in which new physical laws and phenomena may be searched.

 It is also interesting to study whether complementing CEP with a more dynamical axiom, e.g., the one linked to the Neumark extension or Stinespring dilations, can lead to  a more powerful postulate. In the case of the purification postulate such a dynamical postulate can be derived, it is therefore interesting to investigate whether this is also the case for a dynamical version of the CEP, or whether this must be additionally postulated. Determining what form this dynamical postulate should take, and demonstrating that it indeed holds in the theory of non-signalling behaviours, is perhaps the most important direction for follow up research.

\begin{acknowledgments}
MW, TD and KH acknowledge grant Sonata Bis 5 (grant number: 2015/18/E/ST2/00327) from the National Science Center. TD acknowledge Omer Sakarya for his help in numerical simulation.
KH, PH and {\L}P acknowledges ERC grant QOLAPS and Polish Ministry of Science and Higher Education Grant no. IdP2011 000361. MP acknowledges support from European Union’s Horizon 2020 Research and Innovation Programme under the Marie Skłodowska-Curie Action OPERACQC (Grant Agreement No. 661338), and from the Foundational Questions Institute under the Physics of the Observer Programme (Grant No. FQXi-RFP-1601).
RR acknowledges support from the Early Career Scheme (ECS) Grant No. 27210620, the General Research Fund (GRF) Grant No. 17211122, and the Research Impact Fund (RIF) Grant No. R7035-21.  MW, TD, KH, and PH were partially supported by the Foundation for Polish Science through IRAP project co-financed by EU within Smart Growth Operational Programme (contract no. 2018/MAB/5). JHS was supported by the Foundation for
Polish Science through IRAP project co-financed by EU
within Smart Growth Operational Programme (contract
no. 2018/MAB/5). JHS thanks Ana Bel\'en Sainz for helpful discussions. 
\end{acknowledgments}

\section*{Code availability}
The code that supports the theoretical plots and tables within this paper is available from the corresponding author upon reasonable request.

\bibliographystyle{quantum}
\bibliography{References}

\newpage

\appendix
\section{Proofs}
\subsection{Proof of Proposition~\ref{prop:PPtoCEP}}\label{proof:PPtoCEP}
For simplicity in this proof we adopt the diagrammatic notation of \cite{Chiribella2010} using the conventions of \cite{selbyReconstruction}.

Consider some state $s$ and some purification $\sigma^p \in \mathbf{Ext}_P[s]$ and an arbitrary extension $\Sigma \in \mathbf{Ext}[s]$. We want to demonstrate that we can achieve generation, which we will do by showing that there exists a transformation mapping $\sigma^p$ to $\Sigma$, as $\Sigma$ is an arbitrary extension we will therefore have demonstrated that $\sigma^p$ is generating.

To start, let us consider another purification, $\Sigma^p \in \mathbf{Ext}_P[\sigma]$ this time of the state $\Sigma$. It is clear that, as this is pure and an extension of $s$ then it is also a purification of $s$. We therefore have two purifications of $s$ and so one may be tempted to connect them via the essential uniqueness property. However, this will not always be the case, to see this let us examine the systems more carefully. $\sigma^p$ is a state on some purifying system $B$:
\beq
\begin{tikzpicture}
	\begin{pgfonlayer}{nodelayer}
		\node [style=none] (0) at (0, -0.25) {$\sigma^p$};
		\node [style=none] (1) at (-1.5, 0.25) {};
		\node [style=none] (2) at (1.5, 0.25) {};
		\node [style=none] (3) at (0, -1) {};
		\node [style=none] (4) at (-0.75, 0.25) {};
		\node [style=none] (5) at (-0.75, 1.5) {};
		\node [style=none] (6) at (0.75, 0.25) {};
		\node [style={right label}] (7) at (-0.75, 1.25) {$A$};
		\node [style=none] (8) at (0.75, 1.5) {};
		\node [style={right label}] (9) at (0.75, 1.25) {$B$};
	\end{pgfonlayer}
	\begin{pgfonlayer}{edgelayer}
		\draw (1.center) to (2.center);
		\draw (2.center) to (3.center);
		\draw (3.center) to (1.center);
		\draw [qWire] (5.center) to (4.center);
		\draw [qWire] (8.center) to (6.center);
	\end{pgfonlayer}
\end{tikzpicture}}
\eeq
whilst, $\sigma$ could have been an extension on some other system $X$ and the purification of this, $\Sigma^p$ could involve a third system $Y$:
\beq
\begin{tikzpicture}
	\begin{pgfonlayer}{nodelayer}
		\node [style=none] (0) at (0.75, -0.25) {$\Sigma^p$};
		\node [style=none] (1) at (-1.5, 0.25) {};
		\node [style=none] (2) at (3, 0.25) {};
		\node [style=none] (3) at (0.75, -1) {};
		\node [style=none] (4) at (-0.75, 0.25) {};
		\node [style=none] (5) at (-0.75, 1.5) {};
		\node [style=none] (6) at (0.75, 0.25) {};
		\node [style={right label}] (7) at (-0.75, 1.25) {$A$};
		\node [style=none] (8) at (0.75, 1.5) {};
		\node [style={right label}] (9) at (0.75, 1.25) {$X$};
		\node [style={right label}] (10) at (2.25, 1.25) {$Y$};
		\node [style=none] (11) at (2.25, 1.5) {};
		\node [style=none] (12) at (2.25, 0.25) {};
	\end{pgfonlayer}
	\begin{pgfonlayer}{edgelayer}
		\draw (1.center) to (2.center);
		\draw (2.center) to (3.center);
		\draw (3.center) to (1.center);
		\draw [qWire] (5.center) to (4.center);
		\draw [qWire] (8.center) to (6.center);
		\draw [qWire] (11.center) to (12.center);
	\end{pgfonlayer}
\end{tikzpicture}}
\eeq
Unless we are in a highly contrived scenario whereby $B=X\otimes Y$ then we are not able to directly employ the essential uniqueness property. We therefore have to further extend our systems by composing with some extra pure states, $\phi$ and $\chi$, as follows:
\beq
\begin{tikzpicture}
	\begin{pgfonlayer}{nodelayer}
		\node [style=none] (0) at (0, -0.25) {$\sigma^p$};
		\node [style=none] (1) at (-1.5, 0.25) {};
		\node [style=none] (2) at (1.5, 0.25) {};
		\node [style=none] (3) at (0, -1) {};
		\node [style=none] (4) at (-0.75, 0.25) {};
		\node [style=none] (5) at (-0.75, 1.5) {};
		\node [style=none] (6) at (2.5, 0.25) {};
		\node [style={right label}] (7) at (-0.75, 1.25) {$A$};
		\node [style=none] (8) at (2.5, 1.5) {};
		\node [style={right label}] (9) at (2.5, 1.25) {$X$};
		\node [style={right label}] (10) at (4, 1.25) {$Y$};
		\node [style=none] (11) at (4, 1.5) {};
		\node [style=none] (12) at (4, 0.25) {};
		\node [style=none] (13) at (0.75, 1.5) {};
		\node [style=none] (14) at (0.75, 0.25) {};
		\node [style={right label}] (15) at (0.75, 1.25) {$B$};
		\node [style=none] (16) at (3.25, -1) {};
		\node [style=none] (17) at (1.75, 0.25) {};
		\node [style=none] (18) at (2.5, 0.25) {};
		\node [style=none] (19) at (4, 0.25) {};
		\node [style=none] (20) at (4.75, 0.25) {};
		\node [style=none] (21) at (3.25, -0.25) {$\chi$};
	\end{pgfonlayer}
	\begin{pgfonlayer}{edgelayer}
		\draw (1.center) to (2.center);
		\draw (2.center) to (3.center);
		\draw (3.center) to (1.center);
		\draw [qWire] (5.center) to (4.center);
		\draw [qWire, in=90, out=-90, looseness=1.00] (8.center) to (6.center);
		\draw [qWire, in=90, out=-90, looseness=1.00] (11.center) to (12.center);
		\draw [qWire, in=90, out=-90, looseness=1.00] (13.center) to (14.center);
		\draw (17.center) to (20.center);
		\draw (20.center) to (16.center);
		\draw (16.center) to (17.center);
	\end{pgfonlayer}
\end{tikzpicture}}
\qquad\text{and}\qquad
\begin{tikzpicture}
	\begin{pgfonlayer}{nodelayer}
		\node [style=none] (0) at (0.75, -0.25) {$\Sigma^p$};
		\node [style=none] (1) at (-1.5, 0.25) {};
		\node [style=none] (2) at (3, 0.25) {};
		\node [style=none] (3) at (0.75, -1) {};
		\node [style=none] (4) at (-0.75, 0.25) {};
		\node [style=none] (5) at (-0.75, 1.5) {};
		\node [style=none] (6) at (0.75, 0.25) {};
		\node [style={right label}] (7) at (-0.75, 1.25) {$A$};
		\node [style=none] (8) at (2.5, 1.5) {};
		\node [style={right label}] (9) at (2.5, 1.25) {$X$};
		\node [style={right label}] (10) at (4, 1.25) {$Y$};
		\node [style=none] (11) at (4, 1.5) {};
		\node [style=none] (12) at (2.25, 0.25) {};
		\node [style=point] (13) at (4, -0) {$\phi$};
		\node [style=none] (14) at (0.75, 1.5) {};
		\node [style=none] (15) at (4, 0.25) {};
		\node [style={right label}] (16) at (0.75, 1.25) {$B$};
	\end{pgfonlayer}
	\begin{pgfonlayer}{edgelayer}
		\draw (1.center) to (2.center);
		\draw (2.center) to (3.center);
		\draw (3.center) to (1.center);
		\draw [qWire] (5.center) to (4.center);
		\draw [qWire, in=90, out=-90, looseness=1.00] (8.center) to (6.center);
		\draw [qWire, in=90, out=-90, looseness=1.00] (11.center) to (12.center);
		\draw [qWire, in=90, out=-90, looseness=1.00] (14.center) to (15.center);
	\end{pgfonlayer}
\end{tikzpicture}}
\eeq
Then, thanks to our additional assumption that the parallel composite of pure states is pure, then these composite states define further purifications of the original states $s$. Now however, they have the same systems and hence we can use essential uniqueness to conclude that there is a reversible transformation $T$ mapping between them, that is:
\beq
\begin{tikzpicture}
        \begin{pgfonlayer}{nodelayer}
		\node [style=none] (0) at (0, -0.25) {$\sigma^p$};
		\node [style=none] (1) at (-1.5, 0.25) {};
		\node [style=none] (2) at (1.5, 0.25) {};
		\node [style=none] (3) at (0, -1) {};
		\node [style=none] (4) at (-0.75, 0.25) {};
		\node [style=none] (5) at (-0.75, 3.75) {};
		\node [style=none] (6) at (2.5, 0.25) {};
		\node [style={right label}] (7) at (-0.75, 3.5) {$A$};
		\node [style=none] (8) at (2.5, 1.5) {};
		\node [style={right label}] (9) at (2.5, 0.75) {$X$};
		\node [style={right label}] (10) at (4, 0.75) {$Y$};
		\node [style=none] (11) at (4, 1.5) {};
		\node [style=none] (12) at (4, 0.25) {};
		\node [style=none] (13) at (0.75, 1.5) {};
		\node [style=none] (14) at (0.75, 0.25) {};
		\node [style={right label}] (15) at (0.75, 0.75) {$B$};
		\node [style=none] (16) at (3.25, -1) {};
		\node [style=none] (17) at (1.75, 0.25) {};
		\node [style=none] (18) at (2.5, 0.25) {};
		\node [style=none] (19) at (4, 0.25) {};
		\node [style=none] (20) at (4.75, 0.25) {};
		\node [style=none] (21) at (3.25, -0.25) {$\chi$};
		\node [style=none] (22) at (2.5, 2) {$T$};
		\node [style=none] (23) at (0.25, 2.5) {};
		\node [style=none] (24) at (0.25, 1.5) {};
		\node [style=none] (25) at (4.5, 1.5) {};
		\node [style=none] (26) at (4.5, 2.5) {};
		\node [style=none] (27) at (1.75, 2.5) {};
		\node [style=none] (28) at (2.5, 2.5) {};
		\node [style=none] (29) at (1.5, 2.5) {};
		\node [style=none] (30) at (4, 2.5) {};
		\node [style={right label}] (31) at (0.75, 3) {$B$};
		\node [style=none] (32) at (4, 3.75) {};
		\node [style=none] (33) at (2.5, 2.5) {};
		\node [style=none] (34) at (0.75, 2.5) {};
		\node [style=none] (35) at (0.75, 3.75) {};
		\node [style={right label}] (36) at (2.5, 3) {$X$};
		\node [style={right label}] (37) at (4, 3) {$Y$};
		\node [style=none] (38) at (2.5, 3.75) {};
		\node [style=none] (39) at (4, 2.5) {};
	\end{pgfonlayer}
	\begin{pgfonlayer}{edgelayer}
		\draw (1.center) to (2.center);
		\draw (2.center) to (3.center);
		\draw (3.center) to (1.center);
		\draw [qWire] (5.center) to (4.center);
		\draw [qWire, in=90, out=-90, looseness=1.00] (8.center) to (6.center);
		\draw [qWire, in=90, out=-90, looseness=1.00] (11.center) to (12.center);
		\draw [qWire, in=90, out=-90, looseness=1.00] (13.center) to (14.center);
		\draw (17.center) to (20.center);
		\draw (20.center) to (16.center);
		\draw (16.center) to (17.center);
		\draw (23.center) to (24.center);
		\draw (24.center) to (25.center);
		\draw (25.center) to (26.center);
		\draw (26.center) to (23.center);
		\draw [qWire, in=90, out=-90, looseness=1.00] (38.center) to (33.center);
		\draw [qWire, in=90, out=-90, looseness=1.00] (32.center) to (39.center);
		\draw [qWire, in=90, out=-90, looseness=1.00] (35.center) to (34.center);
	\end{pgfonlayer}
\end{tikzpicture}}
\quad =\quad
\begin{tikzpicture}
	\begin{pgfonlayer}{nodelayer}
		\node [style=none] (0) at (0.75, -0.25) {$\Sigma^p$};
		\node [style=none] (1) at (-1.5, 0.25) {};
		\node [style=none] (2) at (3, 0.25) {};
		\node [style=none] (3) at (0.75, -1) {};
		\node [style=none] (4) at (-0.75, 0.25) {};
		\node [style=none] (5) at (-0.75, 1.5) {};
		\node [style=none] (6) at (0.75, 0.25) {};
		\node [style={right label}] (7) at (-0.75, 1.25) {$A$};
		\node [style=none] (8) at (2.5, 1.5) {};
		\node [style={right label}] (9) at (2.5, 1.25) {$X$};
		\node [style={right label}] (10) at (4, 1.25) {$Y$};
		\node [style=none] (11) at (4, 1.5) {};
		\node [style=none] (12) at (2.25, 0.25) {};
		\node [style=point] (13) at (4, -0) {$\phi$};
		\node [style=none] (14) at (0.75, 1.5) {};
		\node [style=none] (15) at (4, 0.25) {};
		\node [style={right label}] (16) at (0.75, 1.25) {$B$};
	\end{pgfonlayer}
	\begin{pgfonlayer}{edgelayer}
		\draw (1.center) to (2.center);
		\draw (2.center) to (3.center);
		\draw (3.center) to (1.center);
		\draw [qWire] (5.center) to (4.center);
		\draw [qWire, in=90, out=-90, looseness=1.00] (8.center) to (6.center);
		\draw [qWire, in=90, out=-90, looseness=1.00] (11.center) to (12.center);
		\draw [qWire, in=90, out=-90, looseness=1.00] (14.center) to (15.center);
	\end{pgfonlayer}
\end{tikzpicture}}
\eeq
Now, we can simply discard the $B$ and $Y$ systems on both sides of this equation to see that we can achieve generation:
\beqa
\input{Diagrams/PPtoCEPProof}
\eeqa
That is, there is a transformation $\tilde{T}$ which maps $\sigma^p$ to $\Sigma$ which completes the proof.

\subsection{Proof of Theorem~\ref{thm:acesstoPMEequivaccesstoMinEns}} \label{proof:acesstoPMEequivaccesstoMinEns}
 To see this, first note that  the set of all PMEs, $S_p$ is a convex set as any two PMEs of $P$, $\cep^1(P), ~\cep^2(P) \in S_P$, their convex combination\footnote{Abstractly we can view PMEs as probability distributions over the set of pure behaviours and hence is a convex set (in particular a simplex) the set $S_P$ is then a subset of this simplex which we will show is closed under convex combinations. } $\lambda~ \cep^1(P) + (1 - \lambda)~\cep^2(P) \in S_P, ~\forall \lambda \in  (0,1)$.  Such a mixture of PMEs is defined as follows: 
Suppose $\{(p_i, P_\texttt{E}^i)\}_i = \cep^1(P)$ and $\{(q_i, P_\texttt{E}^i)\}_i = \cep^2(P)$ are two PMEs of the behaviour $P$, where $V(\cep^1)$ and $V(\cep^2)$ are the set of pure behaviours corresponds to the ensembles. Now define $V_{int} = V(\cep^1) \cap V(\cep^2)$, $V_1 = V(\cep^1) \setminus V(\cep^2)$ and $V_2 = V(\cep^2) \setminus V(\cep^1)$. Then the convex combination of $\cep^1(P)$ and $\cep^2(P)$ is defined as
\ben
\lambda~ \cep^1(P) + (1 - \lambda)~\cep^2(P) 
 := \{(r_i, P_\texttt{E}^i)\}_{ P_\texttt{E}^i \in V(\cep^1) \cup V(\cep^2)},
\een
where $r_i = \lambda p_i + (1 - \lambda)q_i $, $\forall~P_\texttt{E}^i \in V_{int}$, $r_i = \lambda p_i$, $\forall~P_\texttt{E}^i \in V_1$ and $r_i =  (1 - \lambda)q_i $, $\forall~P_\texttt{E}^i \in V_2$. Clearly 
\ben
 \sum_{i: ~P_\texttt{E}^i \in V(\cep^1) \cup V(\cep^2)} r_i P_\texttt{E}^i &=& \sum_{i: ~P_\texttt{E}^i \in V_{int}} (\lambda p_i  + (1 - \lambda)q_i) P_\texttt{E}^i  +  \sum_{i: ~P_\texttt{E}^i \in V_1} \lambda p_i  P_\texttt{E}^i + \sum_{i: ~P_\texttt{E}^i \in V_2} (1 - \lambda)q_i P_\texttt{E}^i ~~~~\\
 &=& \sum_{i: ~P_\texttt{E}^i \in V_{int} \cup V_1}  \lambda p_i P_\texttt{E}^i +  \sum_{i: ~P_\texttt{E}^i \in V_{int} \cup V_2}  (1 - \lambda) q_i P_\texttt{E}^i \\
 &=&  \lambda P + (1 - \lambda) P = P.
\een
Here we use the fact that  $V_{int} \cup V_1 = V(\cep^1)$, $V_{int} \cup V_2 = V(\cep^2)$ and $\sum_{i: ~P_\texttt{E}^i \in V(\cep^1)}  p_i P_\texttt{E}^i = P$ and  $\sum_{i: ~P_\texttt{E}^i \in V(\cep^2)}  q_i P_\texttt{E}^i = P$. The above equation proves that the convex combination of two PMEs of a behaviour $P$ is also an ensemble of $P$, and its members are all pure. Hence, $S_P$ forms a convex set (as we have shown) with only finite number of vertices (as we show below), and all of them are minimal ensembles, as the following lemma proves.

\begin{lemma}[Extremal $\implies$ Minimal]\label{theorem:purification}
      In the set of all pure members ensembles of the behaviour $P$, denoted by $S_P$,  the  ensembles that are extremal  of $S_P$, are minimal. 
      \label{thm:extremal_are_minimal}
 \end{lemma}
\begin{proof}
Suppose by contradiction, this is not true, i.e. there exists a PME, $\cep(P)=\{(p_i,P^i_\texttt{E})\}_{i=1}^{n}$, with $p_i > 0, ~\forall i$, which is extremal in $S_P$,  but is not minimal. Then, there must exists a proper subset ${\cal I} \subset V(\cep)$ such that 
for all $P^j_\texttt{E} \in {\cal{I}}$, there is  some other choices of probabilities $\{q_j\}$, which forms an ensemble $\cm(P) = \{(q_j,P^{j}_\texttt{E})\}_{P^{j}_\texttt{E} \in {\cal I}}$, and it is minimal. Let us now embed the 
distribution $\{q_j\}_{\in {\cal I} }$ which has less than $n$ elements, to obtain new but equivalent distribution with $n$ elements,
by letting $p_i'  := q_i, ~\forall P^i_\texttt{E} \in {\cal I}$ and $p_i': = 0,~ \forall P^i_\texttt{E} \in V(\cep)\setminus {\cal I}$. Let us note that the minimal ensemble $\cm (P)$, is now equivalent to the PME, $\{(p_i',P^i_\texttt{E} )\}_{i=1}^{n}$.

Consider now an ensemble defined as:
\be
\mathscr{N}=\left\{\frac{p_i - p p_i'}{ (1-p)}, P^i_\texttt{E} \right\}_{i=1}^{n} \equiv \{(r_i,P^i_\texttt{E}) \}_{i=1}^{n},
\ee
where we define $p = \frac{p_{min}}{p'_{max}}$ with $p_{min} = \min_i \{ p_i : p_i >0\}$ and $p'_{max} = \max_i \{p'_i\}$. 
Let us first note, that $\mathscr{N}$ is
an ensemble of $P$. indeed, note that
\ben
P = \sum_i p_i P^i_\texttt{E}  = \sum_i (p_i - p p'_i) P^i_\texttt{E}  + p\sum_i p'_i P^i_\texttt{E},  \nonumber \\
= (1-p)\sum_i r_i P^i_\texttt{E}  + p\sum_i p'_i P^i_\texttt{E}, 
\een 
now by assumption $\sum_i p'_i P^i_\texttt{E} = P$, as $\{(p'_i, P^i_\texttt{E})\}_{i=1}^{n} = \cm(P)$. 
Thus:
\be
P = (1-p)\sum_i r_i P^i_\texttt{E}  + p P
\ee
which implies that $\sum_i r_i P^i_E = P$, i.e. $\mathscr{N}$ is an ensemble of $P$, if $0<p<1$.

We will argue now that the latter fact holds. Indeed,
by definition, $p$ is nonzero,  resulting $p_{min}>0$. To see $ p<1$ we will prove that $ p_{min} <  p'_{max}$. If $ p_{min} \geq p'_{max}$, then $1 = \sum_i p_i \geq n p_{min} \geq n p'_{max} \geq n/ |{\cal I} | \Rightarrow |{\cal I} | \geq n$, which  is a contradiction as ${\cal I} $ is a  proper subset of $V(\cep)$. Here we use the fact that $p'_{max} \geq 1/|{\cal I} |$. Hence, $p=\frac{p_{min}}{p'_{max}} < 1$ and  $\mathscr{N}$ is an ensemble of P, and can be denoted as $\mathscr{N}(P)$.
 
Now, we observe that by construction the PME
\be
\ce(P) = p \cm(P) + (1-p)\mathscr{N}(P),
\ee
i.e., $\ce(P)$ is a mixture of two ensembles, that are not equal to each other. This is a contradiction with assumed extremality of the ensemble $\ce(P)$, since the mixture, as shown above, is non-trivial, and the assertion follows. 
\end{proof}

The above theorem proves that all the extremal points in $S_P$, are minimal ensembles, i.e., there is no extremal points in $S_P$ other than the $\cm(P)$. Now we will prove the converse, that no interior point from $S_P$ is a minimal ensemble, i.e., all minimal ensembles are also extremal.
To prove it, we will need the following lemma,   interesting in its own right,  as it characterizes minimal ensembles as those with a unique distribution:

{\lemma The pure members ensemble $\cep(P) = \{(p_i,P^i_\texttt{E})\}_i$ of a behaviour $P$ is minimal, $\cm$, iff the decomposition of this behaviour
into the elements $\{p_i: p_i > 0\}$ is unique, given by corresponding probabilities $p_i$.
\label{lem:unique}
}\\
\begin{proof}
The ``if" direction is trivial: if the elements $\{p_i: p_i > 0\}$, of the decomposition of $\cep(P)$ are unique, then it is not possible to set any probability to zero. Hence, there is no proper subset ${\cal I}\subset V(\cep)$, which forms an ensemble of $P$, with another choice of probabilities.

For the  ``only if" part, suppose $\cm(P) = \{(p_i, P^i_\texttt{E})\}_{i = 1}^m$, is a minimal  ensemble of $P$, we have to prove that the decomposition $\{p_i, p_i > 0\}$, is unique. Assume that the $\{p_i\}$ is not unique, but being a minimal ensemble it should follow
 $\sum_{i=1}^m p_i P^i_\texttt{E} = P$, or in other words the set of following linear equations 
\begin{equation}\label{eqn:set_linear}
\sum_{i=1}^m a_{ki}y_i = c_k, 
\end{equation}

for some $k = 1, \ldots, l'$, has solution in form $y_i = p_i$. Here $c_k$ are the entries of the behaviour $P$,   for the pair $(a,x)$, or in other words the probability of getting $a$, when the input is $x$,  $P(a|x) = c_k$. Similarly the coefficients $\{a_{ki}\}$ are the same  entries for the pair $(a,x)$ of  the pure behaviours  $\{P^i_\texttt{E}(a|x) = a_{ki}\}$. As the behaviour $P$ should follow some equality constraint,  and due to some internal symmetry of it, not all  $l'$  equations in Eq. (\ref{eqn:set_linear}), are linearly independent. Here, by linear independence we mean $\sum_{k\neq k'}\lambda_k (\sum_{i=1}^m a_{ki}y_i - c_k)  \neq \sum_{i=1}^m a_{k'i}y_i - c_k'$, for some $k, k'$ and $\lambda_k$.
Suppose,  there are only $l$ linearly independent equations. 
(There is also a constraint on the $\{y_i\}$, that $\sum_{i=1}^m y_i = 1$ as they represent the probability of the ensemble, but we don't need to consider it separately, as the behaviour $P$ is normalized, so Eq. (\ref{eqn:set_linear}), will take care of it.)
Now the number of linearly independent equations and the number of variables can be in one of the
three orders which we consider separately: 1) $l>m$, 2) $l<m$ 3) $l=m$. Notice first, that it can not be  $l> m$, i.e., for $l$ number of linear equation pertaining $m$ number of variables. Otherwise there would be no solution of the set of equation:
\begin{equation}
\left\{ \sum_i^m a_{ki} y_i = c_k\right\}_{k=1}^{l},
\end{equation}
with variables $y_i$
 but we already have a solution, the initial one: $y_i=p_i > 0, ~\forall i$. 
On the other hand, if $l < m$, then one can always write down any set of $l$, $\{y_i\}_{i = 1}^l$, as a linear functions of the remaining $(m-l)$ $\{y_j\}_{j = l+1}^m$. And in that case one can always set any one (or more) $y_i = 0$ for some $i$,  which violates the  condition of minimal ensembles. Hence we are left with $l =m$.

In this case, we have the same number of linearly independent equation as the number of variables, and in that case the matrix $A =[ a_{ki}]$ is non-singular and invertible, which gives a unique solution of $y_i= p_i >0$ for all $i$.
\end{proof}

We can pass now to prove the extremality of minimal ensembles:
\begin{lemma}[Minimal $\implies$ Extremal]\label{thm:min-ext}
     For a behaviour P, all of its minimal ensembles $\cm(P)$ are extremal in the set $S_P$ of all ensembles of a behaviour $P$.
\end{lemma}
\begin{proof}
Suppose by contradiction, that $\cm(P)$ is not extremal. Then, there exist pure members ensembles $\ce_1(P)$ and  $\ce_2(P)$ such that: 
\be
\cm(P) = \lambda\ce_1(P) + (1-\lambda)\ce_2(P)
\ee
for some $0< \lambda<1$. By the above equality, $V(\ce_1) \subseteq V(\cm)$ and $V(\ce_2) \subseteq V(\cm)$. But by minimality of $\cm$, $V(\ce_1)$ can not be proper subset
of $V(\cm)$, as there are no weights that together with any proper subset of $V(\cm)$ form an ensemble of $P$. Thus $V(\ce_1) = V(\cm)$ and for similar reason
$V(\ce_2) = V(\cm)$. It would mean, that there is an ensemble (let us focus on $\ce_1$) which has different distribution, but the same set of members. It would
mean that the distribution of $\cm$ is not unique: there is another one which together with the same set of members yields an ensemble of $P$. This however
is not possible, since by Lemma \ref{lem:unique}, any minimal ensemble has unique distribution. This proves desired contradiction, hence the assertion follows. 
\end{proof} 

From these two Lemmas \ref{theorem:purification} and \ref{thm:min-ext},  we obtain, that the set $S_P$ of all  pure member ensembles (PMEs)  of $P$ is a convex hull of the set of minimal ensembles $\cm(P)$. And for any behaviour $P$, the set of minimal ensembles is finite, as there are finite number of pure behaviours\footnote{The cardinality of the set of minimal ensembles is bounded by the cardinality of the set of all subsets of pure behaviours which is finite   if the set of pure behaviours is finite.  See section \ref{sec:limitedDim} for an explicit  tighter  upper bound. }
and  corresponding to Lemma \ref{lem:unique} the decomposition of the $p_i$ in minimal ensembles are unique, implies $S_P$ forms a convex polytope.

\subsection{Proof of Corollary~\ref{th:min_puri+coin}} \label{proof:min_puri+coin}
\begin{proof} 
Note that according to the Definition \ref{def:complete_ext}, if ${\cal E}(P)_{AE}(ae|xz)$ is the NSEA of the given behaviour $P_A(a|x)$, then
the only ensembles realized for different choices of input $z$ of the extending party $E$ are the minimal ensembles $\cm(P)$.
If there are $\{\cm_i(P)\}_{i = 1}^N$, $N$ numbers of such minimal ensembles of $P_A$, then there should be $|\mathcal{Z}| = N$, number of distinct inputs in part of the extending system.
Due to Lemma \ref{theorem:purification}  all the extremal points in $S_P$ are minimal ensembles.
From the latter, one can generate any pure members ensemble  $\ce(P)$ by properly mixing the minimal ensembles by using appropriate distribution $\{\tilde{p}(k)\}_{k = 1}^N$, with $\sum_{k = 1}^N \tilde{p}(k) = 1$. For an arbitrary $\ce(P) \in S_P$, the Lemma \ref{theorem:purification}, certifies the existence of at least one such $\{\tilde{p}(k)\}_{k = 1}^N$, which will generate it,
hence, $\ce(P) = \sum_{k = 1}^N \tilde{p}(k) \cm_k(P)$. Each $\cm_k(P) = \{(p(e=i|z = k), P^{ik}_\texttt{E}(a|x))\}_i$, has been obtained from ${\cal E}(P)_{AE}$, by setting the input $z = k$ of the extending party. If the inputs are now chosen probabilistically according to the distribution $\{\tilde{p}(k)\}_{k = 1}^N$, and registering the output $e = i$, then
\ben
\ce(P) &=& 
\sum_{k = 1}^N \tilde{p}(k) \left\{\left(p(e=i|z = k), P^{ik}_\texttt{E}(a|x)\right)\right\} \\
&=& \left\{ \left(q(i), P^{i}_\texttt{E}(a|x)\right)\right\}_{P^{i}_\texttt{E} \in \cup_{k = 1}^N V(\cm_k)},
\een
where $q(i)$ is the probability of getting the pure behaviour $P^{i}_\texttt{E}$, as given in Lemma \ref{theorem:purification}.
\end{proof}

\subsection{Proof of Theorem~\ref{thm:accessequivgeneration}} \label{proof:accessequivgeneration}

\begin{figure}[t]
\begin{center}
\includegraphics[scale=0.6,angle =0]{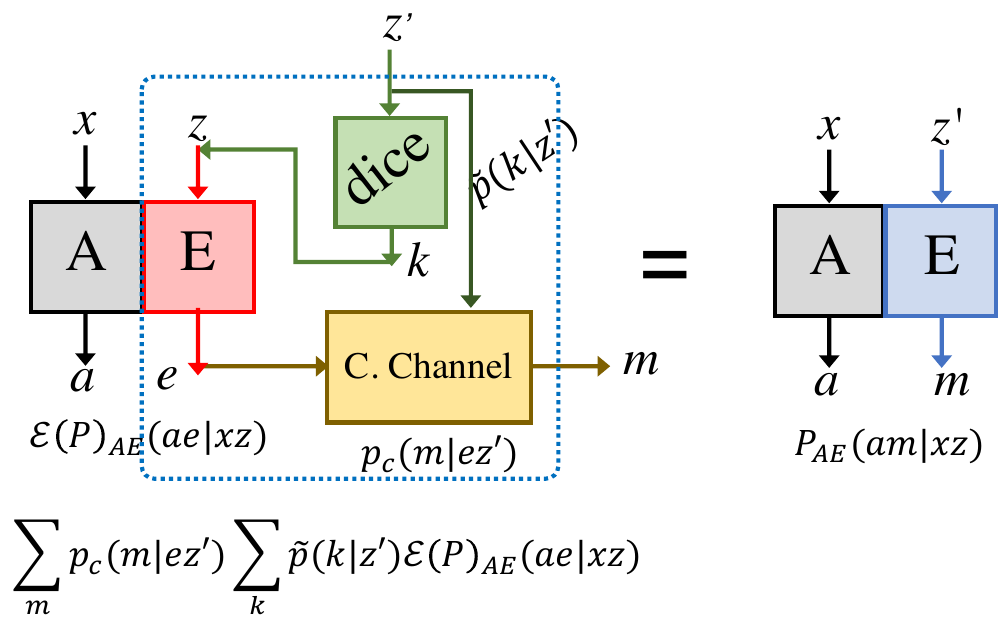}
\end{center}
\caption{Pictorial depiction of Theorem \ref{thm:allensem_anyextension}. 
$E$ holds the additional interfaces of the {NSEA} of behaviour $P_A$. She use an input randomizer (a dice) $\tilde{p}(k|z')$, which can be tuned by the parameter $z'$, and a classical post-processing channel $p_c(m|ez')$, which is also $z'$ dependent. The dice, the extending system and the posy-processing channel together form the new system in part of $E$. The input of the dice $z'$ and the output of the channel $m$, can be considered as the input and output of the new system.
 $E$ only choose some value of input $z'$, generates required randomness and also appropriate conditional probability distribution, resulting different set of mixed ensembles.
}
\label{fig:schematic_arb_extension}
\end{figure}

\begin{lemma}
	(\textsc{access} $\implies$ \textsc{generation}) Access to all ensembles implies access to arbitrary extensions of the extended system.
	\label{thm:cegacae}
\end{lemma}
\begin{proof}
We first follow \cite{Barret-Roberts} (See Eq. (35) there) and observe that for any extension
$P\equiv P(am|xz^\prime)$ of $P(a|x)$ there is
\begin{align}
P(am|xz^\prime) &= P(a|xz^\prime m)P(m|xz^\prime) \\
&=P(a|xz^\prime m)P(m|z^\prime) \\
&\equiv \{P^{ij}(a|x)P(m=i|z^\prime=j)\}_{i,j}
\label{eq:bayes}
\end{align}
In the first of the above equalities we use Bayes rule,
and in the second the non-signalling from $A$ to $E$.
Such obtained equality implies that every bipartite behaviour can be viewed from perspective of the system $E$ as having access to $|z^\prime|$ ensembles of the form $\ce_j(P)=\{(P(m=i|z^\prime=j), P^{ij}(a|x))\}_i$. We argue now,
that from NSEA one can generate any of these ensembles.
Thanks to Theorem \ref{theorem:mixed}, for $|z^\prime|$ inputs $z^\prime=j$ there exists a dice $D_j$ with probability distribution $\tilde{p}(k|z^\prime=j)$ and a classical post-processing channel $C_j$, described with a conditional distribution $p_c(m|e,z^\prime=j)$,
such that when applied on
system $E$ of the NSEA
 $\mathcal{E}(P)(ae|xz)$ they generate
the ensemble $\ce_j$.
Hence from a collection of $\{D_j\}_{j=1}^{|z^\prime|}$ and $\{C_j\}_{j=1}^{|z^\prime|}$ one 
can build via appropriate
``wiring'' a behaviour $P'$ which
upon input $z^\prime=j$ performs
according to a dice $D_j$
and further post-process the output $e$ (and that of $D_j$)
through $C_j$ to give the final output $m$ (see Fig. \ref{fig:schematic_arb_extension}).
In this way one assures
that a new behaviour $P'$ has
access to all  the ensembles 
 $\ce_j$ (and no other), forming
(thanks to Eq. (\ref{eq:bayes}) above)
an extension equivalent to $P$ up to relabelling of inputs in the extending system E. 
\end{proof}

From the above theorem it is clear that the {NSEA} ${\cal E}(P)_{AE}$, of $P_A$ together with access of arbitrary randomness (input randomizer followed by a classical post-processing channel) can generate any collection of ensembles (Theorem \ref{theorem:mixed}) even mixed.  If we consider the whole setup    as a single behaviour, as depicted in Fig. \ref{fig:schematic_arb_extension} (with the  blue rectangular part is in possession of $E$), then it becomes a proper arbitrary extension generating a particular collection of ensembles. 
Note that it can be easily verified that it fulfills all properties of non-signalling behaviour. 

\begin{lemma}
\textsc{generation} $\implies$ \textsc{access}
\end{lemma}
\begin{proof}
The proof for any GPT is given
in Proposition \ref{thm:gen_implies_access}. For the sake of completeness, we show it now using solely arguments from the non-signalling theory. Namely NSEA
is a particular extension, hence
access to arbitrary extension implies access to NSEA. This further
via Theorem \ref{theorem:mixed} implies access to
any (possibly mixed) ensemble. 
\end{proof}

\section{Explicit examples in the theory of no-signaling behavious} \label{app:Examples}

\subsection{Complete extensions of binary input-output behaviours }\label{sec:example-PR}

In section \ref{sec:no-go-discrete}, we proved that the theory of non-signaling behaviours (NS) does not possess the property of purification  as there is not a pure extension of every behaviour.   However, there do  exist some behaviours  for  which  the NSEA  is an extremal,  i.e., pure,  behaviour.  In other words, whilst the purification postulate fails as not every behaviour has a purification, there nonetheless do exist purifications of certain behaviours. 

Here we are going to  show an example of this. Namely,  that if one considers a maximally mixed behaviour with a single binary input and  a single binary output, then it can be extended to an extremal (pure) behaviour in a higher dimensional state space\footnote{Throughout the paper we use the following notation for behaviours, 
\begin{itemize}
\item $P$ -- \textit{italic} represent any generic behaviour.
\item $\mathrm{P}$  -- normal font represent a particular example of a behaviour. 
\end{itemize}}
This pure behaviour is the maximally non-local behaviour equivalent to the Popescu-Rohrlich box \cite{PR}  (up to proper labeling on the extending system)  defined as 
\be \label{eqn:defPR}
\mathrm{P^{PR}}(a,e|x,z)=
\left\{
\bea{l}
\frac12 \quad{\textrm{for }} a\oplus e=x\cdot z, \\
0 \quad{\textrm{otherwise}}.
\eea
\right.
\ee
where $a,e,x,z \in \{0,1\}$.   This is a close analogue to the fact that  in quantum  theory:  a purification of  the maximally mixed state  $\frac{\mathbb{I}_d}{d}$ is the maximally entangled Bell state  (up to local isometry)  \cite{HoroRMP}.

 We will now prove this result. Note that the  maximally mixed behaviour with a single binary input and single binary output is given by
\ben
\mathrm{P_A^m}(a|x)=
\begin{array}{c|c|c}
$\diagbox[width=2em, height=1.8em,innerrightsep=0pt]{$a$}{$x~$}$ & 0 & 1\\ \hline	\\[-1em]
	 0&  1/2 &  1/2  \\ \hline \\[-1em]
	 1 & 1/2 &  1/2
\end{array}
\label{eqn:maximally_mixed}
\een
 here $x$ being the input and $a$ being the output of the behaviour on system $A$.
{This maximally mixed behaviour lies in the ``center'' of the polytope (of the set of behaviours with a single binary input and a single binary output), the extremal points (vertices) of the polytope are
\ben
\mathrm{P}^0_\texttt{E} =
\begin{array}{c|c|c}
$\diagbox[width=1.6em, height=1.6em,innerrightsep=0pt]{$a$}{$x$}$ & 0 & 1\\ \hline	\\[-1em]
	 0&  1 & 1  \\ \hline \\[-1em]
	 1 & 0 & 0
\end{array}, ~~
\mathrm{P}^1_\texttt{E} =
\begin{array}{c|c|c}
$\diagbox[width=1.6em, height=1.6em,innerrightsep=0pt]{$a$}{$x$}$ & 0 & 1\\ \hline	\\[-1em]
	 0&  0 &  0  \\ \hline \\[-1em]
	 1 & 1 &  1
\end{array}, ~~
\mathrm{P}^2_\texttt{E} =
\begin{array}{c|c|c}
$\diagbox[width=1.6em, height=1.6em,innerrightsep=0pt]{$a$}{$x$}$ & 0 & 1\\ \hline	\\[-1em]
	 0&  1 &  0  \\ \hline \\[-1em]
	 1 & 0 &  1
\end{array}, ~~
\mathrm{P}^3_\texttt{E} =
\begin{array}{c|c|c}
$\diagbox[width=1.6em, height=1.6em,innerrightsep=0pt]{$a$}{$x$}$ & 0 & 1\\ \hline	\\[-1em]
	 0&  0 &  1  \\ \hline \\[-1em]
	 1 & 1 &  0
\end{array}. 
\label{eqn:extremal_boxes}
\een
These pure behaviours are 
 are deterministic behaviours as $\mathrm{P}^i_\texttt{E}(a|x) = \delta_{a,g^i(x)} $, for some function $g^i: \{0,1\} \rightarrow \{0,1\}$. 
The polytope has been depicted in Fig. \ref{fig:schematic}, with a blue square, the corners of the square are the extremal (pure) behaviours $\{P_\texttt{E}^i\}$, represented by the black solid circles. The center point of the polytope is the maximally mixed behaviour, $P_A^m$, which is depicted by a white solid circle. 
Any point inside the polytope can be expanded as a convex combination of the vertices.
It is easy to see that, as $\mathrm{P}_A^m$ lies at the intersection of the two diagonals of the square (behaviour polytope), it can be expanded in terms of vertex pairs  $\{\mathrm{P}_\texttt{E}^0, \mathrm{P}_\texttt{E}^1\}$ and $\{\mathrm{P}_\texttt{E}^2, \mathrm{P}_\texttt{E}^3\}$ with equal probabilities. In particular these form the two minimal ensembles of $\mathrm{P}_A^m$ namely
\ben
\cm_0(\mathrm{P}_A^m) &=& \{(1/2, \mathrm{P_\texttt{E}^0});(1/2, \mathrm{P_\texttt{E}^1})\} = \{p(i|0), \mathrm{P}^{i0}_\texttt{E}\}, \label{eqn:PRmin1}\\
\cm_1(\mathrm{P}_A^m) &=& \{(1/2, \mathrm{P}_\texttt{E}^2);(1/2, \mathrm{P}_\texttt{E}^3)\} = \{p(i|1), \mathrm{P}^{i1}_\texttt{E}\},  \label{eqn:PRmin2}
\een
and there are no other minimal ensembles.
Now from the Definition \ref{def:complete_ext}, of  NSEA  to the system $E$, the above two minimal ensembles are obtained in part of system $A$, for two different measurement choices on $E$.
\begin{figure}[t]
\begin{center}
\includegraphics[scale=0.45,angle =0]{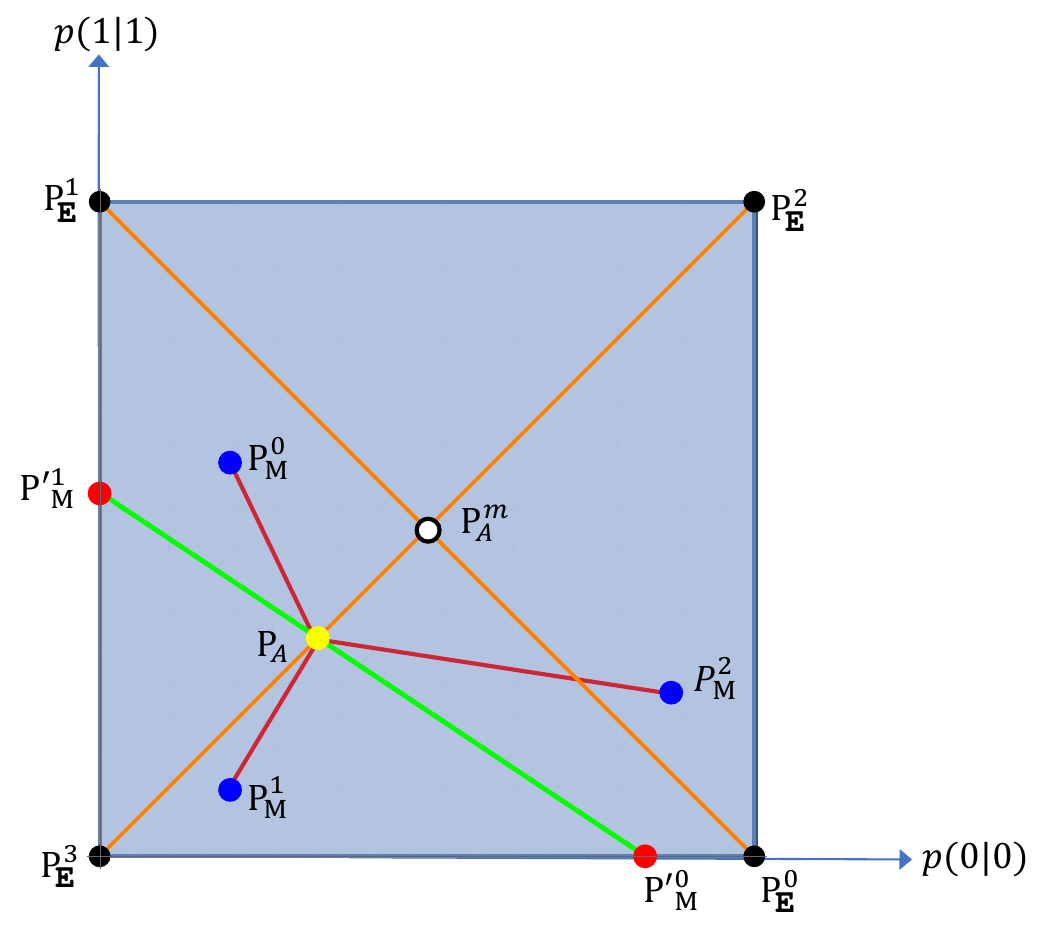}
\end{center}
\caption{Polytope of the set of single party binary input output behaviours. Here the behaviour of consideration given in Eq. (\ref{eqn:our_box}) is the yellow  bullet. The white  bullet is the maximally mixed behaviour.
The black  bullets are  the deterministic, pure or extremal behaviours as given in Eq. (\ref{eqn:extremal_boxes}). The required behaviour can be decomposed as linear combination of the deterministic behaviours as given in Eqs. (\ref{eqn:decomposition}) and (\ref{eqn:choices}). It can also be expanded as convex combination of mixed behaviours which are the red and blue  bullets, these form the mixed ensemble of $P_A$.}
\label{fig:schematic}
\end{figure}
We choose that the first ensemble is  
 obtained by  setting the input $z=0$, and the second one is obtained  by  $z=1$. Labeling the member behaviours of each ensemble with the  output $e$, i.e., 
\ben
\begin{array}{c|c|c|c}
z & e & \text{behaviour on}~ A & \text{Probability} \\
\hline\\[-1em]
 \multirow{2}{*}{0 }    & 0 & \hspace{-1em} \mathrm{P_\texttt{E}^0} & 1/2\\
     & 1 & \hspace{1em}\mathrm{P_\texttt{E}^1} & 1/2\\
     \hline \\[-1em]
  \multirow{2}{*}{1 }    & 0 & \hspace{-1em} \mathrm{P_\texttt{E}^2} & 1/2 \\
  & 1 & \hspace{1em} \mathrm{P_\texttt{E}^3} & 1/2
\end{array}
\een
% We add proper flags $e=0,1,2,3$ to the boxes $\{P_E^i\}_{i = 0}^3$ on this system to distinguish between members of ensemble.
we finally obtain the non-signaling complete extension (NSCE) of the maximally mixed behaviour that presents the same behaviour as the PR box in equation (\ref{eqn:defPR})}: 
\ben
\mathrm{P^{PR}_{AE}}\left(ae|xz\right)=
\begin{array}{cc?cc?cc}
 &\multicolumn{1}{c}{x} & \multicolumn{2}{c}{0}	& \multicolumn{2}{c}{1} \\
z &$\diagbox[width=1.8em, height=1.8em, innerrightsep=0pt]{$e$}{$a$}$  & 0	& 1	&0  & 1\\
	\Cline{1pt}{2-6} \\ [-1em]
 \multirow{2}{*}{0 }  & 0	& 1/2	& 0				& 1/2		& 0\\ 
 & 1	& 0			& 1/2	&0				 	& 1/2 \\
 \Cline{1pt}{2-6}  \\ [-1em]
\multirow{2}{*}{1 }   &0 & 1/2	& 0				&0 				& 1/2\\
& 1 & 0			& 1/2	& 1/2 	& 0 
\end{array}
\label{eqn:PAX_purified_PR}
\een
It is easy to check that if $x\cdot z=0$, the resulting probability distribution on $a$ and $e$ is  perfectly  correlated, while for $x\cdot z=1$  it is perfectly anti-correlated (i.e.,
$e=0$ implies $a=1$ and $e=1$ implies $a=0$). 
%which gives desired perfect anti-correlations. 
Thus in a sense we have derived a PR box solely from the principle of  the complete extension
(CEP).  It is easy to see, that  by  negating $z$ and $e$, one can  instead obtain  another maximally non-local bipartite behaviour. Similarly, all  of  the other  maximally  non-local behaviours (i.e., all non-local vertices of the polytope of two party binary input and binary output behaviours)  can be obtained by proper relabeling of the $z$ and $e$. 

 Note that the  PR box is a vertex in the polytope of two party binary input and binary output behaviours \cite{Barret-Roberts}. Hence, we have the following conclusion:
%which can be viewed as derivation of a PR box from the concept of minimal purification.
{\corollary The PR box is a purification of a maximally mixed behaviour with a single binary input and a single binary output.
}

We have therefore constructed the PR box without any reference to CHSH inequality \cite{CHSH} (in contrast to how it was originally in \cite{PR}), as the non-signaling complete extension (NSCE) of a  maximally mixed  behaviour (\ref{eqn:maximally_mixed}). One  could  argue that the PR box is present in the theory of non-signaling behaviours from the very beginning.  Whilst this is true,  our derivation  is based  solely on the Definition \ref{def:complete_ext},  rather than relying  on the structure of the polytope of non-signaling behaviours.% with two binary inputs and two binary outputs. 

{\rem It is tempting to say that the non-signaling complete extension (NSCE) of two party maximally mixed binary input and binary output behaviours
	is a tensor product of the PR boxes. It is however not the case. This is due to the fact, that one of the valid ensembles of 
	a maximally mixed state $\frac{\mathbb{I}_{AB}}{4} = {\frac 1 2}PR_{AB} + {\frac 1 2} \overline{PR}_{AB}$, where 
	$\overline{PR}_{AB}(ab|xy)=  \frac 12 \delta_{a\oplus b, xy \oplus 1}$ is a non-local behaviour supported on the orthogonal subspace to that of the 
	support of PR. Since this ensemble is clearly minimal, having $2$ members, in definition of NSCE there should be
    the input $z$ which allows the owner of extending system to collapse the system $AB$ into one of these maximally non-local behaviours (each with
    probability half). Suppose now, by contradiction that the NSCE is of the form $PR_{AX_A}\otimes PR_{BX_B}$. It is 
    then clear to see, that in such a behaviour none of direct measurements (choosing the inputs) has outcome behaviour on $AB$ of the form expected by measurement of demanded input $z$.
	However one should consider some other possible ways of measuring system $X_AX_B$ e.g. via wiring.  Yet there is no such action on systems $X_AX_B$, simulating joint outcomes of $z$, since that would lead to the so called {\it non-locality
	swapping}, which is proven to be impossible in Refs. \cite{Barrett, no_ent_swap}.}

By virtue of Theorem \ref{thm:NoGo}, getting a pure behaviour in the higher dimensional state space through the construction of non-signaling complete extension (NSCE) is not always possible for any generic behaviour. 
 Indeed,  if we choose any other behaviour in the polytope of single binary input and single binary output, except for the maximally mixed and the four vertices behaviours, then its  NSCE is not a vertex. 
 For example, let  us consider the following behaviour  
\ben
\mathrm{P_A}=
\begin{array}{c|c|c}
$\diagbox[width=1.6em, height=1.6em, innerrightsep=0pt]{$a$}{$x$}$ & 0 & 1\\ \hline	 \\[-1em]
	 0&  1/3 &  2/3 \\ \hline \\[-1em]
	 1 & 2/3 & 1/3 
\end{array}
\label{eqn:our_box}
\een
%In the polytope of the set of boxes of the above input output,
%there are four  deterministic boxes, which given by
%
%which are also extremal boxes form the vertex of the polytope, as depicted in Fig. \ref{fig:schematic}.
which lies in the polytope of single party binary input and binary output behaviours, and it is represented by the yellow point in Fig.~\ref{fig:schematic}.
 Each behaviour can be expanded in terms of the pure behaviours of the polytope, hence
\ben
\mathrm{P_A} = x_0 \mathrm{P}^0_\texttt{E} + x_1 \mathrm{P}^1_\texttt{E} + x_2 \mathrm{P}^2_\texttt{E} + x_3 \mathrm{P}^3_\texttt{E},
\label{eqn:decomposition}
\een
where $x_i \geq 0, \forall i$, and $\sum_{i =0}^3 x_i = 1$.
The general solutions of Eq. (\ref{eqn:decomposition}) is:
\ben
\begin{cases}\label{eqn:choices}
	x_0=\frac{1}{3}\left(2-3x_3\right),\\
	x_1=\frac{1}{3}\left(2-3x_3\right),\\
	x_2=\frac{1}{3}\left(3x_3-1\right),\\
	x_3=x_3
\end{cases} \hspace{1em} \frac{1}{3} \le x_3 \le \frac{2}{3}
\een
To construct the minimal ensembles of behaviour $P_A$, we have to find out the set of decomposition over the pure points $\{P_\texttt{E}^i\}$, such that 
%From the definition \ref{def:purification-min}, of minimal purification, it is such decomposition of $P_A$ on the pure boxes $ P_i$, 
any proper subset of each choice can not be the ensemble of $P_A$ with another set of probabilities.
This implies that we have to find those solutions of Eq. (\ref{eqn:choices}), where the
 minimal number of $x_i$'s are nonzero. There are two of such choices, given by 
\be
\begin{cases}
	x_0=\frac{1}{3},\\
	x_1=\frac{1}{3},\\
	x_2=0,\\
	x_3=\frac{1}{3}
\end{cases},
\begin{cases}
	x_0=0,\\
	x_1=0,\\
	x_2=\frac{1}{3},\\
	x_3=\frac{2}{3}
\end{cases},
\ee
which, hence, form the two minimal ensembles of $P_A$:% which are
\ben
\cm_0(\mathrm{P_A}) &=& \{(1/3, \mathrm{P_\texttt{E}^0});(1/3, \mathrm{P_\texttt{E}^1});(1/3, \mathrm{P_\texttt{E}^3})\} = \{p(i|0), \mathrm{P}^{i0}_\texttt{E}\}, \label{eqn:min1}\\
\cm_1(\mathrm{P_A}) &=& \{(1/3, \mathrm{P_\texttt{E}^2});(2/3, \mathrm{P_\texttt{E}^3})\} = \{p(i|1), \mathrm{P}^{i1}_\texttt{E}\}.  \label{eqn:min2}
\een
Here we label the ensembles by $\{0,1\}$, according to the inputs $z$ and the members as $\mathrm{P^{00}} = \mathrm{P_\texttt{E}^0}$, $\mathrm{P^{10}} = \mathrm{P}_\texttt{E}^1$, $\mathrm{P^{20}} = \mathrm{P}_\texttt{E}^3$ and $\mathrm{P}^{01} = \mathrm{P}_\texttt{E}^2$, $\mathrm{P^{11}} = \mathrm{P}_\texttt{E}^3$.
%setting the output $e = i$,  
Finally then, the  NSCE of $\mathrm{P_A}$ to system $E$ is given by

\ben
\mathrm{P_{AE}}\left(ae|xz\right)=
\begin{array}{cc?cc?cc}
 &\multicolumn{1}{c}{x} & \multicolumn{2}{c}{0}	& \multicolumn{2}{c}{1} \\
z &$\diagbox[width=1.8em, height=1.8em, innerrightsep=0pt]{$e$}{$a$}$  & 0	& 1	&0  & 1\\
	\Cline{1pt}{2-6} \\ [-1em]
 \multirow{3}{*}{0 }  & 0	& 1/3	& 0				& 1/3		& 0\\
 & 1	& 0			& 1/3	&0				 	& 1/3 \\
 & 2	& 0			& 1/3	& 1/3		& 0  \\
 \Cline{1pt}{2-6}  \\ [-1em]
\multirow{2}{*}{1 }   &0 & 1/3	& 0				&0 				& 1/3\\
& 1 & 0			& 2/3	& 2/3 	& 0 
\end{array}
\label{eqn:PAX_purified}
\een
 which  lies in the polytope of two party  behaviours,   in which  one party has a binary input and output, whereas the other party has a binary input but a ternary output.  Moreover, as we will prove in the following section (Sec.~\ref{sec:CE-notvertex}), this extended behaviour is not pure in the respective behaviour polytope. 
%and the proof has been given in the following sec Sec \ref{sec:CE-notvertex}. 
%In Sec. \ref{} of the supplementary material we provide the explicit construction of CE, and we also proof that it is not a vertex point.}
%Next we will show that this completely extended box $\mathrm{P_{AX}}\left(ae|xz\right)$ is not an extremal box in the set of boxes of system $AX$.

An example of Theorem \ref{th:min_puri+coin}  has been given in Sec. \ref{sec:any-PME},
%i.e., complete extension gives access to any pure members ensemble,  for the box in Eq. (\ref{eqn:our_box}).
where we have shown that any pure members ensemble of $\mathrm{P}_A$, given in Eq. (\ref{eqn:our_box}), can be constructed by taking the convex combination of its two minimal ensembles. This convex combination has been simulated by using an additional randomness (a dice) at the input of the extending party. An arbitrary ensemble of  $\mathrm{P}_A$, can be constructed by passing the output of the extending party through a classical post-processing channel (Theorem \ref{theorem:mixed}) has also been exemplified in Sec. \ref{sec:mixen}. The members of the mixed ensembles have been depicted by the blue dots (for mixed ensemble 1) and by red dots (for mixed ensemble 2). By enumerating the mixed ensembles with the input and the member behaviours with the output of an additional system, an extension of $\mathrm{P}_A$ has been generated. As the mixed ensembles of $\mathrm{P}_A$ are arbitrarily chosen, the extension made out of it is also an arbitrary extension of $\mathrm{P}_A$ --  given in Eq. (\ref{eqn:PAX_purified_mix}) of Sec. \ref{sec:CEgenerate}.
Moreover, for a generic behaviour, the non-signaling  complete extension (NSCE) introduces correlations (possibly non-local) between the extending and the extended system. It is quite clear that, for a pure behaviour (a vertex of a polytope), there will be no correlation due to NSCE, whereas it can inject a maximal amount of correlation when the given behaviour is maximally mixed. For $\mathrm{P}_A$, we calculate the non-local correlation in  Sec. \ref{sec:CEnonlocal}.

Another important aspect of non-signaling complete extensions (NSCE) is that, unlike quantum purifications, the NSCE of an arbitrary behaviour and its  conjugate behaviour are not the same. If ${\cal E}(P)_{AE}(ae|xz)$ is the NSCE of the behaviour $P_A(a|x)$, then the behaviour $P_E(e|z) = \sum_a P_{AE}(ae|xz)$, is the conjugate behaviour of $P_A$. The NSCE of  $P_E(e|z)$, according to the Definition \ref{def:complete_ext}, ${\cal E}(P)_{A'E}(a'e|x'z)$ is not equal to the ${\cal E}(P)_{AE}(ae|xz)$. In section \ref{sec:CEconjugate}, we provide an explicit example in favour of this argument, for the behaviour $\mathrm{P}_A$, given in Eq. (\ref{eqn:our_box}).

\subsubsection{Example: the non-signaling complete extension that is not a vertex}\label{sec:CE-notvertex}
%\jnote{This heading seems like it's  a restatment of the general theorem from earlier, maybe we could add ``An example that...'' or something to make it clear what this section is about. Similarly for the subsequent subsections.}

In this section we will prove that the NSCE of $\mathrm{P}_A$ (given in equation (\ref{eqn:our_box}))
%\jnote{I would suggest using a more distinctive name than $P_A$ to refer to this particular behaviour, it's quite hard to tell when generic statements are being made and when it's particular to a specific behaviour. I have similar problems in places with understanding whether we're talking about a generic non-signalling complete extension (i.e., an extension satisfying generation) vs. a particular one (e.g., the NSEA) I'd recommend generally checking through the paper to sort out this issue.}
, is not a vertex in the higher dimensional behaviour polytope.
%In this section, we will prove that unlike Sec. \ref{subsec:example-PR}, the NSCE,  $\mathrm{P_{AX}}$, of box $\mathrm{P_A}$,  is not extremal in the polytope of the set of all boxes which included $\mathrm{P_{AX}}$.
Consider an arbitrary behaviour $P_{A_1A_2 \ldots A_n}(a_1a_2\ldots a_n|x_1x_2\ldots x_n)$, with $n$ parties $A_1A_2 \ldots A_n$ where the $x_i$ and $a_i$ are respectively the input and output of the $i$th party $A_i$. Suppose there are $m_i$ possible measurement choices of $A_i$, i.e.,  $x_i \in \{1,2, \ldots m_i\}$, and corresponding to each measurement $x_i = j$ the number of possible outcomes is $d_{ij}$, such that  $a_i \in \{1,2, \ldots,d_{ij}\}$. The total number of parameters involved in defining the behaviour is given by Ref.~\cite{Pironio2005lift} as
\ben
t&=&\prod_{i=1}^n \left( \sum_{j=1}^{m_i} d_{ij}\right). 
\een
Among these $t$ parameters not all  of them  are independent as  the behaviour  should obey normalization and non-signaling conditions.  These conditions together imply that 
%Moreover, 
the behaviour must lie within a  particular  polytope of $\mathbb{R}^{t}$. If we construct a vector $v \in \mathbb{R}^{t}$, whose entries are those $t$ probabilities, then the polytope  can be  defined as 
\ben
{\cal P} = \{v~|~{\cal A}v \leq w\}
\een
for some $w \in \mathbb{R}^{s}$, and ${\cal A}$ an $t\times s$ matrix. Here  the condition  ${\cal A}v \leq w$  captures  all  of  the constraints  that the probabilities need to satisfy.

Let us take an arbitrary element $u \in {\cal P}$, and suppose ${\cal A}_u u \leq w_u$ are those inequality constraints among all possible constraints that are satisfied by $u$ with equality \cite{Schrijverbook}, i.e., ${\cal A}_u u = w_u$. Here, ${\cal A}_u$ and $w_u$ are sub-matrices of ${\cal A}$ and $w$ respectively. Then from Theorem 5.7 of Ref. \cite{Schrijverbook}, $u$ will be a pure point of ${\cal P}$ if and only if $\mathrm{rank}\left({\cal A}_u\right)=t$. 

The behaviour $\mathrm{P_{AE}}$, given in Eq. (\ref{eqn:PAX_purified}),  belongs to a polytope which lies in a space of $t = 20$.  We now compute the rank of $\mathcal{A}_u$ to demonstrate that this is not a vertex of the polytope.
%Now we will 
 First we count how many independent equality constraints are acting on $\mathrm{P_{AE}}$.
The $t = 20$ probabilities  are $P_{AE}(00|00), P_{AE}(01|00), \ldots, P_{AE}(12|11)$, i.e., $P_{AE}(ae|xz)$ for $a, x \in \{0,1\}$ and $e \in \{0,1,2\}$ for $z = 0$ and $e \in \{0,1\}$ for $z = 1$.
The equality constraints they need to satisfy are the non-signaling constraint and normalization conditions, namely
\ben
\sum_a P_{AE}(ae|xz) = \sum_a P_{AE}(ae|x'z), ~\forall e,x,x',z, \label{eqn:PAE-nonsingcondition1} \\
\sum_e P_{AE}(ae|xz) = \sum_e P_{AE}(ae|xz'), ~\forall a,x,z,z', \label{eqn:PAE-nonsingcondition2}
\een
and
\ben
\sum_{ae} P_{AE}(ae|xz) = 1, ~\forall x,z. \label{eqn:PAE-normcondition}
\een
There are $5$ from Eq. (\ref{eqn:PAE-nonsingcondition1})\footnote{$\sum_a P_{AE}(ae|xz) = P_E(e|z), ~ e \in \{0,1,2\}$ for $z = 0$ and $e \in \{0,1\}$ for $z = 1$.},  $4$  from Eq. (\ref{eqn:PAE-nonsingcondition2})\footnote{$\sum_e P_{AE}(ae|xz) = P_A(a|x), ~ a,x \in \{0,1\}$}, and 
another $4$  from Eq. (\ref{eqn:PAE-normcondition}), totalling $13$ equality constraints.   Moreover, among the $40$ inequality  constraints of the form $0 \leq P_{AE}(ae|xz) \leq 1,  ~\forall a,e,x,z$, ($2$ inequalities for each of the $20$ probabilities $P_{AE}(ae|xz)$),
%\ben
%0 \leq P_{AE}(ae|xz) \leq 1, ~\forall a,e,x,z
%\een
 only $10$ of them find equality with zeroes. Note that there is no probability which finds equality with $1$, as there is no entry of $1$, in Eq. (\ref{eqn:PAX_purified}). Hence,  the total number of equality constraint  are $ 23$, and the matrix ${\cal A}_u $ for which ${\cal A}_u u = w_u$, is of the dimension $23 \times 20$. To obtain $\mathrm{rank}\left({\cal A}_u\right)$, we need to find out the number of linearly independent 
 equality constraint out of the $23$. 
 The $10$ equality constraint with zero, on the probabilities are all linearly independent, but among those $13$ (non-signaling constraint and normalization conditions) only $9$ of them are linearly independent -- by choosing any $1$ normalization condition, and any $8$ non-signaling condition one can generate the remaining $4$ conditions.
Hence, the total number of the linearly independent constraints, on the behaviour $P_{AE}$ is $9 + 10 = 19$, and rank$({\cal A}_u) = 19 < 20$. Which allows us to state that behaviour $\mathrm{P_{AE}}$ is not a vertex (extreme point) of the given polytope. 

On the other hand the NSCE of the maximally mixed behaviour, the PR box given in Eq. (\ref{eqn:PAX_purified_PR}) is a pure behaviour in the polytope of two binary input output behaviours. This can be shown in the following way: the space where the PR box lives is of $t = 16$ \cite{Barret-Roberts}. Amongst the $16$ parameters there are $8$ equality constraints with zeroes and $4 + 4 = 8$ linearly independent equality comes from the $8$ non-signaling conditions and $4$ normalization condition, hence rank$({\cal A}_u) = 16$, which exactly matches with the dimension of the space.

%We are going to show now that in the polytope of single party binary input and output behaviour, the number of behaviours which can be purified is finite. \jnote{We already know this from our general theorem, what is new here?}
%We have the nice observation.

In the following lines we study particular properties of NSEA in a low dimensional case.

\begin{observation}
	All non-deterministic behaviours in single party, binary input, binary output scenario  have two minimal ensembles. All minimal ensembles of those have either two or three members. 
	\label{obs:num_ens}
\end{observation}
\begin{proof}
The polytope of single party binary input binary output behaviour belongs to set of reals of dimension $d = 2$ and it has been given in Fig. \ref{fig:schematic}.
Suppose, $P$ is any arbitrary behaviour in the polytope, then from the theorem of Carath{\'e}odory \cite{caratheodory1911, Ziegler1995}, the maximal number of pure behaviours  in each minimal ensembles of $P$ is $3$. It is also clear from the figure that any 3 pure behaviours of the polytope form a triangle, and any arbitrary point inside the polytope can be inside at most two overlapping  triangles. Hence, it has at most two minimal ensembles.
\end{proof}

%If any box lies on the line joining any two extremal points, then it has only two members in that particular minimal ensemble
For any arbitrary behaviour inside that polytope,  the minimal ensembles consists of any combination of the following number of members. 
{\corollary	Among single party, binary input, binary output behaviours, only five of them have NSCE that is a purification (NSCE which is a vertex). %\jnote{I think this just proves that the NSEA is only a purification for 5 behaviours, it doesn't seem to necessarily rule out that some other behaviours could have purifications, just that if they did these purifications wouldn't be the NSEA.}
}

%In the following proof we will refer to the single party binary input, binary output boxes as the considered scenario or initial polytope. The cardinality of the input and output of the extended system is fixed to binary.\\
\begin{proof}
Due to the Observation \ref{obs:num_ens} we know that the non-signaling complete extensions of behaviours in considered scenario are bipartite states in one of the following polytopes. Polytope of
\begin{enumerate}
    \item[(i)] one binary input and one unary input with binary outputs: behaviours lying on the edges of the polytope has single minimal ensembles with only two members.
	\item[(ii)] two binary inputs, two binary outputs behaviours: when the initial behaviour has two minimal ensembles and each of the ensembles has two members. 
	\item[(iii)] two binary inputs, one binary output, one binary/ternary output (depending on the corresponding input setting) behaviours: behaviours lying on any one diagonal of the polytope.
	\item[(iv)] two binary inputs, and two ternary outputs behaviours: for any arbitrary behaviour, not belongs to the above sets.
\end{enumerate}
For each local vertex (deterministic box) of the listed polytopes, if we trace out the second party by summation over all of its outcomes, the result is one of the deterministic behaviours of the initial polytope. Due to Theorem~1 of \cite{Barret-Roberts}, we know the form of all non-local vertices in (i,ii,iii) polytopes. In each case after tracing out the extending system (summation over outcomes), the result is the maximally mixed behaviour of the initial system. \\
As we have investigated all the vertices which were suspected of being a purifications of behaviours from single party, one binary input, one binary output scenario and in each case we obtained one of the five states we conclude there are no other behaviours (in the initial polytope) that have purification.
\end{proof}

\subsubsection{Example: NSCE of $\mathrm{P}_A$ gives \textsc{access} to any PME of $\mathrm{P}_A$}\label{sec:any-PME}
In Theorem \ref{th:min_puri+coin}, we state that, the extended system of the NSCE, can access any PMEs of the behaviour $\mathrm{P_A}$, if it is equipped with arbitrary randomness.
Any pure ensemble of $\mathrm{P_A}$, $\ce(\mathrm{P_A}) = \{x_i, \mathrm{P_i}\}_i$, where the $\{x_i\}$ satisfy Eq. (\ref{eqn:choices}), can be written down as convex combination of the minimal ensembles, which is given below
\ben
\ce(\mathrm{P_A}) = \lambda \cm_0(\mathrm{P_A}) + (1 - \lambda) \cm_1(\mathrm{P_A}),
\een
with $\lambda = 2 - 3 x_3 \in [0,1]$, as $x_3 \in [\frac 13, \frac 23]$. If the extending party $X$ chooses to toss a coin $p_t$, (binary output) and feed it to the  input $z$ of her part of the completely extended behaviour with $p_t(0)= \lambda$, and $p_t(1)=1 -  \lambda$, then the extending system has \textsc{access} to any pure ensemble of $\mathrm{P_A}$.
%\textcolor{red}{A schematic diagram of accessing  all pure members ensembles once the complete extension is in hand is given depicted in Fig. \ref{fig:schematic_pure_members_ensemble}.}

\subsubsection{Example: NSCE of $\mathrm{P}_A$ gives \textsc{access} to any mixed ensemble of $\mathrm{P}_A$} \label{sec:mixen}
In this section, we will  explicitly exemplify that the extending system $E$ can access all possible mixed ensemble $\ce_{mix}(\mathrm{P_A}) = \{p_m,\mathrm{P_M^m}\}_m$, of an arbitrary behaviour $\mathrm{P_A}$, (given in Theorem \ref{theorem:mixed}). Here the behaviours $\mathrm{P_M^m}$ are any arbitrary behaviours. \\
Example 1: Suppose $X$ wants to  access the following ensemble of mixed behaviours
\ben
\ce_{mix}(\mathrm{P_A}) = \Big\{\Big(\frac{33}{81},\mathrm{P_M^0}\Big); \Big(\frac{32}{81}, \mathrm{P_M^1} \Big);  \Big(\frac{16}{81}, \mathrm{P_M^2} \Big)\Big\},
\label{eqn:mixed_ensemble}
\een
 in part of system $A$\footnote{Here we use {\it tilde}, on the symbol of ensemble to denote a particular ensemble among the set of ensembles.}, where 
%$
%P_A   = (33/81)P_M^0 +(32/81) P_M^1 + (16/81) P_M^2.
%$
the mixed behaviours (the blue points in Fig. \ref{fig:schematic}), are given by 
\ben
\mathrm{P^0_M} =
\begin{array}{c|c|c}
$\diagbox[width=1.6em, height=1.6em,innerrightsep=0pt]{$a$}{$x$}$ & 0 & 1\\ \hline	\\[-1em]
	 0&  1/5 &  2/5  \\ \hline \\[-1em]
	 1 & 4/5 &  3/5
\end{array}, ~~
\mathrm{P^1_M} =
\begin{array}{c|c|c}
$\diagbox[width=1.6em, height=1.6em,innerrightsep=0pt]{$a$}{$x$}$ & 0 & 1\\ \hline	\\[-1em]
	 0&  1/5 &  9/10  \\ \hline \\[-1em]
	 1 & 4/5 &  1/10
\end{array}, ~~
\mathrm{P^2_M} =
\begin{array}{c|c|c}
$\diagbox[width=1.6em, height=1.6em,innerrightsep=0pt]{$a$}{$x$}$ & 0 & 1\\ \hline	\\[-1em]
	 0&  7/8 &  3/4  \\ \hline \\[-1em]
	 1 & 1/8 &  1/4
\end{array}. \nonumber \\
\label{eqn:mixed_boxes}
\een
%which are depicted as the .
 Each of these mixed behaviours has some decompositions  over the pure behaviours, which are certainly not unique, consider the following minimal one, which are
\ben
\cm(\mathrm{P_M^0}) &=& \Big\{\Big(\frac 25, \mathrm{P_\texttt{E}^1}\Big);\Big(\frac 15, \mathrm{P_\texttt{E}^2}\Big);\Big(\frac 25,\mathrm{P_\texttt{E}^3}\Big)\Big\}, 
%\cm_1(P_M^0) &=& \Big\{\Big(\frac 15, P_E^0\Big);\Big(\frac 35, P_E^1\Big);\Big(\frac 15,P_E^3\Big)\Big\},
\een
\ben
\cm(\mathrm{P_M^1}) &=& \Big\{\Big(\frac{1}{10}, \mathrm{P_\texttt{E}^0}\Big);\Big(\frac {1}{10}, \mathrm{P_\texttt{E}^2}\Big);\Big(\frac 45, \mathrm{P_\texttt{E}^3}\Big)\Big\}, 
%\cm_1(P_M^1) &=& \Big\{\Big(\frac 15, P_E^0\Big);\Big(\frac {1}{10}, P_E^1\Big);\Big(\frac{7}{10},P_E^3\Big)\Big\},
\een
and
\ben
\cm(\mathrm{P_M^2}) &=& \Big\{\Big(\frac 34, \mathrm{P_\texttt{E}^0}\Big);\Big(\frac 18, \mathrm{P_\texttt{E}^1}\Big);\Big(\frac 18, \mathrm{P_\texttt{E}^2}\Big)\Big\}, 
%\cm_1(P_M^2) &=& \Big\{\Big(\frac 58, P_E^0\Big);\Big(\frac 14, P_E^2\Big);\Big(\frac 18,P_E^3\Big)\Big\}.
\een
%One can take any pure ensembles of the mixed boxes $\{P_M^i\}$, by properly mixing the corresponding two minimal ensembles, (see Theorem \ref{th:min_puri+coin}). Here we take only the first minimal ensemble for each mixed boxes and 
Put them into Eq. (\ref{eqn:mixed_ensemble}), the mixed ensemble  then turn out to be the pure one, given by
\ben
\cep(\mathrm{P_A}) = \Big\{\Big( \frac{76}{405}, \mathrm{P_\texttt{E}^0}\Big) ; \Big( \frac{76}{405}, \mathrm{P_\texttt{E}^1}\Big); \Big( \frac{59}{405}, \mathrm{P_\texttt{E}^2}\Big) ; \Big( \frac{194}{405}, \mathrm{P_\texttt{E}^3} \Big)\Big\} = \{r(e), \mathrm{P_\texttt{E}^e}\}
\een
One can check that $\cep(\mathrm{P_A})=  \frac{76}{135}\cm_0(\mathrm{P_A}) +  \frac{59}{135}\cm_1(\mathrm{P_A})$, and 
 $E$ can  access $\cep(\mathrm{P_A})$ by choosing the input according to the  probability distribution, $\{p_t(0) = \frac{76}{135},p_t(1) = \frac{59}{135}\}$. This can be done by feeding the output of a flipped coin  to the input 
$z$ of her part of the NSCE,  ${\cal E}(\mathrm{P})_{AE}(ae|xz)$. 

Once the PME in part of $A$ has been prepared, the prefixed mixed ensembles has been constructed by passing the output $e$ through a classical channel (post-processing channel) 
%part then she pass her output $e$
%\footnote{Recall that there is a one to one map between the output $e \rightarrow P^e_E$.} 
%through a conditional 
$\mathrm{P_c}(m|e)$, which is 
\ben
\mathrm{P_c}(m|e)=
\begin{array}{c|c|c|c|c}
$\diagbox[width=1.7em, height=1.8em, innerrightsep=0pt]{$m$}{$e$}$ & 0 & 1 & 2& 3\\ \hline	\\ [-1em]
	 0& 0 & 33/38 & 33/59 & 33/97\\ \hline \\ [-1em]
	 1 & 4/19 & 0 & 16/59 & 64/97 \\ \hline \\ [-1em]
	 2 & 15/19 & 5/38 & 10/59 & 0 \\ \hline 
\end{array}
\label{eqn:post-processing_channel}
\een
the index $m$ is the flag in part of $E$, different $m$ give the access to different mixed behaviour $\mathrm{P^m_M}$ with probability $p_m$. One can check that $p_m = \sum_e P_c(m|e)r(e)$, and $\mathrm{P^m_M} = \frac{1}{p_m} \sum_e P_c(m|e)r(e)P^e_\texttt{E} $.
Thus we can see that the extending system can be able to access any ensemble of behaviour $\mathrm{P_A}$, by NSCE with arbitrary randomness which will mix the minimal ensembles by mixing the input $z$, and then gluing the output $e$ by a conditional classical channel.

Example 2: $\mathrm{P_A}$ can also be expanded as another mixed ensemble $\ce'_{mix}(\mathrm{P_A}) = \Big\{\Big(\frac{2}{5}, \mathrm{P_M^{'0}}\Big);\Big(\frac{3}{5}, \mathrm{P_M^{'1}}\Big)\Big\}$, (the red points in Fig. \ref{fig:schematic}) where 
\ben
\mathrm{P}^{'0}_M =
\begin{array}{c|c}
	5/6& 1 \\ \hline
	1/6& 0 
\end{array} = \Big\{\Big(\frac 56, \mathrm{P_E^0}\Big);\Big(\frac 16, \mathrm{P_E^3}\Big)\Big\}, \\
\mathrm{P}^{'1}_M =
\begin{array}{c|c}
	0& 4/9 \\ \hline
	1& 5/9 
\end{array} =  \Big\{\Big(\frac 59, \mathrm{P_E^1}\Big);\Big(\frac 49, \mathrm{P_E^3}\Big)\Big\}.
\een
Now the  pure ensemble turn out to $\ce'(\mathrm{P_A}) = \cm_0(\mathrm{P_A})  = \{(1/3, \mathrm{P_\texttt{E}^0});(1/3, \mathrm{P_\texttt{E}^1});(1/3, \mathrm{P_\texttt{E}^3})\} = \{r'(e), \mathrm{P}^e_\texttt{E}\}$. For this, $E$ will chose a completely biased coin,  $p_t(0) = 1, p_t(1) = 0$, and feed its output to the input of the NSCE, and the post-processing channel $\mathrm{P_c}$ is 
\ben
\mathrm{P_c}(m|e)=
\begin{array}{c|c|c|c|c}
$\diagbox[width=1.7em, height=1.8em, innerrightsep=0pt]{$m$}{$e$}$ & 0 & 1 & 2& 3\\ \hline	
	 0& 1 & 0 & - & 1/5\\ \hline
	 1 & 0 & 1 & - & 4/5 \\ \hline 
\end{array}
\label{eqn:pp_channel2}
\een
which will be used to post-process the output of the extending part.
{Here we keep the column for $e = 2$ ``blank" as there is no such incidence that the pure behaviour $P^2_\texttt{E}$ occur. Clearly $p_0 = \sum_e r'(e) \mathrm{P_c}(m|e) = \frac{1}{3} + \frac{1}{3}\times \frac{1}{5} = \frac{2}{5}$ and $\mathrm{P_M^{'0}} = \frac{1}{p_0}\sum_e r'(e) \mathrm{P_c}(m = 0|e)\mathrm{P}^e_\texttt{E} = \frac{5}{6}\mathrm{P}^0_\texttt{E} + \frac{1}{6}\mathrm{P}^3_\texttt{E}$.}

\subsubsection{NSCE can generate any extension}\label{sec:CEgenerate}
(Example of Theorem \ref{thm:allensem_anyextension}).
Numbering these two examples of mixed ensembles with $z' = 0$ and $z' = 1$, we obtain an arbitrary extension of $\mathrm{P_A}$ to the behaviour $\mathrm{P_{AE}}(am|xz')$. Such that  $\{p(m|z'=0),\mathrm{P^{m0}}(a|x)\} = \ce_{mix}(\mathrm{P_A}) $ and $\{p(m|z'=1),\mathrm{P^{m1}}(a|x)\} = \ce'_{mix}(\mathrm{P_A})$. And the arbitrary extended behaviour is
\ben
\mathrm{P^{mix}_{AE}}\left(am|xz'\right)=
\begin{array}{cc?cc?cc}
 &\multicolumn{1}{c}{x} & \multicolumn{2}{c}{0}	& \multicolumn{2}{c}{1} \\ 
z' &$\diagbox[width=1.8em, height=1.8em, innerrightsep=0pt]{$m$}{$a$}$  & 0	& 1	&0  & 1 \\
	\Cline{1pt}{2-6}  \\ [-0.9em]
 \multirow{4}{*}{0 }  & 0	& \frac{11}{135}	& \frac{44}{135}				& \frac{22}{135}		& \frac{11}{45}\\ [4pt]
 & 1	& \frac{32}{405} 	& \frac{128}{405} 	& \frac{16}{45}		& \frac{16}{405} \\ [4pt]
 & 2	 & \frac{14}{81}	& \frac{2}{81}				& \frac{4}{27} 	& \frac{4}{81}\\    [2pt]
 \Cline{1pt}{2-6}  \\ [-0.9em]
\multirow{2}{*}{1 }  & 0&  \frac{1}{3} & \frac{1}{15}	& \frac 25	&0 \\   [4pt]
  & 1 & 0	& \frac 35				&\frac{4}{15}				& \frac 13
\end{array}
\label{eqn:PAX_purified_mix}
\een
We can consider all possible extension of $\mathrm{P_A} \rightarrow \mathrm{P_{AE}}$, which will take care of all possible ensembles of $\mathrm{P_A}$.

%One can check that passing the output $b$, through this post-processing channel $X$ can be able to realize the $\{p_m, P_M^m\}$ in part $A$.
\subsubsection{Quantifying non-locality introduced in NSCE}\label{sec:CEnonlocal}
Here we quantify the amount of non-locality introduced among the extending and the extended system in {the process of the construction of the NSCE}, following Definition \ref{def:complete_ext}.

We have observed the fact that the completely extended behaviour of the maximally mixed single input output  box, has turned out to be the Popescu-Rohrlich box \cite{PR}, which  (under suitable pre and post processing)  can
%and it 
violate any kind of  bipartite  Bell expression maximally. In that case, the maximal amount of non-locality was introduced in {the process of the construction of the NSCE}. 
On the other hand, to quantify the non-locality of the NSCE of the non-maximally mixed single input output  behaviour given in Eq. (\ref{eqn:our_box}), the NSCE is shown in Eq. (\ref{eqn:PAX_purified}),  to have different cardinalities  of outputs.
%. If we forget about the redundant entries in the output of the purifying system then also $e \in \{0,1,2\}$ for $z = 0$, while $a \in \{0,1\}$ for all $x$.
To get rid of this asymmetry in  the extending system  of $X$, we can do two possible surgeries. \\
{\bf Case 1:} One can add one more outputs in the purified system and calculate the Bell like inequality defined by Collins et. al.  (CGLMP) \cite{CollinsGLMP-Bell-hdim} (the acronym is after it finders D. Collins, N. Gissin, N. Linden, S. Massar and S. Popescu).  The behaviour which maximizes the  CGLMP bound has the following form after a local relabeling of the inputs and outputs
\ben
\mathrm{P_{AE}}\left(ae|xz\right)=
\begin{array}{cc?ccc?ccc}
 &\multicolumn{1}{c}{x} & \multicolumn{3}{c}{0}	& \multicolumn{3}{c}{1} \\
z &$\diagbox[width=1.8em, height=1.8em, innerrightsep=0pt]{$e$}{$a$}$  & 0	& 1&2	&0  & 1& 2\\
 \Cline{1pt}{2-8}  \\ [-1em]
\multirow{3}{*}{0 }  &0 & 1/3	& 0			&0	&0 			& 1/3 &0\\
& 1 & 0			& 2/3	&0 & 2/3 	& 0  &0 \\
 & 2 & 0	& 0				&0 				& 0 &0 				& 0\\
 \Cline{1pt}{2-8} \\ [-1em]
 \multirow{3}{*}{1 }  & 0	& 1/3	& 0	&0			& 1/3		& 0 &0\\
 & 1	& 0			& 1/3	&0	&0			 	& 1/3 &0\\ 
 & 2	& 0			& 1/3	&  0 &1/3		& 0 &0 
\end{array}
\label{eqn:PAX_CGLMP}
\een
For this bipartite two inputs and three-output box, the CGLMP bound turns out to be $3$,  which is beyond the quantum limit quoted to be $2.87$ in  Ref. \cite{CollinsGLMP-Bell-hdim}.

\noindent{\bf Case 2:} Another way to calculate the non-locality of this asymmetric behaviour by following the prescription giving in Ref. \cite{Pironio2005lift}. It proposes to merge the extra outcomes in the following way
\ben
\mathrm{P'_{AE}}\left(ae = 1|xz\right) = \mathrm{P_{AE}}\left(ae = 1|xz\right) + \mathrm{P_{AE}}\left(ae = 2|xz\right) \nonumber \\
\een
Hence the behaviour in Eq. (\ref{eqn:PAX_purified}) can be transformed to a  bipartite binary input output box,
\ben
\mathrm{P'_{AE}}\left(ae|xz\right)=
\begin{array}{cc?cc?cc}
& \multicolumn{1}{c}{x} & \multicolumn{2}{c}{0}	& \multicolumn{2}{c}{1} \\
z &$\diagbox[width=1.8em, height=1.8em, innerrightsep=0pt]{$e$}{$a$}$  & 0	& 1	&0  & 1\\
 \Cline{1pt}{2-6}  \\ [-1em]
\multirow{2}{*}{0 }  & 0	& 1/3	 & 0	& 1/3		& 0 \\
 & 1	& 0			& 2/3	&1/3	 & 1/3 \\ 
 \Cline{1pt}{2-6} \\ [-1em]
 \multirow{2}{*}{1 }  &0 & 1/3			&0	&0 			& 1/3 \\
& 1 & 0			& 2/3	 & 2/3 	& 0   \\
\end{array}
\label{eqn:PAX_CGLMP_22}
\een
For this behaviour we have the well known CHSH inequality to quantify the non-locality, and it is $3.33$, which is also beyond the quantum limit. 
However, the amount of non-locality for this NSCE is substantially less than the amount of non-locality present in a PR box.  It therefore seems that it may be possible to quantify the non-locality 
between subsystem and its extending system as a measure of how close a behaviour is to being a vertex  {for NSCE} in the  {theory of non-signaling behaviours}.

Until now, we have given examples in favor of the various properties of NSCE we have discovered so far. Now we want to shed some light on another aspects of NSCE which shows a sharp disparity with the purification principle of the QT. If $|\psi_{AE}\rangle$, is the purification of a quantum state $\rho_A$,  to system $E$, then the same pure state is also the purification of quantum state $\rho_X = \text{tr}_A |\psi_{AE}\rangle\langle \psi_{AE}|$. In the latter section we give an example to show that this is not the case for the NSCE. If we have a behaviour $P_A$, and $P_{AE}$ is its NSCE, then we say the behaviour $P_E = \text{tr}_A P_{AE}$ is the conjugate box. We are going now to construct the NSCE of the conjugate box.

\subsubsection{Complete extension of the conjugate box}\label{sec:CEconjugate}
In this section, we will find the  NSCE of the   conjugate behaviour of the behaviour given in Eq. (\ref{eqn:our_box}).
% If $P_{AX}$ is the complete extension of box $P_A$, then we will say the system $X$, as the conjugate purifying system, and the box belongs to that system 
The conjugate box, $P_E$ can be obtained from Eq. (\ref{eqn:PAX_purified}), by $P_E(e|z) = \sum_a P_{AE}(ae|xz)$, and it is given by
\ben
\mathrm{P_E}(e|z)=
\begin{array}{c|c|c}
$\diagbox[width=1.6em, height=1.6em, innerrightsep=0pt]{$e$}{$z$}$ & 0 & 1\\ \hline	 \\[-1em]
	 0&  1/3 &  1/3 \\ \hline \\[-1em]
	 1 & 1/3 & 2/3 \\ \hline \\[-1em]
	 2 & 1/3 & 0
\end{array}
\label{eqn:purified_box}
\een
This behaviour lies in a $4$ dimensional  behaviour  polytope whose vertices are given by
\ben
\mathrm{P}^0_\texttt{E} =
\begin{array}{c|c|c}
$\diagbox[width=1.6em, height=1.6em, innerrightsep=0pt]{$e$}{$z$}$ & 0 & 1\\ \hline	 \\[-1em]
	 0&  1 &  1 \\ \hline \\[-1em]
	 1 & 0 & 0 \\ \hline \\[-1em]
	 2 & 0 & 0
\end{array}, ~~~
\mathrm{P}^1_\texttt{E} =
\begin{array}{c|c|c}
$\diagbox[width=1.6em, height=1.6em, innerrightsep=0pt]{$e$}{$z$}$ & 0 & 1\\ \hline	 \\[-1em]
	 0&  0 & 0 \\ \hline \\[-1em]
	 1 & 1 & 1 \\ \hline \\[-1em]
	 2 & 0 & 0
\end{array}, ~~~
\mathrm{P}^2_\texttt{E} =
\begin{array}{c|c|c}
$\diagbox[width=1.6em, height=1.6em, innerrightsep=0pt]{$e$}{$z$}$ & 0 & 1\\ \hline	 \\[-1em]
	 0&  0 & 0 \\ \hline \\[-1em]
	 1 & 0 & 0 \\ \hline \\[-1em]
	 2 & 1 & 1
\end{array}, \nonumber \\ \nonumber\\
\mathrm{P}^3_\texttt{E} =
\begin{array}{c|c|c}
$\diagbox[width=1.6em, height=1.6em, innerrightsep=0pt]{$e$}{$z$}$ & 0 & 1\\ \hline	 \\[-1em]
	 0&  1 & 0 \\ \hline \\[-1em]
	 1 & 0 & 1 \\ \hline \\[-1em]
	 2 & 0 & 0
\end{array}, ~~~
\mathrm{P}^4_E =
\begin{array}{c|c|c}
$\diagbox[width=1.6em, height=1.6em, innerrightsep=0pt]{$e$}{$z$}$ & 0 & 1\\ \hline	 \\[-1em]
	 0 & 1 & 0 \\ \hline \\[-1em]
	 1 & 0 & 0 \\ \hline \\[-1em]
	 2 & 0 & 1
\end{array}, ~~~
\mathrm{P}^5_\texttt{E} =
\begin{array}{c|c|c}
$\diagbox[width=1.6em, height=1.6em, innerrightsep=0pt]{$e$}{$z$}$ & 0 & 1\\ \hline	 \\[-1em]
	 0 & 0 & 1 \\ \hline \\[-1em]
	 1 & 1 & 0 \\ \hline \\[-1em]
	 2 & 0 & 0
\end{array}, \nonumber \\ \nonumber \\
\mathrm{P}^6_\texttt{E} =
\begin{array}{c|c|c}
$\diagbox[width=1.6em, height=1.6em, innerrightsep=0pt]{$e$}{$z$}$ & 0 & 1\\ \hline	 \\[-1em]
	 0 & 0 & 1 \\ \hline \\[-1em]
	 1 & 0 & 0 \\ \hline \\[-1em]
	 2 & 1 & 0
\end{array}, ~~~
\mathrm{P}^7_\texttt{E} =
\begin{array}{c|c|c}
$\diagbox[width=1.6em, height=1.6em, innerrightsep=0pt]{$e$}{$z$}$ & 0 & 1\\ \hline	 \\[-1em]
	 0 & 0 & 0 \\ \hline \\[-1em]
	 1 & 0 & 1 \\ \hline \\[-1em]
	 2 & 1 & 0
\end{array}, ~~~
\mathrm{P}^8_\texttt{E} =
\begin{array}{c|c|c}
$\diagbox[width=1.6em, height=1.6em, innerrightsep=0pt]{$e$}{$z$}$ & 0 & 1\\ \hline	 \\[-1em]
	 0 & 0 & 0 \\ \hline \\[-1em]
	 1 & 1 & 0 \\ \hline \\[-1em]
	 2 & 0 & 1
\end{array}. 
\een

To obtain the  NSCE of this box, we need to find the minimal ensembles of $\mathrm{P_E}$, which are
\ben
\cm_0(\mathrm{P_E}) &=& \{(1/3, \mathrm{P_\texttt{E}^0});(1/3,\mathrm{P_\texttt{E}^1});(1/3,\mathrm{P_\texttt{E}^7})\}, \label{eqn:Min1}\\
\cm_1(\mathrm{P_E}) &=& \{(1/3, \mathrm{P_\texttt{E}^1});(1/3,\mathrm{P_\texttt{E}^3});(1/3,\mathrm{P_\texttt{E}^6})\},  \\ \label{eqn:Min2}
\cm_2(\mathrm{P_E}) &=& \{(1/3, \mathrm{P_\texttt{E}^3});(1/3,\mathrm{P_\texttt{E}^5});(1/3,\mathrm{P_\texttt{E}^7})\}.  \label{eqn:Min3}
\een 
Consider the  NSCE  of $\mathrm{P_E}$ to a system $A'$, as $\mathrm{P_{A'E}}(a'e|x'z)$, where $\{p(a'=i|x' = k), \mathrm{P_E}^{ik}(e|z)\} = \cm_k$, the  NSCE of $\mathrm{P_E}$ is
\ben
\mathrm{P_{A'E}}\left(a'e|x'z\right)=
\begin{array}{cc?ccc?ccc?ccc}
 &\multicolumn{1}{c}{x'} & \multicolumn{3}{c}{0}	& \multicolumn{3}{c}{1} & \multicolumn{3}{c}{2} \\
z &$\diagbox[width=1.8em, height=1.8em, innerrightsep=0pt]{$e$}{$a'$}$  & 0	& 1&2 &0  & 1 &2 &0  & 1 &2\\
	\Cline{1pt}{2-11} \\ [-0.9em]
 \multirow{3}{*}{0 }  & 0	& \frac 13	& 0	&0   	&0 & \frac 13	 &0	&\frac 13	& 0 &0\\
 & 1	& 0	& \frac 13	&0	& \frac 13   &0   &0   &0	& \frac 13	&0\\
 & 2	& 0	& 0	& \frac 13	& 0	& 0	& \frac 13 & 0	& 0	& \frac 13\\ [0.15em]
 \Cline{1pt}{2-11}  \\ [-0.9em]
\multirow{2}{*}{1 }   &0 & \frac 13 & 0	& 0	& 0	& 0  & \frac 13 &0  & \frac 13  & 0\\
& 1 & 0 &\frac 13	&\frac 13	&\frac 13	&\frac 13	& 0 &\frac 13 	& 0   & \frac 13\\
\end{array}
\label{eqn:PAX_conjugate_purified}
\een

 One can clearly see  that $\mathrm{P_{AE}} \neq \mathrm{P_{A'E}} $,  for example, by noticing  the mismatch of the cardinality of the inputs of the extended party of the conjugate system.
%\textcolor{red}{ Now we want to calculate the assymetry in the purification of the purified system of the given box.}

\subsection{Dimensionality of the No-signaling complete extension}\label{sec:app:dim}

In this subsection we provide a proof for Theorem \ref{thm:limitedDim} of the main text.

\begin{proof}
Let us recall, that in quantum theory (due to the Schmidt decomposition) purifications can always be found with an extending system with the same dimension as the extended system. We are interested in similarly quantifying the size of the minimal  NSCE. However, there are many possible quantifiers that we could use here, for example, we can measure the size of the  NSCE  using the total number of its outputs, inputs and outputs, or, the dimension of its state space (vector space). The theory of non-signalling behaviours is a discrete convex theory, hence we call its state space a convex polytope, a multidimensional generalization of a convex polyhedron. In this section, we strive to upper bound the dimension of the polytope that contains NSEA of some fixed but arbitrary behaviour $\mathrm{P}$. We identify the dimension of behaviour with the dimension of a polytope it belongs. Although we chose the dimension of NSEA, as a quantifier of its size, other aspects of it, like the number of its inputs and outputs, will also be discussed in the following.

From Theorem 1 of \cite{Pironio2005lift} we know that the dimension of a certain behaviour polytope 
$\mathcal{B}$ is:
\begin{align}\label{eq:dim_def}
    \dim \mathcal{B} = \prod_{i=1}^{n} \left(\sum_{j=1}^{m_i}(v_{ij}-1)+1\right)-1,
\end{align}
where $n$ is the total number of non-signalling parties, $m_i$ is the total number of the inputs of the $i^\mathrm{th}$ party, with $v_{ij}$ being the  number of outputs for the $j^\mathrm{th}$ input. This polytope of $\dim \mathcal{B}$ is contained in the vector space of $\mathbb{R}^t$, where $t=\prod_{i=1}^{n} \sum_{j=1}^{m_i} v_{ij}$ is the total number of outputs.

Suppose now that $\mathrm{P} \in \mathcal{B}$, then there exists a polytope $\tilde{\mathcal{B}}$, such that the NSEA $\mathcal{E}(P) \in \tilde{\mathcal{B}}$. The dimension of $\tilde{\mathcal{B}}$ can be determined by:
\begin{align}
    \dim \tilde{\mathcal{B}} &= \prod_{i=1}^{n+1} \left(\sum_{j=1}^{m_i}(v_{ij}-1)+1\right)-1 \\
    &=\left(\sum_{j=1}^{m_{n+1}}(v_{n+1,j}-1)+1\right) \left(\dim \mathcal{B}+1\right)-1,
\end{align}
where $n+1$ is the index of the extending party, and the last equality is due to expansion of product with respect to $n+1$-th term. 

To obtain the upper bound on $\dim \tilde{\mathcal{B}}$ it is enough now to find upper bounds on the number of inputs and outputs of the extending system. The following considerations have a qualitative character, we are much more interested in the fact that an upper bound on $\dim \tilde{\mathcal{B}}$ exists than in its tightness. The upper bound on the number of outputs can be found via Carath\'{e}odory theorem \cite{Ziegler1995}, i.e., $v_{n+1,j} \le \dim \mathcal{B}+1$, we obtain:
\begin{align}
   &\dim \tilde{\mathcal{B}} \le \left(m_{n+1}\dim \mathcal{B} +1\right) \left(\dim \mathcal{B}+1\right)-1.
\end{align}
It is instructive to notice that the above inequality can be saturated for some generic behaviours. The number of inputs of the extending system is equal to the number of minimal ensembles of the behaviour $\mathrm{P}$. Suppose now that $V$ is the number of vertices of the polytope $\mathcal{B}$. Then, using Carath\'{e}odory theorem again, each choice of $\dim \mathcal{B}+1$ vertices (out of $V$) leads to a minimal ensemble via elimination of the vertices, therefore $m_{n+1} \le \binom{V}{\dim \mathcal{B}+1}$, and hence:
\begin{align}
    &\dim \tilde{\mathcal{B}} < \left(\binom{V}{\dim \mathcal{B}+1}\dim \mathcal{B} +1\right) \left(\dim \mathcal{B}+1\right),
\end{align}
where the last term ($-1$) was neglected. 

The characterisation of the vertices of polytopes of behaviours with two binary inputs and  two binary outputs in bipartite and tripartite scenarios was provided in \cite{Barret-Roberts}. In general, however, finding vertices of a polytope is a hard problem known as the ``vertex enumeration problem''. Using McMullen’s Upper Bound  Theorem \cite{McMullenUpperBound,Ziegler1995,Vertices}, we can, however, still obtain an upper bound for the number of vertices in terms of $\dim \mathcal{B}$ and $t$. Here we use a fact that a polytope 
can be defined as an intersection of a set of halfspaces. Then following \cite{Vertices} (up to notation) we have the subsequent statement. 

Let A be an $m \times t$ matrix of reals and let $b \in \mathbb{R}^m$ be a real vector. Consider a polytope 
\begin{align}\label{def:polyhedron}
    \mathcal{P}=\left\{x \in \mathbb{R}^t: Ax \le b\right\},
\end{align} 
and $V_\mathcal{P}$ be the number of vertices of $\mathcal{P}$ then:
\begin{align}
    V_\mathcal{P} \le \binom{m-t-s}{s}+\binom{m-s-1}{t-s-1},
\end{align}
where $s=\left \lfloor{t/2}\right \rfloor$.

To achieve our goal it is enough to determine the number of rows $m$ of matrix $A$. We divide now all the constraints on the non-signalling polytope into three classes, written in terms of marginal probabilities $p_{ijk}$:
\begin{enumerate}
    \item Probabilistic constraints: $0 \le p_{ijk} \le 1$.
    \item Normalization constraints, i.e, $\sum_k^{v_{ij}} p_{ijk}=1$.
    \item non-signalling constraints (see Section \ref{sec:CE-def}).
\end{enumerate}
It is easy to see that constraints of type $1$ contribute to $2t$ rows of matrix A. In the next step we observe that constraints of type $2$ and $3$ are linear and together reduce the dimension of the space of correlations from $t$ to $\dim \mathcal{B}$, and hence can be encoded in $t-\dim \mathcal{B}$ linearly independent rows of matrix $A$ (irrespectively on the actual number of constraints). The total number of rows is therefore upper bound with $m \le 3t-\dim \mathcal{B}$ (there still is some redundancy because of the relation between constraints of types $1$ and $2$), hence:
\begin{align}
    V \le \binom{2t -\left \lfloor{t/2}\right \rfloor - \dim \mathcal{B} }{\left \lfloor{t/2}\right \rfloor}+ \binom{3t -\left \lfloor{t/2}\right \rfloor - \left(\dim \mathcal{B}+1\right) }{t-\left \lfloor{t/2}\right \rfloor-1}.
\end{align}

Finally, we obtain
\begin{align}
         &\dim \tilde{\mathcal{B}} <\left(\dim \mathcal{B}+1\right) \nonumber\\
         &\times \left(\binom{\binom{2t -\left \lfloor{t/2}\right \rfloor - \dim \mathcal{B} }{\left \lfloor{t/2}\right \rfloor}+ \binom{3t -\left \lfloor{t/2}\right \rfloor - \left(\dim \mathcal{B}+1\right) }{t-\left \lfloor{t/2}\right \rfloor-1}}{\dim \mathrm{B}+1}\dim \mathcal{B} +1\right),
    \end{align}
     where:
     \begin{align}
         &\dim \mathcal{B} = \prod_{i=1}^{n} \left(\sum_{j=1}^{m_i}(v_{ij}-1)+1\right)-1,~~t=\prod_{i=1}^{n} \sum_{j=1}^{m_i} v_{ij}.
     \end{align}

\end{proof}

In Theorem \ref{thm:limitedDim} we were interested only in showing that the dimension of the complete extension is bounded from above, and hence our result is very loose.  In particular the number of candidates for a minimal ensemble is very inefficient upper bound on the number of minimal ensembles, although obtained with simple combinatorics.  Therefore, the first place for improvement is to find tighter upper bound on the number of minimal ensembles. A more rigorous treatment can show that many candidates lead to the same minimal ensemble and moreover that some of them are not valid. Furthermore, the upper bound on the number of vertices that we used is very general and does not incorporate, for example, symmetries of the non-signalling polytope.

\subsection{Complete extensions of three-cycle contextual behaviour}\label{sec:spekkens}

\textcolor{black}{In the previous sections, we have considered the NSCE of an arbitrary behavior which is inherently of the form $P_A(a|x)$, 
 i.e.,  a conditional probability distribution consist of only single set of input $x$ and output $a$. The polytope of those behaviors are completely determined by the constraints probability distributions satisfy, like non-signaling and normalization condition. Here, 
within the theory of non-signalling behaviours (NS) we will study contextuality scenarios, and, hence, investigate NSCE of contextual behaviours. The polytope of contextual behaviors are not simply the non-signalling polytopes but rather they are termed as no-disturbance polytopes. And while considering the minimal ensembles of a contextual behavior we will focus on the pure members of the no-disturbance polytopes. } In particular we are going to focus on the three-cycle contextuality scenario \cite{abramsky2011sheaf,Cabello2013-noncontext} (aka Specker's triangle \cite{specker1990logik,liang2011specker}).

First we will introduce this contextuality scenario in standard notation before showing how to recast it within NS. The three-cycle scenario consists of a triple of binary observables $\{X_0, X_1, X_2\}$ with outcomes $\{a,b,c\}$ valued in $\{-1,1\}$, together with the constraint that these are pairwise compatible -- that is, any pair $\{X_i,X_j\}$ can be jointly measured without disturbance, but are globally incompatible -- that is, there is no way to measure all three without disturbance. Such a setup can be realised in quantum theory \cite{kunjwal2014quantum} provided that the observables are not taken to correspond to projective measurements. This compatibility structure can be captured by the set of maximal contexts  \cite{abramsky2011sheaf,Cabello2013-noncontext}
\begin{equation}
\mathscr{C}_3 = \{\{X_0,X_1\}, \{X_1,X_2\},\{X_2,X_0\}\}.
\end{equation}
The behaviour of a particular realisation of this scenario is captured by the joint distributions over the outcomes of the compatible observables, i.e., $p(a,b)$; $p(b,c)$; and $p(c,a)$, which can be represented in the following table:
\ben
\begin{array}{c|c|c?c|c|c?c|c|c}
$\diagbox[width=1.8em, height=1.8em, innerrightsep=0pt]{$a$}{$b$}$ & $+1$ & $-1 $ & $\diagbox[width=1.8em, height=1.8em, innerrightsep=0pt]{$b$}{$c$}$ & $+1$ & $-1 $ & $\diagbox[width=1.8em, height=1.8em, innerrightsep=0pt]{$c$}{$a$}$ & $+1$ & $-1 $ \\
\Cline{0.5pt}{1-9} \\ [-1em]
$+1$ & $$ & $$ & $+1$ & $$ & $$ & $+1$ & $$ & $$ \\ \hline
$-1$ & $$ & $$ & $-1$ & $$ & $$ & $-1$ & $$ & $$  
\end{array}
\een
Any such behaviour must satisfy certain no-disturbance constraints\footnote{
As it should follow the no-disturbance condition like the non-signaling one in non-locality, which is
\ben
p(a) = \sum_{b = -1}^1 p(a,b) = \sum_{c = -1}^1 p(c,a) \\
p(b) = \sum_{a = -1}^1 p(a,b) = \sum_{c = -1}^1 p(b,c) \\
p(c) = \sum_{b = -1}^1 p(b,c) = \sum_{a = -1}^1 p(c,a) 
\een
}
 \cite{abramsky2011sheaf,Cabello2013-noncontext}, which together define the no-disturbance polytope. Formally defining contextuality is beyond the scope of this paper, but, for our purposes it suffices to note that a behaviour is non-contextual if it is a convex combination of the eight deterministic vertices of the no-disturbance polytope
 \ben
{\cal N}^{\cal C}_0 = 
\begin{array}{c|c?c|c?c|c}
 $1$ & $0$ & $1$ & $0$ &  $1$ & $0$ \\ \hline
 $0$ & $0$ & $0$ & $0$ &  $0$ & $0$ 
\end{array}~~~
{\cal N}^{\cal C}_1 = 
\begin{array}{c|c?c|c?c|c}
 $1$ & $0$ & $0$ & $1$ &  $0$ & $0$ \\ \hline
 $0$ & $0$ & $0$ & $0$ &  $1$ & $0$ 
\end{array} \nonumber
\een
\ben
{\cal N}^{\cal C}_2 = 
\begin{array}{c|c?c|c?c|c}
 $0$ & $1$ & $0$ & $0$ &  $1$ & $0$ \\ \hline
 $0$ & $0$ & $1$ & $0$ &  $0$ & $0$ 
\end{array}~~~
{\cal N}^{\cal C}_3 = 
\begin{array}{c|c?c|c?c|c}
 $0$ & $0$ & $1$ & $0$ &  $0$ & $1$ \\ \hline
 $1$ & $0$ & $0$ & $0$ &  $0$ & $0$ 
\end{array}~~~
{\cal N}^{\cal C}_4 = 
\begin{array}{c|c?c|c?c|c}
 $0$ & $1$ & $0$ & $0$ &  $0$ & $0$ \\ \hline
 $0$ & $0$ & $0$ & $1$ &  $1$ & $0$ 
\end{array} \nonumber
\een
\ben
{\cal N}^{\cal C}_5 = 
\begin{array}{c|c?c|c?c|c}
 $0$ & $0$ & $0$ & $1$ &  $0$ & $0$ \\ \hline
 $1$ & $0$ & $0$ & $0$ &  $0$ & $1$ 
\end{array}~~~
{\cal N}^{\cal C}_6 = 
\begin{array}{c|c?c|c?c|c}
 $0$ & $0$ & $0$ & $0$ &  $0$ & $1$ \\ \hline
 $0$ & $1$ & $1$ & $0$ &  $0$ & $0$ 
\end{array} ~~~
{\cal N}^{\cal C}_7 = 
\begin{array}{c|c?c|c?c|c}
 $0$ & $0$ & $0$ & $0$ &  $0$ & $0$ \\ \hline
 $0$ & $1$ & $0$ & $1$ &  $0$ & $1$ 
\end{array} \nonumber
\een
otherwise, it is a contextual behaviour. 
Any other vertex of the no-disturbance polytope is therefore contextual, indeed, there are four of these which are
\ben
{\cal C}_0 = 
\begin{array}{c|c?c|c?c|c}
 $1/2$ & $0$ & $1/2$ & $0$ &  $0$ & $1/2$ \\ \hline
 $0$ & $1/2$ & $0$ & $1/2$ &  $1/2$ & $0$ 
\end{array} ~~~~~~
{\cal C}_1 = 
\begin{array}{c|c?c|c?c|c}
 $1/2$ & $0$ & $0$ & $1/2$ &  $1/2$ & $0$ \\ \hline
 $0$ & $1/2$ & $1/2$ & $0$ &  $0$ & $1/2$ 
\end{array}\nonumber
\een
\ben
{\cal C}_{2} = 
\begin{array}{c|c?c|c?c|c}
 $0$ & $1/2$ & $1/2$ & $0$ &  $1/2$ & $0$ \\ \hline
 $1/2$ & $0$ & $0$ & $1/2$ &  $0$ & $1/2$ 
\end{array}~~~~~~
{\cal C}_3 = 
\begin{array}{c|c?c|c?c|c}
 $0$ & $1/2$ & $0$ & $1/2$ &  $0$ & $1/2$ \\ \hline
 $1/2$ & $0$ & $1/2$ & $0$ &  $1/2$ & $0$ 
\end{array}\nonumber.
\een

 Note that  the dimension of the no-disturbance polytope is $6$, hence, according to the Carath\'{e}odory theorem \cite{Ziegler1995}, the set of minimal ensembles of an arbitrary behaviour inside the polytope has at most $7$ elements. 

We will now focus on a particular class of behaviours within this polytope, namely those lying on an isotropic line -- a mixture of a contextual vertex, say ${\cal C}_0$ and the maximally mixed behaviour\footnote{
The maximally mixed behaviour $\mathrm{M}$ is given by 
\ben
M =
\begin{array}{c|c?c|c?c|c}
 \frac{1}{4} &  \frac{1}{4} & \frac{1}{4} & \frac{1}{4} &  \frac{1}{4} &  \frac{1}{4} \\ [2pt] \hline \\ [-0.9em]
 \frac{1}{4} &  \frac{1}{4} &  \frac{1}{4} &  \frac{1}{4} &  \frac{1}{4}  &  \frac{1}{4}
\end{array}
\een}
%Here, we are trying to find the  NSEA  of a behaviour 
%which is in the isotropic line, i.e., a mixture of a contextual behaviour say ${\cal C}_0$ (Speken's triangle \jnote{Add citation}) and the maximally mixed behaviour
which is a non-contextual behaviour. We denote an element on this isotropic line (parameterised by $\lambda \in (0,1)$) as $\mathrm{P}_\lambda$ which can be explicitly written as
\ben
\mathrm{P}_\lambda &=& (1 - \lambda){\cal C}_0 + \lambda M  \nonumber \\ &=& 
\begin{array}{c|c?c|c?c|c}
 \frac{2-\lambda}{4} & \frac{\lambda}{4} & \frac{2-\lambda}{4} & \frac{\lambda}{4} &  \frac{\lambda}{4} & \frac{2-\lambda}{4} \\ [2pt] \hline \\ [-0.9em]
 \frac{\lambda}{4} & \frac{2-\lambda}{4} & \frac{\lambda}{4} & \frac{2-\lambda}{4} &  \frac{2-\lambda}{4} & \frac{\lambda}{4} 
\end{array}
\label{contextual_box}
\een

The minimal ensembles of $\mathrm{P}_\lambda$, have been found by applying numerical techniques  to obtain   the solution of a set of linear equations, and they  are as follows.

%{\bf Having} 
 For all $\lambda \in (0,1)$ we have the following 
\ben
v &=& \Big[~\frac{\lambda}{4},  ~~~~ \frac{\lambda}{4}, ~1 - \frac{\lambda}{2}\Big] \\
{\cal M}_1(\mathrm{P}) &=& \Big[ {\cal N^C}_2, ~{\cal N^C}_5, ~ ~{\cal C}_0 ~\Big]
\een
\ben
v &=& \Big[~\frac{\lambda}{4},  ~~~~\frac{\lambda}{4}, ~1 - \lambda, ~~~\frac{\lambda}{2}~\Big] \\
{\cal M}_2 &=& \Big[ {\cal N^C}_3, ~{\cal N^C}_4, ~~ {\cal C}_0,  ~~~{\cal C}_1~\Big] \\ 
{\cal M}_3 &=& \Big[ {\cal N^C}_1, ~{\cal N^C}_6, ~~ {\cal C}_0,  ~~~{\cal C}_{2}~\Big] \\
{\cal M}_4 &=& \Big[ {\cal N^C}_0, ~{\cal N^C}_7, ~~ {\cal C}_0,  ~~~{\cal C}_{3}~\Big] 
\een
\ben
v &=& \Big[1- \frac{3\lambda}{4},  ~\frac{\lambda}{4}, ~~~\frac{\lambda}{4},  ~~~ \frac{\lambda}{4}~\Big] \\
{\cal M}_5 &=& \Big[ ~~~~{\cal C}_0, ~~~{\cal C}_1, ~~~ {\cal C}_{2}, ~~ {\cal C}_{3} ~\Big]
\een
\ben
v &=& \Big[~\frac{\lambda}{4},  ~~~~\frac{\lambda}{4}, ~~~~\frac{\lambda}{4}, ~~~~\frac{\lambda}{4}, ~1 - \lambda \Big] \\
{\cal M}_6 &=& \Big[ {\cal N^C}_1, ~{\cal N^C}_2, ~ {\cal N^C}_3,  ~{\cal N^C}_7, ~ {\cal C}_0~\Big] \\ 
{\cal M}_7 &=& \Big[ {\cal N^C}_0, ~{\cal N^C}_4, ~ {\cal N^C}_5,  ~{\cal N^C}_6, ~{\cal C}_{0}~\Big] 
\een
%{\bf Having} 
 In $\lambda \in (0, \frac 23)$, we additionally have 
%all the above $7$ ensemble and the following
\ben
v &=& \Big[~\frac{\lambda}{4},  ~~~~\frac{\lambda}{4}, ~~~~\frac{\lambda}{4}, ~~~~\frac{\lambda}{4}, ~~~~\frac{\lambda}{4}, ~~~~\frac{\lambda}{4}, 1 - \frac{3\lambda}{2} \Big] \hspace{2em}\\
{\cal M}_8 &=& \Big[ {\cal N^C}_0, {\cal N^C}_1, ~ {\cal N^C}_3,  \,\,{\cal N^C}_4, \, {\cal N^C}_6,  \,{\cal N^C}_7, ~\, {\cal C}_0~~\Big] 
\een
 Whilst in  $\lambda  \in ( \frac 23, 1)$, the behaviour has another $5$ minimal ensembles.  However,  in this range of $\lambda$, $\mathrm{P}_\lambda$ is non-contextual (see Theorem 1 in \cite{Cabello2013-noncontext})  and we are interested in exploring the case in which $\mathrm{P}_\lambda$ displays contextuality so we will not go into detail about these. 

 To study the NSEAs for contextuality scenarios we must first recast the scenario into the theory of non-signalling behaviours. In particular, rather than describing the scenario by a triple of bipartite distributions (i.e., $p(a,b)$, $p(b,c)$, and $p(c,a)$) we will instead describe the scenario by a conditional probability distribution. We do this by associating the input $x\in \{0,1,2\}$  to the maximal contexts, i.e., $0 \sim \{X_0,X_1\}$, $1 \sim \{X_1,X_2\}$, and $2 \sim \{X_2,X_0\}$; and associating an outcome $a'\in\{0,1,2,3\}$ to the outcome of the bipartite distributions $\{+1,-1\}\times\{+1,-1\}$. This defines an embedding of the no-disturbance polytope into the polytope of single input output behaviours (with the relevant dimensions). 

In our particular example of interest, namely $\mathrm{P}_\lambda$ using this embedding we can associate the behaviour of the contextuality scenario $\mathrm{P}_\lambda$ to the single input output behaviour
\ben\label{box_interest2}
\mathrm{P'_A}_\lambda = 
\begin{array}{c|c|c|c}
$\diagbox[width=1.8em, height=1.8em, innerrightsep=0pt]{$a'$}{$x$}$ & 0 & 1 & 2  
\\ \Cline{0.5pt}{1-4} \\ [-0.8em]
$0$ & \frac{2-\lambda}{4}  & \frac{2-\lambda}{4} & \frac{\lambda}{4} \\ [0.7ex] \Cline{0.5pt}{1-4} \\ [-0.8em]
$1$ & \frac{\lambda}{4} & \frac{\lambda}{4} & \frac{2-\lambda}{4} \\ [0.7ex] \Cline{0.5pt}{1-4} \\ [-0.8em]
$2$ & \frac{\lambda}{4} & \frac{\lambda}{4} & \frac{2-\lambda}{4} \\ [0.7ex] \Cline{0.5pt}{1-4} \\ [-0.8em]
$3$ & \frac{2-\lambda}{4}  & \frac{2-\lambda}{4} & \frac{\lambda}{4} 
\end{array}
\een

% To find out the  NSEA  of the 3-cycle contextual behaviour given in Eq. (\ref{contextual_box}), we will now map it in the following way:
% Consider $\mathrm{P'}(a'|x = 0) = p(a,b)$ with a one to one correspondence between  $a' \in \{0,1,2,3\}$ and $a,b \in \{+1,-1\}\times  \{+1,-1\}$. Similarly $\mathrm{P'}(a'|x = 1) = p(b,c)$  and  $\mathrm{P'}(a'|x = 2) = p(c,a)$.
% Hence the single input output behaviour now has the following form
% %\end{minipage}
% %\begin{minipage}{0.5\textwidth}
% \ben\label{box_interest2}
% \mathrm{P'_A} = 
% \begin{array}{c|c|c|c}
% $\diagbox[width=1.8em, height=1.8em, innerrightsep=0pt]{$a'$}{$x$}$ & 0 & 1 & 2  
% \\ \Cline{0.5pt}{1-4} \\ [-0.8em]
% $0$ & \frac{2-\lambda}{4}  & \frac{2-\lambda}{4} & \frac{\lambda}{4} \\ [0.7ex] \Cline{0.5pt}{1-4} \\ [-0.8em]
% $1$ & \frac{\lambda}{4} & \frac{\lambda}{4} & \frac{2-\lambda}{4} \\ [0.7ex] \Cline{0.5pt}{1-4} \\ [-0.8em]
% $2$ & \frac{\lambda}{4} & \frac{\lambda}{4} & \frac{2-\lambda}{4} \\ [0.7ex] \Cline{0.5pt}{1-4} \\ [-0.8em]
% $3$ & \frac{2-\lambda}{4}  & \frac{2-\lambda}{4} & \frac{\lambda}{4} 
% \end{array}
% \een

 Now we are in the position that we can construct the NSCE  of ${P'_A}_\lambda$, in particular, focusing on the contextual case in which $\lambda \in (0, \frac 23)$. Note however, that we want to construct the NSCE  not for the minimal ensembles of the behaviour within the single input output polytope, but for the ensembles within the embedded no-disturbance polytope. The NSCE of this will however still live inside the polytope of no signalling behaviours with a pair of inputs and outputs (of suitable dimensions).

In particular, one can show that the NSEA  of ${P'_A}_\lambda$ in $\lambda \in (0, \frac 23)$ is given by 
\ben
\mathrm{P'_{AE}}_\lambda\left(ae|xz\right)= ~~
\begin{array}{cc?cccc?cccc?cccc}
 &\multicolumn{1}{c}{x} & \multicolumn{4}{c}{0}	& \multicolumn{4}{c}{1} & \multicolumn{4}{c}{2} \\
z &$\diagbox[width=1.8em, height=1.8em, innerrightsep=0pt]{$e$}{$a$}$  & 0	& 1 &2 &3	&0  & 1 &2 &3 &0 &1 &2 &3\\
	\Cline{1pt}{2-14} \\ [-0.9em]
 \multirow{3}{*}{0 }  & 0 & 0 & \frac{\lambda}{4} & 0 & 0 & 0 & 0 & \frac{\lambda}{4}  & 0 & \frac{\lambda}{4}  & 0 & 0 &0 \\ 
 & 1	& 0	&0 & \frac{\lambda}{4} & 0 & 0	 & \frac{\lambda}{4}	& 0& 0 &0 &0 &0 & \frac{\lambda}{4}\\
 & 2	& \frac{2-\lambda}{4}	&0 & 0 & \frac{2-\lambda}{4}  & \frac{2-\lambda}{4}	& 0	& 0& \frac{2-\lambda}{4} &0 & \frac{2-\lambda}{4} &\frac{2-\lambda}{4} & 0 \\ [2pt]
 \Cline{1pt}{2-14}  \\ [-0.9em]
 
\multirow{4}{*}{1 }  & 0 & 0 & 0 & \frac{\lambda}{4}  & 0  & \frac{\lambda}{4} & 0 & 0  & 0 & 0 &  \frac{\lambda}{4}   & 0 &0 \\ 
& 1 &  0 & \frac{\lambda}{4}  & 0 & 0  & 0 & 0  & 0 & \frac{\lambda}{4}  & 0 & 0 &  \frac{\lambda}{4}    &0 \\
& 2 & \frac{1 -\lambda}{2}  &  0 & 0 & \frac{1 - \lambda}{2}  & \frac{1 - \lambda}{2}  & 0 & 0  & \frac{1 - \lambda}{2} & 0 & \frac{1 - \lambda}{2} &  \frac{1 - \lambda}{2}   &0 \\
& 3 & \frac{\lambda}{4}  &  0 & 0 & \frac{\lambda}{4}  & 0 & \frac{\lambda}{4}     & \frac{ \lambda}{4} & 0 & \frac{\lambda}{4}  & 0 &  0  &\frac{\lambda}{4} \\ [2pt]
 \Cline{1pt}{2-14}  \\ [-0.9em]
 
 \multirow{4}{*}{2 }  & 0  & \frac{\lambda}{4}  & 0 & 0 & 0 & 0 & \frac{\lambda}{4}  & 0  & 0 & 0 & 0 &  \frac{\lambda}{4}    &0 \\ 
& 1 & 0 & 0 &  0 & \frac{\lambda}{4}    & 0 & 0  & \frac{\lambda}{4} & 0  & 0  &  \frac{\lambda}{4}  & 0  &0 \\
& 2 & \frac{1 -\lambda}{2}  &  0 & 0 & \frac{1 - \lambda}{2}  & \frac{1 - \lambda}{2}  & 0 & 0  & \frac{1 - \lambda}{2} & 0 & \frac{1 - \lambda}{2} &  \frac{1 - \lambda}{2}   &0 \\
& 3 &  0 & \frac{\lambda}{4}    & \frac{\lambda}{4} & 0 & \frac{\lambda}{4} & 0     & 0 & \frac{\lambda}{4} & \frac{\lambda}{4}  & 0 &  0  &\frac{\lambda}{4} \\ [2pt]
 \Cline{1pt}{2-14}  \\ [-0.9em]
 
 \multirow{4}{*}{3 }  & 0 & \frac{\lambda}{4}  & 0 & 0  & 0 & \frac{\lambda}{4}  & 0   & 0 & 0 &  \frac{\lambda}{4}  & 0  &0 & 0\\ 
& 1  & 0 & 0  & 0 & \frac{\lambda}{4}  & 0   & 0 & 0 &  \frac{\lambda}{4}  & 0  &0 & 0 & \frac{\lambda}{4}\\
& 2 & \frac{1 -\lambda}{2}  &  0 & 0 & \frac{1 - \lambda}{2}  & \frac{1 - \lambda}{2}  & 0 & 0  & \frac{1 - \lambda}{2} & 0 & \frac{1 - \lambda}{2} &  \frac{1 - \lambda}{2}   &0 \\
& 3 &  0 & \frac{\lambda}{4}    & \frac{\lambda}{4} & 0  & 0 & \frac{\lambda}{4}      & \frac{\lambda}{4} & 0&  0 & \frac{\lambda}{4}   &\frac{\lambda}{4}  & 0  \\ [2pt]
 \Cline{1pt}{2-14}  \\ [-0.9em]

  \multirow{4}{*}{4 }  & 0 & \frac{4 - 3\lambda}{8}  & 0 & 0  & \frac{4 - 3\lambda}{8} & \frac{4 - 3\lambda}{8}  & 0   & 0 & \frac{4 - 3\lambda}{8} &  0  & \frac{4 - 3\lambda}{8}  &\frac{4 - 3\lambda}{8} & 0\\ 
& 1  & \frac{\lambda}{8} & 0  & 0 & \frac{\lambda}{8}  & 0   & \frac{\lambda}{8} & \frac{\lambda}{8} &  0  & \frac{\lambda}{8}  &0 & 0 & \frac{\lambda}{8}\\
& 2 &0  & \frac{\lambda}{8}  & \frac{\lambda}{8}  &0    & \frac{\lambda}{8} & 0 & 0 & \frac{\lambda}{8} &   \frac{\lambda}{8}  &0 & 0 & \frac{\lambda}{8}\\
& 3 &  0 & \frac{\lambda}{8}    & \frac{\lambda}{8} & 0  & 0 & \frac{\lambda}{8}      & \frac{\lambda}{8} & 0&  0 & \frac{\lambda}{8}   &\frac{\lambda}{8}  & 0  \\ [2pt]
 \Cline{1pt}{2-14}  \\ [-0.9em]
 
  \multirow{5}{*}{5 }  & 0  & \frac{\lambda}{4}  & 0 & 0 & 0 & 0 & \frac{\lambda}{4}  & 0  & 0 & 0 & 0 &  \frac{\lambda}{4}    &0\\ 
& 1 & 0 & \frac{\lambda}{4} & 0 & 0 & 0 & 0 & \frac{\lambda}{4}  & 0 & \frac{\lambda}{4}  & 0 & 0 &0\\
& 2 & 0 & 0 & \frac{\lambda}{4}  & 0  & \frac{\lambda}{4} & 0 & 0  & 0 & 0 &  \frac{\lambda}{4}   & 0 &0 \\
& 3  & 0 & 0  & 0 & \frac{\lambda}{4}  & 0   & 0 & 0 &  \frac{\lambda}{4}  & 0  &0 & 0 & \frac{\lambda}{4}  \\ 
& 4 & \frac{1 -\lambda}{2}  &  0 & 0 & \frac{1 - \lambda}{2}  & \frac{1 - \lambda}{2}  & 0 & 0  & \frac{1 - \lambda}{2} & 0 & \frac{1 - \lambda}{2} &  \frac{1 - \lambda}{2}   &0 \\ [2pt]
 \Cline{1pt}{2-14}  \\ [-0.9em]

 \multirow{5}{*}{6 }  & 0 & \frac{\lambda}{4}  & 0 & 0  & 0 & \frac{\lambda}{4}  & 0   & 0 & 0 &  \frac{\lambda}{4}  & 0  &0 & 0\\ 
 & 1 &  0 & \frac{\lambda}{4}  & 0 & 0  & 0 & 0  & 0 & \frac{\lambda}{4}  & 0 & 0 &  \frac{\lambda}{4}    &0  \\
 & 2	& 0	&0 & \frac{\lambda}{4} & 0 & 0	 & \frac{\lambda}{4}	& 0& 0 &0 &0 &0 & \frac{\lambda}{4}\\
 & 3 & 0 & 0 &  0 & \frac{\lambda}{4}    & 0 & 0  & \frac{\lambda}{4} & 0  & 0  &  \frac{\lambda}{4}  & 0  &0  \\
 & 4 & \frac{1 -\lambda}{2}  &  0 & 0 & \frac{1 - \lambda}{2}  & \frac{1 - \lambda}{2}  & 0 & 0  & \frac{1 - \lambda}{2} & 0 & \frac{1 - \lambda}{2} &  \frac{1 - \lambda}{2}   &0 \\ [2pt]
 \Cline{1pt}{2-14}  \\ [-0.9em]
%  \end{array}
%\een
%\ben 
% \begin{array}{cc?cccc?cccc?cccc}
  \multirow{7}{*}{7 }  & 0 & \frac{\lambda}{4}  & 0 & 0  & 0 & \frac{\lambda}{4}  & 0   & 0 & 0 &  \frac{\lambda}{4}  & 0  &0 & 0\\ 
  & 1  & \frac{\lambda}{4}  & 0 & 0 & 0 & 0 & \frac{\lambda}{4}  & 0  & 0 & 0 & 0 &  \frac{\lambda}{4}    &0\\ 
  & 2 & 0 & 0 & \frac{\lambda}{4}  & 0  & \frac{\lambda}{4} & 0 & 0  & 0 & 0 &  \frac{\lambda}{4}   & 0 &0 \\
 & 3 &  0 & \frac{\lambda}{4}  & 0 & 0  & 0 & 0  & 0 & \frac{\lambda}{4}  & 0 & 0 &  \frac{\lambda}{4}    &0  \\
  & 4 & 0 & 0 &  0 & \frac{\lambda}{4}    & 0 & 0  & \frac{\lambda}{4} & 0  & 0  &  \frac{\lambda}{4}  & 0  &0  \\
  & 5  & 0 & 0  & 0 & \frac{\lambda}{4}  & 0   & 0 & 0 &  \frac{\lambda}{4}  & 0  &0 & 0 & \frac{\lambda}{4}  \\ 
 & 6 & \frac{2 -3\lambda}{4}  &  0 & 0 &  \frac{2 -3\lambda}{4}  &  \frac{2 -3\lambda}{4}  & 0 & 0  &  \frac{2 -3\lambda}{4} & 0 &  \frac{2 -3\lambda}{4}  &   \frac{2 -3\lambda}{4}  &0 \\ [2pt]
 \Cline{1pt}{2-14}  \\ [-0.9em]
\end{array}
\label{eqn:3cyclePAX1}
\een

One can then observe that non-locality is a special case of contextuality in which the no-disturbance polytope coincides with a non-signalling polytope -- in such a case the NSCE that we construct for the contextuality scenario will be precisely that which we would construct in the non-signalling scenario.

Note that as the no-disturbance polytope is a discrete theory, that theorem \ref{thm:NoGo} applies and so in general the NSEAs that we construct will not be purifications. However, it would be interesting to study the cases in which they are.

\subsection{Complete extension of the behaviours lying on the isotropic line }\label{sec:Bell-Tsirelsen}

% Complete extension can also be constructed for a box, consisting of more than one non-signaling subsystems.
% In this section, we will find the CE of the Tsirelson's box \cite{Tsirelson-bound}. This behaviour reaches the quantum limit in violating CHSH inequality \cite{CHSH}, and describes statistics that can be achieved with quantum state and measurements on bipartite binary input and binary output scenario. 

% To this end we have to apply numerical approach, as our search space of possible minimal ensembles is quite large. We first make suitable observation, which makes the space to be searched smaller:

The aim of this section is to construct explicitly the  NSCE of the isotropic behaviour, that is, a mixture of the $PR$ and anti-$PR$ box.
We therefore focus on the polytope of behaviours, $P_{AB}(ab|xy)$, with two binary inputs $x,y \in \{0,1\}$, two binary outputs $a,b \in \{0,1\}$, and satisfying the non-signaling condition. There are $24$ extremal \textcolor{blue}{(pure)} behaviours (vertices)
of this polytope, among which $16$ are 
local behaviours given by
\be 
\mathrm{L}_{\alpha\beta\gamma\delta}(ab|xy) = \left\{ \begin{array}{ll}
         1 & \mbox{if $a= \alpha x\oplus b$,} \\
         	  & ~~~\mbox{$b = \gamma y \oplus \delta$} \\
        0 & \mbox{otherwise}.\end{array} \right.
\ee
with $\alpha,\beta,\gamma$ and $\delta \in \{0,1\}$. And another $8$ non-local behaviours, which are
\be 
\mathrm{B}_{rst}(ab|xy) = \left\{ \begin{array}{ll}
         1/2 & \mbox{if $a\oplus b = xy\oplus rx \oplus sy \oplus t$} \\
        0 & \mbox{otherwise}.\end{array} \right.
\ee
with $r,s,t$ taking values either $0$ or $1$ \cite{Barret-Roberts}. The triple $(r,s,t)$ enumerates the non-local behaviours in the polytope of behaviours with
two binary inputs and two binary outputs.
Hence, the isotropic behaviour can be formulated as, $B(\eta)_{AB}(ab|xy) = \eta B_{000}(ab|xy) + (1 - \eta) B_{111}(ab|xy)$, where $\eta \in (0,1)$, and  where  $B_{000}$ and $B_{111}$  are  the $PR$ and anti-$PR$ boxes respectively. 
One can easily check that $B(\frac 12) = P^m_{AB}(ab|xy)$, is a maximally mixed behaviour in the polytope of two input two output behaviours (the one taking value $P(ab|xy)=\frac 14$ for all $a,b,x,y\in\{0,1\}$).  All the behaviours  $B(\eta)$, for $\eta \in (1,\frac 12)$ can be transformed into to the behaviours $B(\eta)$, for $\eta \in (\frac 12, 0)$ by local relabeling of inputs and outputs.  Thus, in our investigation we will consider only the behaviours lying on the isotropic line from the PR box to maximally mixed behaviour, i.e., for  $B(\eta)$, where $ \eta \in [\frac 12, 1)$.

We know from the definition of  NSCE, Definition \ref{def:NSEA_SI},  %\ref{def:complete_ext_main_text})
 that, corresponding to each input of the extending party, there is a minimal ensemble ${\cal M}(B(\eta))$, where the members of the ensembles are enumerated  by  the outputs of the extending system. Hence, finding all possible minimal ensembles of a behaviour is sufficient for the construction of  NSCE.

\subsubsection{Minimal ensembles for isotropic behaviours}
     Our aim is to find the minimal ensembles for the behaviours lying on  the  isotropic line, but, in this section we first focus on the minimal ensembles for a particular behaviour, the Bell Tsirelson behaviour \cite{Tsirelson-bound}. 
     The Bell Tsirelson behaviour, lying on the isotropic line for $\eta = \frac{2+\sqrt{2}}{4}$, reaches the quantum limit in violating the CHSH inequality \cite{CHSH}. 
     The  challenge  in finding the minimal ensembles is that   the Bell Tsirelson behaviour is  specified by an irrational number, however,  despite this,  we are able to find all of its minimal ensembles  analytically, due to the following observations and a theorem. 

 {\observation All the pure members ensembles of the behaviours $B(\eta)$, in the isotropic line, contains the PR box, $B_{000}$ as one of its member element, for  all  $\eta \in (\frac 34, 1]$. 
 \label{obs:b000}
 }
 \begin{proof}
 By direct inspection of the $24$ vertices of the polytope of behaviours with two binary inputs and two binary outputs we observe that $PR$-box is the only behaviour, which has value $4$
 of the CHSH functional which reads
 \begin{equation}
     \beta(P) = \sum_{x,y}\sum_{(a,b) \in \mbox{supp}(B_{000})} P(ab|xy)
 \end{equation}
 Moreover, for all other vertices the functional achieves values $\leq 3$. It is also clear that the functional $\beta$ is linear (in fact it is equal to $2\<P|B_{000}\>$ where $\<.|.\>$ denotes Euclidean scalar product of vectors (see Lemma 4 of \cite{Grudka_contextuality}). Let us then suppose
 that there exists an ensemble $\{p_i,V_i\}_i$ of a behaviour $P$ from the isotropic line $(3/4,1]$, which does not have $PR$ in its set of members (here $V_i$ are the vertices of the polytope). We have then $\beta(P)=\beta(\sum_ip_iV_i)=\sum_i p_i \beta(V_i)\leq \sum_i p_i 3\leq 3$. This contradicts the fact that
 all behaviours from the considered set satisfy $\beta(P)>3$ i.e. they violate CHSH inequality.
 \end{proof}
  Let  $S_{\cm}(\eta)$ denote the set of all minimal ensembles of the isotropic behaviour $B(\eta)$.

% % {\it Proof.}  It is easy to see, that if $\alpha' >\alpha$ one can generate an ensemble of $B(\alpha')$ from $\cm(B(\alpha))$
% % by admixing more of $B_{000}$, which by definition belongs to it, and keep the other members of ensemble unchanged,
% % obtaining an ensemble of $B(\alpha')$.
% %  \blacksquare~

% % {\observation Let $S_\cm(\eta)=S_\cm(\eta')$ and for some $\eta< \eta'' <\eta'$ there is $\cm_1 \in S(\eta'')$ and $\cm_1 \notin S_\cm(\eta')$. Then for any $\cm \in S_\cm(\eta')$ there is $V(\cm(\eta) \neq V(\cm(\eta'))$.
% % }
% % \begin{proof}
% % Suppose by contradiction that there is $\cm_1=\{q_i,V_i\}$ of $S(\eta)$ that do not belong to $S_\cm(\eta')$, but there
% % exists $\cm_2 \in S(\eta')$ such that $\cm_2 = \{r_i, V_i\}$
% % i.e. with the same set of members as $\cm_1$. Since $B(\eta'')=
% % p B(\eta) + (1-p) B(\eta')$, we have $B(\eta'') = p \sum_i r_i V_i +(1-p) \sum_i r_i V_$
% % \end{proof}

 {\lemma Let $1>\eta'>\eta> \frac 34$, then for any minimal ensemble from $S_\cm(\eta)$, there exists a unique minimal ensemble of $S_\cm(\eta')$ with the same set of members.
 \label{lem:similar}
 }
 \begin{proof}
 Since $\eta, \eta' \in (\frac 34, 1)$ by observation \ref{obs:b000}, all the minimal ensembles of $B(\eta)$ must contain $B_{000}$ as its members elements. 
% % We can then use Observation \ref{obs:monot}, to see that for any $\cm_1 \in S_\cm(\eta)$
% % there exists $\ce \in S_\cm(\eta')$ such that $V(\cm_1)=V(\ce)$.
% % We will show now, that $\ce$ is minimal.
 Now, from lemma \ref{lem:unique}, for the set $V(\cm_1)$ there exists unique weights $\{p_i >0\}_{i=0}^n$ such that 
 \begin{equation}
     \sum_{V_i\in V(\cm_1)} p_i V_i = B(\eta),
 \end{equation}
 where $V_0=B_{000}$ and $V_i$ for $i=\in\{1,...,n\}$ are vertices of the polytope
 of behaviours with two binary inputs and two binary outputs.
 Now, from $\eta'>\eta$ 
% %and convexity of the interval $[B(\eta),B_{000}]$,
 there exists unique weight $1>q>0$ such, that $B(\eta')=qB(\eta)+(1-q)B_{000}$. Hence, for the set $\{V_i\}_{i=0}^n$ there exist unique weights $\{r_i\}_{i=0}^n$
 given by equations $r_0 = q p_0 + (1-q)$ and $r_i=p_i(1-q)$ for $i\in\{1,...,n\}$, such that $\sum_i r_i V_i = B(\eta')$. Hence,
 by lemma \ref{lem:unique}, the ensemble $\{r_i, V_i\}_{i=0}^n$ is minimal ensemble of $B(\eta')$, and the assertion follows.
 \end{proof}

 We will  now adopt  the  notation, that ${\cal V}$ is
 the set of the sets of members of minimal ensembles:
 \begin{equation}
     {\mathcal V}(P) :=  \bigcup_{{\mathcal M} \in S_{\mathcal M}(P)}  \{ V({\mathcal M}(P))\}.
 \end{equation}

% % Now we define similarity of the set of minimal ensembles:
% %  {\definition[Similarity of two sets of minimal ensembles]
% %   We will say that a set $S_\cm(P_1)$ is similar to $S_\cm(P_2)$, if corresponding to each minimal ensemble $\cm(P_1)$ there exists a unique minimal ensemble $\cm(P_2)$, such that $V(\cm(P_1)) = V(\cm(P_2)$. We write then $S_{\cm}(P_1)\rightarrow S_{\cm}(P_2)$.
% %  \label{def:isomorphism_minens}
% %  }

 From lemma \ref{lem:similar} we have the immediate corollary:
 {\corollary Let $1>\eta'>\eta> \frac 34$, and $B(\eta) = \eta B_{000}+ (1-\eta) B_{111}$ then, there is
 ${\mathcal V}(P(\eta)) \subseteq {\mathcal V}(P(\eta'))$
 \label{cor:inclusion}
 }

% % {\observation Let \sout{minimal} ensemble $\cm(B(\eta))$,
% %  contain $B_{000}$, and $\eta' > \eta$. Then, for each  such \sout{minimal}  ensemble $\cm(B(\eta)) \in S_{\cm}(\eta)$
% % there is a \sout{minimal} ensemble $\cm(B(\eta')) \in S_{\cm}(\eta')$ such that $V(\cm(B(\eta))) = V(\cm(B(\eta')))$.
% % \label{obs:monot}
% % }\\
% % \begin{proof}
% % It is enough to observe that $B(\eta') = \gamma PR + (1-\gamma) B(\eta)$. Consider then any ensemble of $B(\eta)$ equal  $\{(p_0,B_{000}),(p_1,V_1),\ldots,(p_n,V_n)\}$. Note that we use here Observation \ref{obs:b000} i.e. the fact that any ensemble of behaviour from considered part of isotropic line contains the $B_{000}$. Then the ensemble $\{(\gamma p_0 +(1-\gamma), B_{000}),((1-\gamma)p_1,V_1),\ldots,((1-\gamma)p_n,V_n)\}$ is an ensemble of $B(\eta')$ with the same set of members.
% % \end{proof}

% %  Now we define the isomorphism of the set of minimal ensembles:
% %  {\definition[Isomorphism between the set of minimal ensembles]
% %  Suppose, $B(\eta)$ and $B(\eta')$ are two behaviours in the isotropic line, with $\eta \neq \eta'$. We will say the set of all minimal ensembles of these two behaviours are isomorphic if corresponds to each minimal ensembles of $\cm(B(\eta))$ there exist a minimal ensemble $\cm(B(\eta'))$, such that $V(\cm(B(\eta)) = V(\cm(B(\eta'))$. And for this two isomorphic set of minimal ensembles we can write $S_{\cm}(\eta)=S_{\cm}(\eta')$.
% %  \label{def:isomorphism_minens}
% %  }

 {\theorem Let $\frac 34 < \eta' < \eta < \eta'' < 1$. If ${\mathcal V}(B(\eta')) = {\mathcal V}(B(\eta''))\equiv {\mathcal V}$,  then we also have
 ${\mathcal V}(\eta)  = {\mathcal V}$.
 }
 \begin{proof}
 Since $\eta', \eta'' \in (\frac 34, 1)$, from 
 the Corollary \ref{cor:inclusion}, we have
% %Observation \ref{obs:monot},  
% %and the Definition \ref{def:isomorphism_minens},  
 \begin{equation}
     {\mathcal V}(B(\eta')) \subseteq {\mathcal V}(B(\eta)) \subseteq
     {\mathcal V}(B(\eta''))
 \end{equation}
 However by assumption there is 
 $ {\mathcal V}(B(\eta'')) \subseteq {\mathcal V}(B(\eta'))$, hence
 ${\mathcal V}(B(\eta)) \subseteq {\mathcal V}(B(\eta'))$, and the assertion follows. 
 \end{proof}

 The above theorem holds for the
 range of the isotropic parameter $(\frac 34,1)$. Moreover, due to this theorem we are able to find out the minimal ensembles of all non-local $B(\eta)$ for $\eta \in (\frac 12, 1)$.
 However, numerical investigate indicates that it holds true also for the $( \frac 12, \frac 34]$, hence we conjecture as follows:
 \begin{conjecture}
 Let $\frac 12 < \eta' < \eta < \eta'' < 1$. If ${\mathcal V}(B(\eta')) = {\mathcal V}(B(\eta''))\equiv {\mathcal V}$,  then we also have
 ${\mathcal V}(\eta)  = {\mathcal V}$.
 \label{conj:half}
 \end{conjecture}

The above observation and theorem help us to find out the  NSEA  of the Bell-Tsirelson's box. There are a total of $354$ minimal ensembles of  $B(\eta)$, for $\eta = \frac{2+\sqrt{2}}{4}$.  Surprisingly,  the  NSEA  of a noisy PR box leads to an extending system $E$ embedded in a  vector space  of dimension $2,837$ (taking into account the normalization constraints it effectively lives in  a  $2,483$ dimensional space) and the entire NSEA, is lying in a polytope which is embedded in a vector space of dimension $45,392$ taking into account the normalization
and non-signaling constraints effectively lives in space of dimension $22,355$. 
The list of these ensembles are given in Sec. \ref{appen:Bell_Tsirelson_box}.

\begin{figure*}[h]
\begin{center}
\includegraphics[scale=0.5,angle =0]{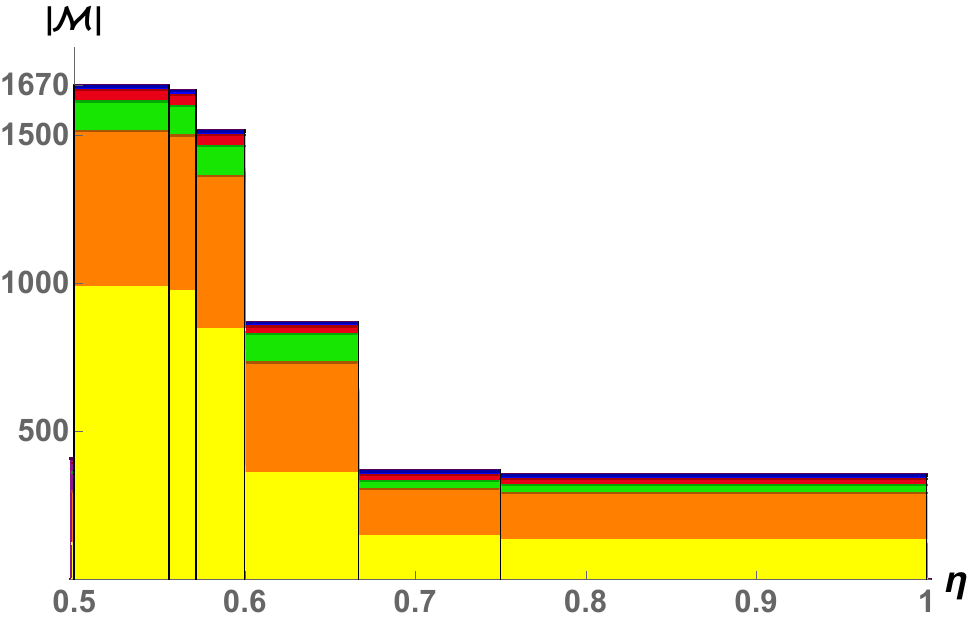}
\includegraphics[scale=0.5,angle =0]{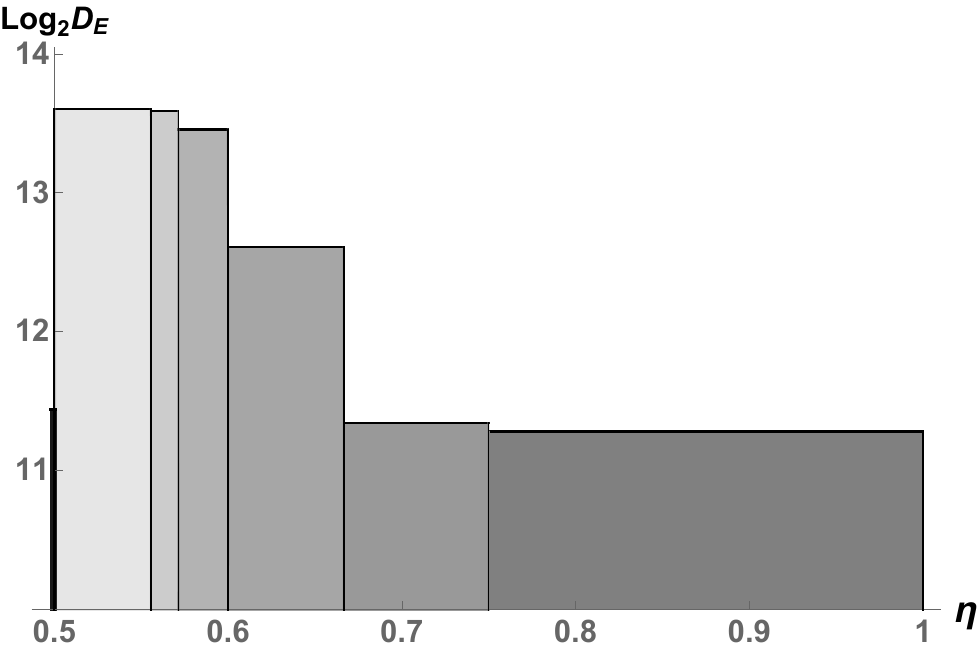}
\end{center}
\caption{Panel a) Histogram plot of the total number of inputs $|{\cal Z}| = |{\cal M}|$ of the extending system of the  NSEA  of the behaviours along the isotropic line. Different color in each column of the histogram shows the the number of inputs having different numbers of outputs. The height of the yellow color in each column represents the number of inputs with $9$ outputs, orange represents the same for $8$ outputs, green is for $7$ outputs, red is for $6$ outputs and so on. The inputs of the extending system for various values of outputs are given in Table \ref{tab:min_ensemb_iso}. Panel b) Histogram plot of the memory required in the  extending system to  store the information about the minimal ensembles of the behaviours along the isotropic line. Here $D_E$ represents the dimension of the extending system of  NSEA.  All the plots have been given from $\eta = 0$ to $\eta = \frac 12$, i.e., from the PR box to the maximally mixed box.}
\label{fig:memory_fig}
\end{figure*}
\subsubsection{Dimension of the extending party for the behaviours lying on the isotropic line}

In the  previous  section, we discussed how one can construct the  NSCE  of the  Bell-Tsirelson box. In this section, due to the large number of minimal ensembles for a generic $B(\eta)$, we will discuss some of the statistics of the  NSCE of   more general  isotropic behaviours. 
%$B(\eta)$ lying on the isotropic line. 

%, from $\eta \in (\frac 12, 1)$ it is completely symmetric.   

For any value of $\eta \in [\frac 12, 1)$, we have provided an estimate of the 
%non-signaling complete extension (NSCE) 
 NSCE for the bipartite behaviour $B(\eta)$ for $\eta \in (\frac 12 ,1)$. This is indeed an  NSCE of this behaviour 
%given 
 provided  that conjecture \ref{conj:half} holds true. %Due to the large size of the NSCE we will make an study for the dimension of the extending system. 
The number of elements in each of the minimal ensembles is bounded between $2$  and  
%to the 
$\dim {\cal B} + 1$, which is $9$, for two input and two output binary behaviours. In the entire range of $\eta \in (\frac 12, 1)$, we have numerically calculated the total number of inputs (number of minimal ensembles $|\cal{M}|$) of the extending system,  which have an  equal number of  outputs $v_j \in \{2,3,\ldots, 9\}$.  This  is  given in table \ref{tab:min_ensemb_iso}.  The Bell-Tsirelson box, $\eta = \frac{2+\sqrt{2}}{4}$, lying in the range $\eta \in (\frac 34,1)$, is given in the last column of Table \ref{tab:min_ensemb_iso}, having total  number of inputs as $354$. Among these $354$ inputs there are $130$ inputs having $9$ outputs, $160$ inputs having $8$ outputs and so on. The minimal ensembles associated with each input of the Bell-Tsirelson box have been given in Sec. \ref{appen:Bell_Tsirelson_box} of the supplemental material.
Here, ${\cal M}$ represent the number of minimal ensembles of a {behaviour} and the length of the column represent the number of pure behaviours in each of the minimal ensembles.

\begin{table}[h]
    \centering
\renewcommand{\arraystretch}{1.5} % Default value: 1
\begin{tabular}{|c|c|c|c|c|c|c|c|}%{|p{0.2\textwidth}|p{0.32\textwidth}|p{0.4\textwidth}|}
\hline 
\diagbox[width=2.5em, height=2.5em, innerrightsep=1pt]{$v_j$}{$\eta$} & $~~~\frac 12 ~~~$ & $(\frac 12,\frac 59]$ & $(\frac 59,\frac 47]$ & $(\frac 47,\frac 35]$ & $(\frac 35,\frac 23]$ & $(\frac 23,\frac 34]$ & $(\frac 34, 1)$ \\
\hline
2 & 4 & 1 & 1 & 1 & 1 & 1 & 1 \\
\hline 
3 & 0 & 3 & 3 & 3 & 3 & 3 & 3 \\
\hline
4 & 12 & 0 & 0 & 0 & 0 & 0 & 0 \\
\hline
5 & 0 & 12 & 12 & 12 & 12 & 12 & 12 \\
\hline
6 & 32 & 38 & 38 & 38 & 26 & 20 & 20 \\
\hline 
7 & 64 & 100 & 100 & 100 & 96 & 28 & 28 \\
\hline 
8 & 176 & 528 & 528 & 520 & 376 & 160 & 160 \\
\hline 
9 & 120 & 988 & 972 & 844 & 356 & 144 & 130 \\
\hline
total & 408 & 1670 & 1654 & 1518 & 870 & 368 & 354 \\
\hline
\end{tabular}
 \caption{Table shows the number of inputs (number of minimal ensembles $|\cal M|$), of the extending system  which have  same number of outputs $v_j$, for various values of $v_j \in \{2,3,\ldots, 9\}$, in the entire range of the parameter $\eta$.}
    \label{tab:min_ensemb_iso}
\end{table}
The number of inputs of the extending system for various values of $\eta$, has been given in figure \ref{fig:memory_fig}(a). The color represent the number of inputs among the total number of inputs having equal number of outputs. The yellow color stands for the set of inputs having $9$ outputs, orange is for $8$, green is for $7$ and red is for $6$ and so on. We have also plotted the memory required by the extending system, to store the information about the all possible minimal ensembles of the behaviours lying on the isotropic line, in figure \ref{fig:memory_fig}(b). Here $D_E$ represent the dimension of the extending system. 

\begin{table}[h]
    \centering
    \renewcommand{\arraystretch}{1.5}
    \begin{tabular}{|c|c|c|c|c|c|c|c|}
\hline 
\diagbox[width=3.5em, height=2em, innerrightsep=1pt]{dim}{$\eta$} & $~~~\frac 12 ~~~$ & $(\frac 12,\frac 59]$ & $(\frac 59,\frac 47]$ & $(\frac 47,\frac 35]$ & $(\frac 35,\frac 23]$ & $(\frac 23,\frac 34]$ & $(\frac 34, 1)$ \\
\hline
$E$ & 2776 & 12445 & 12317 & 11237 & 6241 & 2595 & 2483 \\
\hline
$NSCE$ & 24992 & 112013 & 110861 & 101141 & 56177 & 23363 & 22355 \\
\hline 
\end{tabular}
    \caption{Table shows the dimension of the extending system as well as the total dimension of the NSCE, of  the behaviours lying on the isotropic line.}
    \label{tab:dimension}
\end{table}
\renewcommand{\arraystretch}{1}

The dimension of the extending system as well as the total dimension of the  NSCE, as given in Eq. (\ref{eq:dim_def}), for various values of $\eta \in (\frac 12, 1)$ has been enlisted in Table \ref{tab:dimension}. We have observed that the dimension of the  NSCE is a maximum when the behaviour is in the vicinity of the maximally mixed behaviour. The system size is considerably lower in the range of our interest i.e., when $\eta \in (
\frac 34, 1)$,  that is, when  the behaviour $B(\eta)$ starts showing non-classical features.

\subsubsection{Minimal ensembles for non-local isotropic behaviours}\label{appen:Bell_Tsirelson_box}

Here we present the minimal ensembles of the Bell-Tsirelson box, which lies on the isotropic line
\begin{equation}
  B(\eta) = \eta \mathrm{B}_{000} + (1 - \eta) \mathrm{B}_{001},
\end{equation}
for $\eta = \frac{2+\sqrt{2}}{4}$.
This behaviour lies in a polytope of dimension $ d = 8$ \cite{Barret-Roberts}, and, according to the theorem of Carath{\'e}odory \cite{caratheodory1911}, the minimal ensembles of $B(\eta)$, consists of  at most $d+1$ pure behaviours.
Hence, we present only those minimal ensembles having at most $9$ elements.

We group the ensembles by the intervals of  the  parameter $\eta$ for
which they are valid minimal ensembles. They are minimal, as otherwise
we would find their subsets which would be minimal and also valid for
higher $\alpha$. The interval of the parameter $\alpha$ is the  due to  
the requirement that the $\mathrm{B}_{000}$ coefficient must be greater or
equal $0$.

We also group the ensembles by the vector of coefficients corresponding to the behaviours of the decomposition which summed up give the $B(\alpha)$. 

% %%%%%%%%%%%%%%%%%%%%%%%%%%%%%%%%%%%%%%%%%%%%%%%%%%
\begin{enumerate}
    
\item {Having $\eta \in \left(0, 1\right).$}

\[v = \left [ \eta, \quad 1- \eta \right ]\]
\vspace{-2em}
\[\mathcal{M}_{1} = [\mathrm{B}_{000}, \mathrm{B}_{001}] \]

\item {Having $\eta  \in \left(\frac{1}{4}, 1\right).$}

\begin{equation*}
    v = \left[ \frac{4 \eta}{3} - \frac{1}{3}, \quad \frac{1}{6}- \frac{\eta}{6}, \quad \frac{1}{6} - \frac{\eta}{6}, \quad \frac{1}{6} - \frac{\eta}{6}, \quad \frac{1}{6} - \frac{\eta}{6}, \quad \frac{1}{6} - \frac{\eta}{6}, \quad \frac{1}{6} - \frac{\eta}{6}, \quad \frac{1}{6} - \frac{\eta}{6}, \quad \frac{1}{6} - \frac{\eta}{6}\right ]
\end{equation*}
\vspace{-1em}
\[\mathcal{M}_{2} = [\mathrm{B}_{000}, L_{0001}, L_{0011}, L_{0100}, L_{0110}, L_{1001}, L_{1010}, L_{1100}, L_{1111}] \]

\item {Having $\eta \in \left(\frac{1}{3}, 1\right).$}

\[v = \left [ \frac{3 \eta}{2} - \frac{1}{2}, \quad \frac{1}{4} - \frac{\eta}{4} , \quad \frac{1}{4} - \frac{\eta}{4}, \quad \frac{1}{4} - \frac{\eta}{4}, \quad \frac{1}{4} - \frac{\eta}{4}, \quad \frac{1}{4} - \frac{\eta}{4}, \quad \frac{1}{4} - \frac{\eta}{4}\right ]\]
\vspace{-1em}
\begin{eqnarray*}
\mathcal{M}_{3} = [\mathrm{B}_{000}, L_{0000}, L_{0001}, L_{0100}, L_{0110}, L_{1001}, L_{1111}] \\
\mathcal{M}_{4} = [\mathrm{B}_{000}, L_{0001}, L_{0011}, L_{0100}, L_{0101}, L_{1010}, L_{1100}] \\
\mathcal{M}_{5} = [\mathrm{B}_{000}, L_{0001}, L_{0011}, L_{0110}, L_{0111}, L_{1010}, L_{1100}] \\
\mathcal{M}_{6} = [\mathrm{B}_{000}, L_{0001}, L_{0110}, L_{1000}, L_{1001}, L_{1100}, L_{1111}] \\
\mathcal{M}_{7} = [\mathrm{B}_{000}, L_{0001}, L_{0110}, L_{1010}, L_{1011}, L_{1100}, L_{1111}] \\
\mathcal{M}_{8} = [\mathrm{B}_{000}, L_{0010}, L_{0011}, L_{0100}, L_{0110}, L_{1001}, L_{1111}] \\
\mathcal{M}_{9} = [\mathrm{B}_{000}, L_{0011}, L_{0100}, L_{1001}, L_{1010}, L_{1100}, L_{1101}] \\
\mathcal{M}_{10} = [\mathrm{B}_{000}, L_{0011}, L_{0100}, L_{1001}, L_{1010}, L_{1110}, L_{1111}]
\end{eqnarray*}

\[v = \left [ \frac{3 \eta}{2} - \frac{1}{2}, \quad \frac{1}{2} - \frac{\eta}{2}, \quad \frac{1}{4} - \frac{\eta}{4}, \quad \frac{1}{4} - \frac{\eta}{4}, \quad \frac{1}{4} - \frac{\eta}{4}, \quad \frac{1}{4} - \frac{\eta}{4}\right ]\]
\vspace{-1em}
\begin{eqnarray*}
\mathcal{M}_{11} = [\mathrm{B}_{000}, \mathrm{B}_{010}, L_{0001}, L_{0011}, L_{0100}, L_{0110}] \\
\mathcal{M}_{12} = [\mathrm{B}_{000}, \mathrm{B}_{011}, L_{1001}, L_{1010}, L_{1100}, L_{1111}] \\
\mathcal{M}_{13} = [\mathrm{B}_{000}, \mathrm{B}_{100}, L_{0001}, L_{0100}, L_{1001}, L_{1100}]\\ 
\mathcal{M}_{14} = [\mathrm{B}_{000}, \mathrm{B}_{101}, L_{0011}, L_{0110}, L_{1010}, L_{1111}] \\
\mathcal{M}_{15} = [\mathrm{B}_{000}, \mathrm{B}_{110}, L_{0011}, L_{0110}, L_{1001}, L_{1100}] \\
\mathcal{M}_{16} = [\mathrm{B}_{000}, \mathrm{B}_{111}, L_{0001}, L_{0100}, L_{1010}, L_{1111}] 
\end{eqnarray*}

\item {Having $\eta \in \left(\frac{2}{5}, 1\right).$}

\begin{equation*}
v = \left [ \frac{5 \eta}{3} - \frac{2}{3}, \quad \frac{1}{3} - \frac{\eta}{3} , \quad \frac{1}{6} - \frac{\eta}{6}, \quad \frac{1}{6} - \frac{\eta}{6}, \quad \frac{1}{6} - \frac{\eta}{6}, \quad \frac{1}{6} - \frac{\eta}{6}, \quad \frac{1}{6} - \frac{\eta}{6}, \quad \frac{1}{6} - \frac{\eta}{6}, \quad \frac{1}{3} - \frac{\eta}{3} \right ]
\end{equation*}
%\vspace{-1.5em}
\begin{eqnarray*}
\mathcal{M}_{17} = [\mathrm{B}_{000}, \mathrm{B}_{110}, L_{0001}, L_{0011}, \mathrm{L}_{0101}, \mathrm{L}_{0110}, \mathrm{L}_{1010}, \mathrm{L}_{1011}, \mathrm{L}_{1100}] \\
\mathcal{M}_{18} = [\mathrm{B}_{000}, \mathrm{B}_{111}, \mathrm{L}_{0001}, \mathrm{L}_{0010}, \mathrm{L}_{0100}, \mathrm{L}_{0110}, \mathrm{L}_{1000}, \mathrm{L}_{1001}, \mathrm{L}_{1111}] 
\end{eqnarray*}

\[v = \left [ \frac{5 \eta}{3} - \frac{2}{3}, \quad \frac{1}{3} - \frac{\eta}{3}, \quad \frac{1}{6} - \frac{\eta}{6}, \quad \frac{1}{6} - \frac{\eta}{6}, \quad \frac{1}{6} - \frac{\eta}{6}, \quad \frac{1}{6} - \frac{\eta}{6}, \quad \frac{1}{3} - \frac{\eta}{3}, \quad \frac{1}{6} - \frac{\eta}{6}, \quad \frac{1}{6} - \frac{\eta}{6}\right ]\]
\vspace{-1.5em}
\begin{eqnarray*}
\mathcal{M}_{19} = [\mathrm{B}_{000}, \mathrm{B}_{110}, \mathrm{L}_{0000}, \mathrm{L}_{0011}, \mathrm{L}_{0100}, \mathrm{L}_{0110}, \mathrm{L}_{1001}, \mathrm{L}_{1110}, \mathrm{L}_{1111}] \\
\mathcal{M}_{20} = [\mathrm{B}_{000}, \mathrm{B}_{111}, \mathrm{L}_{0001}, \mathrm{L}_{0011}, \mathrm{L}_{0100}, \mathrm{L}_{0111}, \mathrm{L}_{1010}, \mathrm{L}_{1100}, \mathrm{L}_{1101}] 
\end{eqnarray*}

\[v = \left [ \frac{5 \eta}{3} - \frac{2}{3}, \quad \frac{1}{3} - \frac{\eta}{3}, \quad  \frac{1}{6} - \frac{\eta}{6}, \quad  \frac{1}{6} - \frac{\eta}{6}, \quad  \frac{1}{6} - \frac{\eta}{6}, \quad  \frac{1}{6} - \frac{\eta}{6}, \quad \frac{1}{3} - \frac{\eta}{3}, \quad \frac{1}{3} - \frac{\eta}{3}\right ]\]
\vspace{-1.5em}
\begin{eqnarray*}
\mathcal{M}_{21} = [\mathrm{B}_{000}, \mathrm{B}_{011}, \mathrm{L}_{0000}, \mathrm{L}_{0010}, \mathrm{L}_{0100}, \mathrm{L}_{0110}, \mathrm{L}_{1001}, \mathrm{L}_{1111}] \\
\mathcal{M}_{22} = [\mathrm{B}_{000}, \mathrm{B}_{011}, \mathrm{L}_{0001}, \mathrm{L}_{0011}, \mathrm{L}_{0101}, \mathrm{L}_{0111}, \mathrm{L}_{1010}, \mathrm{L}_{1100}] 
\end{eqnarray*}

\[v = \left [ \frac{5 \eta}{3} - \frac{2}{3}, \quad \frac{1}{3} - \frac{\eta}{3}, \quad  \frac{1}{6} - \frac{\eta}{6}, \quad  \frac{1}{6} - \frac{\eta}{6}, \quad  \frac{1}{6} - \frac{\eta}{6}, \quad \frac{1}{3} - \frac{\eta}{3}, \quad  \frac{1}{6} - \frac{\eta}{6}, \quad \frac{1}{3} - \frac{\eta}{3}\right ]\]
\vspace{-1.5em}
\[\mathcal{M}_{23} = [\mathrm{B}_{000}, \mathrm{B}_{011}, \mathrm{L}_{0010}, \mathrm{L}_{0110}, \mathrm{L}_{1000}, \mathrm{L}_{1001}, \mathrm{L}_{1100}, \mathrm{L}_{1111}] \]

\[v = \left [ \frac{5 \eta}{3} - \frac{2}{3}, \quad \frac{1}{3} - \frac{\eta}{3}, \quad  \frac{1}{6} - \frac{\eta}{6}, \quad  \frac{1}{6} - \frac{\eta}{6}, \quad  \frac{1}{6} - \frac{\eta}{6}, \quad \frac{1}{3} - \frac{\eta}{3}, \quad \frac{1}{3} - \frac{\eta}{3}, \quad  \frac{1}{6} - \frac{\eta}{6}\right ]\]
\vspace{-1.5em}
\[\mathcal{M}_{24} = [\mathrm{B}_{000}, \mathrm{B}_{011}, \mathrm{L}_{0011}, \mathrm{L}_{0111}, \mathrm{L}_{1001}, \mathrm{L}_{1010}, \mathrm{L}_{1100}, \mathrm{L}_{1101}] \]

\[v = \left [ \frac{5 \eta}{3} - \frac{2}{3}, \quad \frac{1}{3} - \frac{\eta}{3}, \quad  \frac{1}{6} - \frac{\eta}{6}, \quad  \frac{1}{6} - \frac{\eta}{6}, \quad \frac{1}{3} - \frac{\eta}{3}, \quad  \frac{1}{6} - \frac{\eta}{6}, \quad  \frac{1}{6} - \frac{\eta}{6}, \quad  \frac{1}{6} - \frac{\eta}{6}, \quad  \frac{1}{6} - \frac{\eta}{6}\right ]\]
\vspace{-1.5em}
\begin{eqnarray*}
\mathcal{M}_{25} = [\mathrm{B}_{000}, \mathrm{B}_{110}, \mathrm{L}_{0000}, \mathrm{L}_{0001}, \mathrm{L}_{0110}, \mathrm{L}_{1001}, \mathrm{L}_{1011}, \mathrm{L}_{1100}, \mathrm{L}_{1111}] \\
\mathcal{M}_{26} = [\mathrm{B}_{000}, \mathrm{B}_{111}, \mathrm{L}_{0010}, \mathrm{L}_{0011}, \mathrm{L}_{0100}, \mathrm{L}_{1001}, \mathrm{L}_{1010}, \mathrm{L}_{1101}, \mathrm{L}_{1111}] 
\end{eqnarray*}

\[v = \left [ \frac{5 \eta}{3} - \frac{2}{3}, \quad \frac{1}{3} - \frac{\eta}{3}, \quad  \frac{1}{6} - \frac{\eta}{6}, \quad  \frac{1}{6} - \frac{\eta}{6}, \quad \frac{1}{3} - \frac{\eta}{3}, \quad  \frac{1}{6} - \frac{\eta}{6}, \quad  \frac{1}{6} - \frac{\eta}{6}, \quad \frac{1}{3} - \frac{\eta}{3}\right ]\]
\vspace{-1.5em}
\begin{eqnarray*}
\mathcal{M}_{27} = [\mathrm{B}_{000}, \mathrm{B}_{011}, \mathrm{L}_{0000}, \mathrm{L}_{0100}, \mathrm{L}_{1001}, \mathrm{L}_{1010}, \mathrm{L}_{1110}, \mathrm{L}_{1111}] \\
\mathcal{M}_{28} = [\mathrm{B}_{000}, \mathrm{B}_{101}, \mathrm{L}_{0000}, \mathrm{L}_{0001}, \mathrm{L}_{0110}, \mathrm{L}_{1000}, \mathrm{L}_{1001}, \mathrm{L}_{1111}]\\
\mathcal{M}_{29} = [\mathrm{B}_{000}, \mathrm{B}_{101}, \mathrm{L}_{0000}, \mathrm{L}_{0001}, \mathrm{L}_{0110}, \mathrm{L}_{1010}, \mathrm{L}_{1011}, \mathrm{L}_{1111}] \\
\mathcal{M}_{30} = [\mathrm{B}_{000}, \mathrm{B}_{101}, \mathrm{L}_{0010}, \mathrm{L}_{0011}, \mathrm{L}_{0110}, \mathrm{L}_{1000}, \mathrm{L}_{1001}, \mathrm{L}_{1111}] 
\end{eqnarray*}

\[v = \left [ \frac{5 \eta}{3} - \frac{2}{3}, \quad \frac{1}{3} - \frac{\eta}{3}, \quad  \frac{1}{6} - \frac{\eta}{6}, \quad  \frac{1}{6} - \frac{\eta}{6}, \quad \frac{1}{3} - \frac{\eta}{3}, \quad  \frac{1}{6} - \frac{\eta}{6}, \quad \frac{1}{3} - \frac{\eta}{3}, \quad  \frac{1}{6} - \frac{\eta}{6}\right ]\]
\vspace{-1em}
\[\mathcal{M}_{31} = [\mathrm{B}_{000}, \mathrm{B}_{011}, \mathrm{L}_{0001}, \mathrm{L}_{0101}, \mathrm{L}_{1010}, \mathrm{L}_{1011}, \mathrm{L}_{1100}, \mathrm{L}_{1111}] \]

\[v = \left [ \frac{5 \eta}{3} - \frac{2}{3}, \quad \frac{1}{3} - \frac{\eta}{3}, \quad  \frac{1}{6} - \frac{\eta}{6}, \quad  \frac{1}{6} - \frac{\eta}{6}, \quad \frac{1}{3} - \frac{\eta}{3}, \quad \frac{1}{3} - \frac{\eta}{3}, \quad  \frac{1}{6} - \frac{\eta}{6}, \quad  \frac{1}{6} - \frac{\eta}{6}\right ]\]
\vspace{-1em}
\begin{eqnarray*}
\mathcal{M}_{32} = [\mathrm{B}_{000}, \mathrm{B}_{100}, \mathrm{L}_{0000}, \mathrm{L}_{0001}, \mathrm{L}_{0100}, \mathrm{L}_{1001}, \mathrm{L}_{1110}, \mathrm{L}_{1111}] \\
\mathcal{M}_{33} = [\mathrm{B}_{000}, \mathrm{B}_{100}, \mathrm{L}_{0010}, \mathrm{L}_{0011}, \mathrm{L}_{0100}, \mathrm{L}_{1001}, \mathrm{L}_{1100}, \mathrm{L}_{1101}] \\
\mathcal{M}_{34} = [\mathrm{B}_{000}, \mathrm{B}_{100}, \mathrm{L}_{0010}, \mathrm{L}_{0011}, \mathrm{L}_{0100}, \mathrm{L}_{1001}, \mathrm{L}_{1110}, \mathrm{L}_{1111}] 
\end{eqnarray*}

\[v = \left [ \frac{5 \eta}{3} - \frac{2}{3}, \quad \frac{1}{3} - \frac{\eta}{3}, \quad  \frac{1}{6} - \frac{\eta}{6}, \quad \frac{1}{3} - \frac{\eta}{3}, \quad  \frac{1}{6} - \frac{\eta}{6}, \quad \frac{1}{3} - \frac{\eta}{3}, \quad  \frac{1}{6} - \frac{\eta}{6}, \quad  \frac{1}{6} - \frac{\eta}{6}\right ]\]
%\vspace{-2em}
\[\mathcal{M}_{35} = [\mathrm{B}_{000}, \mathrm{B}_{010}, \mathrm{L}_{0000}, \mathrm{L}_{0001}, \mathrm{L}_{0100}, \mathrm{L}_{0110}, \mathrm{L}_{1011}, \mathrm{L}_{1111}] \]

\[v = \left [ \frac{5 \eta}{3} - \frac{2}{3}, \quad \frac{1}{3} - \frac{\eta}{3}, \quad  \frac{1}{6} - \frac{\eta}{6}, \quad \frac{1}{3} - \frac{\eta}{3}, \quad \frac{1}{3} - \frac{\eta}{3}, \quad  \frac{1}{6} - \frac{\eta}{6}, \quad  \frac{1}{6} - \frac{\eta}{6}, \quad  \frac{1}{6} - \frac{\eta}{6}\right ]\]
\vspace{-1em}
\begin{eqnarray*}
\mathcal{M}_{36} = [\mathrm{B}_{000}, \mathrm{B}_{010}, \mathrm{L}_{0001}, \mathrm{L}_{0011}, \mathrm{L}_{0100}, \mathrm{L}_{0101}, \mathrm{L}_{1010}, \mathrm{L}_{1110}] \\
\mathcal{M}_{37} = [\mathrm{B}_{000}, \mathrm{B}_{010}, \mathrm{L}_{0010}, \mathrm{L}_{0011}, \mathrm{L}_{0100}, \mathrm{L}_{0110}, \mathrm{L}_{1001}, \mathrm{L}_{1101}] 
\end{eqnarray*}

\[v = \left [ \frac{5 \eta}{3} - \frac{2}{3}, \quad \frac{1}{3} - \frac{\eta}{3}, \quad \frac{1}{3} - \frac{\eta}{3}, \quad  \frac{1}{6} - \frac{\eta}{6}, \quad  \frac{1}{6} - \frac{\eta}{6}, \quad  \frac{1}{6} - \frac{\eta}{6}, \quad  \frac{1}{6} - \frac{\eta}{6}, \quad  \frac{1}{6} - \frac{\eta}{6}, \quad  \frac{1}{6} - \frac{\eta}{6}\right ]\]
\vspace{-1em}
\begin{eqnarray*}
\mathcal{M}_{38} = [\mathrm{B}_{000}, \mathrm{B}_{110}, \mathrm{L}_{0011}, \mathrm{L}_{0100}, \mathrm{L}_{0101}, \mathrm{L}_{1001}, \mathrm{L}_{1010}, \mathrm{L}_{1100}, \mathrm{L}_{1110}] \\
\mathcal{M}_{39} = [\mathrm{B}_{000}, \mathrm{B}_{111}, \mathrm{L}_{0001}, \mathrm{L}_{0110}, \mathrm{L}_{0111}, \mathrm{L}_{1000}, \mathrm{L}_{1010}, \mathrm{L}_{1100}, \mathrm{L}_{1111}] 
\end{eqnarray*}

\[v = \left [ \frac{5 \eta}{3} - \frac{2}{3}, \quad \frac{1}{3} - \frac{\eta}{3}, \quad \frac{1}{3} - \frac{\eta}{3}, \quad  \frac{1}{6} - \frac{\eta}{6}, \quad  \frac{1}{6} - \frac{\eta}{6}, \quad  \frac{1}{6} - \frac{\eta}{6}, \quad  \frac{1}{6} - \frac{\eta}{6}, \quad \frac{1}{3} - \frac{\eta}{3}\right ]\]
\vspace{-1em}
\begin{eqnarray*}
\mathcal{M}_{40} &=& [\mathrm{B}_{000}, \mathrm{B}_{011}, \mathrm{B}_{100}, \mathrm{L}_{0001}, \mathrm{L}_{0111}, \mathrm{L}_{1001}, \mathrm{L}_{1010}, \mathrm{L}_{1100}] \\
\mathcal{M}_{41} &=&[\mathrm{B}_{000}, \mathrm{B}_{011}, \mathrm{B}_{101}, \mathrm{L}_{0000}, \mathrm{L}_{0110}, \mathrm{L}_{1001}, \mathrm{L}_{1010}, \mathrm{L}_{1111}] \\
\mathcal{M}_{42} &=& [\mathrm{B}_{000}, \mathrm{B}_{011}, \mathrm{B}_{110}, \mathrm{L}_{0011}, \mathrm{L}_{0101}, \mathrm{L}_{1001}, \mathrm{L}_{1010}, \mathrm{L}_{1100}] \\
\mathcal{M}_{43} &=& [\mathrm{B}_{000}, \mathrm{B}_{011}, \mathrm{B}_{111}, \mathrm{L}_{0010}, \mathrm{L}_{0100}, \mathrm{L}_{1001}, \mathrm{L}_{1010}, \mathrm{L}_{1111}] \\
\mathcal{M}_{44} &=& [\mathrm{B}_{000}, \mathrm{B}_{100}, \mathrm{B}_{110}, \mathrm{L}_{0001}, \mathrm{L}_{0110}, \mathrm{L}_{1001}, \mathrm{L}_{1011}, \mathrm{L}_{1100}] \\
\mathcal{M}_{45} &=& [\mathrm{B}_{000}, \mathrm{B}_{100}, \mathrm{L}_{0001}, \mathrm{L}_{0100}, \mathrm{L}_{0101}, \mathrm{L}_{1010}, \mathrm{L}_{1011}, \mathrm{L}_{1100}]\\
\mathcal{M}_{46}  &=&[\mathrm{B}_{000}, \mathrm{B}_{100}, \mathrm{L}_{0001}, \mathrm{L}_{0110}, \mathrm{L}_{0111}, \mathrm{L}_{1000}, \mathrm{L}_{1001}, \mathrm{L}_{1100}] \\
\mathcal{M}_{47}  &=&[\mathrm{B}_{000}, \mathrm{B}_{100}, \mathrm{L}_{0001}, \mathrm{L}_{0110}, \mathrm{L}_{0111}, \mathrm{L}_{1010}, \mathrm{L}_{1011}, \mathrm{L}_{1100}] \\
\mathcal{M}_{48}  &=&[ \mathrm{B}_{000}, \mathrm{B}_{101}, \mathrm{B}_{111}, \mathrm{L}_{0001}, \mathrm{L}_{0110}, \mathrm{L}_{1000}, \mathrm{L}_{1010}, \mathrm{L}_{1111}] 
\end{eqnarray*}

\[v = \left [ \frac{5 \eta}{3} - \frac{2}{3}, \quad \frac{1}{3} - \frac{\eta}{3}, \quad \frac{1}{3} - \frac{\eta}{3}, \quad  \frac{1}{6} - \frac{\eta}{6}, \quad  \frac{1}{6} - \frac{\eta}{6}, \quad \frac{1}{3} - \frac{\eta}{3}, \quad  \frac{1}{6} - \frac{\eta}{6}, \quad  \frac{1}{6} - \frac{\eta}{6}\right ]\]
\begin{eqnarray*}
\mathcal{M}_{49} &=& [\mathrm{B}_{000}, \mathrm{B}_{010}, \mathrm{B}_{100}, \mathrm{L}_{0001}, \mathrm{L}_{0011}, \mathrm{L}_{0100}, \mathrm{L}_{1001}, \mathrm{L}_{1110}] \\
\mathcal{M}_{50} &=& [\mathrm{B}_{000}, \mathrm{B}_{010}, \mathrm{B}_{101}, \mathrm{L}_{0001}, \mathrm{L}_{0011}, \mathrm{L}_{0110}, \mathrm{L}_{1000}, \mathrm{L}_{1111}]\\
\mathcal{M}_{51} &=& [\mathrm{B}_{000}, \mathrm{B}_{010}, \mathrm{B}_{110}, \mathrm{L}_{0001}, \mathrm{L}_{0011}, \mathrm{L}_{0110}, \mathrm{L}_{1011}, \mathrm{L}_{1100}] \\
\mathcal{M}_{52} &=& [\mathrm{B}_{000}, \mathrm{B}_{010}, \mathrm{B}_{111}, \mathrm{L}_{0001}, \mathrm{L}_{0011}, \mathrm{L}_{0100}, \mathrm{L}_{1010}, \mathrm{L}_{1101}] \\
\mathcal{M}_{53} &=& [\mathrm{B}_{000}, \mathrm{B}_{011}, \mathrm{B}_{100}, \mathrm{L}_{0010}, \mathrm{L}_{0100}, \mathrm{L}_{1001}, \mathrm{L}_{1100}, \mathrm{L}_{1111}] \\
\mathcal{M}_{54} &=& [\mathrm{B}_{000}, \mathrm{B}_{011}, \mathrm{B}_{101}, \mathrm{L}_{0011}, \mathrm{L}_{0101}, \mathrm{L}_{1010}, \mathrm{L}_{1100}, \mathrm{L}_{1111}] \\
\mathcal{M}_{55} &=& [\mathrm{B}_{000}, \mathrm{B}_{011}, \mathrm{B}_{110}, \mathrm{L}_{0000}, \mathrm{L}_{0110}, \mathrm{L}_{1001}, \mathrm{L}_{1100}, \mathrm{L}_{1111}] \\
\mathcal{M}_{56} &=& [\mathrm{B}_{000}, \mathrm{B}_{011}, \mathrm{B}_{111}, \mathrm{L}_{0001}, \mathrm{L}_{0111}, \mathrm{L}_{1010}, \mathrm{L}_{1100}, \mathrm{L}_{1111}] \\
\mathcal{M}_{57} &=& [\mathrm{B}_{000}, \mathrm{B}_{100}, \mathrm{B}_{110}, \mathrm{L}_{0011}, \mathrm{L}_{0100}, \mathrm{L}_{1001}, \mathrm{L}_{1100}, \mathrm{L}_{1110}] \\
\mathcal{M}_{58} &=& [\mathrm{B}_{000}, \mathrm{B}_{100}, \mathrm{B}_{111}, \mathrm{L}_{0001}, \mathrm{L}_{0010}, \mathrm{L}_{0100}, \mathrm{L}_{1001}, \mathrm{L}_{1111}] \\
%\end{eqnarray*}
%\begin{eqnarray*}
\mathcal{M}_{59} &=& [\mathrm{B}_{000}, \mathrm{B}_{101}, \mathrm{B}_{110}, \mathrm{L}_{0000}, \mathrm{L}_{0011}, \mathrm{L}_{0110}, \mathrm{L}_{1001}, \mathrm{L}_{1111}] \\
\mathcal{M}_{60} &=& [\mathrm{B}_{000}, \mathrm{B}_{101}, \mathrm{B}_{111}, \mathrm{L}_{0011}, \mathrm{L}_{0100}, \mathrm{L}_{1010}, \mathrm{L}_{1101}, \mathrm{L}_{1111}] \\
\mathcal{M}_{61} &=& [\mathrm{B}_{000}, \mathrm{B}_{101}, \mathrm{L}_{0011}, \mathrm{L}_{0100}, \mathrm{L}_{0101}, \mathrm{L}_{1010}, \mathrm{L}_{1100}, \mathrm{L}_{1101}] \\
\mathcal{M}_{62} &=& [\mathrm{B}_{000}, \mathrm{B}_{101}, \mathrm{L}_{0011}, \mathrm{L}_{0100}, \mathrm{L}_{0101}, \mathrm{L}_{1010}, \mathrm{L}_{1110}, \mathrm{L}_{1111}] \\
\mathcal{M}_{63}&=&  [\mathrm{B}_{000}, \mathrm{B}_{101}, \mathrm{L}_{0011}, \mathrm{L}_{0110}, \mathrm{L}_{0111}, \mathrm{L}_{1010}, \mathrm{L}_{1100}, \mathrm{L}_{1101}] 
\end{eqnarray*}

\vspace{-1em}
\[v = \left [ \frac{5 \eta}{3} - \frac{2}{3}, \quad \frac{1}{3} - \frac{\eta}{3}, \quad \frac{1}{3} - \frac{\eta}{3}, \quad  \frac{1}{6} - \frac{\eta}{6}, \quad \frac{1}{3} - \frac{\eta}{3}, \quad  \frac{1}{6} - \frac{\eta}{6}, \quad  \frac{1}{6} - \frac{\eta}{6}, \quad  \frac{1}{6} - \frac{\eta}{6}\right ]\]
\[\mathcal{M}_{64} = [\mathrm{B}_{000}, \mathrm{B}_{010}, \mathrm{L}_{0001}, \mathrm{L}_{0011}, \mathrm{L}_{0110}, \mathrm{L}_{0111}, \mathrm{L}_{1000}, \mathrm{L}_{1100}] \]

\vspace{-1em}
\[v = \left [ \frac{5 \eta}{3} - \frac{2}{3}, \quad \frac{1}{3} - \frac{\eta}{3}, \quad \frac{1}{3} - \frac{\eta}{3}, \quad \frac{1}{3} - \frac{\eta}{3}, \quad  \frac{1}{6} - \frac{\eta}{6}, \quad  \frac{1}{6} - \frac{\eta}{6}, \quad  \frac{1}{6} - \frac{\eta}{6}, \quad  \frac{1}{6} - \frac{\eta}{6}\right ]\]
\vspace{-1.5em}
\begin{eqnarray*}
\mathcal{M}_{65} &=& [\mathrm{B}_{000}, \mathrm{B}_{010}, \mathrm{B}_{100}, \mathrm{L}_{0001}, \mathrm{L}_{0100}, \mathrm{L}_{0110}, \mathrm{L}_{1011}, \mathrm{L}_{1100}] \\
\mathcal{M}_{66} &=& [\mathrm{B}_{000}, \mathrm{B}_{010}, \mathrm{B}_{101}, \mathrm{L}_{0011}, \mathrm{L}_{0100}, \mathrm{L}_{0110}, \mathrm{L}_{1010}, \mathrm{L}_{1101}] \\
\mathcal{M}_{67} &=& [\mathrm{B}_{000}, \mathrm{B}_{010}, \mathrm{B}_{110}, \mathrm{L}_{0011}, \mathrm{L}_{0100}, \mathrm{L}_{0110}, \mathrm{L}_{1001}, \mathrm{L}_{1110}] \\
\mathcal{M}_{68}  &=&[\mathrm{B}_{000}, \mathrm{B}_{010}, \mathrm{B}_{111}, \mathrm{L}_{0001}, \mathrm{L}_{0100}, \mathrm{L}_{0110}, \mathrm{L}_{1000}, \mathrm{L}_{1111}] \\
\mathcal{M}_{69} &=& [\mathrm{B}_{000}, \mathrm{B}_{010}, \mathrm{L}_{0001}, \mathrm{L}_{0110}, \mathrm{L}_{1000}, \mathrm{L}_{1011}, \mathrm{L}_{1100}, \mathrm{L}_{1111}] \\
\mathcal{M}_{70}  &=& [\mathrm{B}_{000}, \mathrm{B}_{010}, \mathrm{L}_{0011}, \mathrm{L}_{0100}, \mathrm{L}_{1001}, \mathrm{L}_{1010}, \mathrm{L}_{1101}, \mathrm{L}_{1110}] \\
\mathcal{M}_{71} &=& [\mathrm{B}_{000}, \mathrm{B}_{100}, \mathrm{B}_{111}, \mathrm{L}_{0001}, \mathrm{L}_{0100}, \mathrm{L}_{0111}, \mathrm{L}_{1010}, \mathrm{L}_{1100}] \\
\mathcal{M}_{72}  &=& [\mathrm{B}_{000}, \mathrm{B}_{101}, \mathrm{B}_{110}, \mathrm{L}_{0011}, \mathrm{L}_{0101}, \mathrm{L}_{0110}, \mathrm{L}_{1010}, \mathrm{L}_{1100}] 
\end{eqnarray*}

\vspace{-1em}
\[v = \left [ \frac{5 \eta}{3} - \frac{2}{3}, \quad \frac{1}{3} - \frac{\eta}{3}, \quad \frac{1}{3} - \frac{\eta}{3}, \quad \frac{1}{3} - \frac{\eta}{3}, \quad \frac{1}{3} - \frac{\eta}{3}, \quad \frac{1}{3} - \frac{\eta}{3}\right ]\]
\vspace{-1.5em}
\begin{eqnarray*}
\mathcal{M}_{73} &=& [\mathrm{B}_{000}, \mathrm{B}_{010}, \mathrm{B}_{100}, \mathrm{B}_{111}, \mathrm{L}_{0001}, \mathrm{L}_{0100}] \\
\mathcal{M}_{74} &=& [\mathrm{B}_{000}, \mathrm{B}_{010}, \mathrm{B}_{101}, \mathrm{B}_{110}, \mathrm{L}_{0011}, \mathrm{L}_{0110}] \\
\mathcal{M}_{75} &=& [\mathrm{B}_{000}, \mathrm{B}_{011}, \mathrm{B}_{100}, \mathrm{B}_{110}, \mathrm{L}_{1001}, \mathrm{L}_{1100}] \\
\mathcal{M}_{76} &=& [\mathrm{B}_{000}, \mathrm{B}_{011}, \mathrm{B}_{101}, \mathrm{B}_{111}, \mathrm{L}_{1010}, \mathrm{L}_{1111}] 
\end{eqnarray*}

\item {Having $\eta \in \left(\frac{3}{7}, 1\right).$}
\[v = \left [ \frac{7 \eta}{4} - \frac{3}{4}, \quad \frac{1}{4} - \frac{\eta}{4}, \quad \frac{1}{4} - \frac{\eta}{4}, \quad \frac{1}{8} - \frac{\eta}{8}, \quad \frac{1}{8} - \frac{\eta}{8}, \quad \frac{1}{4} - \frac{\eta}{4}, \quad \frac{1}{8} - \frac{\eta}{8}, \quad \frac{1}{4} - \frac{\eta}{4}, \quad \frac{3}{8} - \frac{3 \eta}{8}\right ]\]
\vspace{-1.5em}
\[\mathcal{M}_{77} = [\mathrm{B}_{000}, \mathrm{B}_{011}, \mathrm{B}_{101}, \mathrm{L}_{0000}, \mathrm{L}_{0010}, \mathrm{L}_{0110}, \mathrm{L}_{1000}, \mathrm{L}_{1001}, \mathrm{L}_{1111}] \]

\[v = \left [ \frac{7 \eta}{4} - \frac{3}{4}, \quad \frac{1}{4} - \frac{\eta}{4}, \quad \frac{1}{4} - \frac{\eta}{4}, \quad \frac{1}{8} - \frac{\eta}{8}, \quad \frac{1}{8} - \frac{\eta}{8}, \quad \frac{1}{4} - \frac{\eta}{4}, \quad \frac{1}{4} - \frac{\eta}{4}, \quad \frac{1}{8} - \frac{\eta}{8}, \quad \frac{3}{8} - \frac{3 \eta}{8}\right ]\]
\begin{eqnarray*}
\mathcal{M}_{78} = [\mathrm{B}_{000}, \mathrm{B}_{011}, \mathrm{B}_{110}, \mathrm{L}_{0001}, \mathrm{L}_{0011}, \mathrm{L}_{0101}, \mathrm{L}_{1010}, \mathrm{L}_{1011}, \mathrm{L}_{1100}] \\
\mathcal{M}_{79} = [\mathrm{B}_{000}, \mathrm{B}_{101}, \mathrm{B}_{111}, \mathrm{L}_{0001}, \mathrm{L}_{0010}, \mathrm{L}_{0110}, \mathrm{L}_{1000}, \mathrm{L}_{1001}, \mathrm{L}_{1111}] 
\end{eqnarray*}

\[v = \left [ \frac{7 \eta}{4} - \frac{3}{4}, \quad \frac{1}{4} - \frac{\eta}{4}, \quad \frac{1}{4} - \frac{\eta}{4}, \quad \frac{1}{8} - \frac{\eta}{8}, \quad \frac{1}{8} - \frac{\eta}{8}, \quad \frac{1}{4} - \frac{\eta}{4}, \quad \frac{3}{8} - \frac{3 \eta}{8}, \quad \frac{1}{8} - \frac{\eta}{8}, \quad \frac{1}{4} - \frac{\eta}{4}\right ]\]
%\vspace{-1em}
\[\mathcal{M}_{80} = [\mathrm{B}_{000}, \mathrm{B}_{011}, \mathrm{B}_{100}, \mathrm{L}_{0000}, \mathrm{L}_{0010}, \mathrm{L}_{0100}, \mathrm{L}_{1001}, \mathrm{L}_{1110}, \mathrm{L}_{1111}] \]

\[v = \left [ \frac{7 \eta}{4} - \frac{3}{4}, \quad \frac{1}{4} - \frac{\eta}{4}, \quad \frac{1}{4} - \frac{\eta}{4}, \quad \frac{1}{8} - \frac{\eta}{8}, \quad \frac{1}{8} - \frac{\eta}{8}, \quad \frac{1}{4} - \frac{\eta}{4}, \quad \frac{3}{8} - \frac{3 \eta}{8}, \quad \frac{1}{4} - \frac{\eta}{4}, \quad \frac{1}{8} - \frac{\eta}{8}\right ]\]
%\vspace{-1em}
\begin{eqnarray*}
\mathcal{M}_{81} = [\mathrm{B}_{000}, \mathrm{B}_{011}, \mathrm{B}_{111}, \mathrm{L}_{0001}, \mathrm{L}_{0011}, \mathrm{L}_{0111}, \mathrm{L}_{1010}, \mathrm{L}_{1100}, \mathrm{L}_{1101}] \\
\mathcal{M}_{82} = [\mathrm{B}_{000}, \mathrm{B}_{100}, \mathrm{B}_{110}, \mathrm{L}_{0000}, \mathrm{L}_{0011}, \mathrm{L}_{0100}, \mathrm{L}_{1001}, \mathrm{L}_{1110}, \mathrm{L}_{1111}] 
\end{eqnarray*}

\[v = \left [ \frac{7 \eta}{4} - \frac{3}{4}, \quad \frac{1}{4} - \frac{\eta}{4}, \quad \frac{1}{4} - \frac{\eta}{4}, \quad \frac{1}{8} - \frac{\eta}{8}, \quad \frac{1}{4} - \frac{\eta}{4}, \quad \frac{3}{8} - \frac{3 \eta}{8}, \quad \frac{1}{8} - \frac{\eta}{8}, \quad \frac{1}{8} - \frac{\eta}{8}, \quad \frac{1}{4} - \frac{\eta}{4}\right ]\]
%\vspace{-1em} 
\begin{eqnarray*}
\mathcal{M}_{83} = [\mathrm{B}_{000}, \mathrm{B}_{010}, \mathrm{B}_{101}, \mathrm{L}_{0000}, \mathrm{L}_{0001}, \mathrm{L}_{0110}, \mathrm{L}_{1000}, \mathrm{L}_{1011}, \mathrm{L}_{1111}] \\
\mathcal{M}_{84} = [\mathrm{B}_{000}, \mathrm{B}_{010}, \mathrm{B}_{111}, \mathrm{L}_{0010}, \mathrm{L}_{0011}, \mathrm{L}_{0100}, \mathrm{L}_{1001}, \mathrm{L}_{1010}, \mathrm{L}_{1101}] 
\end{eqnarray*}
\[v = \left [ \frac{7 \eta}{4} - \frac{3}{4}, \quad \frac{1}{4} - \frac{\eta}{4}, \quad \frac{1}{4} - \frac{\eta}{4}, \quad \frac{1}{8} - \frac{\eta}{8}, \quad \frac{1}{4} - \frac{\eta}{4}, \quad \frac{3}{8} - \frac{3 \eta}{8}, \quad \frac{1}{4} - \frac{\eta}{4}, \quad \frac{1}{8} - \frac{\eta}{8}, \quad \frac{1}{8} - \frac{\eta}{8}\right ]\]
%\vspace{-1em}
\begin{eqnarray*}
\mathcal{M}_{85} = [\mathrm{B}_{000}, \mathrm{B}_{010}, \mathrm{B}_{100}, \mathrm{L}_{0010}, \mathrm{L}_{0011}, \mathrm{L}_{0100}, \mathrm{L}_{1001}, \mathrm{L}_{1101}, \mathrm{L}_{1110}] \\
\mathcal{M}_{86} = [\mathrm{B}_{000}, \mathrm{B}_{010}, \mathrm{B}_{110}, \mathrm{L}_{0000}, \mathrm{L}_{0001}, \mathrm{L}_{0110}, \mathrm{L}_{1011}, \mathrm{L}_{1100}, \mathrm{L}_{1111}] 
\end{eqnarray*}
\[v = \left [ \frac{7 \eta}{4} - \frac{3}{4}, \quad \frac{1}{4} - \frac{\eta}{4}, \quad \frac{1}{4} - \frac{\eta}{4}, \quad \frac{1}{4} - \frac{\eta}{4}, \quad \frac{1}{8} - \frac{\eta}{8}, \quad \frac{1}{8} - \frac{\eta}{8}, \quad \frac{1}{8} - \frac{\eta}{8}, \quad \frac{1}{4} - \frac{\eta}{4}, \quad \frac{3}{8} - \frac{3 \eta}{8}\right ]\]
%\vspace{-1em}
\begin{eqnarray*}
\mathcal{M}_{87} = [\mathrm{B}_{000}, \mathrm{B}_{011}, \mathrm{B}_{111}, \mathrm{L}_{0010}, \mathrm{L}_{0100}, \mathrm{L}_{0110}, \mathrm{L}_{1000}, \mathrm{L}_{1001}, \mathrm{L}_{1111}] \\
\mathcal{M}_{88} = [\mathrm{B}_{000}, \mathrm{B}_{100}, \mathrm{B}_{110}, \mathrm{L}_{0001}, \mathrm{L}_{0101}, \mathrm{L}_{0110}, \mathrm{L}_{1010}, \mathrm{L}_{1011}, \mathrm{L}_{1100}] 
\end{eqnarray*}
\[v = \left [ \frac{7 \eta}{4} - \frac{3}{4}, \quad \frac{1}{4} - \frac{\eta}{4}, \quad \frac{1}{4} - \frac{\eta}{4}, \quad \frac{1}{4} - \frac{\eta}{4}, \quad \frac{1}{8} - \frac{\eta}{8}, \quad \frac{1}{8} - \frac{\eta}{8}, \quad \frac{1}{4} - \frac{\eta}{4}, \quad \frac{1}{8} - \frac{\eta}{8}, \quad \frac{3}{8} - \frac{3 \eta}{8}\right ]\]
%\vspace{-1em}
\[\mathcal{M}_{89} = [\mathrm{B}_{000}, \mathrm{B}_{011}, \mathrm{B}_{100}, \mathrm{L}_{0001}, \mathrm{L}_{0101}, \mathrm{L}_{0111}, \mathrm{L}_{1010}, \mathrm{L}_{1011}, \mathrm{L}_{1100}] \]

\[v = \left [ \frac{7 \eta}{4} - \frac{3}{4}, \quad \frac{1}{4} - \frac{\eta}{4}, \quad \frac{1}{4} - \frac{\eta}{4}, \quad \frac{1}{4} - \frac{\eta}{4}, \quad \frac{1}{8} - \frac{\eta}{8}, \quad \frac{1}{8} - \frac{\eta}{8}, \quad \frac{3}{8} - \frac{3 \eta}{8}, \quad \frac{1}{8} - \frac{\eta}{8}, \quad \frac{1}{4} - \frac{\eta}{4}\right ]\]
\vspace{-1em}
\begin{eqnarray*}
\mathcal{M}_{90} = [\mathrm{B}_{000}, \mathrm{B}_{011}, \mathrm{B}_{110}, \mathrm{L}_{0000}, \mathrm{L}_{0100}, \mathrm{L}_{0110}, \mathrm{L}_{1001}, \mathrm{L}_{1110}, \mathrm{L}_{1111}] \\
\mathcal{M}_{91} = [\mathrm{B}_{000}, \mathrm{B}_{101}, \mathrm{B}_{111}, \mathrm{L}_{0011}, \mathrm{L}_{0100}, \mathrm{L}_{0111}, \mathrm{L}_{1010}, \mathrm{L}_{1100}, \mathrm{L}_{1101}] 
\end{eqnarray*}
\[v = \left [ \frac{7 \eta}{4} - \frac{3}{4}, \quad \frac{1}{4} - \frac{\eta}{4}, \quad \frac{1}{4} - \frac{\eta}{4}, \quad \frac{1}{4} - \frac{\eta}{4}, \quad \frac{1}{8} - \frac{\eta}{8}, \quad \frac{1}{8} - \frac{\eta}{8}, \quad \frac{3}{8} - \frac{3 \eta}{8}, \quad \frac{1}{4} - \frac{\eta}{4}, \quad \frac{1}{8} - \frac{\eta}{8}\right ]\]
%\vspace{-1em}
\[\mathcal{M}_{92} = [\mathrm{B}_{000}, \mathrm{B}_{011}, \mathrm{B}_{101}, \mathrm{L}_{0011}, \mathrm{L}_{0101}, \mathrm{L}_{0111}, \mathrm{L}_{1010}, \mathrm{L}_{1100}, \mathrm{L}_{1101}] \]

\[v = \left [ \frac{7 \eta}{4} - \frac{3}{4}, \quad \frac{1}{4} - \frac{\eta}{4}, \quad \frac{1}{4} - \frac{\eta}{4}, \quad \frac{1}{4} - \frac{\eta}{4}, \quad \frac{1}{8} - \frac{\eta}{8}, \quad \frac{3}{8} - \frac{3 \eta}{8}, \quad \frac{1}{8} - \frac{\eta}{8}, \quad \frac{1}{8} - \frac{\eta}{8}, \quad \frac{1}{4} - \frac{\eta}{4}\right ]\]
%\vspace{-1em}
\[\mathcal{M}_{93} = [\mathrm{B}_{000}, \mathrm{B}_{101}, \mathrm{B}_{110}, \mathrm{L}_{0000}, \mathrm{L}_{0001}, \mathrm{L}_{0110}, \mathrm{L}_{1001}, \mathrm{L}_{1011}, \mathrm{L}_{1111}] \]

\[v = \left [ \frac{7 \eta}{4} - \frac{3}{4}, \quad \frac{1}{4} - \frac{\eta}{4}, \quad \frac{1}{4} - \frac{\eta}{4}, \quad \frac{1}{4} - \frac{\eta}{4}, \quad \frac{1}{8} - \frac{\eta}{8}, \quad \frac{3}{8} - \frac{3 \eta}{8}, \quad \frac{1}{4} - \frac{\eta}{4}, \quad \frac{1}{8} - \frac{\eta}{8}, \quad \frac{1}{8} - \frac{\eta}{8}\right ]\]
%\vspace{-1.5em}
\[\mathcal{M}_{94} = [\mathrm{B}_{000}, \mathrm{B}_{100}, \mathrm{B}_{111}, \mathrm{L}_{0010}, \mathrm{L}_{0011}, \mathrm{L}_{0100}, \mathrm{L}_{1001}, \mathrm{L}_{1101}, \mathrm{L}_{1111}] \]

\vspace{-1em}
\[v = \left [ \frac{7 \eta}{4} - \frac{3}{4}, \quad \frac{1}{4} - \frac{\eta}{4}, \quad \frac{1}{4} - \frac{\eta}{4}, \quad \frac{1}{4} - \frac{\eta}{4}, \quad \frac{1}{4} - \frac{\eta}{4}, \quad \frac{1}{4} - \frac{\eta}{4}, \quad \frac{1}{4} - \frac{\eta}{4}, \quad \frac{1}{4} - \frac{\eta}{4}\right ]\]
%\vspace{-1.5em}
\begin{eqnarray*}
\mathcal{M}_{95} = [\mathrm{B}_{000}, \mathrm{B}_{010}, \mathrm{B}_{100}, \mathrm{B}_{110}, \mathrm{L}_{0001}, \mathrm{L}_{0110}, \mathrm{L}_{1011}, \mathrm{L}_{1100}] \\
\mathcal{M}_{96} = [\mathrm{B}_{000}, \mathrm{B}_{010}, \mathrm{B}_{100}, \mathrm{B}_{110}, \mathrm{L}_{0011}, \mathrm{L}_{0100}, \mathrm{L}_{1001}, \mathrm{L}_{1110}] \\
\mathcal{M}_{97} = [\mathrm{B}_{000}, \mathrm{B}_{010}, \mathrm{B}_{101}, \mathrm{B}_{111}, \mathrm{L}_{0001}, \mathrm{L}_{0110}, \mathrm{L}_{1000}, \mathrm{L}_{1111}] \\
\mathcal{M}_{98} = [\mathrm{B}_{000}, \mathrm{B}_{010}, \mathrm{B}_{101}, \mathrm{B}_{111}, \mathrm{L}_{0011}, \mathrm{L}_{0100}, \mathrm{L}_{1010}, \mathrm{L}_{1101}] \\
\mathcal{M}_{99} = [\mathrm{B}_{000}, \mathrm{B}_{011}, \mathrm{B}_{100}, \mathrm{B}_{111}, \mathrm{L}_{0001}, \mathrm{L}_{0111}, \mathrm{L}_{1010}, \mathrm{L}_{1100}] \\
\mathcal{M}_{100} = [\mathrm{B}_{000}, \mathrm{B}_{011}, \mathrm{B}_{100}, \mathrm{B}_{111}, \mathrm{L}_{0010}, \mathrm{L}_{0100}, \mathrm{L}_{1001}, \mathrm{L}_{1111}] \\
\mathcal{M}_{101} = [\mathrm{B}_{000}, \mathrm{B}_{011}, \mathrm{B}_{101}, \mathrm{B}_{110}, \mathrm{L}_{0000}, \mathrm{L}_{0110}, \mathrm{L}_{1001}, \mathrm{L}_{1111}] \\
\mathcal{M}_{102} = [\mathrm{B}_{000}, \mathrm{B}_{011}, \mathrm{B}_{101}, \mathrm{B}_{110}, \mathrm{L}_{0011}, \mathrm{L}_{0101}, \mathrm{L}_{1010}, \mathrm{L}_{1100}] 
\end{eqnarray*}
\[v = \left [ \frac{7 \eta}{4} - \frac{3}{4}, \quad \frac{1}{4} - \frac{\eta}{4}, \quad \frac{1}{4} - \frac{\eta}{4}, \quad \frac{3}{8} - \frac{3 \eta}{8}, \quad \frac{1}{8} - \frac{\eta}{8}, \quad \frac{1}{4} - \frac{\eta}{4}, \quad \frac{1}{8} - \frac{\eta}{8}, \quad \frac{1}{8} - \frac{\eta}{8}, \quad \frac{1}{4} - \frac{\eta}{4}\right ]\]
%\vspace{-1em}
\[\mathcal{M}_{103} = [\mathrm{B}_{000}, \mathrm{B}_{100}, \mathrm{B}_{111}, \mathrm{L}_{0001}, \mathrm{L}_{0110}, \mathrm{L}_{0111}, \mathrm{L}_{1000}, \mathrm{L}_{1010}, \mathrm{L}_{1100}] \]

\[v = \left [ \frac{7 \eta}{4} - \frac{3}{4}, \quad \frac{1}{4} - \frac{\eta}{4}, \quad \frac{1}{4} - \frac{\eta}{4}, \quad \frac{3}{8} - \frac{3 \eta}{8}, \quad \frac{1}{8} - \frac{\eta}{8}, \quad \frac{1}{4} - \frac{\eta}{4}, \quad \frac{1}{4} - \frac{\eta}{4}, \quad \frac{1}{8} - \frac{\eta}{8}, \quad \frac{1}{8} - \frac{\eta}{8}\right ]\]
%\vspace{-1.5em}
\[\mathcal{M}_{104} = [\mathrm{B}_{000}, \mathrm{B}_{101}, \mathrm{B}_{110}, \mathrm{L}_{0011}, \mathrm{L}_{0100}, \mathrm{L}_{0101}, \mathrm{L}_{1010}, \mathrm{L}_{1100}, \mathrm{L}_{1110}] \]

\[v = \left [ \frac{7 \eta}{4} - \frac{3}{4}, \quad \frac{1}{4} - \frac{\eta}{4}, \quad \frac{1}{4} - \frac{\eta}{4}, \quad \frac{3}{8} - \frac{3 \eta}{8}, \quad \frac{1}{4} - \frac{\eta}{4}, \quad \frac{1}{8} - \frac{\eta}{8}, \quad \frac{1}{8} - \frac{\eta}{8}, \quad \frac{1}{8} - \frac{\eta}{8}, \quad \frac{1}{4} - \frac{\eta}{4}\right ]\]
%\vspace{-1.5em}
\begin{eqnarray*}
\mathcal{M}_{105} = [\mathrm{B}_{000}, \mathrm{B}_{010}, \mathrm{B}_{100}, \mathrm{L}_{0001}, \mathrm{L}_{0110}, \mathrm{L}_{0111}, \mathrm{L}_{1000}, \mathrm{L}_{1011}, \mathrm{L}_{1100}] \\
\mathcal{M}_{106} = [\mathrm{B}_{000}, \mathrm{B}_{010}, \mathrm{B}_{110}, \mathrm{L}_{0011}, \mathrm{L}_{0100}, \mathrm{L}_{0101}, \mathrm{L}_{1001}, \mathrm{L}_{1010}, \mathrm{L}_{1110}] 
\end{eqnarray*}

\[v = \left [ \frac{7 \eta}{4} - \frac{3}{4}, \quad \frac{1}{4} - \frac{\eta}{4}, \quad \frac{1}{4} - \frac{\eta}{4}, \quad \frac{3}{8} - \frac{3 \eta}{8}, \quad \frac{1}{4} - \frac{\eta}{4}, \quad \frac{1}{8} - \frac{\eta}{8}, \quad \frac{1}{4} - \frac{\eta}{4}, \quad \frac{1}{8} - \frac{\eta}{8}, \quad \frac{1}{8} - \frac{\eta}{8}\right ]\]
%\vspace{-1.5em}
\begin{eqnarray*}
\mathcal{M}_{107} = [\mathrm{B}_{000}, \mathrm{B}_{010}, \mathrm{B}_{101}, \mathrm{L}_{0011}, \mathrm{L}_{0100}, \mathrm{L}_{0101}, \mathrm{L}_{1010}, \mathrm{L}_{1101}, \mathrm{L}_{1110}] \\
\mathcal{M}_{108} = [\mathrm{B}_{000}, \mathrm{B}_{010}, \mathrm{B}_{111}, \mathrm{L}_{0001}, \mathrm{L}_{0110}, \mathrm{L}_{0111}, \mathrm{L}_{1000}, \mathrm{L}_{1100}, \mathrm{L}_{1111}] 
\end{eqnarray*}

\vspace{1em}
\item {Having $\eta \in \left(\frac{4}{9}, 1\right).$}

\[v = \left [ \frac{9 \eta}{5} - \frac{4}{5}, \quad \frac{1}{5} - \frac{\eta}{5}, \quad \frac{1}{5} - \frac{\eta}{5}, \quad \frac{1}{5} - \frac{\eta}{5}, \quad \frac{1}{5} - \frac{\eta}{5}, \quad \frac{1}{5} - \frac{\eta}{5}, \quad \frac{1}{5} - \frac{\eta}{5}, \quad \frac{1}{5} - \frac{\eta}{5}, \quad \frac{2}{5} - \frac{2 \eta}{5}\right ]\]
%\vspace{-1.5em}
\begin{eqnarray*}
\mathcal{M}_{109} = [\mathrm{B}_{000}, \mathrm{B}_{011}, \mathrm{B}_{100}, \mathrm{B}_{110}, \mathrm{L}_{0001}, \mathrm{L}_{0101}, \mathrm{L}_{1010}, \mathrm{L}_{1011}, \mathrm{L}_{1100}] \\
\mathcal{M}_{110} = [\mathrm{B}_{000}, \mathrm{B}_{011}, \mathrm{B}_{101}, \mathrm{B}_{111}, \mathrm{L}_{0010}, \mathrm{L}_{0110}, \mathrm{L}_{1000}, \mathrm{L}_{1001}, \mathrm{L}_{1111}] 
\end{eqnarray*}

\[v = \left [ \frac{9 \eta}{5} - \frac{4}{5}, \quad \frac{1}{5} - \frac{\eta}{5}, \quad \frac{1}{5} - \frac{\eta}{5}, \quad \frac{1}{5} - \frac{\eta}{5}, \quad \frac{1}{5} - \frac{\eta}{5}, \quad \frac{1}{5} - \frac{\eta}{5}, \quad \frac{2}{5} - \frac{2 \eta}{5}, \quad \frac{1}{5} - \frac{\eta}{5}, \quad \frac{1}{5} - \frac{\eta}{5}\right ]\]
\vspace{-1.5em}
\begin{eqnarray*}
\mathcal{M}_{111} = [\mathrm{B}_{000}, \mathrm{B}_{010}, \mathrm{B}_{100}, \mathrm{B}_{111}, \mathrm{L}_{0010}, \mathrm{L}_{0011}, \mathrm{L}_{0100}, \mathrm{L}_{1001}, \mathrm{L}_{1101}] \\
\mathcal{M}_{112} = [\mathrm{B}_{000}, \mathrm{B}_{010}, \mathrm{B}_{101}, \mathrm{B}_{110}, \mathrm{L}_{0000}, \mathrm{L}_{0001}, \mathrm{L}_{0110}, \mathrm{L}_{1011}, \mathrm{L}_{1111}] \\
\mathcal{M}_{113} = [\mathrm{B}_{000}, \mathrm{B}_{011}, \mathrm{B}_{100}, \mathrm{B}_{110}, \mathrm{L}_{0000}, \mathrm{L}_{0100}, \mathrm{L}_{1001}, \mathrm{L}_{1110}, \mathrm{L}_{1111}] \\
\mathcal{M}_{114} = [\mathrm{B}_{000}, \mathrm{B}_{011}, \mathrm{B}_{101}, \mathrm{B}_{111}, \mathrm{L}_{0011}, \mathrm{L}_{0111}, \mathrm{L}_{1010}, \mathrm{L}_{1100}, \mathrm{L}_{1101}] 
\end{eqnarray*}

%\vspace{-1.5em}
\[v = \left [ \frac{9 \eta}{5} - \frac{4}{5}, \quad \frac{1}{5} - \frac{\eta}{5}, \quad \frac{1}{5} - \frac{\eta}{5}, \quad \frac{1}{5} - \frac{\eta}{5}, \quad \frac{2}{5} - \frac{2 \eta}{5}, \quad \frac{1}{5} - \frac{\eta}{5}, \quad \frac{1}{5} - \frac{\eta}{5}, \quad \frac{1}{5} - \frac{\eta}{5}, \quad \frac{1}{5} - \frac{\eta}{5}\right ]\]
%\vspace{-1.5em}
\begin{eqnarray*}
\mathcal{M}_{115} = [\mathrm{B}_{000}, \mathrm{B}_{010}, \mathrm{B}_{100}, \mathrm{B}_{111}, \mathrm{L}_{0001}, \mathrm{L}_{0110}, \mathrm{L}_{0111}, \mathrm{L}_{1000}, \mathrm{L}_{1100}] \\
\mathcal{M}_{116} = [\mathrm{B}_{000}, \mathrm{B}_{010}, \mathrm{B}_{101}, \mathrm{B}_{110}, \mathrm{L}_{0011}, \mathrm{L}_{0100}, \mathrm{L}_{0101}, \mathrm{L}_{1010}, \mathrm{L}_{1110}] 
\end{eqnarray*}

\item {Having $\eta \in \left(\frac{1}{2}, 1\right).$}

\[v = \left [ 2 \eta - 1, \quad \frac{1}{3} - \frac{\eta}{3}, \quad \frac{1}{3} - \frac{\eta}{3}, \quad  \frac{1}{6} - \frac{\eta}{6}, \quad  \frac{1}{6} - \frac{\eta}{6}, \quad  \frac{1}{6} - \frac{\eta}{6}, \quad  \frac{1}{6} - \frac{\eta}{6}, \quad \frac{1}{3} - \frac{\eta}{3}, \quad \frac{1}{3} - \frac{\eta}{3}\right ]\]
%\vspace{-1.5em}
\begin{eqnarray*}
\mathcal{M}_{117} = [\mathrm{B}_{000}, \mathrm{B}_{100}, \mathrm{B}_{110}, \mathrm{L}_{0000}, \mathrm{L}_{0011}, \mathrm{L}_{0100}, \mathrm{L}_{0111}, \mathrm{L}_{1001}, \mathrm{L}_{1110}] \\
\mathcal{M}_{118} = [\mathrm{B}_{000}, \mathrm{B}_{100}, \mathrm{B}_{110}, \mathrm{L}_{0001}, \mathrm{L}_{0010}, \mathrm{L}_{0101}, \mathrm{L}_{0110}, \mathrm{L}_{1011}, \mathrm{L}_{1100}] \\
\mathcal{M}_{119} = [\mathrm{B}_{000}, \mathrm{B}_{101}, \mathrm{B}_{111}, \mathrm{L}_{0000}, \mathrm{L}_{0011}, \mathrm{L}_{0100}, \mathrm{L}_{0111}, \mathrm{L}_{1010}, \mathrm{L}_{1101}] \\
\mathcal{M}_{120} = [\mathrm{B}_{000}, \mathrm{B}_{101}, \mathrm{B}_{111}, \mathrm{L}_{0001}, \mathrm{L}_{0010}, \mathrm{L}_{0101}, \mathrm{L}_{0110}, \mathrm{L}_{1000}, \mathrm{L}_{1111}]
\end{eqnarray*}

\vspace{-1em}
\[v = \left [ 2 \eta - 1, \quad \frac{1}{3} - \frac{\eta}{3}, \quad \frac{1}{3} - \frac{\eta}{3}, \quad  \frac{1}{6} - \frac{\eta}{6}, \quad  \frac{1}{6} - \frac{\eta}{6}, \quad  \frac{1}{6} - \frac{\eta}{6}, \quad \frac{1}{3} - \frac{\eta}{3}, \quad  \frac{1}{6} - \frac{\eta}{6}, \quad \frac{1}{3} - \frac{\eta}{3}\right ]\]
%\vspace{-1.5em}
\begin{eqnarray*}
\mathcal{M}_{121} = [\mathrm{B}_{000}, \mathrm{B}_{100}, \mathrm{B}_{110}, \mathrm{L}_{0011}, \mathrm{L}_{0111}, \mathrm{L}_{1000}, \mathrm{L}_{1001}, \mathrm{L}_{1100}, \mathrm{L}_{1110}] \\
\mathcal{M}_{122} = [\mathrm{B}_{000}, \mathrm{B}_{101}, \mathrm{B}_{111}, \mathrm{L}_{0011}, \mathrm{L}_{0111}, \mathrm{L}_{1000}, \mathrm{L}_{1010}, \mathrm{L}_{1100}, \mathrm{L}_{1101}] 
\end{eqnarray*}

\vspace{-1em}
\[v = \left [ 2 \eta - 1, \quad \frac{1}{3} - \frac{\eta}{3}, \quad \frac{1}{3} - \frac{\eta}{3}, \quad  \frac{1}{6} - \frac{\eta}{6}, \quad  \frac{1}{6} - \frac{\eta}{6}, \quad  \frac{1}{6} - \frac{\eta}{6}, \quad \frac{1}{3} - \frac{\eta}{3}, \quad \frac{1}{3} - \frac{\eta}{3}, \quad  \frac{1}{6} - \frac{\eta}{6}\right ]\]
%\vspace{-1.5em}
\begin{eqnarray*}
\mathcal{M}_{123} = [\mathrm{B}_{000}, \mathrm{B}_{100}, \mathrm{B}_{110}, \mathrm{L}_{0001}, \mathrm{L}_{0101}, \mathrm{L}_{1010}, \mathrm{L}_{1011}, \mathrm{L}_{1100}, \mathrm{L}_{1110}] \\
\mathcal{M}_{124} = [\mathrm{B}_{000}, \mathrm{B}_{100}, \mathrm{B}_{110}, \mathrm{L}_{0010}, \mathrm{L}_{0110}, \mathrm{L}_{1001}, \mathrm{L}_{1011}, \mathrm{L}_{1100}, \mathrm{L}_{1101}] 
\end{eqnarray*}

%\vspace{-1em}
\[v = \left [ 2 \eta - 1, \quad \frac{1}{3} - \frac{\eta}{3}, \quad \frac{1}{3} - \frac{\eta}{3}, \quad  \frac{1}{6} - \frac{\eta}{6}, \quad  \frac{1}{6} - \frac{\eta}{6}, \quad \frac{1}{3} - \frac{\eta}{3}, \quad  \frac{1}{6} - \frac{\eta}{6}, \quad  \frac{1}{6} - \frac{\eta}{6}, \quad \frac{1}{3} - \frac{\eta}{3}\right ]\]\vspace{-1.5em}
\begin{eqnarray*}
\mathcal{M}_{125} = [\mathrm{B}_{000}, \mathrm{B}_{010}, \mathrm{B}_{111}, \mathrm{L}_{0000}, \mathrm{L}_{0001}, \mathrm{L}_{0100}, \mathrm{L}_{1010}, \mathrm{L}_{1011}, \mathrm{L}_{1101}] \\
\mathcal{M}_{126} = [\mathrm{B}_{000}, \mathrm{B}_{010}, \mathrm{B}_{111}, \mathrm{L}_{0010}, \mathrm{L}_{0011}, \mathrm{L}_{0100}, \mathrm{L}_{1000}, \mathrm{L}_{1001}, \mathrm{L}_{1101}] \\
\mathcal{M}_{127} = [\mathrm{B}_{000}, \mathrm{B}_{010}, \mathrm{B}_{111}, \mathrm{L}_{0010}, \mathrm{L}_{0011}, \mathrm{L}_{0100}, \mathrm{L}_{1010}, \mathrm{L}_{1011}, \mathrm{L}_{1101}] \\
\mathcal{M}_{128} = [\mathrm{B}_{000}, \mathrm{B}_{011}, \mathrm{B}_{110}, \mathrm{L}_{0000}, \mathrm{L}_{0001}, \mathrm{L}_{0101}, \mathrm{L}_{1010}, \mathrm{L}_{1011}, \mathrm{L}_{1100}] \\
\mathcal{M}_{129} = [\mathrm{B}_{000}, \mathrm{B}_{011}, \mathrm{B}_{110}, \mathrm{L}_{0010}, \mathrm{L}_{0011}, \mathrm{L}_{0101}, \mathrm{L}_{1000}, \mathrm{L}_{1001}, \mathrm{L}_{1100}] \\
\mathcal{M}_{130} = [\mathrm{B}_{000}, \mathrm{B}_{011}, \mathrm{B}_{110}, \mathrm{L}_{0010}, \mathrm{L}_{0011}, \mathrm{L}_{0101}, \mathrm{L}_{1010}, \mathrm{L}_{1011}, \mathrm{L}_{1100}] \\
\mathcal{M}_{131} = [\mathrm{B}_{000}, \mathrm{B}_{101}, \mathrm{B}_{111}, \mathrm{L}_{0001}, \mathrm{L}_{0101}, \mathrm{L}_{1000}, \mathrm{L}_{1010}, \mathrm{L}_{1110}, \mathrm{L}_{1111}] \\
\mathcal{M}_{132} = [\mathrm{B}_{000}, \mathrm{B}_{101}, \mathrm{B}_{111}, \mathrm{L}_{0010}, \mathrm{L}_{0110}, \mathrm{L}_{1000}, \mathrm{L}_{1001}, \mathrm{L}_{1101}, \mathrm{L}_{1111}] 
\end{eqnarray*}

%\vspace{-1.5em}
\[v = \left [ 2 \eta - 1, \quad \frac{1}{3} - \frac{\eta}{3}, \quad \frac{1}{3} - \frac{\eta}{3}, \quad  \frac{1}{6} - \frac{\eta}{6}, \quad  \frac{1}{6} - \frac{\eta}{6}, \quad \frac{1}{3} - \frac{\eta}{3}, \quad  \frac{1}{6} - \frac{\eta}{6}, \quad \frac{1}{3} - \frac{\eta}{3}, \quad  \frac{1}{6} - \frac{\eta}{6}\right ]\]
%\vspace{-1.5em}
\begin{eqnarray*}
\mathcal{M}_{133} = [\mathrm{B}_{000}, \mathrm{B}_{100}, \mathrm{B}_{110}, \mathrm{L}_{0000}, \mathrm{L}_{0100}, \mathrm{L}_{1001}, \mathrm{L}_{1011}, \mathrm{L}_{1110}, \mathrm{L}_{1111}] \\
\mathcal{M}_{134} = [\mathrm{B}_{000}, \mathrm{B}_{101}, \mathrm{B}_{111}, \mathrm{L}_{0000}, \mathrm{L}_{0100}, \mathrm{L}_{1010}, \mathrm{L}_{1011}, \mathrm{L}_{1101}, \mathrm{L}_{1111}] 
\end{eqnarray*}

\vspace{-1em}
\[v = \left [ 2 \eta - 1, \quad \frac{1}{3} - \frac{\eta}{3}, \quad \frac{1}{3} - \frac{\eta}{3}, \quad  \frac{1}{6} - \frac{\eta}{6}, \quad  \frac{1}{6} - \frac{\eta}{6}, \quad \frac{1}{3} - \frac{\eta}{3}, \quad \frac{1}{3} - \frac{\eta}{3}, \quad  \frac{1}{6} - \frac{\eta}{6}, \quad  \frac{1}{6} - \frac{\eta}{6}\right ]\]
%\vspace{-1.5em}
\begin{eqnarray*}
\mathcal{M}_{135} = [\mathrm{B}_{000}, \mathrm{B}_{010}, \mathrm{B}_{110}, \mathrm{L}_{0000}, \mathrm{L}_{0001}, \mathrm{L}_{0110}, \mathrm{L}_{1011}, \mathrm{L}_{1100}, \mathrm{L}_{1101}] \\
\mathcal{M}_{136} = [\mathrm{B}_{000}, \mathrm{B}_{010}, \mathrm{B}_{110}, \mathrm{L}_{0000}, \mathrm{L}_{0001}, \mathrm{L}_{0110}, \mathrm{L}_{1011}, \mathrm{L}_{1110}, \mathrm{L}_{1111}] \\
\mathcal{M}_{137} = [\mathrm{B}_{000}, \mathrm{B}_{010}, \mathrm{B}_{110}, \mathrm{L}_{0010}, \mathrm{L}_{0011}, \mathrm{L}_{0110}, \mathrm{L}_{1011}, \mathrm{L}_{1100}, \mathrm{L}_{1101}] \\
\mathcal{M}_{138} = [\mathrm{B}_{000}, \mathrm{B}_{011}, \mathrm{B}_{111}, \mathrm{L}_{0000}, \mathrm{L}_{0001}, \mathrm{L}_{0111}, \mathrm{L}_{1010}, \mathrm{L}_{1100}, \mathrm{L}_{1101}] \\
\mathcal{M}_{139} = [\mathrm{B}_{000}, \mathrm{B}_{011}, \mathrm{B}_{111}, \mathrm{L}_{0000}, \mathrm{L}_{0001}, \mathrm{L}_{0111}, \mathrm{L}_{1010}, \mathrm{L}_{1110}, \mathrm{L}_{1111}] \\
\mathcal{M}_{140} = [\mathrm{B}_{000}, \mathrm{B}_{011}, \mathrm{B}_{111}, \mathrm{L}_{0010}, \mathrm{L}_{0011}, \mathrm{L}_{0111}, \mathrm{L}_{1010}, \mathrm{L}_{1100}, \mathrm{L}_{1101}] 
\end{eqnarray*}
\[v = \left [ 2 \eta - 1, \quad \frac{1}{3} - \frac{\eta}{3}, \quad \frac{1}{3} - \frac{\eta}{3}, \quad  \frac{1}{6} - \frac{\eta}{6}, \quad \frac{1}{3} - \frac{\eta}{3}, \quad  \frac{1}{6} - \frac{\eta}{6}, \quad \frac{1}{3} - \frac{\eta}{3}, \quad  \frac{1}{6} - \frac{\eta}{6}, \quad  \frac{1}{6} - \frac{\eta}{6}\right ]\]
\vspace{-1.5em}
\begin{eqnarray*}
\mathcal{M}_{141} = [\mathrm{B}_{000}, \mathrm{B}_{100}, \mathrm{B}_{111}, \mathrm{L}_{0000}, \mathrm{L}_{0001}, \mathrm{L}_{0100}, \mathrm{L}_{0111}, \mathrm{L}_{1010}, \mathrm{L}_{1110}] \\
\mathcal{M}_{142} = [\mathrm{B}_{000}, \mathrm{B}_{101}, \mathrm{B}_{110}, \mathrm{L}_{0000}, \mathrm{L}_{0011}, \mathrm{L}_{0100}, \mathrm{L}_{0101}, \mathrm{L}_{1010}, \mathrm{L}_{1110}] 
\end{eqnarray*}
\[v = \left [ 2 \eta - 1, \quad \frac{1}{3} - \frac{\eta}{3}, \quad \frac{1}{3} - \frac{\eta}{3}, \quad  \frac{1}{6} - \frac{\eta}{6}, \quad \frac{1}{3} - \frac{\eta}{3}, \quad \frac{1}{3} - \frac{\eta}{3}, \quad  \frac{1}{6} - \frac{\eta}{6}, \quad  \frac{1}{6} - \frac{\eta}{6}, \quad  \frac{1}{6} - \frac{\eta}{6}\right ]\]
\begin{eqnarray*}
\mathcal{M}_{143} = [\mathrm{B}_{000}, \mathrm{B}_{100}, \mathrm{B}_{111}, \mathrm{L}_{0001}, \mathrm{L}_{0010}, \mathrm{L}_{0100}, \mathrm{L}_{0101}, \mathrm{L}_{1011}, \mathrm{L}_{1111}] \\
\mathcal{M}_{144} = [\mathrm{B}_{000}, \mathrm{B}_{101}, \mathrm{B}_{110}, \mathrm{L}_{0010}, \mathrm{L}_{0011}, \mathrm{L}_{0101}, \mathrm{L}_{0110}, \mathrm{L}_{1000}, \mathrm{L}_{1100}] 
\end{eqnarray*}
\[v = \left [ 2 \eta - 1, \quad \frac{1}{3} - \frac{\eta}{3}, \quad \frac{1}{3} - \frac{\eta}{3}, \quad \frac{1}{3} - \frac{\eta}{3}, \quad  \frac{1}{6} - \frac{\eta}{6}, \quad  \frac{1}{6} - \frac{\eta}{6}, \quad  \frac{1}{6} - \frac{\eta}{6}, \quad  \frac{1}{6} - \frac{\eta}{6}, \quad \frac{1}{3} - \frac{\eta}{3}\right ]\]
\begin{eqnarray*}
\mathcal{M}_{145} = [\mathrm{B}_{000}, \mathrm{B}_{010}, \mathrm{B}_{110}, \mathrm{L}_{0011}, \mathrm{L}_{0100}, \mathrm{L}_{0101}, \mathrm{L}_{1000}, \mathrm{L}_{1001}, \mathrm{L}_{1110}] \\
\mathcal{M}_{146} = [\mathrm{B}_{000}, \mathrm{B}_{010}, \mathrm{B}_{110}, \mathrm{L}_{0011}, \mathrm{L}_{0100}, \mathrm{L}_{0101}, \mathrm{L}_{1010}, \mathrm{L}_{1011}, \mathrm{L}_{1110}] \\
\mathcal{M}_{147} = [\mathrm{B}_{000}, \mathrm{B}_{010}, \mathrm{B}_{110}, \mathrm{L}_{0011}, \mathrm{L}_{0110}, \mathrm{L}_{0111}, \mathrm{L}_{1000}, \mathrm{L}_{1001}, \mathrm{L}_{1110}] \\
\mathcal{M}_{148} = [\mathrm{B}_{000}, \mathrm{B}_{011}, \mathrm{B}_{111}, \mathrm{L}_{0010}, \mathrm{L}_{0100}, \mathrm{L}_{0101}, \mathrm{L}_{1000}, \mathrm{L}_{1001}, \mathrm{L}_{1111}] \\
\mathcal{M}_{149} = [\mathrm{B}_{000}, \mathrm{B}_{011}, \mathrm{B}_{111}, \mathrm{L}_{0010}, \mathrm{L}_{0100}, \mathrm{L}_{0101}, \mathrm{L}_{1010}, \mathrm{L}_{1011}, \mathrm{L}_{1111}] \\
\mathcal{M}_{150} = [\mathrm{B}_{000}, \mathrm{B}_{011}, \mathrm{B}_{111}, \mathrm{L}_{0010}, \mathrm{L}_{0110}, \mathrm{L}_{0111}, \mathrm{L}_{1000}, \mathrm{L}_{1001}, \mathrm{L}_{1111}] 
\end{eqnarray*}

\vspace{-1.5em}
\[v = \left [ 2 \eta - 1, \quad \frac{1}{3} - \frac{\eta}{3}, \quad \frac{1}{3} - \frac{\eta}{3}, \quad \frac{1}{3} - \frac{\eta}{3}, \quad  \frac{1}{6} - \frac{\eta}{6}, \quad  \frac{1}{6} - \frac{\eta}{6}, \quad \frac{1}{3} - \frac{\eta}{3}, \quad  \frac{1}{6} - \frac{\eta}{6}, \quad  \frac{1}{6} - \frac{\eta}{6}\right ]\]
\begin{eqnarray*}
\mathcal{M}_{151} = [\mathrm{B}_{000}, \mathrm{B}_{010}, \mathrm{B}_{111}, \mathrm{L}_{0001}, \mathrm{L}_{0100}, \mathrm{L}_{0101}, \mathrm{L}_{1000}, \mathrm{L}_{1110}, \mathrm{L}_{1111}] \\
\mathcal{M}_{152} = [\mathrm{B}_{000}, \mathrm{B}_{010}, \mathrm{B}_{111}, \mathrm{L}_{0001}, \mathrm{L}_{0110}, \mathrm{L}_{0111}, \mathrm{L}_{1000}, \mathrm{L}_{1100}, \mathrm{L}_{1101}] \\
\mathcal{M}_{153} = [\mathrm{B}_{000}, \mathrm{B}_{010}, \mathrm{B}_{111}, \mathrm{L}_{0001}, \mathrm{L}_{0110}, \mathrm{L}_{0111}, \mathrm{L}_{1000}, \mathrm{L}_{1110}, \mathrm{L}_{1111}] \\
\mathcal{M}_{154} = [\mathrm{B}_{000}, \mathrm{B}_{011}, \mathrm{B}_{110}, \mathrm{L}_{0000}, \mathrm{L}_{0100}, \mathrm{L}_{0101}, \mathrm{L}_{1001}, \mathrm{L}_{1110}, \mathrm{L}_{1111}] \\
\mathcal{M}_{155} = [\mathrm{B}_{000}, \mathrm{B}_{011}, \mathrm{B}_{110}, \mathrm{L}_{0000}, \mathrm{L}_{0110}, \mathrm{L}_{0111}, \mathrm{L}_{1001}, \mathrm{L}_{1100}, \mathrm{L}_{1101}] \\
\mathcal{M}_{156} = [\mathrm{B}_{000}, \mathrm{B}_{011}, \mathrm{B}_{110}, \mathrm{L}_{0000}, \mathrm{L}_{0110}, \mathrm{L}_{0111}, \mathrm{L}_{1001}, \mathrm{L}_{1110}, \mathrm{L}_{1111}] \\
\mathcal{M}_{157} = [\mathrm{B}_{000}, \mathrm{B}_{100}, \mathrm{B}_{111}, \mathrm{L}_{0001}, \mathrm{L}_{0010}, \mathrm{L}_{0110}, \mathrm{L}_{0111}, \mathrm{L}_{1000}, \mathrm{L}_{1100}] \\
\mathcal{M}_{158} = [\mathrm{B}_{000}, \mathrm{B}_{101}, \mathrm{B}_{110}, \mathrm{L}_{0000}, \mathrm{L}_{0001}, \mathrm{L}_{0101}, \mathrm{L}_{0110}, \mathrm{L}_{1011}, \mathrm{L}_{1111}] 
\end{eqnarray*}
\[v = \left [ 2 \eta - 1, \quad \frac{1}{3} - \frac{\eta}{3}, \quad \frac{1}{3} - \frac{\eta}{3}, \quad \frac{1}{3} - \frac{\eta}{3}, \quad  \frac{1}{6} - \frac{\eta}{6}, \quad \frac{1}{3} - \frac{\eta}{3}, \quad  \frac{1}{6} - \frac{\eta}{6}, \quad  \frac{1}{6} - \frac{\eta}{6}, \quad  \frac{1}{6} - \frac{\eta}{6}\right ]\]
\begin{eqnarray*}
\mathcal{M}_{159} = [\mathrm{B}_{000}, \mathrm{B}_{100}, \mathrm{B}_{111}, \mathrm{L}_{0010}, \mathrm{L}_{0011}, \mathrm{L}_{0100}, \mathrm{L}_{0111}, \mathrm{L}_{1001}, \mathrm{L}_{1101}] \\
\mathcal{M}_{160} = [\mathrm{B}_{000}, \mathrm{B}_{101}, \mathrm{B}_{110}, \mathrm{L}_{0000}, \mathrm{L}_{0011}, \mathrm{L}_{0110}, \mathrm{L}_{0111}, \mathrm{L}_{1001}, \mathrm{L}_{1101}] 
\end{eqnarray*}

\[v = \left [ 2 \eta - 1, \quad \frac{1}{3} - \frac{\eta}{3}, \quad \frac{1}{3} - \frac{\eta}{3}, \quad \frac{1}{3} - \frac{\eta}{3}, \quad \frac{1}{3} - \frac{\eta}{3}, \quad  \frac{1}{6} - \frac{\eta}{6}, \quad  \frac{1}{6} - \frac{\eta}{6}, \quad  \frac{1}{6} - \frac{\eta}{6}, \quad  \frac{1}{6} - \frac{\eta}{6}\right ]\]
\begin{eqnarray*}
\mathcal{M}_{161} = [\mathrm{B}_{000}, \mathrm{B}_{100}, \mathrm{B}_{111}, \mathrm{L}_{0001}, \mathrm{L}_{0111}, \mathrm{L}_{1000}, \mathrm{L}_{1010}, \mathrm{L}_{1100}, \mathrm{L}_{1110}] \\
\mathcal{M}_{162} = [\mathrm{B}_{000}, \mathrm{B}_{100}, \mathrm{B}_{111}, \mathrm{L}_{0010}, \mathrm{L}_{0100}, \mathrm{L}_{1001}, \mathrm{L}_{1011}, \mathrm{L}_{1101}, \mathrm{L}_{1111}] \\
\mathcal{M}_{163} = [\mathrm{B}_{000}, \mathrm{B}_{101}, \mathrm{B}_{110}, \mathrm{L}_{0000}, \mathrm{L}_{0110}, \mathrm{L}_{1001}, \mathrm{L}_{1011}, \mathrm{L}_{1101}, \mathrm{L}_{1111}] \\
\mathcal{M}_{164} = [\mathrm{B}_{000}, \mathrm{B}_{101}, \mathrm{B}_{110}, \mathrm{L}_{0011}, \mathrm{L}_{0101}, \mathrm{L}_{1000}, \mathrm{L}_{1010}, \mathrm{L}_{1100}, \mathrm{L}_{1110}] 
\end{eqnarray*}

\[v = \left [ 2 \eta - 1, \quad \frac{1}{2} - \frac{\eta}{2}, \quad \frac{1}{4} - \frac{\eta}{4}, \quad \frac{1}{4} - \frac{\eta}{4}, \quad \frac{1}{4} - \frac{\eta}{4}, \quad \frac{1}{4} - \frac{\eta}{4}, \quad \frac{1}{4} - \frac{\eta}{4}, \quad \frac{1}{4} - \frac{\eta}{4}\right ]\]
\begin{eqnarray*}
\mathcal{M}_{165} = [\mathrm{B}_{000}, \mathrm{B}_{010}, \mathrm{L}_{0000}, \mathrm{L}_{0001}, \mathrm{L}_{0100}, \mathrm{L}_{0110}, \mathrm{L}_{1011}, \mathrm{L}_{1101}] \\
\mathcal{M}_{166} = [\mathrm{B}_{000}, \mathrm{B}_{010}, \mathrm{L}_{0001}, \mathrm{L}_{0011}, \mathrm{L}_{0100}, \mathrm{L}_{0101}, \mathrm{L}_{1000}, \mathrm{L}_{1110}] \\
\mathcal{M}_{167} = [\mathrm{B}_{000}, \mathrm{B}_{010}, \mathrm{L}_{0001}, \mathrm{L}_{0011}, \mathrm{L}_{0110}, \mathrm{L}_{0111}, \mathrm{L}_{1000}, \mathrm{L}_{1110}] \\
\mathcal{M}_{168} = [\mathrm{B}_{000}, \mathrm{B}_{010}, \mathrm{L}_{0001}, \mathrm{L}_{0110}, \mathrm{L}_{1000}, \mathrm{L}_{1011}, \mathrm{L}_{1100}, \mathrm{L}_{1101}] \\
\mathcal{M}_{169} = [\mathrm{B}_{000}, \mathrm{B}_{010}, \mathrm{L}_{0001}, \mathrm{L}_{0110}, \mathrm{L}_{1000}, \mathrm{L}_{1011}, \mathrm{L}_{1110}, \mathrm{L}_{1111}] \\
\mathcal{M}_{170} = [\mathrm{B}_{000}, \mathrm{B}_{010}, \mathrm{L}_{0010}, \mathrm{L}_{0011}, \mathrm{L}_{0100}, \mathrm{L}_{0110}, \mathrm{L}_{1011}, \mathrm{L}_{1101}] \\
\mathcal{M}_{171} = [\mathrm{B}_{000}, \mathrm{B}_{010}, \mathrm{L}_{0011}, \mathrm{L}_{0100}, \mathrm{L}_{1000}, \mathrm{L}_{1001}, \mathrm{L}_{1101}, \mathrm{L}_{1110}] \\
\mathcal{M}_{172} = [\mathrm{B}_{000}, \mathrm{B}_{010}, \mathrm{L}_{0011}, \mathrm{L}_{0100}, \mathrm{L}_{1010}, \mathrm{L}_{1011}, \mathrm{L}_{1101}, \mathrm{L}_{1110}] \\
\mathcal{M}_{173} = [\mathrm{B}_{000}, \mathrm{B}_{011}, \mathrm{L}_{0000}, \mathrm{L}_{0001}, \mathrm{L}_{0101}, \mathrm{L}_{0111}, \mathrm{L}_{1010}, \mathrm{L}_{1100}] \\
\mathcal{M}_{174} = [\mathrm{B}_{000}, \mathrm{B}_{011}, \mathrm{L}_{0000}, \mathrm{L}_{0010}, \mathrm{L}_{0100}, \mathrm{L}_{0101}, \mathrm{L}_{1001}, \mathrm{L}_{1111}] \\
\mathcal{M}_{175} = [\mathrm{B}_{000}, \mathrm{B}_{011}, \mathrm{L}_{0000}, \mathrm{L}_{0010}, \mathrm{L}_{0110}, \mathrm{L}_{0111}, \mathrm{L}_{1001}, \mathrm{L}_{1111}] \\
\mathcal{M}_{176} = [\mathrm{B}_{000}, \mathrm{B}_{011}, \mathrm{L}_{0000}, \mathrm{L}_{0111}, \mathrm{L}_{1001}, \mathrm{L}_{1010}, \mathrm{L}_{1100}, \mathrm{L}_{1101}] \\
\mathcal{M}_{177} = [\mathrm{B}_{000}, \mathrm{B}_{011}, \mathrm{L}_{0000}, \mathrm{L}_{0111}, \mathrm{L}_{1001}, \mathrm{L}_{1010}, \mathrm{L}_{1110}, \mathrm{L}_{1111}] \\
\mathcal{M}_{178} = [\mathrm{B}_{000}, \mathrm{B}_{011}, \mathrm{L}_{0010}, \mathrm{L}_{0011}, \mathrm{L}_{0101}, \mathrm{L}_{0111}, \mathrm{L}_{1010}, \mathrm{L}_{1100}] \\
\mathcal{M}_{179} = [\mathrm{B}_{000}, \mathrm{B}_{011}, \mathrm{L}_{0010}, \mathrm{L}_{0101}, \mathrm{L}_{1000}, \mathrm{L}_{1001}, \mathrm{L}_{1100}, \mathrm{L}_{1111}] \\
\mathcal{M}_{180} = [\mathrm{B}_{000}, \mathrm{B}_{011}, \mathrm{L}_{0010}, \mathrm{L}_{0101}, \mathrm{L}_{1010}, \mathrm{L}_{1011}, \mathrm{L}_{1100}, \mathrm{L}_{1111}] \\
\mathcal{M}_{181} = [\mathrm{B}_{000}, \mathrm{B}_{100}, \mathrm{L}_{0000}, \mathrm{L}_{0001}, \mathrm{L}_{0100}, \mathrm{L}_{0111}, \mathrm{L}_{1001}, \mathrm{L}_{1110}] \\
\mathcal{M}_{182} = [\mathrm{B}_{000}, \mathrm{B}_{100}, \mathrm{L}_{0001}, \mathrm{L}_{0010}, \mathrm{L}_{0100}, \mathrm{L}_{0101}, \mathrm{L}_{1011}, \mathrm{L}_{1100}] \\
\mathcal{M}_{183} = [\mathrm{B}_{000}, \mathrm{B}_{100}, \mathrm{L}_{0001}, \mathrm{L}_{0010}, \mathrm{L}_{0110}, \mathrm{L}_{0111}, \mathrm{L}_{1011}, \mathrm{L}_{1100}] \\
\mathcal{M}_{184} = [\mathrm{B}_{000}, \mathrm{B}_{100}, \mathrm{L}_{0001}, \mathrm{L}_{0111}, \mathrm{L}_{1000}, \mathrm{L}_{1001}, \mathrm{L}_{1100}, \mathrm{L}_{1110}] \\
\mathcal{M}_{185} = [\mathrm{B}_{000}, \mathrm{B}_{100}, \mathrm{L}_{0001}, \mathrm{L}_{0111}, \mathrm{L}_{1010}, \mathrm{L}_{1011}, \mathrm{L}_{1100}, \mathrm{L}_{1110}] \\
\mathcal{M}_{186} = [\mathrm{B}_{000}, \mathrm{B}_{100}, \mathrm{L}_{0010}, \mathrm{L}_{0011}, \mathrm{L}_{0100}, \mathrm{L}_{0111}, \mathrm{L}_{1001}, \mathrm{L}_{1110}] \\
\mathcal{M}_{187} = [\mathrm{B}_{000}, \mathrm{B}_{100}, \mathrm{L}_{0010}, \mathrm{L}_{0100}, \mathrm{L}_{1001}, \mathrm{L}_{1011}, \mathrm{L}_{1100}, \mathrm{L}_{1101}] \\
\mathcal{M}_{188} = [\mathrm{B}_{000}, \mathrm{B}_{100}, \mathrm{L}_{0010}, \mathrm{L}_{0100}, \mathrm{L}_{1001}, \mathrm{L}_{1011}, \mathrm{L}_{1110}, \mathrm{L}_{1111}] \\
\mathcal{M}_{189} = [\mathrm{B}_{000}, \mathrm{B}_{101}, \mathrm{L}_{0000}, \mathrm{L}_{0001}, \mathrm{L}_{0101}, \mathrm{L}_{0110}, \mathrm{L}_{1000}, \mathrm{L}_{1111}] \\
\mathcal{M}_{190} = [\mathrm{B}_{000}, \mathrm{B}_{101}, \mathrm{L}_{0000}, \mathrm{L}_{0011}, \mathrm{L}_{0100}, \mathrm{L}_{0101}, \mathrm{L}_{1010}, \mathrm{L}_{1101}] \\
\mathcal{M}_{191} = [\mathrm{B}_{000}, \mathrm{B}_{101}, \mathrm{L}_{0000}, \mathrm{L}_{0011}, \mathrm{L}_{0110}, \mathrm{L}_{0111}, \mathrm{L}_{1010}, \mathrm{L}_{1101}] \\
\mathcal{M}_{192} = [\mathrm{B}_{000}, \mathrm{B}_{101}, \mathrm{L}_{0000}, \mathrm{L}_{0110}, \mathrm{L}_{1000}, \mathrm{L}_{1001}, \mathrm{L}_{1101}, \mathrm{L}_{1111}] \\
\mathcal{M}_{193} = [\mathrm{B}_{000}, \mathrm{B}_{101}, \mathrm{L}_{0000}, \mathrm{L}_{0110}, \mathrm{L}_{1010}, \mathrm{L}_{1011}, \mathrm{L}_{1101}, \mathrm{L}_{1111}] \\
\mathcal{M}_{194} = [\mathrm{B}_{000}, \mathrm{B}_{101}, \mathrm{L}_{0010}, \mathrm{L}_{0011}, \mathrm{L}_{0101}, \mathrm{L}_{0110}, \mathrm{L}_{1000}, \mathrm{L}_{1111}] \\
\mathcal{M}_{195} = [\mathrm{B}_{000}, \mathrm{B}_{101}, \mathrm{L}_{0011}, \mathrm{L}_{0101}, \mathrm{L}_{1000}, \mathrm{L}_{1010}, \mathrm{L}_{1100}, \mathrm{L}_{1101}] \\
\mathcal{M}_{196} = [\mathrm{B}_{000}, \mathrm{B}_{101}, \mathrm{L}_{0011}, \mathrm{L}_{0101}, \mathrm{L}_{1000}, \mathrm{L}_{1010}, \mathrm{L}_{1110}, \mathrm{L}_{1111}] \\
\mathcal{M}_{197} = [\mathrm{B}_{000}, \mathrm{B}_{110}, \mathrm{L}_{0000}, \mathrm{L}_{0001}, \mathrm{L}_{0101}, \mathrm{L}_{0110}, \mathrm{L}_{1011}, \mathrm{L}_{1100}] \\
\mathcal{M}_{198} = [\mathrm{B}_{000}, \mathrm{B}_{110}, \mathrm{L}_{0000}, \mathrm{L}_{0011}, \mathrm{L}_{0100}, \mathrm{L}_{0101}, \mathrm{L}_{1001}, \mathrm{L}_{1110}] \\
\mathcal{M}_{199} = [\mathrm{B}_{000}, \mathrm{B}_{110}, \mathrm{L}_{0000}, \mathrm{L}_{0011}, \mathrm{L}_{0110}, \mathrm{L}_{0111}, \mathrm{L}_{1001}, \mathrm{L}_{1110}] \\
\mathcal{M}_{200} = [\mathrm{B}_{000}, \mathrm{B}_{110}, \mathrm{L}_{0000}, \mathrm{L}_{0110}, \mathrm{L}_{1001}, \mathrm{L}_{1011}, \mathrm{L}_{1100}, \mathrm{L}_{1101}] \\
\mathcal{M}_{201} = [\mathrm{B}_{000}, \mathrm{B}_{110}, \mathrm{L}_{0000}, \mathrm{L}_{0110}, \mathrm{L}_{1001}, \mathrm{L}_{1011}, \mathrm{L}_{1110}, \mathrm{L}_{1111}] \\
\mathcal{M}_{202} = [\mathrm{B}_{000}, \mathrm{B}_{110}, \mathrm{L}_{0010}, \mathrm{L}_{0011}, \mathrm{L}_{0101}, \mathrm{L}_{0110}, \mathrm{L}_{1011}, \mathrm{L}_{1100}] \\
\mathcal{M}_{203} = [\mathrm{B}_{000}, \mathrm{B}_{110}, \mathrm{L}_{0011}, \mathrm{L}_{0101}, \mathrm{L}_{1000}, \mathrm{L}_{1001}, \mathrm{L}_{1100}, \mathrm{L}_{1110}] \\
\mathcal{M}_{204} = [\mathrm{B}_{000}, \mathrm{B}_{110}, \mathrm{L}_{0011}, \mathrm{L}_{0101}, \mathrm{L}_{1010}, \mathrm{L}_{1011}, \mathrm{L}_{1100}, \mathrm{L}_{1110}] 
\end{eqnarray*}
\begin{eqnarray*}
\mathcal{M}_{205} = [\mathrm{B}_{000}, \mathrm{B}_{111}, \mathrm{L}_{0000}, \mathrm{L}_{0001}, \mathrm{L}_{0100}, \mathrm{L}_{0111}, \mathrm{L}_{1010}, \mathrm{L}_{1101}] \\
\mathcal{M}_{206} = [\mathrm{B}_{000}, \mathrm{B}_{111}, \mathrm{L}_{0001}, \mathrm{L}_{0010}, \mathrm{L}_{0100}, \mathrm{L}_{0101}, \mathrm{L}_{1000}, \mathrm{L}_{1111}] \\
\mathcal{M}_{207} = [\mathrm{B}_{000}, \mathrm{B}_{111}, \mathrm{L}_{0001}, \mathrm{L}_{0010}, \mathrm{L}_{0110}, \mathrm{L}_{0111}, \mathrm{L}_{1000}, \mathrm{L}_{1111}] \\
\mathcal{M}_{208} = [\mathrm{B}_{000}, \mathrm{B}_{111}, \mathrm{L}_{0001}, \mathrm{L}_{0111}, \mathrm{L}_{1000}, \mathrm{L}_{1010}, \mathrm{L}_{1100}, \mathrm{L}_{1101}] \\
\mathcal{M}_{209} = [\mathrm{B}_{000}, \mathrm{B}_{111}, \mathrm{L}_{0001}, \mathrm{L}_{0111}, \mathrm{L}_{1000}, \mathrm{L}_{1010}, \mathrm{L}_{1110}, \mathrm{L}_{1111}] \\
\mathcal{M}_{210} = [\mathrm{B}_{000}, \mathrm{B}_{111}, \mathrm{L}_{0010}, \mathrm{L}_{0011}, \mathrm{L}_{0100}, \mathrm{L}_{0111}, \mathrm{L}_{1010}, \mathrm{L}_{1101}] \\
\mathcal{M}_{211} = [\mathrm{B}_{000}, \mathrm{B}_{111}, \mathrm{L}_{0010}, \mathrm{L}_{0100}, \mathrm{L}_{1000}, \mathrm{L}_{1001}, \mathrm{L}_{1101}, \mathrm{L}_{1111}] \\
\mathcal{M}_{212} = [\mathrm{B}_{000}, \mathrm{B}_{111}, \mathrm{L}_{0010}, \mathrm{L}_{0100}, \mathrm{L}_{1010}, \mathrm{L}_{1011}, \mathrm{L}_{1101}, \mathrm{L}_{1111}] \\
\end{eqnarray*}

\vspace{-2em}
\[v = \left [ 2 \eta - 1, \quad \frac{1}{2} - \frac{\eta}{2}, \quad \frac{1}{2} - \frac{\eta}{2}, \quad \frac{1}{4} - \frac{\eta}{4}, \quad \frac{1}{4} - \frac{\eta}{4}, \quad \frac{1}{4} - \frac{\eta}{4}, \quad \frac{1}{4} - \frac{\eta}{4}\right ]\]
\vspace{-1.5em}
\begin{eqnarray*}
\mathcal{M}_{213} = [\mathrm{B}_{000}, \mathrm{B}_{010}, \mathrm{B}_{100}, \mathrm{L}_{0001}, \mathrm{L}_{0100}, \mathrm{L}_{1011}, \mathrm{L}_{1110}] \\
\mathcal{M}_{214} = [\mathrm{B}_{000}, \mathrm{B}_{010}, \mathrm{B}_{101}, \mathrm{L}_{0011}, \mathrm{L}_{0110}, \mathrm{L}_{1000}, \mathrm{L}_{1101}] \\
\mathcal{M}_{215} = [\mathrm{B}_{000}, \mathrm{B}_{010}, \mathrm{B}_{110}, \mathrm{L}_{0011}, \mathrm{L}_{0110}, \mathrm{L}_{1011}, \mathrm{L}_{1110}] \\
\mathcal{M}_{216} = [\mathrm{B}_{000}, \mathrm{B}_{010}, \mathrm{B}_{111}, \mathrm{L}_{0001}, \mathrm{L}_{0100}, \mathrm{L}_{1000}, \mathrm{L}_{1101}] \\
\mathcal{M}_{217} = [\mathrm{B}_{000}, \mathrm{B}_{011}, \mathrm{B}_{100}, \mathrm{L}_{0010}, \mathrm{L}_{0111}, \mathrm{L}_{1001}, \mathrm{L}_{1100}] \\
\mathcal{M}_{218} = [\mathrm{B}_{000}, \mathrm{B}_{011}, \mathrm{B}_{101}, \mathrm{L}_{0000}, \mathrm{L}_{0101}, \mathrm{L}_{1010}, \mathrm{L}_{1111}] \\
\mathcal{M}_{219} = [\mathrm{B}_{000}, \mathrm{B}_{011}, \mathrm{B}_{110}, \mathrm{L}_{0000}, \mathrm{L}_{0101}, \mathrm{L}_{1001}, \mathrm{L}_{1100}] \\
\mathcal{M}_{220} = [\mathrm{B}_{000}, \mathrm{B}_{011}, \mathrm{B}_{111}, \mathrm{L}_{0010}, \mathrm{L}_{0111}, \mathrm{L}_{1010}, \mathrm{L}_{1111}] \\
\mathcal{M}_{221} = [\mathrm{B}_{000}, \mathrm{B}_{100}, \mathrm{B}_{110}, \mathrm{L}_{1001}, \mathrm{L}_{1011}, \mathrm{L}_{1100}, \mathrm{L}_{1110}] \\
\mathcal{M}_{222} = [\mathrm{B}_{000}, \mathrm{B}_{100}, \mathrm{B}_{111}, \mathrm{L}_{0001}, \mathrm{L}_{0010}, \mathrm{L}_{0100}, \mathrm{L}_{0111}] \\
\mathcal{M}_{223} = [\mathrm{B}_{000}, \mathrm{B}_{101}, \mathrm{B}_{110}, \mathrm{L}_{0000}, \mathrm{L}_{0011}, \mathrm{L}_{0101}, \mathrm{L}_{0110}] \\
\mathcal{M}_{224} = [\mathrm{B}_{000}, \mathrm{B}_{101}, \mathrm{B}_{111}, \mathrm{L}_{1000}, \mathrm{L}_{1010}, \mathrm{L}_{1101}, \mathrm{L}_{1111}] 
\end{eqnarray*}

%\vspace{-1em}
\[v = \left [ 2 \eta - 1, \quad \frac{1}{2} - \frac{\eta}{2}, \quad \frac{1}{2} - \frac{\eta}{2}, \quad \frac{1}{2} - \frac{\eta}{2}, \quad \frac{1}{2} - \frac{\eta}{2}\right ]\]
\vspace{-1em}
\begin{eqnarray*}
\mathcal{M}_{225} = [\mathrm{B}_{000}, \mathrm{L}_{0000}, \mathrm{L}_{0001}, \mathrm{L}_{0100}, \mathrm{L}_{0101}] \\
\mathcal{M}_{226} = [\mathrm{B}_{000}, \mathrm{L}_{0000}, \mathrm{L}_{0001}, \mathrm{L}_{0110}, \mathrm{L}_{0111}] \\
\mathcal{M}_{227} = [\mathrm{B}_{000}, \mathrm{L}_{0000}, \mathrm{L}_{0100}, \mathrm{L}_{1001}, \mathrm{L}_{1101}] \\
\mathcal{M}_{228} = [\mathrm{B}_{000}, \mathrm{L}_{0001}, \mathrm{L}_{0101}, \mathrm{L}_{1000}, \mathrm{L}_{1100}] \\
\mathcal{M}_{229} = [\mathrm{B}_{000}, \mathrm{L}_{0010}, \mathrm{L}_{0011}, \mathrm{L}_{0100}, \mathrm{L}_{0101}] \\
\mathcal{M}_{230} = [\mathrm{B}_{000}, \mathrm{L}_{0010}, \mathrm{L}_{0011}, \mathrm{L}_{0110}, \mathrm{L}_{0111}] \\
\mathcal{M}_{231} = [\mathrm{B}_{000}, \mathrm{L}_{0010}, \mathrm{L}_{0110}, \mathrm{L}_{1011}, \mathrm{L}_{1111}] \\
\mathcal{M}_{232} = [\mathrm{B}_{000}, \mathrm{L}_{0011}, \mathrm{L}_{0111}, \mathrm{L}_{1010}, \mathrm{L}_{1110}] \\
\mathcal{M}_{233} = [\mathrm{B}_{000}, \mathrm{L}_{1000}, \mathrm{L}_{1001}, \mathrm{L}_{1100}, \mathrm{L}_{1101}] \\
\mathcal{M}_{234} = [\mathrm{B}_{000}, \mathrm{L}_{1000}, \mathrm{L}_{1001}, \mathrm{L}_{1110}, \mathrm{L}_{1111}] \\
\mathcal{M}_{235} = [\mathrm{B}_{000}, \mathrm{L}_{1010}, \mathrm{L}_{1011}, \mathrm{L}_{1100}, \mathrm{L}_{1101}] \\
\mathcal{M}_{236} = [\mathrm{B}_{000}, \mathrm{L}_{1010}, \mathrm{L}_{1011}, \mathrm{L}_{1110}, \mathrm{L}_{1111}] 
\end{eqnarray*}

\vspace{-1.em}
\[v = \left [ 2 \eta - 1, \quad 1 - \eta , \quad 1 - \eta \right ]\]
\vspace{-1.5em}
\begin{eqnarray*}
\mathcal{M}_{237} = [\mathrm{B}_{000}, \mathrm{B}_{010}, \mathrm{B}_{011}] \\
\mathcal{M}_{238} = [\mathrm{B}_{000}, \mathrm{B}_{100}, \mathrm{B}_{101}] \\
\mathcal{M}_{239} = [\mathrm{B}_{000}, \mathrm{B}_{110}, \mathrm{B}_{111}] 
\end{eqnarray*}

\item {Having $\eta \in \left(\frac{5}{9}, 1\right).$}

\[v = \left [ \frac{9 \eta}{4} - \frac{5}{4}, \quad \frac{1}{4} - \frac{\eta}{4}, \quad \frac{1}{4} - \frac{\eta}{4}, \quad \frac{1}{4} - \frac{\eta}{4}, \quad \frac{1}{4} - \frac{\eta}{4}, \quad \frac{1}{4} - \frac{\eta}{4}, \quad \frac{1}{4} - \frac{\eta}{4}, \quad \frac{1}{4} - \frac{\eta}{4}, \quad \frac{1}{2} - \frac{\eta}{2}\right ]\]

\begin{eqnarray*}
\mathcal{M}_{240} = [\mathrm{B}_{000}, \mathrm{B}_{010}, \mathrm{B}_{100}, \mathrm{B}_{110}, \mathrm{L}_{0011}, \mathrm{L}_{0111}, \mathrm{L}_{1000}, \mathrm{L}_{1001}, \mathrm{L}_{1110}] \\
\mathcal{M}_{241} = [\mathrm{B}_{000}, \mathrm{B}_{010}, \mathrm{B}_{101}, \mathrm{B}_{111}, \mathrm{L}_{0000}, \mathrm{L}_{0100}, \mathrm{L}_{1010}, \mathrm{L}_{1011}, \mathrm{L}_{1101}] 
\end{eqnarray*}
\[v = \left [ \frac{9 \eta}{4} - \frac{5}{4}, \quad \frac{1}{4} - \frac{\eta}{4}, \quad \frac{1}{4} - \frac{\eta}{4}, \quad \frac{1}{4} - \frac{\eta}{4}, \quad \frac{1}{4} - \frac{\eta}{4}, \quad \frac{1}{4} - \frac{\eta}{4}, \quad \frac{1}{2} - \frac{\eta}{2}, \quad \frac{1}{4} - \frac{\eta}{4}, \quad \frac{1}{4} - \frac{\eta}{4}\right ]\]

\begin{eqnarray*}
\mathcal{M}_{242} = [\mathrm{B}_{000}, \mathrm{B}_{010}, \mathrm{B}_{100}, \mathrm{B}_{110}, \mathrm{L}_{0010}, \mathrm{L}_{0110}, \mathrm{L}_{1011}, \mathrm{L}_{1100}, \mathrm{L}_{1101}] \\
\mathcal{M}_{243} = [\mathrm{B}_{000}, \mathrm{B}_{010}, \mathrm{B}_{101}, \mathrm{B}_{111}, \mathrm{L}_{0001}, \mathrm{L}_{0101}, \mathrm{L}_{1000}, \mathrm{L}_{1110}, \mathrm{L}_{1111}] \\
\mathcal{M}_{244} = [\mathrm{B}_{000}, \mathrm{B}_{011}, \mathrm{B}_{100}, \mathrm{B}_{111}, \mathrm{L}_{0000}, \mathrm{L}_{0001}, \mathrm{L}_{0111}, \mathrm{L}_{1010}, \mathrm{L}_{1110}] \\
\mathcal{M}_{245} = [\mathrm{B}_{000}, \mathrm{B}_{011}, \mathrm{B}_{101}, \mathrm{B}_{110}, \mathrm{L}_{0010}, \mathrm{L}_{0011}, \mathrm{L}_{0101}, \mathrm{L}_{1000}, \mathrm{L}_{1100}] 
\end{eqnarray*}

\[v = \left [ \frac{9 \eta}{4} - \frac{5}{4}, \quad \frac{1}{4} - \frac{\eta}{4}, \quad \frac{1}{4} - \frac{\eta}{4}, \quad \frac{1}{4} - \frac{\eta}{4}, \quad \frac{1}{2} - \frac{\eta}{2}, \quad \frac{1}{4} - \frac{\eta}{4}, \quad \frac{1}{4} - \frac{\eta}{4}, \quad \frac{1}{4} - \frac{\eta}{4}, \quad \frac{1}{4} - \frac{\eta}{4}\right ]\]

\begin{eqnarray*}
\mathcal{M}_{246} = [\mathrm{B}_{000}, \mathrm{B}_{011}, \mathrm{B}_{100}, \mathrm{B}_{111}, \mathrm{L}_{0010}, \mathrm{L}_{0100}, \mathrm{L}_{0101}, \mathrm{L}_{1011}, \mathrm{L}_{1111}] \\
\mathcal{M}_{247} = [\mathrm{B}_{000}, \mathrm{B}_{011}, \mathrm{B}_{101}, \mathrm{B}_{110}, \mathrm{L}_{0000}, \mathrm{L}_{0110}, \mathrm{L}_{0111}, \mathrm{L}_{1001}, \mathrm{L}_{1101}] 
\end{eqnarray*}

\item {Having $\eta \in \left(\frac{4}{7}, 1\right).$}

\[v = \left [ \frac{7 \eta}{3} - \frac{4}{3}, \quad \frac{1}{3} - \frac{\eta}{3}, \quad \frac{1}{3} - \frac{\eta}{3}, \quad  \frac{1}{6} - \frac{\eta}{6}, \quad  \frac{1}{6} - \frac{\eta}{6}, \quad \frac{1}{3} - \frac{\eta}{3}, \quad  \frac{1}{6} - \frac{\eta}{6}, \quad \frac{1}{3} - \frac{\eta}{3}, \quad \frac{1}{2} - \frac{\eta}{2}\right ]\]

\begin{eqnarray*}
\mathcal{M}_{248} = [\mathrm{B}_{000}, \mathrm{B}_{010}, \mathrm{B}_{111}, \mathrm{L}_{0000}, \mathrm{L}_{0010}, \mathrm{L}_{0100}, \mathrm{L}_{1010}, \mathrm{L}_{1011}, \mathrm{L}_{1101}] \\
\mathcal{M}_{249} = [\mathrm{B}_{000}, \mathrm{B}_{100}, \mathrm{B}_{110}, \mathrm{L}_{0000}, \mathrm{L}_{0011}, \mathrm{L}_{0111}, \mathrm{L}_{1000}, \mathrm{L}_{1001}, \mathrm{L}_{1110}] 
\end{eqnarray*}

\[v = \left [ \frac{7 \eta}{3} - \frac{4}{3}, \quad \frac{1}{3} - \frac{\eta}{3}, \quad \frac{1}{3} - \frac{\eta}{3}, \quad  \frac{1}{6} - \frac{\eta}{6}, \quad  \frac{1}{6} - \frac{\eta}{6}, \quad \frac{1}{3} - \frac{\eta}{3}, \quad \frac{1}{3} - \frac{\eta}{3}, \quad  \frac{1}{6} - \frac{\eta}{6}, \quad \frac{1}{2} - \frac{\eta}{2}\right ]\]

\[\mathcal{M}_{250} = [\mathrm{B}_{000}, \mathrm{B}_{010}, \mathrm{B}_{100}, \mathrm{L}_{0001}, \mathrm{L}_{0011}, \mathrm{L}_{0111}, \mathrm{L}_{1000}, \mathrm{L}_{1001}, \mathrm{L}_{1110}] \]

\[v = \left [ \frac{7 \eta}{3} - \frac{4}{3}, \quad \frac{1}{3} - \frac{\eta}{3}, \quad \frac{1}{3} - \frac{\eta}{3}, \quad  \frac{1}{6} - \frac{\eta}{6}, \quad  \frac{1}{6} - \frac{\eta}{6}, \quad \frac{1}{3} - \frac{\eta}{3}, \quad \frac{1}{2} - \frac{\eta}{2}, \quad  \frac{1}{6} - \frac{\eta}{6}, \quad \frac{1}{3} - \frac{\eta}{3}\right ]\]

\begin{eqnarray*}
\mathcal{M}_{251} = [\mathrm{B}_{000}, \mathrm{B}_{010}, \mathrm{B}_{110}, \mathrm{L}_{0000}, \mathrm{L}_{0010}, \mathrm{L}_{0110}, \mathrm{L}_{1011}, \mathrm{L}_{1100}, \mathrm{L}_{1101}] \\
\mathcal{M}_{252} = [\mathrm{B}_{000}, \mathrm{B}_{101}, \mathrm{B}_{111}, \mathrm{L}_{0001}, \mathrm{L}_{0010}, \mathrm{L}_{0101}, \mathrm{L}_{1000}, \mathrm{L}_{1110}, \mathrm{L}_{1111}] 
\end{eqnarray*}

\[v = \left [ \frac{7 \eta}{3} - \frac{4}{3}, \quad \frac{1}{3} - \frac{\eta}{3}, \quad \frac{1}{3} - \frac{\eta}{3}, \quad  \frac{1}{6} - \frac{\eta}{6}, \quad  \frac{1}{6} - \frac{\eta}{6}, \quad \frac{1}{3} - \frac{\eta}{3}, \quad \frac{1}{2} - \frac{\eta}{2}, \quad \frac{1}{3} - \frac{\eta}{3}, \quad  \frac{1}{6} - \frac{\eta}{6}\right ]\]

\[\mathcal{M}_{253} = [\mathrm{B}_{000}, \mathrm{B}_{010}, \mathrm{B}_{101}, \mathrm{L}_{0001}, \mathrm{L}_{0011}, \mathrm{L}_{0101}, \mathrm{L}_{1000}, \mathrm{L}_{1110}, \mathrm{L}_{1111}] \]

\[v = \left [ \frac{7 \eta}{3} - \frac{4}{3}, \quad \frac{1}{3} - \frac{\eta}{3}, \quad \frac{1}{3} - \frac{\eta}{3}, \quad  \frac{1}{6} - \frac{\eta}{6}, \quad \frac{1}{3} - \frac{\eta}{3}, \quad \frac{1}{2} - \frac{\eta}{2}, \quad  \frac{1}{6} - \frac{\eta}{6}, \quad  \frac{1}{6} - \frac{\eta}{6}, \quad \frac{1}{3} - \frac{\eta}{3}\right ]\]

\[\mathcal{M}_{254} = [\mathrm{B}_{000}, \mathrm{B}_{100}, \mathrm{B}_{111}, \mathrm{L}_{0000}, \mathrm{L}_{0001}, \mathrm{L}_{0111}, \mathrm{L}_{1000}, \mathrm{L}_{1010}, \mathrm{L}_{1110}] \]

\[v = \left [ \frac{7 \eta}{3} - \frac{4}{3}, \quad \frac{1}{3} - \frac{\eta}{3}, \quad \frac{1}{3} - \frac{\eta}{3}, \quad  \frac{1}{6} - \frac{\eta}{6}, \quad \frac{1}{3} - \frac{\eta}{3}, \quad \frac{1}{2} - \frac{\eta}{2}, \quad \frac{1}{3} - \frac{\eta}{3}, \quad  \frac{1}{6} - \frac{\eta}{6}, \quad  \frac{1}{6} - \frac{\eta}{6}\right ]\]

\[\mathcal{M}_{255} = [\mathrm{B}_{000}, \mathrm{B}_{101}, \mathrm{B}_{110}, \mathrm{L}_{0010}, \mathrm{L}_{0011}, \mathrm{L}_{0101}, \mathrm{L}_{1000}, \mathrm{L}_{1100}, \mathrm{L}_{1110}] \]

\[v = \left [ \frac{7 \eta}{3} - \frac{4}{3}, \quad \frac{1}{3} - \frac{\eta}{3}, \quad \frac{1}{3} - \frac{\eta}{3}, \quad \frac{1}{3} - \frac{\eta}{3}, \quad  \frac{1}{6} - \frac{\eta}{6}, \quad  \frac{1}{6} - \frac{\eta}{6}, \quad  \frac{1}{6} - \frac{\eta}{6}, \quad \frac{1}{3} - \frac{\eta}{3}, \quad \frac{1}{2} - \frac{\eta}{2}\right ]\]

\[\mathcal{M}_{256} = [\mathrm{B}_{000}, \mathrm{B}_{010}, \mathrm{B}_{101}, \mathrm{L}_{0000}, \mathrm{L}_{0100}, \mathrm{L}_{0110}, \mathrm{L}_{1010}, \mathrm{L}_{1011}, \mathrm{L}_{1101}] \]

\[v = \left [ \frac{7 \eta}{3} - \frac{4}{3}, \quad \frac{1}{3} - \frac{\eta}{3}, \quad \frac{1}{3} - \frac{\eta}{3}, \quad \frac{1}{3} - \frac{\eta}{3}, \quad  \frac{1}{6} - \frac{\eta}{6}, \quad  \frac{1}{6} - \frac{\eta}{6}, \quad \frac{1}{3} - \frac{\eta}{3}, \quad  \frac{1}{6} - \frac{\eta}{6}, \quad \frac{1}{2} - \frac{\eta}{2}\right ]\]

\begin{eqnarray*}
\mathcal{M}_{257} = [\mathrm{B}_{000}, \mathrm{B}_{010}, \mathrm{B}_{110}, \mathrm{L}_{0011}, \mathrm{L}_{0101}, \mathrm{L}_{0111}, \mathrm{L}_{1000}, \mathrm{L}_{1001}, \mathrm{L}_{1110}] \\
\mathcal{M}_{258} = [\mathrm{B}_{000}, \mathrm{B}_{101}, \mathrm{B}_{111}, \mathrm{L}_{0000}, \mathrm{L}_{0100}, \mathrm{L}_{0111}, \mathrm{L}_{1010}, \mathrm{L}_{1011}, \mathrm{L}_{1101}] 
\end{eqnarray*}
\[v = \left [ \frac{7 \eta}{3} - \frac{4}{3}, \quad \frac{1}{3} - \frac{\eta}{3}, \quad \frac{1}{3} - \frac{\eta}{3}, \quad \frac{1}{3} - \frac{\eta}{3}, \quad  \frac{1}{6} - \frac{\eta}{6}, \quad  \frac{1}{6} - \frac{\eta}{6}, \quad \frac{1}{2} - \frac{\eta}{2}, \quad  \frac{1}{6} - \frac{\eta}{6}, \quad \frac{1}{3} - \frac{\eta}{3}\right ]\]

\[\mathcal{M}_{259} = [\mathrm{B}_{000}, \mathrm{B}_{010}, \mathrm{B}_{100}, \mathrm{L}_{0010}, \mathrm{L}_{0100}, \mathrm{L}_{0110}, \mathrm{L}_{1011}, \mathrm{L}_{1100}, \mathrm{L}_{1101}] \]

\[v = \left [ \frac{7 \eta}{3} - \frac{4}{3}, \quad \frac{1}{3} - \frac{\eta}{3}, \quad \frac{1}{3} - \frac{\eta}{3}, \quad \frac{1}{3} - \frac{\eta}{3}, \quad  \frac{1}{6} - \frac{\eta}{6}, \quad  \frac{1}{6} - \frac{\eta}{6}, \quad \frac{1}{2} - \frac{\eta}{2}, \quad \frac{1}{3} - \frac{\eta}{3}, \quad  \frac{1}{6} - \frac{\eta}{6}\right ]\]
\begin{eqnarray*}
\mathcal{M}_{260} = [\mathrm{B}_{000}, \mathrm{B}_{010}, \mathrm{B}_{111}, \mathrm{L}_{0001}, \mathrm{L}_{0101}, \mathrm{L}_{0111}, \mathrm{L}_{1000}, \mathrm{L}_{1110}, \mathrm{L}_{1111}] \\
\mathcal{M}_{261} = [\mathrm{B}_{000}, \mathrm{B}_{100}, \mathrm{B}_{110}, \mathrm{L}_{0010}, \mathrm{L}_{0101}, \mathrm{L}_{0110}, \mathrm{L}_{1011}, \mathrm{L}_{1100}, \mathrm{L}_{1101}] 
\end{eqnarray*}
\[v = \left [ \frac{7 \eta}{3} - \frac{4}{3}, \quad \frac{1}{3} - \frac{\eta}{3}, \quad \frac{1}{3} - \frac{\eta}{3}, \quad \frac{1}{3} - \frac{\eta}{3}, \quad  \frac{1}{6} - \frac{\eta}{6}, \quad \frac{1}{2} - \frac{\eta}{2}, \quad  \frac{1}{6} - \frac{\eta}{6}, \quad  \frac{1}{6} - \frac{\eta}{6}, \quad \frac{1}{3} - \frac{\eta}{3}\right ]\]

\begin{eqnarray*}
\mathcal{M}_{262} = [\mathrm{B}_{000}, \mathrm{B}_{011}, \mathrm{B}_{100}, \mathrm{L}_{0000}, \mathrm{L}_{0001}, \mathrm{L}_{0111}, \mathrm{L}_{1001}, \mathrm{L}_{1010}, \mathrm{L}_{1110}] \\
\mathcal{M}_{263} = [\mathrm{B}_{000}, \mathrm{B}_{011}, \mathrm{B}_{110}, \mathrm{L}_{0010}, \mathrm{L}_{0011}, \mathrm{L}_{0101}, \mathrm{L}_{1000}, \mathrm{L}_{1011}, \mathrm{L}_{1100}] 
\end{eqnarray*}
\[v = \left [ \frac{7 \eta}{3} - \frac{4}{3}, \quad \frac{1}{3} - \frac{\eta}{3}, \quad \frac{1}{3} - \frac{\eta}{3}, \quad \frac{1}{3} - \frac{\eta}{3}, \quad  \frac{1}{6} - \frac{\eta}{6}, \quad \frac{1}{2} - \frac{\eta}{2}, \quad \frac{1}{3} - \frac{\eta}{3}, \quad  \frac{1}{6} - \frac{\eta}{6}, \quad  \frac{1}{6} - \frac{\eta}{6}\right ]\]
\begin{eqnarray*}
\mathcal{M}_{264} = [\mathrm{B}_{000}, \mathrm{B}_{011}, \mathrm{B}_{101}, \mathrm{L}_{0010}, \mathrm{L}_{0011}, \mathrm{L}_{0101}, \mathrm{L}_{1000}, \mathrm{L}_{1100}, \mathrm{L}_{1111}] \\
\mathcal{M}_{265} = [\mathrm{B}_{000}, \mathrm{B}_{011}, \mathrm{B}_{111}, \mathrm{L}_{0000}, \mathrm{L}_{0001}, \mathrm{L}_{0111}, \mathrm{L}_{1010}, \mathrm{L}_{1101}, \mathrm{L}_{1110}] 
\end{eqnarray*}

\[v = \left [ \frac{7 \eta}{3} - \frac{4}{3}, \quad \frac{1}{3} - \frac{\eta}{3}, \quad \frac{1}{3} - \frac{\eta}{3}, \quad \frac{1}{3} - \frac{\eta}{3}, \quad \frac{1}{3} - \frac{\eta}{3}, \quad \frac{1}{3} - \frac{\eta}{3}, \quad \frac{1}{3} - \frac{\eta}{3}, \quad \frac{1}{3} - \frac{\eta}{3}\right ]\]
\begin{eqnarray*}
\mathcal{M}_{266} = [\mathrm{B}_{000}, \mathrm{B}_{010}, \mathrm{B}_{100}, \mathrm{B}_{111}, \mathrm{L}_{0001}, \mathrm{L}_{0111}, \mathrm{L}_{1000}, \mathrm{L}_{1110}] \\
\mathcal{M}_{267} = [\mathrm{B}_{000}, \mathrm{B}_{010}, \mathrm{B}_{100}, \mathrm{B}_{111}, \mathrm{L}_{0010}, \mathrm{L}_{0100}, \mathrm{L}_{1011}, \mathrm{L}_{1101}] \\
\mathcal{M}_{268} = [\mathrm{B}_{000}, \mathrm{B}_{010}, \mathrm{B}_{101}, \mathrm{B}_{110}, \mathrm{L}_{0000}, \mathrm{L}_{0110}, \mathrm{L}_{1011}, \mathrm{L}_{1101}] \\
\mathcal{M}_{269} = [\mathrm{B}_{000}, \mathrm{B}_{010}, \mathrm{B}_{101}, \mathrm{B}_{110}, \mathrm{L}_{0011}, \mathrm{L}_{0101}, \mathrm{L}_{1000}, \mathrm{L}_{1110}] \\
\mathcal{M}_{270} = [\mathrm{B}_{000}, \mathrm{B}_{011}, \mathrm{B}_{100}, \mathrm{B}_{110}, \mathrm{L}_{0000}, \mathrm{L}_{0111}, \mathrm{L}_{1001}, \mathrm{L}_{1110}] \\
\mathcal{M}_{271} = [\mathrm{B}_{000}, \mathrm{B}_{011}, \mathrm{B}_{100}, \mathrm{B}_{110}, \mathrm{L}_{0010}, \mathrm{L}_{0101}, \mathrm{L}_{1011}, \mathrm{L}_{1100}] \\
\mathcal{M}_{272} = [\mathrm{B}_{000}, \mathrm{B}_{011}, \mathrm{B}_{101}, \mathrm{B}_{111}, \mathrm{L}_{0000}, \mathrm{L}_{0111}, \mathrm{L}_{1010}, \mathrm{L}_{1101}] \\
\mathcal{M}_{273} = [\mathrm{B}_{000}, \mathrm{B}_{011}, \mathrm{B}_{101}, \mathrm{B}_{111}, \mathrm{L}_{0010}, \mathrm{L}_{0101}, \mathrm{L}_{1000}, \mathrm{L}_{1111}] 
\end{eqnarray*}
\[v = \left [ \frac{7 \eta}{3} - \frac{4}{3}, \quad \frac{1}{3} - \frac{\eta}{3}, \quad \frac{1}{3} - \frac{\eta}{3}, \quad \frac{1}{2} - \frac{\eta}{2}, \quad  \frac{1}{6} - \frac{\eta}{6}, \quad \frac{1}{3} - \frac{\eta}{3}, \quad  \frac{1}{6} - \frac{\eta}{6}, \quad  \frac{1}{6} - \frac{\eta}{6}, \quad \frac{1}{3} - \frac{\eta}{3}\right ]\]
\begin{eqnarray*}
\mathcal{M}_{274} = [\mathrm{B}_{000}, \mathrm{B}_{011}, \mathrm{B}_{101}, \mathrm{L}_{0000}, \mathrm{L}_{0110}, \mathrm{L}_{0111}, \mathrm{L}_{1001}, \mathrm{L}_{1010}, \mathrm{L}_{1101}] \\
\mathcal{M}_{275} = [\mathrm{B}_{000}, \mathrm{B}_{011}, \mathrm{B}_{111}, \mathrm{L}_{0010}, \mathrm{L}_{0100}, \mathrm{L}_{0101}, \mathrm{L}_{1000}, \mathrm{L}_{1011}, \mathrm{L}_{1111}] 
\end{eqnarray*}
\[v = \left [ \frac{7 \eta}{3} - \frac{4}{3}, \quad \frac{1}{3} - \frac{\eta}{3}, \quad \frac{1}{3} - \frac{\eta}{3}, \quad \frac{1}{2} - \frac{\eta}{2}, \quad  \frac{1}{6} - \frac{\eta}{6}, \quad \frac{1}{3} - \frac{\eta}{3}, \quad \frac{1}{3} - \frac{\eta}{3}, \quad  \frac{1}{6} - \frac{\eta}{6}, \quad  \frac{1}{6} - \frac{\eta}{6}\right ]\]
\begin{eqnarray*}
\mathcal{M}_{276} = [\mathrm{B}_{000}, \mathrm{B}_{011}, \mathrm{B}_{100}, \mathrm{L}_{0010}, \mathrm{L}_{0100}, \mathrm{L}_{0101}, \mathrm{L}_{1011}, \mathrm{L}_{1100}, \mathrm{L}_{1111}] \\
\mathcal{M}_{277} = [\mathrm{B}_{000}, \mathrm{B}_{011}, \mathrm{B}_{110}, \mathrm{L}_{0000}, \mathrm{L}_{0110}, \mathrm{L}_{0111}, \mathrm{L}_{1001}, \mathrm{L}_{1101}, \mathrm{L}_{1110}] 
\end{eqnarray*}
\[v = \left [ \frac{7 \eta}{3} - \frac{4}{3}, \quad \frac{1}{3} - \frac{\eta}{3}, \quad \frac{1}{3} - \frac{\eta}{3}, \quad \frac{1}{2} - \frac{\eta}{2}, \quad \frac{1}{3} - \frac{\eta}{3}, \quad  \frac{1}{6} - \frac{\eta}{6}, \quad  \frac{1}{6} - \frac{\eta}{6}, \quad  \frac{1}{6} - \frac{\eta}{6}, \quad \frac{1}{3} - \frac{\eta}{3}\right ]\]
\[\mathcal{M}_{278} = [\mathrm{B}_{000}, \mathrm{B}_{101}, \mathrm{B}_{110}, \mathrm{L}_{0000}, \mathrm{L}_{0110}, \mathrm{L}_{0111}, \mathrm{L}_{1001}, \mathrm{L}_{1011}, \mathrm{L}_{1101}] \]

\[v = \left [ \frac{7 \eta}{3} - \frac{4}{3}, \quad \frac{1}{3} - \frac{\eta}{3}, \quad \frac{1}{3} - \frac{\eta}{3}, \quad \frac{1}{2} - \frac{\eta}{2}, \quad \frac{1}{3} - \frac{\eta}{3}, \quad  \frac{1}{6} - \frac{\eta}{6}, \quad \frac{1}{3} - \frac{\eta}{3}, \quad  \frac{1}{6} - \frac{\eta}{6}, \quad  \frac{1}{6} - \frac{\eta}{6}\right ]\]
\[\mathcal{M}_{279} = [\mathrm{B}_{000}, \mathrm{B}_{100}, \mathrm{B}_{111}, \mathrm{L}_{0010}, \mathrm{L}_{0100}, \mathrm{L}_{0101}, \mathrm{L}_{1011}, \mathrm{L}_{1101}, \mathrm{L}_{1111}] \]

\item {Having $\eta \in \left(\frac{3}{5}, 1\right).$}

\[v = \left [ \frac{5 \eta}{2} - \frac{3}{2}, \quad \frac{1}{2} - \frac{\eta}{2}, \quad \frac{1}{4} - \frac{\eta}{4}, \quad \frac{1}{4} - \frac{\eta}{4}, \quad \frac{1}{4} - \frac{\eta}{4}, \quad \frac{1}{4} - \frac{\eta}{4}, \quad \frac{1}{4} - \frac{\eta}{4}, \quad \frac{1}{4} - \frac{\eta}{4}, \quad \frac{1}{2} - \frac{\eta}{2}\right ]\]
\begin{eqnarray*}
\mathcal{M}_{280} = [\mathrm{B}_{000}, \mathrm{B}_{110}, \mathrm{L}_{0000}, \mathrm{L}_{0011}, \mathrm{L}_{0101}, \mathrm{L}_{0111}, \mathrm{L}_{1000}, \mathrm{L}_{1001}, \mathrm{L}_{1110}] \\
\mathcal{M}_{281} = [\mathrm{B}_{000}, \mathrm{B}_{111}, \mathrm{L}_{0000}, \mathrm{L}_{0010}, \mathrm{L}_{0100}, \mathrm{L}_{0111}, \mathrm{L}_{1010}, \mathrm{L}_{1011}, \mathrm{L}_{1101}] 
\end{eqnarray*}
\[v = \left [ \frac{5 \eta}{2} - \frac{3}{2}, \quad \frac{1}{2} - \frac{\eta}{2}, \quad \frac{1}{4} - \frac{\eta}{4}, \quad \frac{1}{4} - \frac{\eta}{4}, \quad \frac{1}{4} - \frac{\eta}{4}, \quad \frac{1}{4} - \frac{\eta}{4}, \quad \frac{1}{2} - \frac{\eta}{2}, \quad \frac{1}{4} - \frac{\eta}{4}, \quad \frac{1}{4} - \frac{\eta}{4}\right ]\]
\begin{eqnarray*}
\mathcal{M}_{282} = [\mathrm{B}_{000}, \mathrm{B}_{110}, \mathrm{L}_{0000}, \mathrm{L}_{0010}, \mathrm{L}_{0101}, \mathrm{L}_{0110}, \mathrm{L}_{1011}, \mathrm{L}_{1100}, \mathrm{L}_{1101}] \\
\mathcal{M}_{283} = [\mathrm{B}_{000}, \mathrm{B}_{111}, \mathrm{L}_{0001}, \mathrm{L}_{0010}, \mathrm{L}_{0101}, \mathrm{L}_{0111}, \mathrm{L}_{1000}, \mathrm{L}_{1110}, \mathrm{L}_{1111}] 
\end{eqnarray*}
\[v = \left [ \frac{5 \eta}{2} - \frac{3}{2}, \quad \frac{1}{2} - \frac{\eta}{2}, \quad \frac{1}{4} - \frac{\eta}{4}, \quad \frac{1}{4} - \frac{\eta}{4}, \quad \frac{1}{4} - \frac{\eta}{4}, \quad \frac{1}{4} - \frac{\eta}{4}, \quad \frac{1}{2} - \frac{\eta}{2}, \quad \frac{1}{2} - \frac{\eta}{2}\right ]\]
\begin{eqnarray*}
\mathcal{M}_{284} = [\mathrm{B}_{000}, \mathrm{B}_{010}, \mathrm{L}_{0000}, \mathrm{L}_{0010}, \mathrm{L}_{0100}, \mathrm{L}_{0110}, \mathrm{L}_{1011}, \mathrm{L}_{1101}] \\
\mathcal{M}_{285} = [\mathrm{B}_{000}, \mathrm{B}_{010}, \mathrm{L}_{0001}, \mathrm{L}_{0011}, \mathrm{L}_{0101}, \mathrm{L}_{0111}, \mathrm{L}_{1000}, \mathrm{L}_{1110}] 
\end{eqnarray*}
\[v = \left [ \frac{5 \eta}{2} - \frac{3}{2}, \quad \frac{1}{2} - \frac{\eta}{2}, \quad \frac{1}{4} - \frac{\eta}{4}, \quad \frac{1}{4} - \frac{\eta}{4}, \quad \frac{1}{4} - \frac{\eta}{4}, \quad \frac{1}{2} - \frac{\eta}{2}, \quad \frac{1}{4} - \frac{\eta}{4}, \quad \frac{1}{2} - \frac{\eta}{2}\right ]\]
\[\mathcal{M}_{286} = [\mathrm{B}_{000}, \mathrm{B}_{010}, \mathrm{L}_{0010}, \mathrm{L}_{0110}, \mathrm{L}_{1000}, \mathrm{L}_{1011}, \mathrm{L}_{1100}, \mathrm{L}_{1101}] \]

\[v = \left [ \frac{5 \eta}{2} - \frac{3}{2}, \quad \frac{1}{2} - \frac{\eta}{2}, \quad \frac{1}{4} - \frac{\eta}{4}, \quad \frac{1}{4} - \frac{\eta}{4}, \quad \frac{1}{4} - \frac{\eta}{4}, \quad \frac{1}{2} - \frac{\eta}{2}, \quad \frac{1}{2} - \frac{\eta}{2}, \quad \frac{1}{4} - \frac{\eta}{4}\right ]\]
\[\mathcal{M}_{287} = [\mathrm{B}_{000}, \mathrm{B}_{010}, \mathrm{L}_{0000}, \mathrm{L}_{0100}, \mathrm{L}_{1010}, \mathrm{L}_{1011}, \mathrm{L}_{1101}, \mathrm{L}_{1110}] \]

\[v = \left [ \frac{5 \eta}{2} - \frac{3}{2}, \quad \frac{1}{2} - \frac{\eta}{2}, \quad \frac{1}{4} - \frac{\eta}{4}, \quad \frac{1}{4} - \frac{\eta}{4}, \quad \frac{1}{2} - \frac{\eta}{2}, \quad \frac{1}{4} - \frac{\eta}{4}, \quad \frac{1}{4} - \frac{\eta}{4}, \quad \frac{1}{4} - \frac{\eta}{4}, \quad \frac{1}{4} - \frac{\eta}{4}\right ]\]
\begin{eqnarray*}
\mathcal{M}_{288} = [\mathrm{B}_{000}, \mathrm{B}_{110}, \mathrm{L}_{0010}, \mathrm{L}_{0011}, \mathrm{L}_{0101}, \mathrm{L}_{1000}, \mathrm{L}_{1011}, \mathrm{L}_{1100}, \mathrm{L}_{1110}] \\
\mathcal{M}_{289} = [\mathrm{B}_{000}, \mathrm{B}_{111}, \mathrm{L}_{0000}, \mathrm{L}_{0001}, \mathrm{L}_{0111}, \mathrm{L}_{1000}, \mathrm{L}_{1010}, \mathrm{L}_{1101}, \mathrm{L}_{1110}] 
\end{eqnarray*}
\[v = \left [ \frac{5 \eta}{2} - \frac{3}{2}, \quad \frac{1}{2} - \frac{\eta}{2}, \quad \frac{1}{4} - \frac{\eta}{4}, \quad \frac{1}{4} - \frac{\eta}{4}, \quad \frac{1}{2} - \frac{\eta}{2}, \quad \frac{1}{4} - \frac{\eta}{4}, \quad \frac{1}{4} - \frac{\eta}{4}, \quad \frac{1}{2} - \frac{\eta}{2}\right ]\]
\begin{eqnarray*}
\mathcal{M}_{290} = [\mathrm{B}_{000}, \mathrm{B}_{010}, \mathrm{L}_{0011}, \mathrm{L}_{0111}, \mathrm{L}_{1000}, \mathrm{L}_{1001}, \mathrm{L}_{1101}, \mathrm{L}_{1110}] \\
\mathcal{M}_{291} = [\mathrm{B}_{000}, \mathrm{B}_{100}, \mathrm{L}_{0000}, \mathrm{L}_{0001}, \mathrm{L}_{0111}, \mathrm{L}_{1000}, \mathrm{L}_{1001}, \mathrm{L}_{1110}] \\
\mathcal{M}_{292} = [\mathrm{B}_{000}, \mathrm{B}_{100}, \mathrm{L}_{0000}, \mathrm{L}_{0001}, \mathrm{L}_{0111}, \mathrm{L}_{1010}, \mathrm{L}_{1011}, \mathrm{L}_{1110}] \\
\mathcal{M}_{293} = [\mathrm{B}_{000}, \mathrm{B}_{100}, \mathrm{L}_{0010}, \mathrm{L}_{0011}, \mathrm{L}_{0111}, \mathrm{L}_{1000}, \mathrm{L}_{1001}, \mathrm{L}_{1110}] 
\end{eqnarray*}

\[v = \left [ \frac{5 \eta}{2} - \frac{3}{2}, \quad \frac{1}{2} - \frac{\eta}{2}, \quad \frac{1}{4} - \frac{\eta}{4}, \quad \frac{1}{4} - \frac{\eta}{4}, \quad \frac{1}{2} - \frac{\eta}{2}, \quad \frac{1}{4} - \frac{\eta}{4}, \quad \frac{1}{2} - \frac{\eta}{2}, \quad \frac{1}{4} - \frac{\eta}{4}\right ]\]
\[\mathcal{M}_{294} = [\mathrm{B}_{000}, \mathrm{B}_{010}, \mathrm{L}_{0001}, \mathrm{L}_{0101}, \mathrm{L}_{1000}, \mathrm{L}_{1011}, \mathrm{L}_{1110}, \mathrm{L}_{1111}] \]

\[v = \left [ \frac{5 \eta}{2} - \frac{3}{2}, \quad \frac{1}{2} - \frac{\eta}{2}, \quad \frac{1}{4} - \frac{\eta}{4}, \quad \frac{1}{4} - \frac{\eta}{4}, \quad \frac{1}{2} - \frac{\eta}{2}, \quad \frac{1}{2} - \frac{\eta}{2}, \quad \frac{1}{4} - \frac{\eta}{4}, \quad \frac{1}{4} - \frac{\eta}{4}\right ]\]
\begin{eqnarray*}
\mathcal{M}_{295} = [\mathrm{B}_{000}, \mathrm{B}_{101}, \mathrm{L}_{0000}, \mathrm{L}_{0001}, \mathrm{L}_{0101}, \mathrm{L}_{1000}, \mathrm{L}_{1110}, \mathrm{L}_{1111}] \\
\mathcal{M}_{296} = [\mathrm{B}_{000}, \mathrm{B}_{101}, \mathrm{L}_{0010}, \mathrm{L}_{0011}, \mathrm{L}_{0101}, \mathrm{L}_{1000}, \mathrm{L}_{1100}, \mathrm{L}_{1101}] \\
\mathcal{M}_{297} = [\mathrm{B}_{000}, \mathrm{B}_{101}, \mathrm{L}_{0010}, \mathrm{L}_{0011}, \mathrm{L}_{0101}, \mathrm{L}_{1000}, \mathrm{L}_{1110}, \mathrm{L}_{1111}] 
\end{eqnarray*}

\[v = \left [ \frac{5 \eta}{2} - \frac{3}{2}, \quad \frac{1}{2} - \frac{\eta}{2}, \quad \frac{1}{4} - \frac{\eta}{4}, \quad \frac{1}{2} - \frac{\eta}{2}, \quad \frac{1}{4} - \frac{\eta}{4}, \quad \frac{1}{2} - \frac{\eta}{2}, \quad \frac{1}{4} - \frac{\eta}{4}, \quad \frac{1}{4} - \frac{\eta}{4}\right ]\]
\[\mathcal{M}_{298} = [\mathrm{B}_{000}, \mathrm{B}_{011}, \mathrm{L}_{0000}, \mathrm{L}_{0010}, \mathrm{L}_{0100}, \mathrm{L}_{0101}, \mathrm{L}_{1011}, \mathrm{L}_{1111}] \]

\[v = \left [ \frac{5 \eta}{2} - \frac{3}{2}, \quad \frac{1}{2} - \frac{\eta}{2}, \quad \frac{1}{2} - \frac{\eta}{2}, \quad \frac{1}{4} - \frac{\eta}{4}, \quad \frac{1}{4} - \frac{\eta}{4}, \quad \frac{1}{4} - \frac{\eta}{4}, \quad \frac{1}{4} - \frac{\eta}{4}, \quad \frac{1}{4} - \frac{\eta}{4}, \quad \frac{1}{4} - \frac{\eta}{4}\right ]\]
\begin{eqnarray*}
\mathcal{M}_{299} = [\mathrm{B}_{000}, \mathrm{B}_{110}, \mathrm{L}_{0000}, \mathrm{L}_{0110}, \mathrm{L}_{0111}, \mathrm{L}_{1001}, \mathrm{L}_{1011}, \mathrm{L}_{1101}, \mathrm{L}_{1110}] \\
\mathcal{M}_{300} = [\mathrm{B}_{000}, \mathrm{B}_{111}, \mathrm{L}_{0010}, \mathrm{L}_{0100}, \mathrm{L}_{0101}, \mathrm{L}_{1000}, \mathrm{L}_{1011}, \mathrm{L}_{1101}, \mathrm{L}_{1111}] 
\end{eqnarray*}
\[v = \left [ \frac{5 \eta}{2} - \frac{3}{2}, \quad \frac{1}{2} - \frac{\eta}{2}, \quad \frac{1}{2} - \frac{\eta}{2}, \quad \frac{1}{4} - \frac{\eta}{4}, \quad \frac{1}{4} - \frac{\eta}{4}, \quad \frac{1}{4} - \frac{\eta}{4}, \quad \frac{1}{4} - \frac{\eta}{4}, \quad \frac{1}{2} - \frac{\eta}{2}\right ]\]
\begin{eqnarray*}
\mathcal{M}_{301} &=& [\mathrm{B}_{000}, \mathrm{B}_{010}, \mathrm{B}_{100}, \mathrm{L}_{0001}, \mathrm{L}_{0111}, \mathrm{L}_{1000}, \mathrm{L}_{1011}, \mathrm{L}_{1110}] \\
\mathcal{M}_{302} &=& [\mathrm{B}_{000}, \mathrm{B}_{010}, \mathrm{B}_{101}, \mathrm{L}_{0000}, \mathrm{L}_{0110}, \mathrm{L}_{1000}, \mathrm{L}_{1011}, \mathrm{L}_{1101}] \\
\mathcal{M}_{303} &=& [\mathrm{B}_{000}, \mathrm{B}_{010}, \mathrm{B}_{110}, \mathrm{L}_{0011}, \mathrm{L}_{0101}, \mathrm{L}_{1000}, \mathrm{L}_{1011}, \mathrm{L}_{1110}] \\
\mathcal{M}_{304} &=& [\mathrm{B}_{000}, \mathrm{B}_{010}, \mathrm{B}_{111}, \mathrm{L}_{0010}, \mathrm{L}_{0100}, \mathrm{L}_{1000}, \mathrm{L}_{1011}, \mathrm{L}_{1101}] \\
\mathcal{M}_{305} &=& [\mathrm{B}_{000}, \mathrm{B}_{100}, \mathrm{B}_{110}, \mathrm{L}_{0000}, \mathrm{L}_{0111}, \mathrm{L}_{1001}, \mathrm{L}_{1011}, \mathrm{L}_{1110}] \\
\mathcal{M}_{306} &=& [\mathrm{B}_{000}, \mathrm{B}_{101}, \mathrm{B}_{111}, \mathrm{L}_{0000}, \mathrm{L}_{0111}, \mathrm{L}_{1000}, \mathrm{L}_{1010}, \mathrm{L}_{1101}] \\
\mathcal{M}_{307} &=& [\mathrm{B}_{000}, \mathrm{B}_{101}, \mathrm{L}_{0000}, \mathrm{L}_{0100}, \mathrm{L}_{0101}, \mathrm{L}_{1010}, \mathrm{L}_{1011}, \mathrm{L}_{1101}]\\
\mathcal{M}_{308} &=& [\mathrm{B}_{000}, \mathrm{B}_{101}, \mathrm{L}_{0000}, \mathrm{L}_{0110}, \mathrm{L}_{0111}, \mathrm{L}_{1000}, \mathrm{L}_{1001}, \mathrm{L}_{1101}]\\
\mathcal{M}_{309} &=& [\mathrm{B}_{000}, \mathrm{B}_{101}, \mathrm{L}_{0000}, \mathrm{L}_{0110}, \mathrm{L}_{0111}, \mathrm{L}_{1010}, \mathrm{L}_{1011}, \mathrm{L}_{1101}] 
\end{eqnarray*}
\[v = \left [ \frac{5 \eta}{2} - \frac{3}{2}, \quad \frac{1}{2} - \frac{\eta}{2}, \quad \frac{1}{2} - \frac{\eta}{2}, \quad \frac{1}{4} - \frac{\eta}{4}, \quad \frac{1}{4} - \frac{\eta}{4}, \quad \frac{1}{2} - \frac{\eta}{2}, \quad \frac{1}{4} - \frac{\eta}{4}, \quad \frac{1}{4} - \frac{\eta}{4}\right ]\]
\begin{eqnarray*}
\mathcal{M}_{310} &=& [\mathrm{B}_{000}, \mathrm{B}_{010}, \mathrm{B}_{100}, \mathrm{L}_{0010}, \mathrm{L}_{0100}, \mathrm{L}_{1011}, \mathrm{L}_{1101}, \mathrm{L}_{1110}] \\
\mathcal{M}_{311} &=& [\mathrm{B}_{000}, \mathrm{B}_{010}, \mathrm{B}_{101}, \mathrm{L}_{0011}, \mathrm{L}_{0101}, \mathrm{L}_{1000}, \mathrm{L}_{1101}, \mathrm{L}_{1110}] \\
\mathcal{M}_{312} &=& [\mathrm{B}_{000}, \mathrm{B}_{010}, \mathrm{B}_{110}, \mathrm{L}_{0000}, \mathrm{L}_{0110}, \mathrm{L}_{1011}, \mathrm{L}_{1101}, \mathrm{L}_{1110}] \\
\mathcal{M}_{313} &=& [\mathrm{B}_{000}, \mathrm{B}_{010}, \mathrm{B}_{111}, \mathrm{L}_{0001}, \mathrm{L}_{0111}, \mathrm{L}_{1000}, \mathrm{L}_{1101}, \mathrm{L}_{1110}] \\
\mathcal{M}_{314} &=& [\mathrm{B}_{000}, \mathrm{B}_{011}, \mathrm{B}_{100}, \mathrm{L}_{0000}, \mathrm{L}_{0010}, \mathrm{L}_{0111}, \mathrm{L}_{1001}, \mathrm{L}_{1110}] \\
\mathcal{M}_{315} &=& [\mathrm{B}_{000}, \mathrm{B}_{011}, \mathrm{B}_{101}, \mathrm{L}_{0000}, \mathrm{L}_{0010}, \mathrm{L}_{0101}, \mathrm{L}_{1000}, \mathrm{L}_{1111}] \\
\mathcal{M}_{316} &=& [\mathrm{B}_{000}, \mathrm{B}_{011}, \mathrm{B}_{110}, \mathrm{L}_{0000}, \mathrm{L}_{0010}, \mathrm{L}_{0101}, \mathrm{L}_{1011}, \mathrm{L}_{1100}] \\
\mathcal{M}_{317} &=& [\mathrm{B}_{000}, \mathrm{B}_{011}, \mathrm{B}_{111}, \mathrm{L}_{0000}, \mathrm{L}_{0010}, \mathrm{L}_{0111}, \mathrm{L}_{1010}, \mathrm{L}_{1101}] \\
\mathcal{M}_{318} &=& [\mathrm{B}_{000}, \mathrm{B}_{011}, \mathrm{L}_{0000}, \mathrm{L}_{0001}, \mathrm{L}_{0101}, \mathrm{L}_{0111}, \mathrm{L}_{1010}, \mathrm{L}_{1110}]\\
\mathcal{M}_{319} &=& [\mathrm{B}_{000}, \mathrm{B}_{011}, \mathrm{L}_{0000}, \mathrm{L}_{0010}, \mathrm{L}_{0110}, \mathrm{L}_{0111}, \mathrm{L}_{1001}, \mathrm{L}_{1101}]\\
\mathcal{M}_{320} &=& [\mathrm{B}_{000}, \mathrm{B}_{100}, \mathrm{B}_{110}, \mathrm{L}_{0010}, \mathrm{L}_{0101}, \mathrm{L}_{1011}, \mathrm{L}_{1100}, \mathrm{L}_{1110}] \\
\mathcal{M}_{321} &=& [\mathrm{B}_{000}, \mathrm{B}_{100}, \mathrm{B}_{111}, \mathrm{L}_{0001}, \mathrm{L}_{0010}, \mathrm{L}_{0111}, \mathrm{L}_{1000}, \mathrm{L}_{1110}] \\
\mathcal{M}_{322} &=& [\mathrm{B}_{000}, \mathrm{B}_{100}, \mathrm{L}_{0010}, \mathrm{L}_{0100}, \mathrm{L}_{0101}, \mathrm{L}_{1011}, \mathrm{L}_{1100}, \mathrm{L}_{1101}] \\
\mathcal{M}_{323} &=& [\mathrm{B}_{000}, \mathrm{B}_{100}, \mathrm{L}_{0010}, \mathrm{L}_{0100}, \mathrm{L}_{0101}, \mathrm{L}_{1011}, \mathrm{L}_{1110}, \mathrm{L}_{1111}] \\
\mathcal{M}_{324} &=& [\mathrm{B}_{000}, \mathrm{B}_{100}, \mathrm{L}_{0010}, \mathrm{L}_{0110}, \mathrm{L}_{0111}, \mathrm{L}_{1011}, \mathrm{L}_{1100}, \mathrm{L}_{1101}] \\
\mathcal{M}_{325} &=& [\mathrm{B}_{000}, \mathrm{B}_{101}, \mathrm{B}_{110}, \mathrm{L}_{0000}, \mathrm{L}_{0011}, \mathrm{L}_{0101}, \mathrm{L}_{1000}, \mathrm{L}_{1110}] \\
\mathcal{M}_{326} &=& [\mathrm{B}_{000}, \mathrm{B}_{101}, \mathrm{B}_{111}, \mathrm{L}_{0010}, \mathrm{L}_{0101}, \mathrm{L}_{1000}, \mathrm{L}_{1101}, \mathrm{L}_{1111}] 
\end{eqnarray*}
\[v = \left [ \frac{5 \eta}{2} - \frac{3}{2}, \quad \frac{1}{2} - \frac{\eta}{2}, \quad \frac{1}{2} - \frac{\eta}{2}, \quad \frac{1}{4} - \frac{\eta}{4}, \quad \frac{1}{2} - \frac{\eta}{2}, \quad \frac{1}{4} - \frac{\eta}{4}, \quad \frac{1}{4} - \frac{\eta}{4}, \quad \frac{1}{4} - \frac{\eta}{4}\right ]\]
\[\mathcal{M}_{327} = [\mathrm{B}_{000}, \mathrm{B}_{011}, \mathrm{L}_{0010}, \mathrm{L}_{0011}, \mathrm{L}_{0101}, \mathrm{L}_{0111}, \mathrm{L}_{1000}, \mathrm{L}_{1100}] \]

\[v = \left [ \frac{5 \eta}{2} - \frac{3}{2}, \quad \frac{1}{2} - \frac{\eta}{2}, \quad \frac{1}{2} - \frac{\eta}{2}, \quad \frac{1}{2} - \frac{\eta}{2}, \quad \frac{1}{4} - \frac{\eta}{4}, \quad \frac{1}{4} - \frac{\eta}{4}, \quad \frac{1}{4} - \frac{\eta}{4}, \quad \frac{1}{4} - \frac{\eta}{4}\right ]\]
\begin{eqnarray*}
\mathcal{M}_{328} &=& [\mathrm{B}_{000}, \mathrm{B}_{011}, \mathrm{B}_{100}, \mathrm{L}_{0010}, \mathrm{L}_{0101}, \mathrm{L}_{0111}, \mathrm{L}_{1011}, \mathrm{L}_{1100}] \\
\mathcal{M}_{329} &=& [\mathrm{B}_{000}, \mathrm{B}_{011}, \mathrm{B}_{101}, \mathrm{L}_{0000}, \mathrm{L}_{0101}, \mathrm{L}_{0111}, \mathrm{L}_{1010}, \mathrm{L}_{1101}] \\
\mathcal{M}_{330} &=& [\mathrm{B}_{000}, \mathrm{B}_{011}, \mathrm{B}_{110}, \mathrm{L}_{0000}, \mathrm{L}_{0101}, \mathrm{L}_{0111}, \mathrm{L}_{1001}, \mathrm{L}_{1110}] \\
\mathcal{M}_{331} &=& [\mathrm{B}_{000}, \mathrm{B}_{011}, \mathrm{B}_{111}, \mathrm{L}_{0010}, \mathrm{L}_{0101}, \mathrm{L}_{0111}, \mathrm{L}_{1000}, \mathrm{L}_{1111}] \\
\mathcal{M}_{332} &=& [\mathrm{B}_{000}, \mathrm{B}_{011}, \mathrm{L}_{0000}, \mathrm{L}_{0111}, \mathrm{L}_{1001}, \mathrm{L}_{1010}, \mathrm{L}_{1101}, \mathrm{L}_{1110}]\\ 
\mathcal{M}_{333} &=& [\mathrm{B}_{000}, \mathrm{B}_{011}, \mathrm{L}_{0010}, \mathrm{L}_{0101}, \mathrm{L}_{1000}, \mathrm{L}_{1011}, \mathrm{L}_{1100}, \mathrm{L}_{1111}] \\
\mathcal{M}_{334} &=& [\mathrm{B}_{000}, \mathrm{B}_{100}, \mathrm{B}_{111}, \mathrm{L}_{0010}, \mathrm{L}_{0100}, \mathrm{L}_{0111}, \mathrm{L}_{1011}, \mathrm{L}_{1101}] \\
\mathcal{M}_{335} &=& [\mathrm{B}_{000}, \mathrm{B}_{101}, \mathrm{B}_{110}, \mathrm{L}_{0000}, \mathrm{L}_{0101}, \mathrm{L}_{0110}, \mathrm{L}_{1011}, \mathrm{L}_{1101}] 
\end{eqnarray*}

\[v = \left [ \frac{5 \eta}{2} - \frac{3}{2}, \quad \frac{1}{2} - \frac{\eta}{2}, \quad \frac{1}{2} - \frac{\eta}{2}, \quad \frac{1}{2} - \frac{\eta}{2}, \quad \frac{1}{2} - \frac{\eta}{2}, \quad \frac{1}{2} - \frac{\eta}{2}\right ]\]
\begin{eqnarray*}
\mathcal{M}_{336} = [\mathrm{B}_{000}, \mathrm{B}_{010}, \mathrm{B}_{100}, \mathrm{B}_{110}, \mathrm{L}_{1011}, \mathrm{L}_{1110}] \\
\mathcal{M}_{337} = [\mathrm{B}_{000}, \mathrm{B}_{010}, \mathrm{B}_{101}, \mathrm{B}_{111}, \mathrm{L}_{1000}, \mathrm{L}_{1101}] \\
\mathcal{M}_{338} = [\mathrm{B}_{000}, \mathrm{B}_{011}, \mathrm{B}_{100}, \mathrm{B}_{111}, \mathrm{L}_{0010}, \mathrm{L}_{0111}] \\
\mathcal{M}_{339} = [\mathrm{B}_{000}, \mathrm{B}_{011}, \mathrm{B}_{101}, \mathrm{B}_{110}, \mathrm{L}_{0000}, \mathrm{L}_{0101}] 
\end{eqnarray*}

\item {Having $\eta \in \left(\frac{2}{3}, 1\right).$}

\[v = \left [ 3 \eta - 2, \quad \frac{1}{2} - \frac{\eta}{2}, \quad \frac{1}{2} - \frac{\eta}{2}, \quad \frac{1}{2} - \frac{\eta}{2}, \quad \frac{1}{2} - \frac{\eta}{2}, \quad \frac{1}{2} - \frac{\eta}{2}, \quad \frac{1}{2} - \frac{\eta}{2}\right ]\]
\begin{eqnarray*}
\mathcal{M}_{340} = [\mathrm{B}_{000}, \mathrm{L}_{0000}, \mathrm{L}_{0001}, \mathrm{L}_{0101}, \mathrm{L}_{0111}, \mathrm{L}_{1000}, \mathrm{L}_{1110}] \\
\mathcal{M}_{341} = [\mathrm{B}_{000}, \mathrm{L}_{0000}, \mathrm{L}_{0010}, \mathrm{L}_{0100}, \mathrm{L}_{0101}, \mathrm{L}_{1011}, \mathrm{L}_{1101}] \\
\mathcal{M}_{342} = [\mathrm{B}_{000}, \mathrm{L}_{0000}, \mathrm{L}_{0010}, \mathrm{L}_{0110}, \mathrm{L}_{0111}, \mathrm{L}_{1011}, \mathrm{L}_{1101}] \\
\mathcal{M}_{343} = [\mathrm{B}_{000}, \mathrm{L}_{0000}, \mathrm{L}_{0111}, \mathrm{L}_{1000}, \mathrm{L}_{1001}, \mathrm{L}_{1101}, \mathrm{L}_{1110}] \\
\mathcal{M}_{344} = [\mathrm{B}_{000}, \mathrm{L}_{0000}, \mathrm{L}_{0111}, \mathrm{L}_{1010}, \mathrm{L}_{1011}, \mathrm{L}_{1101}, \mathrm{L}_{1110}] \\
\mathcal{M}_{345} = [\mathrm{B}_{000}, \mathrm{L}_{0010}, \mathrm{L}_{0011}, \mathrm{L}_{0101}, \mathrm{L}_{0111}, \mathrm{L}_{1000}, \mathrm{L}_{1110}] \\
\mathcal{M}_{346} = [\mathrm{B}_{000}, \mathrm{L}_{0010}, \mathrm{L}_{0101}, \mathrm{L}_{1000}, \mathrm{L}_{1011}, \mathrm{L}_{1100}, \mathrm{L}_{1101}] \\
\mathcal{M}_{347} = [\mathrm{B}_{000}, \mathrm{L}_{0010}, \mathrm{L}_{0101}, \mathrm{L}_{1000}, \mathrm{L}_{1011}, \mathrm{L}_{1110}, \mathrm{L}_{1111}] 
\end{eqnarray*}

\[v = \left [ 3 \eta - 2, \quad 1 - \eta , \quad \frac{1}{2} - \frac{\eta}{2}, \quad \frac{1}{2} - \frac{\eta}{2}, \quad \frac{1}{2} - \frac{\eta}{2}, \quad \frac{1}{2} - \frac{\eta}{2}\right ]\]
\begin{eqnarray*}
\mathcal{M}_{348} = [\mathrm{B}_{000}, \mathrm{B}_{010}, \mathrm{L}_{1000}, \mathrm{L}_{1011}, \mathrm{L}_{1101}, \mathrm{L}_{1110}] \\
\mathcal{M}_{349} = [\mathrm{B}_{000}, \mathrm{B}_{011}, \mathrm{L}_{0000}, \mathrm{L}_{0010}, \mathrm{L}_{0101}, \mathrm{L}_{0111}] \\
\mathcal{M}_{350} = [\mathrm{B}_{000}, \mathrm{B}_{100}, \mathrm{L}_{0010}, \mathrm{L}_{0111}, \mathrm{L}_{1011}, \mathrm{L}_{1110}] \\
\mathcal{M}_{351} = [\mathrm{B}_{000}, \mathrm{B}_{101}, \mathrm{L}_{0000}, \mathrm{L}_{0101}, \mathrm{L}_{1000}, \mathrm{L}_{1101}] \\
\mathcal{M}_{352} = [\mathrm{B}_{000}, \mathrm{B}_{110}, \mathrm{L}_{0000}, \mathrm{L}_{0101}, \mathrm{L}_{1011}, \mathrm{L}_{1110}] \\
\mathcal{M}_{353} = [\mathrm{B}_{000}, \mathrm{B}_{111}, \mathrm{L}_{0010}, \mathrm{L}_{0111}, \mathrm{L}_{1000}, \mathrm{L}_{1101}] 
\end{eqnarray*}

\item {Having $\eta \in \left(\frac{3}{4}, 1\right).$}

\[v = \left [ 4 \eta - 3, \quad \frac{1}{2} - \frac{\eta}{2}, \quad \frac{1}{2} - \frac{\eta}{2}, \quad \frac{1}{2} - \frac{\eta}{2}, \quad \frac{1}{2} - \frac{\eta}{2}, \quad \frac{1}{2} - \frac{\eta}{2}, \quad \frac{1}{2} - \frac{\eta}{2}, \quad \frac{1}{2} - \frac{\eta}{2}, \quad \frac{1}{2} - \frac{\eta}{2}\right ]\]
\[\mathcal{M}_{354} = \left [\mathrm{B}_{000}, \mathrm{L}_{0000}, \mathrm{L}_{0010}, \mathrm{L}_{0101}, \mathrm{L}_{0111}, \mathrm{L}_{1000}, \mathrm{L}_{1011}, \mathrm{L}_{1101}, \mathrm{L}_{1110} \right] \]

\end{enumerate}

\newpage

\twocolumn

\section*{List of Symbols and abbreviations:}

\begin{itemize}\itemsep0em 
    \item[]{{PP}: Purification postulate.}
    \item[]{{CEP}: Complete extension postulate.}
    \item[]{{A}: Access property.}
    \item[]{{G}: Generation property.}
    \item[]{{EU}: Essential uniqueness.}
    \item[]{{ONSEA}: Overcomplete non-signaling extension with access.}
    \item[]{{NSEA}: Complete non-signaling extension with access.}
    \item[]{{NSCE}:  Non-signaling complete extension.}
    \item[]{{GPT}:  Generalised probabilistic theory.}
    \item[]{{$ \mathcal{G} $ }: A generalized probabilistic theory.}
    %%%%%%%%%%%%%%%%%%%
    \item[]{{$\textsf{Syst}[\mathcal{G}]$ }: Systems associated with $\mathcal{G}$.}
     \item[]{{$  \otimes $}: Composition rule in a GPT.}
    \item[]{{$ \times $}:  Cartesian product. }
    \item[]{{$ V_A $}: A finite dimensional real vector space associated to  system $A$.}
    \item[]{{$ V^*_A $}: Dual space of $ V_A $.}
    \item[]{{$ \mathds{R}^\Lambda $}: \textcolor{black}{ Vector space of real valued functions for some finite sample space $\Lambda$ in classical} }
    \item[]{{$  $} \textcolor{black}{~theory}.}
    \item[]{{$ \Omega_A $}: Convex set within the vector space $ V_A $, represents the state space of  system $A$.}
    \item[]{{$ \Omega_A^* $}: subset of dual vectors in $V_A^*$ which evaluate to (a subset of) the unit interval on $\Omega_A$.}
    \item[]{{$ {\cal E}_A $}: Convex set within the dual space $ V^*_A $, represents the effect space of  system $A$.}
    \item[]{{$ K_A $}: Convex cone associated with $\Omega_A$.}
    \item[]{{$ e $}: An instance of an effect in the effect space $ {\cal E}_A$.}
    \item[]{{$ e(\Omega_A) $}:  Image of $e$ when its domain is restricted to $\Omega_A$.  }
    %Probability associated with the effect $e$ acting on $ e(\Omega_A)$.}
    \item[]{{$ u_A $}: Unit effect on $\Omega_A$.}
    \item[]{{$ \mathcal{T}_A^B $}: Space of transformations from system $A$ to system $B$.}
    \item[]{{$ M $}: \textcolor{black}{Measurement, belongs to $\mathcal{T}_A^B$.}}
    \item[]{{$ \Delta_I $}: Classical systems contained in every GPT,  where $I$ denotes the set of classical }
    \vspace{-0.9em}  
    \item[]{{$  $}  (deterministic) states.  }
    \item[]{{$ \delta_i $}: a vertex of $\Delta_I$.}
    \item[]{{$ \epsilon_i $}: a vertex of $\Delta_I^*$.}
    \item[]{{$ \mathds{1}_{\Delta_I} $}: Identity transformation for system $\Delta_I$.}
    \item[]{{$ \omega_A $}: \textcolor{black}{Arbitrary state of system $A$}.}
    \item[]{{$ \epsilon_{AB} $}: \textcolor{black}{A non-signalling extension of $ \omega_A $ to system $B$.}}
    \item[]{{$ E_{\omega_A} $}: \textcolor{black}{The set of all states that purify $\omega_A$.}}
    \item[]{{$ \mathrm{Tr}_B $}: \textcolor{black}{Partial trace over system $B$, of a composite system.}}
    \item[]{{$ T (T')$}: \textcolor{black}{A system type in $\textsf{Syst}[\mathcal{G}]$.}}
    % \item[]{{$  $}: .}
    \item[]{{$ s $}: An arbitrary state in a GPT.}
    \item[]{{$ s(T) $}: \textcolor{black}{An arbitrary state of type $T$.}}
    \item[]{{$ \{(p_i, s_i)\} $}: An ensemble of states.}
    \item[]{{$ \mathbf{Ens}[s] $}: The set of all possible ensembles for a state $s$.}
    \item[]{{$ \mathbf{Ens}_P[s] $}: The set of all pure ensembles of  $s$.}
    \item[]{{$ \mathbf{Face}[s] $}: Face of a state $s$.}
    \item[]{{$ \mathbf{Ext}[s] $}: The set of extensions of $s$.}
    \item[]{{$ \mathbf{Ext}_P[s] $}: The set of all pure extensions of $s$.}
    \item[]{{$ \mathbf{Ext}_{class}[s] $}: A class of extension of  $s$, where the extending system is classical.}
    \item[]{{$  \mathsf{Vert}[\Delta_I] $}: The set of vertices of $\Delta_I$.}
    %\item[]{{$ Sumbols $}: \textcolor{GrT}{From Section III (John can you update this)}.}
    \item[]{{$ \sigma^p $}: \textcolor{black}{A purification of system $s$, belongs to $ \mathbf{Ext}_P[s] $}.}
    \item[]{{$ \Sigma $}: \textcolor{black}{Arbitrary extension of system $s$, belongs to $ \mathbf{Ext}[s] $}.}
    \item[]{{$ T_{1 \rightarrow 2} $}: \textcolor{black}{A reversible transformation in the extending system}.}
    \item[]{{$ p_A^* $}: Alice's cheating probability in integer-commitment.}
    \item[]{{$ p_B^*  $}: Bob's cheating probability in integer-commitment.}
    %\item[]{{$  $}: .}
    %\item[]{{$  $}: .}
    \item[]{{$ \mathrm{Tr}_B $}: Partial trace on system $B$ of a composite quantum state $\rho_{AB}$.}
    \item[]{{$ \mathrm{Tr}_{\neq A_i} $}: Partial trace on all systems  despite of system $A_i$ of a composite state $\rho_{A_1A_2\dots A_N}$.}
    \item[]{{$ V $}: Number of the vertices within a theory.}
    \item[]{{$ \aleph_0$}:  The cardinality of the set of natural numbers.}
    \item[]{{$ \mathfrak{c}$}:  The cardinality of the continuum.}
    \item[]{{$ \mathcal{P}$}:  A partition of a composite system into mutually non-signaling subsystems.}
    \item[]{{$ \mathcal{S}_i^\mathcal{P}$}:  An $i^\mathrm{th}$ system of a partition $\mathcal{P}$.}
    % \item[]{{$ Symbols $}: \textcolor{GrT}{From section V (Marek can you do this)}.}
    % \item[]{{$  $}: .}
    % \item[]{{$  $}: .}
    \item[]{{$\rho_A$}: Arbitrary quantum state of system $A$.}
    \item[]{{$\ket{\psi_{AE}}$}: A pure quantum state of the composite system A and E.}
    \item[]{{$\Theta_E$}: Quantum channel in part of system E.}
    \item[]{{$\{(p_i,\rho_A^i)\}$}: Ensemble of a quantum state.}
    \item[]{{$ P_A $}: A behaviour of system $A$.}
    \item[]{{$ P_{AE} $}: A composite behaviour of system $A$ and $E$.}
    \item[]{{$ \{p_i, P_A^i\} $}: Ensemble of the behaviour of system $A$.}
    \item[]{{$ p_{\cal A|X}(a|x) $}: A conditional probability distribution.}
    \item[]{{$ p_{\cal AE|XZ}(ae|xz) $}: A bipartite conditional probability distribution.}
    \item[]{{$ \cal X $}: The set of all input $\{x\}$, of the conditional probability distribution $p_{\cal A|X}(a|x)$.}
    \item[]{{$ \cal Z $}: The set of all input $\{z\}$, of the bipartite conditional probability distribution }
    \vspace{-0.7em}  
    \item[]{{$\mathbf{    }$}~   $ p_{\cal AE|XZ}(ae|xz) $. } 
    \item[]{{$ \cal A $}: The set of all output $\{a\}$, of the conditional probability distribution $ p_{\cal A|X}(a|x) $.}
    \item[]{{$ \cal E $}: The set of all output $\{e\}$, of the bipartite conditional probability distribution }
    \vspace{-0.7em}  
    \item[]{{$\mathbf{    }$}~   $p_{\cal AE|XZ}(ae|xz)$. }
    \item[]{{$ \mathscr{P}_E $}: Classical pre(post)-processing channel acting on the inputs(outputs) of system $E$.}
    \item[]{{$ P_\texttt{E} $}: A behaviour which is an extremal point.}
    \item[]{{$ \ce(P) $}: An arbitrary ensemble of the behaviour $P$.}
    \item[]{{$ V(\ce) $}: The set of members of the ensemble $\ce$.}
    \item[]{{$ \cep(P) $}: Pure members ensemble of the behaviour $P$.}
    \item[]{{$ \cm(P) $}: Minimal ensemble of the behaviour $P$.}
    \item[]{{$ {\cal E}(P)_{AE} $}: Non-signaling  Extension with Access of behaviour $P_A$.}
    \item[]{{$ D_j $}: A dice, characterized by the probability distribution $\tilde{p}(k|z^\prime=j)$.}
    \item[]{{$ C_j $}:  Classical post-processing channel, characterized by conditional distribution }
    \vspace{-0.7em}  
    \item[]{{$\mathbf{    }$}~   $p_c(m|e,z^\prime=j)$. }
    \item[]{{$ \mathcal{B} $}: \textcolor{black}{ A behaviour polytope.}}
    \item[]{{$ \dim \mathcal{B} $}: Dimension of a behaviour polytope.}
    \item[]{{$ P^{PR} $}: Popescu-Rohrlich box.}
    \item[]{{$ P^m  $}: Maximally mixed behaviour.}
    \item[]{{$ P_{A'} $}: Conjugate behaviour of behaviour $P_A$.}
    \item[]{{$ \mathrm{L}_{\alpha\beta\gamma\delta} $}: Local vertices of the polytope of binary input out behaviours, $\alpha,\beta,\gamma,\delta \in \{0,1\}$.}
    \item[]{{$ \mathrm{B}_{rst} $}: Non-local vertices of the polytope of binary input out behaviours, $r,s,t \in \{0,1\}$.}
    \item[]{{$ B(\eta) $}: Arbitrary behaviour on the line joining PR box and anti-PR box.}
    \item[]{{$ \beta(P) $}: Linear map on the behaviour $P$, computing the CHSH functional.}
    \item[]{{$ {\mathcal V}(P) $}: Set of the sets of members of minimal ensembles.}
    \item[]{{$ \mathscr{C}_3 $}:  The set of maximal contexts for the three-cycle contextuality  scenario.}
    \item[]{{$ {\cal N}^{\cal C}_j $}: $j^\mathrm{th}$ non-contextual vertex of the no-disturbance polytope.}
    \item[]{{$ {\cal C}_j $}: $j^\mathrm{th}$ contextual vertex of the no-disturbance polytope.}
\end{itemize}

\onecolumn

\end{document}